\documentclass{article}

\usepackage{euscript}

\usepackage{latexsym}
\usepackage{amsmath, amssymb,graphics}
\usepackage[all]{xy}

\usepackage{fullpage}

\numberwithin{equation}{section}

\def\gg {\mathfrak{g}}

\renewcommand{\(}{\begin{equation*}}
\renewcommand{\)}{\end{equation*}}
\newcommand{\bea}{\begin{eqnarray*}}
\newcommand{\eea}{\end{eqnarray*}}
\newcommand{\R}{{\mathbb R}}
\newcommand{\C}{{\mathbb C}}
\newcommand{\Z}{{\mathbb Z}}
\newcommand{\Q}{{\mathbb Q}}

\newcommand{\cA}{\ensuremath{\mathcal A}}

\newcommand{\cE}{\ensuremath{\mathcal E}}

\newcommand{\cL}{\ensuremath{\mathcal L}}

\newcommand{\cR}{\ensuremath{\mathcal R}}

\def\H{\ensuremath{\ES{H}}}
\def\i{\ensuremath{\dot\imath}}

\def\L{\ensuremath{{\cal L}}}

\def\H{\ensuremath{\ES{H}}}
\def\i{\ensuremath{\dot\imath}}

\def\L{\ensuremath{{\cal L}}}

\newcommand{\N}{\ensuremath{\mathcal N}}
\newcommand{\beq}{\begin{equation}}
\newcommand{\eeq}{\end{equation}}

\numberwithin{equation}{section}

\renewcommand{\(}{\begin{equation}}
\renewcommand{\)}{\end{equation}}

\newcommand{\CC}{{\mathbb C}}
\newcommand{\RR}{{\mathbb R}}
\newcommand{\ZZ}{{\mathbb Z}}

\newcommand{\CP}{\CC \text{P}}
\newcommand{\RP}{\RR \text{P}}

\def\H{{\mathbb H}}
\def\N{{\mathbb N}}
\def\R{{\mathbb R}}
\def\Z{{\mathbb Z}}
\def\Q{{\mathbb Q}}
\def\C{{\mathbb C}}

\def\1{{\bf 1}}

\def\<{\langle}
\def\>{\rangle}

\newcommand{\ttsmat}[4]{\big({ \textstyle {#1 \atop #3}{#2 \atop #4}}\big)}

\numberwithin{equation}{section}

\renewcommand{\(}{\begin{equation}}
\renewcommand{\)}{\end{equation}}

\begin{document}

\begin{titlepage}


\vspace{2em}
\def\thefootnote{\fnsymbol{footnote}}


\begin{center}
{\Large\bf Geometry of 
Spin and 
Spin${}^{c}$ 
structures \\
in the M-theory partition function}
\end{center}
\vspace{1em}

\begin{center}
Hisham Sati 
\footnote{e-mail: {\tt
hsati@math.umd.edu}
\\
Current address: Department of Mathematics, University of Pittsburgh, Pittsburgh, PA 15260.
}
\end{center}

\begin{center}
Department of Mathematics\\
University of Maryland\\
College Park, MD 20742 
\end{center}

\vspace{0em}
\begin{abstract}
\noindent

We study the effects of having multiple Spin structures on the
partition function of the spacetime fields in M-theory. 
This leads to a potential anomaly which appears in the 
eta-invariants upon variation of the Spin structure.  
The main sources of such spaces 
are manifolds with nontrivial fundamental group, which
are also important in realistic models.  
We extend the discussion to the Spin${}^c$ case and find 
the phase of the partition function, and revisit 
the quantization condition for the $C$-field in this case.
 In type IIA string theory
in ten dimensions, 
the mod 2 index of the Dirac operator is the obstruction 
to having a well-defined partition function. We geometrically 
characterize manifolds with and without such an anomaly
and extend to the case of 
nontrivial fundamental group. 
The lift to KO-theory gives the $\alpha$-invariant, which in 
general depends on the Spin structure.  
This reveals 
many interesting connections to positive scalar curvature 
manifolds
and constructions related to the Gromov-Lawson-Rosenberg
conjecture. 
In the twelve-dimensional theory bounding M-theory, we 
study similar geometric questions, including choices of metrics and
obtaining elements of K-theory in ten dimensions by pushforward
in K-theory on the disk fiber. We interpret the latter in terms of the families 
index theorem for Dirac operators on the M-theory circle and disk. 
This involves superconnections, eta-forms,
and infinite-dimensional bundles, and 
gives elements in Deligne cohomology in lower dimensions. 
We illustrate our discussion with many examples throughout.

\end{abstract}

\vfill

\end{titlepage}
\setcounter{footnote}{0}
\renewcommand{\thefootnote}{\arabic{footnote}}

\pagebreak
\renewcommand{\thepage}{\arabic{page}}

\tableofcontents

\section{Introduction and Summary}


Index theorems play an important role in characterizing anomalies
and zero modes in theories in physics. The Atiyah-Singer (AS)
index theorem in even dimensions, in the absence of boundaries,  
involves invariants which do not depend on the choice of geometric 
notions such as a connection or a Spin structure. However, the 
situation in the presence of a boundary is drastically different, as the 
contribution from the boundary depends on geometric and analytical 
quantities in a precise way. M-theory is described by this setting \cite{Flux}
of the Atiyah-Patodi-Singer (APS) index theorem \cite{APSI} \cite{APSII}
\cite{APSIII}.

\vspace{3mm}
The definition and dynamics of M-theory remain a mystery. While
there is mounting evidence that this theory exists, there is yet no
intrinsic understanding of this theory that does not use the
`corners of the moduli space' such as ten-dimensional superstring
theories or eleven-dimensional supergravity. A semi-classical
approach to M-theory involves `continuing' the eleven-dimensional
supergravity action to the quantum regime. Of interest is the
topological part $I_{\rm top}$, comprised of the Chern-Simons term
\cite{CJS} and the one-loop term \cite{VW} \cite{DLM}. $I_{\rm top}$ was shown by
Witten \cite{Flux} to be written as a sum of the index of a Dirac operator
coupled to an $E_8$ bundle and a Rarita-Schwinger operator. Part of
the motivation was first the fact that the cohomology class of the degree four field $G_4$ in M-theory
contains, as a summand, a characteristic class of an $E_8$ bundle, and second, 
existence of the $E_8 \times E_8$ heterotic string theory on the
boundary \cite{HW}.


\vspace{3mm} The dimensional reduction of this $E_8$ gauge theory
from the total space $Y^{11}$
of a circle bundle to ten dimensions was
performed by \cite{DMW} where the partition function was matched
(in a limit) with the corresponding K-theoretic partition function of the
Ramond-Ramond fields in type IIA string theory.
Many subtleties are involved in this quantum equivalence. 
The partition function of the $C$-field $C_3$ and the Rarita-Schwinger
field $\psi_1$ can be factorized into a modulus and a phase, the latter being given by
\cite{DMW}
\(
\Phi = \exp \left[ 2\pi i \left( \frac{1}{2} \overline{\eta}_{E_8}
+ 
 \frac{1}{4} \overline{\eta}_{RS}
\right)\right]=
\exp \left[
2\pi i\left( \frac{1}{4}(h_{E_8}+\eta_{E_8})
+\frac{1}{8}(h_{RS}+\eta_{RS})\right)\right]\;,
\label{Phi}
\)
where $\overline{\eta}_V=\frac{1}{2}(\eta_V + {\rm dim~Ker}D_V)=
\frac{1}{2}(\eta_V + h_V)$ is the reduced eta invariant
of the Dirac operator $D_V$ coupled to the bundle $V$, with $V$ being
the $E_8$ bundle or the Rarita-Schwinger bundle (i.e. RS$=TY^{11}- 3 \mathcal{O}$, 
with $\mathcal{O}$ a trivial real line bundle).

\vspace{3mm}
Taking $Y^{11}= X^{10} \times S^1$, the $C$-field,  being a pullback from 
$X^{10}$,  has an orientation-reversing symmetry.
Under this symmetry, the Chern-Simons term reverses sign, and the phase
of the partition function is complex-conjugated. The Dirac operator changes sign
under reflection of one coordinate, so the nonzero eigenvalues appear in pairs 
$(\lambda, -\lambda)$, so
that the eta invariants are zero \cite{DMW}.
The M-theory/type IIA connection via $E_8$ gauge theory 
was further studied
in \cite{MS} and incorporated into the description of the twisted
K-theory partition function in the presence of the Neveu-Schwarz (NS)
H-field $H_3$ (or its `potential', the $B$-field), as well as in \cite{Sgerbe}
in relation to gerbes.
In this paper we  
 explore effects related to Spin and Spin${}^c$ structures.

\vspace{3mm}
In the above works both the ten- and eleven-dimensional manifolds 
are taken to be Spin. In this case we study the dependence of the phase of the 
partition function on the choice of Spin structure (starting in section \ref{sec spin}).
 We then work out an extension to  the ${\rm Spin}^c$ case (starting in section
 \ref{nonproj} and then more extensively in section \ref{sec spinc}). 
 We study how Spin and ${\rm Spin}^c$ 
structures in eleven dimensions and ten dimensions are related, especially 
in the context of the partition function. First, we look at the case
when the ten-dimensional manifold is ${\rm Spin}^c$ and then when both
 the ten- and eleven-dimensional 
manifolds
are ${\rm Spin}^c$.

\paragraph{Outline.}
This paper is composed of two intermixed parts: exposition and original results.
The first part is written with an emphasis on a contextual point of view 
(and hence is not traditional) and involves:
\begin{itemize}
\item Recasting physical results in a mathematical (and invariant) language, 
thus highlighting otherwise hidden structures. 
\item Bringing into physics results from mathematics to give constructive descriptions 
of structures appearing from the physics side. 
\end{itemize}
We hope this makes the paper more self-contained. Thus, this paper can be 
used by physicists interested in seeing how geometric ideas can be used in an effective
way in constructing and studying the partition function as well as by mathematicians who
are interested in applications and examples. 
The second part on original research, which is the larger part of the paper,
has as main points the following
\begin{enumerate}
\item {\it Highlight the importance of Spin structures in M-theory}. 
When multiple Spin structures exist, the different Spin
structures may lead to different answers for analytical and 
geometric entities, and hence for physical entities (section \ref{sec spin}). 
For example, the mod 2 index of the Dirac operator in 
dimensionally-reduced M-theory to ten dimensions
admits a lift to KO-theory whose value depends on the 
choice of Spin structure (section \ref{sec geo}). 
This 
suggests that when calculating the {\it full} partition function,
which includes moduli of the metric, one should also 
take 
contributions from the corresponding 
different Spin structures into account in the sum. 
Furthermore, considering the bounding (as opposed to the nonbounding)
 Spin structure on the M-theory
circle leads to many interesting 
connections to global analysis: eta-forms, adiabatic limits,  and superconnections
(section \ref{fam}).

\item {\it Study the partition function on manifolds which are not simply connected.} 
Manifolds with multiple Spin structures are non-simply connected. The main 
examples are spherical space forms $S^n/\Gamma$, which in even dimensions can only be
 projective spaces. Representations of the finite group $\Gamma$ 
 provide possibly multiple Spin structures. At the level of bundles this 
 can be encoded by flat vector bundles and their holonomy. 
 We identify the fields in ten and eleven dimensions which could support
 nontrivial fundamental group. 
 This is the subject of section \ref{sec spin} and then sections 
 \ref{pi 1 10} and \ref{fund eleven}.
 Framed manifolds, considered at the end of section \ref{fund eleven},
provide interesting examples.
 We also do the same for 
 the second homotopy groups, which is relevant for Spin${}^c$ structures,
 in section \ref{sec spinc} and section \ref{pie2}.

\item {\it Study the relation between Spin and Spin${}^c$ structures on
one hand on $Y^{11}$ and the corresponding structures on $X^{10}$
on the other hand}.
We consider type IIA string theory as obtained from M-theory via quotient 
by a circle action. The spaces are Spin or Spin${}^c$ depending on whether
the circle action lifts to the Spin bundle. We start with an eleven-dimensional 
Spin manifold.  For an action of even type, or projectable 
action, the ten-dimensional quotient is Spin. For an action of odd type,  
or nonprojectable action, the ten-dimensional quotient is Spin${}^c$. 
We study both cases in detail in section \ref{sec spin} 
(especially sections \ref{proj} and \ref{nonproj}) and section
 \ref{sec spinc}.
Important examples in eleven dimensions 
include contact manifolds and spherical space forms.
 Important examples in 
ten dimensions include K\"ahler manifolds. 

\item {\it Study the dependence of the phase of the partition function on the Spin structure}.
The phase of the partition function in M-theory is given in terms of 
exponentiated eta invariants in equation (\ref{Phi}). 
These are geometric/analytical as opposed to topological invariants.
We ask whether this depends on the 
choice of Spin structure. We perform the analysis for both operators. 
In order to illustrate the point, we work out examples in eleven and seven dimensions  
showing explicitly how the eta function depends on the choice of Spin
structure. We also extract the number of zero modes of the corresponding 
Dirac operators, as these appear in the phase in section \ref{dep pha}.

\item {\it Extend the description of  the phase of the partition function to the Spin${}^c$ case}. We do this in section \ref{sec spinc}.
We motivate the case for Spin${}^c$ 
in ten, eleven, and twelve dimensions
in section \ref{nonproj} and section \ref{why c}. 
Important examples in odd dimensions are 
contact manifolds and spherical space forms, the motivation for both of which 
we highlight
in section \ref{sec rel}. In 
even dimensions, prominent examples are K\"ahler
manifolds and coboundaries of spherical space forms. 
We also study situations of dependence on almost complex structures, which 
 are closely related to Spin${}^c$ structures
 (section \ref{sec eve}). This makes connection to obstructions 
 and generalized
  cohomology in section \ref{cobo}. 
The phase in the adiabatic limit, i.e. when the volume of the base ten-dimensional 
manifold is
taken to be large with respect to the volume of the (circle) fiber, turns out to admit an 
expression which 
is much more involved than that of the Spin case. 
We do this in section \ref{bdry case}, building on discussions
from section \ref{sec apsc}.
This leads to nontrivial congruences
that depend on the M-theory circle bundle, the line bundle associated to the 
${\rm Spin}^c$-structure, the degree four characteristic class $a$ of the $E_8$ 
bundle $E$, and the Pontrjagin classes of the ten-dimensional manifold $X^{10}$. 
We illustrate how the M-theory circle can be used to define the Spin${}^c$ structure. 
We also consider the partition functions of the M2-brane and the M5-brane in 
the Spin${}^c$ case in section \ref{sec eff c}.

\vspace{3mm}
The main feature of our discussion is that we write the phase of the 
eleven-dimensional expression as an integral of quantities in ten 
dimensions. Thus that makes a departure from the discussion of \cite{DMW}
in two respects: We are looking at the phase in eleven dimensions,
involving the eta invariant, instead of the phase in twelve dimensions,
involving the index, and the expression is a ten-dimensional integral. 
This makes connection to the treatment in \cite{MS} using eta-forms.

\item {\it Highlight the importance of geometry, especially Ricci and scalar 
curvatures, in the eta invariant and the mod 2 index appearing in the 
construction of the partition function}. 
We do this starting in section \ref{sec sca}.
The mod 2 index appears in the 
K-theoretic anomaly of the Ramond-Ramond fields in ten dimensions
\cite{DMW}. In section \ref{mod 2 ind}
we characterize the mod 2 index
in ten dimensions from topological, analytical, and geometric angles. 
The mod 2 index in dimensions 2 and 10 can be viewed as an instance of 
the $\widehat{A}$-genus and admits a lift to KO-theory called the $\alpha$-invariant, which 
depends on the choice of Spin structure on the ten-dimensional manifold. 
Requiring this invariant to vanish is relating to positive scalar curvature (psc)
metrics via the Gromov-Lawson-Rosenberg conjecture. We use the 
notion of a Clifford-linear Dirac operator \cite{LM} which, in a sense,
unifies all Spin representations. We give many examples that are important
in type IIA string theory (starting earlier in section \ref{sec abs}). 
The kernel of the $\alpha$ invariant, which is the lift of the $\widehat{A}$-genus, 
 determines the situations for which the mod 2 index of the Dirac operator 
 vanishes, and hence for which the partition function in ten dimensions 
 is anomaly-free. Total spaces of $\H P^2$ bundles
give representatives in each cobordism class of ker $\alpha$.
We do this also for the non-simply connected case. The discussion 
(in section \ref{mod 2 ind}) then
naturally relates structures in ten dimensions to structures in 
two dimensions. 
As far as the geometry of the bounding twelve-manifold
is involved, we investigate how that is related to the physical problem 
in the context of the Atiyah-Patodi-Singer index theorem, and find in 
section \ref{sec typ} 
that
there are preferred metrics that take into account 
the fibration structure between 
type IIA string theory and M-theory.

\item {\it Extend the constructions to the family case}. 
We model the physical setting in terms of families of Dirac operators on the 
M-theory circle (i.e. the eleventh direction) parametrized by the ten-dimensional
base, home to type IIA string theory. This leads to several interesting 
consequences:

\textbullet~ In section \ref{sec fam 10} we explain physically the role of the Quillen and Bismut superconnections.
Because of nonchirality, type IIA is a natural 
setting for the superconnection formalism.
The superconnection represents the geometric 
description of the pushforward map in K-theory.
Thus, starting from bundles on the fiber in the M-theory
compactification, we can produce K-theory elements in 
the base. 
They lead to a sort of family version of the dilatino supersymmetry transformation. 
We also characterize the corresponding infinite-dimensional bundles.

\textbullet~ In section \ref{sub fam} we consider a generalization of usual 
dimensional reduction mechanisms. The Scherk-Schwarz dimensional 
reduction uses an abelian rigid symmetry
of the equations of motion to generalize the reduction ansatz by allowing a 
{\it linear} dependence on the coordinate of the fiber. For example
$v(x, z)=m z + v(x)$. 
What we consider is a generalization of this so that the dependence 
on the fiber coordinate is not necessarily linear, and also allows
the base spacetime derivatives of the fiber vector $\nabla_\mu v_z$
 to be nonzero in general. 
The generalization of the second type has been considered in 
\cite{ALT} where it was related to metric torsion. The general case
we consider will also involve torsion in an intimate way, which 
will essentially be the RR 2-form $F_2$.

\textbullet~ The adiabatic limit (physically in our interpretation: the semiclassical limit) of the eta 
invariant is given in terms of the Bismut-Cheeger eta-forms. We characterize these
physically and provide some examples in section \ref{equiv e}. 

\textbullet~ The eta-forms lead to gerbes and Deligne cohomology classes on the base.
Varying the dimension of the fiber (mostly either one- or two-dimensional) we get
gerbes which are either of even or odd degree. We characterize these physically
in section \ref{form del}. We also consider the new feature of taking into 
account contributions from the kernels of vertical Dirac operators. 

\textbullet~ Starting in section \ref{sub fam},
we consider the pushforward of bundles in eleven dimensions
to bundles in ten (and lower) dimensions at the level of K-theory. 
Several new features arise, including the appearance of
infinite-dimensional bundles.

\item {\it Highlight consequences for the fields in ten and eleven dimensions.}
We focus on the M-theory circle. From the point of view of the ten-dimensional 
type IIA string theory, this means that we are considering the Ramond-Ramond 
(RR) 1-form potential, corresponding to the connection on the circle bundle, and its
curvature 2-form RR field strength $F_2$. From a homological point of view, 
the object that carries a charge with respect to $F_2$ is the D0-brane \cite{Pol}. 
Such branes carry K-theory charges \cite{MM}. On the other hand, such charges are
associated with the Kaluza-Klein (Fourier) modes for the reduction on the circle.
We argue in section \ref{form del}
that generation of such modes corresponds to the Adams operations
on the corresponding line bundle. This operation is an automorphism in K-theory 
which raises the charge of the brane by $k$ units, which in turn corresponds to the
$k$th Kaluza-Klein mode.  

\textbullet ~The $B$-field in ten dimensions
is essentially the integration of the $C$-field over the circle fiber. 
In section \ref{B har} we characterize 
this as the harmonic representative of the Spin${}^c$ structure in the adiabatic
limit. This is analogous to the characterization of the $C$-field as (essentially)
a harmonic 
representative of the  String class in \cite{tcu}.

\textbullet ~The Ramond-Ramond fields: the K-theory class in ten dimensions
corresponding to the 
family of Dirac operators can be obtained. This suggests that one can set 
up a formalism which relates
M-theory and type IIA string theory in K-theory directly without having to 
pass through cohomology.

\textbullet~ The dilaton plays an important role in our discussion
starting in sections \ref{circle actions}, 
\ref{proj}, \ref{nonproj} and then in sections \ref{sec fam 10}
and \ref{equiv e}. First, it represents 
the volume of the circle fiber, so that its size measures the string coupling. 
We use this to give the eigenvalues of the dimensionally reduced 
Dirac operators in the weak string coupling limit. 
Second, in the expression of the eta-forms, a scalar parameter appears, which
we identify essentially with the dilaton. This suggests the interpretation of the integral 
defining the eta-fom to be an integral over all possible sizes of the M-theory circle, and 
hence over all values of the string coupling constants (at least formally).


\textbullet~The dilatino -- the superpartner of the dilaton --
also has an interesting description in our setting (section \ref{sec fam 10}). 
Its supersymmetry transformations, which is essentially the difference
between the Lie derivative with respect to the eleventh direction
and the covariant derivative, admit a generalization to 
the families case. It becomes essentially the difference between 
the Bismut superconnection and the Quillen superconnection. 

\textbullet~The $C$-field in M-theory admits a shifted quantization 
condition in the Spin case \cite{Flux}. We extend this to the Spin${}^c$ case
in section \ref{sec eff c}.
M-theory admits a parity symmetry. We consider the effect
of this symmetry on Spin structures and eta forms in section \ref{ana eff} and 
\ref{equiv e}.

\textbullet~ We highlight with many concrete examples throughout.
We hope that such examples
could be useful for future investigations, especially in relation to
realistic dimensional 
reduction to lower dimensions. 
\end{enumerate}

\paragraph{Spin structures.}
A {\it Spin structure} on a $n$-dimensional 
manifold $M$ consists of a Spin$(n)$ principal bundle 
$P_{\rm Spin}(M)$ over $M$ together with a two-fold covering 
map $\rho : {\rm Spin}(n) \to SO(n)$ such that the diagram 
\(
\xymatrix{
P_{\rm Spin}(M) \times {\rm Spin}(n) 
\ar[rr]
\ar[dd]^{\varphi \times \rho}
&&
P_{\rm Spin}(M)
\ar[dd]^\varphi
\ar[dr]
\\
&& & M
\\
P_{SO}(M) \times SO(n) 
\ar[rr] 
&&
P_{SO}(M)
\ar[ur]
}
\label{spin str def}
\)
commutes.
The horizontal arrows are given by principal bundle structure. 
The set of Spin structures on a manifold $M$ is a torsor over 
$H^1(M;\Z_2)$.  Spin${}^c$ structures are complex analogs of Spin 
structures and are defined and studied in section 
\ref{basic spin c}.
Extensive discussions on Spin and Spin${}^c$ structures
can be found in \cite{LM}.

\paragraph{Spin vs. Spin${}^c$ and ten vs. eleven dimensions.}
Consider the circle bundle $S^1 \to Y^{11} \to X^{10}$ with projection $\pi$. 
Let $\mathcal{L}$ be the complex line bundle associated to $\pi$ with
first Chern class $e=c_1(\mathcal{L})$.  
Since $TY^{11}\cong \pi^* (TX^{10}) \oplus TS^1$, the tangent bundle 
along the fibers is trivial with trivialization provided by
 the vector field generating the 
$S^1$-action on $Y^{11}$. 

\begin{enumerate}
\item Hence a Spin structure on $X^{10}$ -- whenever 
it exists, i.e. when $w_2(X^{10})=0$-- induces a Spin structure on $Y^{11}$
which we denote $\sigma_1$. 
\item If $w_2(X^{10})=e$ mod 2, i.e. $X^{10}$ admits a Spin${}^c$ structure, 
then $TX^{10}\oplus \mathcal{L}$ admits a  Spin structure. The choice of such 
a Spin structure gives a Spin structure on the disk bundle $\mathbb{D}\mathcal{L}$. 
Denote by $\sigma_2$ the restriction of the latter 
to the sphere bundle $S\mathcal{L}=Y^{11}$. 
\item If $w_2(X^{10})=0$, i.e. if $X^{10}$ is Spin, and if $e=0$ mod 2, then 
$\sigma_1$ and $\sigma_2$ are {\it different} Spin structures on $Y^{11}$
since the restriction of $\sigma_1$ to the fiber $S^1$ is the nontrivial 
Spin structure, which does not extend over $\mathbb{D}^2$, whereas the 
restriction of $\sigma_2$ does by construction.
\end{enumerate}
We will consider Spin structures and Spin${}^c$ structures extensively in section 
\ref{sec spin} and section \ref{sec spinc}, respectively.

\paragraph{Reality check 1: Multiple Spin structures 
and the standard model from string theory.}
We will see in section \ref{multi spin} that the main source of multiple Spin 
structures is spaces with nontrivial fundamental group. Such spaces do 
appear in string theory in trying to connect to four-dimensional physics.
In fact, in a sense, they are even more desirable. 
Heterotic vacua (related to other string theories and to M-theory via dualities)
on a Calabi-Yau manifold $M$ require  
a nontrivial fundamental group $\pi_1(M)\neq 1$
in order to produce the standard model gauge group, as only
such manifolds will allow Wilson lines 
\cite{W0} \cite{Se1} \cite{EO} \cite{BOS1}.
For example, 
torus fibered Calabi-Yau threefolds with $\pi_1(M)=\Z_2$ 
admit stable, holomorphic vector bundles with structure group
$SU(5)$ \cite{DOTW1} \cite{DOTW2} \cite{DOTW3}.
Torus fibered Calabi-Yau threefolds with $\pi_1=\Z_2 \times \Z_2$
admit
stable, holomorphic vector bundles with structure group $SU(4)$
\cite{OPR1} \cite{OPR2} \cite{DOPR}. 
Examples with $\pi_1(M)=\Z_3 \times \Z_3$ are given in 
\cite{BOPR}.
In all cases the groups are
broken down to the standard model group.

\paragraph{Reality check 2: Multiple Spin structures and cosmology.}
There is a great deal of activity in trying to figure out the 
topology of the spatial part of the universe. Theoretically, isotropy 
and homogeneity suggest that the universe be 
of constant sectional curvature. Observation suggests that the 
universe is almost flat, but not quite, i.e. is probably curved 
with a very small curvature. Recent WMAP data suggests that
the universe is a spherical space form \cite{LWRLU} or
a a hyperbolic space form with ``thorned topology" 
\cite{ALST}. In both cases, the spaces have nontrivial fundamental 
groups, and this is nonabelian --the group $E_8$-- in the 
first case. There is no definite answer on whether or not 
the universe is not simply connected, but the above results/suggestions
seem reasonable. If this is the case, then the universe will 
admit multiple Spin structures, and the question of distinguishing 
them becomes of great importance.

\section{M-Theory on Eleven-Dimensional Spin Manifolds}
\label{sec spin}

\subsection{Multiple Spin structures}
\label{multi spin}

 If a manifold is not simply connected then it may have more than one 
Spin structure \cite{LM} (see section \ref{spinpie}). The number of Spin structures
is given by $|H^1(M;\Z_2)|$. In particular, a simply connected manifold 
has at most one Spin structure.  

\vspace{3mm}
Such a multiplicity of Spin structures occurs and is important on the 
string worldsheet \cite{SW}; it explains the need for the GSO projection.
In the case of chiral bosons, there is a partition function for each choice 
of Spin structure. Different partition functions are given by theta
functions, with the ones corresponding to different Spin structures 
permuted by $SL(2;\Z)$ \cite{W-duality}. Similarly for chiral $p$-forms 
\cite{W-duality}. 
 What we
are considering here are Spin structures in spacetime. Such a situation
can and does occur in realistic scenarios. For example (see also 
the end of the introduction above)
\begin{enumerate}
\item Every compact, locally irreducible, Ricci-flat manifold of non-generic 
holonomy is a Spin manifold provided that it is simply connected. 
In the non-simply connected case, $M$ may have several Spin structures. 
\item A Calabi-Yau manifold $K$ with $SU(3)$ holonomy always has a Spin structure
even when it is not simply connected, in which case it has several Spin structures. 
\end{enumerate}

\noindent The physical significance of multiple Spin structures over spacetime seems
 far less
clear than the case of the worldsheets. 
In fact, in \cite{Mc1} a way was designed to actually avoid having multiple
such structures. On the other hand, we saw in the introduction that manifolds with 
nontrivial fundamental group are essential for model building. One goal 
of this article is, to some extent, to try to fill this gap by highlighting the
significance of 
multiple Spin structures in spacetime and providing ways of studying them
in the context of partition functions in M-theory and string theory. That is, we study
the consequences of allowing multiple Spin structures. We first consider several 
examples which will be useful later in the paper.

\paragraph{Spin structures on the circle.}
The simplest example of a manifold with more than one 
Spin structure is the circle, which has two inequivalent Spin structures (see \cite{LM}). 
 The bundle of orthonormal frames is $P_{SO}(S^1)\cong S^1$. 
  Since both $S^1$ and $SO(2)$ 
can be identified with $U(1)$, and Spin$(1)=\Z_2$, these structures are given by the maps
\(
\xymatrix{
U(1) \times \Z_2 
\ar[r]^{~~~{\rm pr}_1} 
&
U(1) 
\ar[r]^{\rm id} 
&
U(1)
}
\)
and
\(
\xymatrix{
U(1) \ar[rr]^{\rm square} &&
U(1) \ar[r]^{\rm id} &U(1)
}\;.
\)
A Spin structure on $S^1$ is a real line bundle $L$ on $S^1$ together with
an isomorphism $L\otimes L \cong TS^1$. Two choices are possible for $L$:
\begin{enumerate}
\item {\it Periodic}: $S^1_P=(S^1, L_P)$.

\item {\it Antiperiodic} (or {\it M\"obius}): $S^1_A=(S^1, L_A)$. 
\end{enumerate}
The above cases can be looked at from the point of view of cobordism
(see section \ref{cobo} for 
more on Spin cobordism).
Start with the case when the circle is a boundary of a two-disk. 
A frame for the disk is formed by the tangent vector to the boundary $S^1$
and the unit normal vector. Going around the boundary once gives
a loop in the frame bundle of the disk whose lift to the Spin structure does
not close up.
Since $S^1=\partial \mathbb{D}^2$ and 
$\mathbb{D}^2$ admits a unique Spin structure, $S^1$ as a boundary of 
the disk gives a two-fold connected covering of $S^1$ which corresponds to the 
trivial element in $\Omega_1^{\rm Spin}\cong \Z_2$.
The two-fold disconnected covering of $S^1$ is formed by two copies of 
$S^1$ with opposite orientations obtained by circle inversion. This covering 
cannot be represented as a boundary of a Spin manifold, and hence represents
the generator $\sigma$ of $\Omega_1^{\rm Spin}$. 
Twice this generator, i.e. $2\sigma$, 
is zero in Spin cobordism, since it bounds two cylinders, as 
two copies of the trivial structure on the disk bound one
cylinder. 

\vspace{3mm}
Since $S^1\cong SO(2)$, $S^1 \cong {\rm Spin}(2)$, the exact sequence
$
0 \to \Z_2 \to {\rm Spin}(2) \to SO(2) \to 0
$
takes $z \in {\rm Spin}(2)$ to $z^2$. 
For every $n \geq 1$, $N=\binom{n}{2}$, 
$
P_{{SO}(N)} (SO(n))=SO(n) \times SO(N)
$
and the two coverings are
\begin{eqnarray}
P^1_{{\rm Spin}(N)} (SO(n)) &=& SO(n) \times {\rm Spin}(N)
\nonumber\\
P^2_{{\rm Spin}(N)} (SO(n)) &=& ({\rm Spin}(n) \times {\rm Spin(N)})/\Z_2\;,
\end{eqnarray}
where $\Z_2$ acts on ${\rm Spin}(n) \times {\rm Spin}(N)$ by the map
$(\varepsilon_1, \varepsilon_2) \mapsto (-\varepsilon_1, -\varepsilon_2)$. 
Thus, for $n=2$, $N=1$,
\begin{eqnarray}
P^1_{{\rm Spin}(1)} (SO(2)) &=& SO(2) \times {\rm Spin}(1)
\nonumber\\
P^2_{{\rm Spin}(1)} (SO(2)) &=& ({\rm Spin}(2) \times {\rm Spin}(1))/\Z_2\;,
\end{eqnarray}
which translates into 
\(
P^1_{\Z_2} (S^1) = S^1 \times \Z_2\;,
\qquad \quad
P^2_{\Z_2} (S^1) = (S^1 \times \Z_2)/\Z_2\;,
\)
where $\Z_2$ acts on $S^1 \times \Z_2$ by the map
$(w, \pm 1) \mapsto (-w, \mp1)$. 
The bundle projection maps onto $S^1$ are given by 
$
P^1_{\Z_2} (S^1) \to  S^1 \times \Z_2
$,
$(z, \pm) \mapsto  \pm z$,
and 
$
P^2_{\Z_2} (S^1) \to  (S^1 \times \Z_2)/\Z_2
$,
$(w, \pm) \mapsto  w^2$.
The first total space of the bundle is not connected. The second total space
is connected, since $S^1\times \Z_2)/\Z_2=\{ [w, \varepsilon] \}$, with
$[w, \varepsilon ]=\{ (w, \varepsilon ), (-w, -\varepsilon ) \}$. 
The properties of (and terminology for) the two Spin structures on
 the circle are summarized in the following
table. 
\(
\begin{tabular}{|c||c|c|c|c|c|}
\hline
Trivial & bounding & non-supersymmetric & Neveu-Schwarz & even & periodic
\\
\hline
\hline
Nontrivial & nonbounding & supersymmetric & Ramond & odd & antiperiodic
\\
\hline
\end{tabular}
\nonumber
\)

\paragraph{Spin structures on Riemann surfaces.}
Let $\Sigma_g$ be a closed connected and orientable Riemann surface of genus $g$. 
Let $S^1 \buildrel{i}\over{\to} U(\Sigma_g) \buildrel{\pi}\over{\to}\Sigma_g$ be the 
unit tangent bundle (i.e. the sphere bundle of the tangent bundle). The cohomological 
definition of a Spin structure
says that there is a cohomology class $\xi \in H^1(\Sigma_g;\Z_2)$ that restricts to 
a generator of $H^1(S^1;\Z_2) \cong \Z_2$ on each fiber $S^1$. The short 
exact sequence
\(
0 \longrightarrow H^1(\Sigma_g;\Z_2) \buildrel{\pi^*}\over{\longrightarrow} H^1(U(\Sigma_g);
\Z_2) \buildrel{i^*}\over{\longrightarrow} H^1(S^1;\Z_2) \longrightarrow 0
\)
establishes that the set ${\rm Spin}(\Sigma_g)$ of Spin structures on $\Sigma_g$
is in one-to-one correspondence with $H^1(\Sigma_g;\Z_2)\cong \Z_2^{2g}$
\cite{At}.
The correspondence is not a group isomorphism and ${\rm Spin}(\Sigma_g)$ is 
a nontrivial coset of $H^1(\Sigma_g;\Z_2)$ in $H^1(U(\Sigma_g);\Z_2)$.
Let $\xi_1, \cdots, \xi_{2g}$ be a basis for $H^1(\Sigma_{g};\Z)$ and let
$z$ denote the extra element in $H^1(U(\Sigma_g);\Z)$. A basis 
for $H^1(U(\Sigma_g);\Z)$ is given by $\xi_1, \cdots, \xi_{2g}, z$. Then 
the mod 2 reduction $r_2(z) \in H^1(U(\Sigma_g);\Z_2)$ is a particular 
Spin structure and the set of Spin structures is given by 
\(
{\rm Spin}(\Sigma_g)=\left\{
\sum_{i=1}^{2g} 
x_i r_2(\xi_i) + r_2(z) ~\vert~ {\rm all}~x_i \in \Z_2
\right\}\;.
\)

\paragraph{Spin structures on spheres.}
There is a unique Spin structure on the $n$-sphere $S^n$, $n>1$,  given by \cite{DT}
\(
{\rm Spin}(n+1) \to SO(n+1) \to S^n\;.
\)

\paragraph{Spin structures on projective spaces.} Consider the projective space 
$\mathbb{K}P^n$ over the field $\mathbb{K}=\R, \C$, or $\mathbb{H}$.
The total Stiefel-Whitney class is
\(
w(\mathbb{K}P^n)=1+ w_1 + w_2 + \cdots = (1+x)^{n+1}\;,
\)
where $x$ is the generator of $H^*(\mathbb{K}P^n;\Z_2)$ with 
${\rm dim}_{\mathbb K} x=1$. 

\noindent $\circ$ $\mathbb{K}=\R$: Requiring $w_1=0=w_2$ is equivalent to 
$n+1\equiv \frac{1}{2}n(n+1) \equiv 0$ (mod 2). This means 
that $\R P^n$ is Spin iff $n\equiv 3$ (mod 4). Consider the latter case,
$\R P^{2k+1}$: 
These real projective spaces are orientable with fundamental group
(see section \ref{spinpie})
\(
\pi_1( \R P^{2k+1})=
\left\{
\begin{tabular}{ll}
$\Z_2$, & for $k$ odd,
\\
$\Z_2 \times \Z_2$, & for $k$ even,
\end{tabular}
\right.
\)
so that

1. $\R P^{4l+1}$ has $w_1=0$ and $w_2 \neq 0$ for $l=1,2,3,\cdots$.

2. $\R P^{4l-1}$ has $w_1=0$ and $w_2=0$ and $\pi_1=\Z_2$, so has two 
inequivalent Spin structures
\(
\pi^{\pm} : {\rm Spin}(4l)/{\Z_2^\pm} \to SO(4l)/\Z_2\;,
\)
where $\pi^{\pm}$ are the obvious projections and the action of 
Spin$(4l-1)$ in Spin$(4l)/\Z_2^\pm$ is obtained from the natural 
action of Spin$(4l-1)$ in Spin$(4l)$ by passing to the quotient.
The two inequivalent Spin structures are related by an orientation-reversing 
isometry.

\noindent $\circ$ $\mathbb{K}=\C$: Requiring $w_1=0=w_2$ is equivalent to 
$n+1\equiv 0$ (mod 2). This means that $\C P^n$ is Spin
iff $n$ is odd.

\noindent $\circ$ $\mathbb{K}=\mathbb{H}$: The condition is vacuous for dimension
reasons, so that the quaternionic projective space
$\mathbb{H}P^n$ is Spin for all $n$. 

\paragraph{Spin structures on Lie groups.}
Let $G$ be a simply connected Lie group with a left invariant metric. 
The elements of the Lie algebra $\frak{g}={\rm Lie}(G)$ can be regarded as left 
invariant vector fields. The choice of an oriented orthonormal 
basis of $\frak{g}$ gives a trivialization of the frame bundle
$P_{SO}(G)=G \times SO(n)$. The unique Spin structure can be written 
as $P_{\rm Spin}(G)=G \times {\rm Spin}(n)$, with $\varphi = {\rm id} \times \rho$ 
in (\ref{spin str def}). Spinor fields on the group $G$ are then
just maps $G \to \Delta_n$, corresponding to Spin representations $\Delta_n$
of the group 
$G$.  

\vspace{3mm}
If $G$ is not simply connected then the situation is different.
The group manifold $SO(n)$ is not simply connected and so 
has two distinct Spin structures. Corresponding 
to the (trivial) principal frame bundle $P(SO(n))=SO(n) \times SO(N)$, $N=\frac{1}{2}n(n-1)$,
are two coverings
\begin{eqnarray}
Q_1 (SO(n))&=&SO(n) \times {\rm Spin}(N)
\nonumber\\
Q_2 (SO(n))&=&\left({\rm Spin}(n) \times {\rm Spin}(N)  \right)/\Z_2\;,
\end{eqnarray} 
where $\Z_2$ acts on ${\rm Spin}(n) \times {\rm Spin}(N)$
by the map $(g, h) \mapsto (-g, -h)$.

\paragraph{Spin structures on Ricci-flat $n$-dimensional 
Spin manifolds.}
We review the description in \cite{Mc2} (and in more precise form in \cite{MoS}).
Consider a Ricci-flat $n$-manifold $M$.
By the Cheeger-Gromoll theorem \cite{CG}, $M$ can be expressed as $M=\widetilde{M}/\Gamma$,
where $\widetilde{M}$ is the universal Riemannian cover of $M$ and $\Gamma$ is a finite
group of isometries acting freely on $\widetilde{M}$. 
Let $f\in \Gamma$ and $\tilde{f}$ be its lift to a bundle automorphism 
$
\tilde{f} : P_{SO}(\widetilde{M}) \to P_{SO}(\widetilde{M}) 
$
of $P_{SO}(\widetilde{M})$, and let $\hat{f}$ be the Spin lift 
\(
\hat{f} : P_{\rm Spin}(\widetilde{M}) \to 
P_{\rm Spin}(\widetilde{M}) 
\) 
covering $\tilde{f}$. 
Denote by $-\hat{f}$ the composite of $\hat{f}$ with $-1$, where 
$-1$ is the action by $-1 \in P_{\rm Spin}(M)$ on $P_{\rm Spin}(\widetilde{M})$.
Then $P_{\rm Spin}(\widetilde{M})/\Gamma$ is a principal
Spin$(n)$  bundle over $\widetilde{M}/\Gamma=M$, and so is 
a Spin structure over $M$. 
There is an ambiguity in the way $\Gamma$ acts on ${\rm Spin}(\widetilde{M})$:
Each $f$ of even order can be represented by either $\hat{f}$ or by 
$-\hat{f}$ (similarly for order $p$: $\hat{f}^p=1$).
These various actions by $\Gamma$ produce different Spin structures when the
quotient ${\rm Spin}(\widetilde{M})/\Gamma$ is taken. The resulting Spin structures
are in one-to-one correspondence with Hom$(\Gamma, \Z_2)=H^1(M;\Z_2)$.
Consider the following examples \cite{Mc2}.

\vspace{3mm}
\noindent {\it Example 1: $G_2$-manifolds.}
Let $M$ be a Joyce manifold, i.e. a compact 7-manifold with holonomy group
$G_2$ and fundamental group isomorphic to $\Z_2$. So $M=\widetilde{M}/\Z_2$,
with $\Z_2$ generated by an involution $f$. Then $\widetilde{M}$ has a unique 
Spin structure $P_{\rm Spin}(\widetilde{M})$ but $M$ has two distinct Spin structures
$P_{\rm Spin}(\widetilde{M})/\{1, \hat{f}\}$ and 
$P_{\rm Spin}(\widetilde{M})/\{1, -\hat{f}\}$.

\vspace{3mm}
\noindent  {\it Example 2: Calabi-Yau threefolds.} 
Let $M$ be a Calabi-Yau 3-fold of the form $\widetilde{M}/(\Z_2 \times \Z_2)$,
with $\Gamma=\Z_2 \times \Z_2$ generated by a pair of commuting 
involutions $(a,b)$. 
Then $M$ has {\it four} distinct Spin structures, given by 
$P_{\rm Spin}(\widetilde{M})/\Gamma'$,
with $\Gamma'$ one of the finite groups 
$\{ 1, \hat{a}, \hat{b}, \hat{a}\hat{b}\}$,
$\{ 1, -\hat{a}, \hat{b}, -\hat{a}\hat{b}\}$,
$\{ 1, \hat{a}, -\hat{b}, \hat{a}\hat{b}\}$,
$\{ 1, -\hat{a}, -\hat{b}, \hat{a}\hat{b}\}$.

\subsection{Spin structures and the fundamental group}
\label{spinpie}
We have seen that the main source of multiple Spin structures is
spaces with nontrivial fundamental group. We study this in more detail here.
We start with (classes of) examples and then consider the general case. 

\paragraph{Flat manifolds.}
Flat manifolds are of course solutions to both 
M-theory in eleven dimensions 
and to type IIA string theory in ten dimensions.  

\vspace{2mm}
\noindent {\it 1. The $n$-torus.} The $n$-torus $T^n$ admits $2^n$ Spin structures
\cite{Fr1}. Consider $T^n=\R^n/\Gamma$ with $\Gamma$ a lattice in 
$\R^n$. Let $b_1, \cdots, b_n$ be a $\Z$-basis for $\Gamma$
and $b_1^*, \cdots, b_n^*$ be a dual basis for the dual lattice
$\Gamma^*$. 
Spin structures on $T^n$ can be classified by $n$-tuples 
$(\delta_1, \dots, \delta_n)$ where each 
$\delta_j \in \{0,1\}$ indicates the twist of the Spin structure in
the direction $b_j$. 

\vspace{2mm}
\noindent {\it 2. Torus bundles over flat manifolds.}
Such manifolds are classical solutions to supergravity. 
There are many examples of these toral extensions 
which are Spin. However, 
this is not automatic and there 
also exist examples which are not Spin \cite{AuS}.

\vspace{2mm}
\noindent {\it 3. Flat manifolds with cyclic holonomy.}
Let $M^n$ be an oriented $n$-dimensional
 flat Riemannian manifold. The fundamental group
 $\Gamma=\pi_1(M^n)$ determines a short exact sequence
 $
 0 \to \Z^n \to \Gamma \buildrel{p}\over{\longrightarrow}
 F \to 0\
 $, 
where $\Z^n$ is a torsion-free abelian group of rank $n$
and $F$ is a finite group which is isomorphic to 
the holonomy group of $M^n$. 
The universal cover of $M^n$ is $\R^n$, so that 
$\Gamma$ is isomorphic to a discrete cocompact 
subgroup of the isometry group 
${\rm Isom}(\R^n)=\R^n \ltimes SO(n)$. 
The existence of a Spin structure on $M^n$ is equivalent to 
the existence of a homomorphism $\epsilon : \Gamma \to {\rm Spin}(n)$
such that $\rho \epsilon =p$, where $\rho: {\rm Spin}(n) \to SO(n)$ is 
the covering map. Any flat manifold with holonomy of odd order
admits a Spin structure \cite{Vas}. In fact, every finite group 
is the holonomy group of a Spin flat manifold \cite{DSSz}.
If the fundamental groups 
$\Gamma_i$, $i=1,2$ of two flat oriented $n$-manifolds 
$M_i$ are isomorphic then $M_1$ has 
a Spin structure iff $M_2$ has a Spin structure \cite{HiS}.
Multiple Spin structures are possible here. For example, 
if $\Gamma$ is of the so-called diagonal type and 
$M^n$ has holonomy $\Z_2$ then there are $2^n$ 
Spin structures (as in the case of the $n$-torus)
\cite{MP}.

\subsubsection{Spherical space forms}
\label{sph sp}
A classical result of Hopf characterizes a spherical space form as follows:
A Riemannian manifold $M$ of dimension
$n \geq 2$ is a connected complete manifold of positive 
constant curvature iff $M$ is isometric to $S^n/\Gamma$, where 
$\Gamma$ is a finite subgroup of $O(n+1)$ which acts freely 
on the sphere $S^n$.
 Any fixed-point-free map $g: S^n \to S^n$ is homotopic to the
antipodal map. Hence deg$(g)=(-1)^{n+1}$. Thus if $n$ is even,
the composite of two fixed-point free maps has a fixed point. This means
that the only group which acts freely on an even-dimensional sphere
is $\Z_2$, so the only spherical space forms in even dimensions 
are the sphere itself $S^{2m}$, corresponding to the trivial element 
in $\Z_2$, and the real projective space $\R P^{2m}$, corresponding 
to the nontrivial element in $\Z_2$.
However, the case $n$ odd is much more interesting. 

\vspace{3mm}
The main examples of spherical space forms are 
Clifford-Klein manifolds. These are complete Riemannian 
manifolds with constant sectional curvature equal to $+1$,
and are of the form $S^n/\Gamma$, where $\Gamma$ is a finite group
acting freely and orthogonally on $S^n$, as described above. 
Equivalently, they
are given by an orthogonal representation $\varrho : \Gamma \to O(n+1)$
with $\varrho (g)$ having no $+1$ eigenvalue for all nontrivial 
elements $g \in \Gamma$. The representation 
$\varrho$ is conjugate to a unitary fixed-point-free representation 
$\varrho : \Gamma \to U(k) \subseteq SO(2k) \subset O(2k)$.
The classification of Clifford-Klein 
manifolds $M(\Gamma, \varrho)$
is thus a completely algebraic question in group
representation theory, whose solution is given completely 
by Wolf \cite{Wo}.
More precisely, the classification of spherical space forms can be reduced to the 
determination of all finite groups having fixed-point-free real orthogonal 
representations and hence free unitary representations.

\vspace{3mm}
Since $S^n$ is a universal $\Gamma$-covering space of 
$M^n$, and since $\pi_1(M^n)=\Gamma$ and $\pi_i(M^n)\cong \pi_i(S^n)$
 for $i \neq n$, the integral cohomology groups of $M^n$ are 
\(
H^i(M^n;\Z)=
\left\{
\begin{array}{ll}
\Z, & {\rm if~} i=0 {\rm ~or~} i=n,\\
H^i(\Gamma;\Z), & {\rm if~} 0<i<n,\\
0, & {\rm otherwise}.
\end{array}
\right.
\)
 If $\Gamma$ acts freely on $S^{2n-1}$ then $\Gamma$ has a free
resolution of length (or period) $2n$ \cite{DM}. A free resolution of period $n$
of $\Gamma$ is an exact sequence
\(
0 
\longrightarrow
 \Z
 \buildrel{\mu}\over{\longrightarrow} 
 F_{n-1} 
 \longrightarrow
  \cdots 
  \longrightarrow
   F_1 
   \longrightarrow
    F_0 
    \buildrel{\varepsilon}\over{\longrightarrow}
\Z 
\longrightarrow 
0
\)
of $\Z \Gamma$-modules with the $F_i$ finitely generated and free,
with $\Gamma$ acting trivially on the two $\Z$ terms. 
The fact that $\Gamma$ has a free resolution of period $2n$ implies that 
$H^{2n+1}(\Gamma;\Z)=\Z/|\Gamma|$, which gives that the abelian 
subgroups of $\Gamma$ are cyclic \cite{DM}. For example:
\begin{enumerate}
\item The cyclic groups $\Z_n$ act freely and orthogonally 
on $S^1$. 

\item The binary dihedral groups 
$
Q(4k)=\{ x,y~|~x^{2k}=1, x^k=y^2, yxy^{-1}=x^{-1}\}
$
are subgroups of the unit quaternions $S^3$ and so act freely
and orthogonally on $S^3$ via quaternionic multiplication.
When $k$ is a power of 2, $Q(4k)$ is a generalized quaternion 
group. 
\end{enumerate}
A finite group $\Gamma$ is called a {\it spherical space form group} if $\Gamma$ satisfies 
either of the two equivalent conditions \cite{DM}: 

(i). The cohomology of $\Gamma$ is periodic. 

(ii). The group $\Gamma$ acts without fixed points on some sphere.

\vspace{3mm}
\noindent Let $\Gamma$ be a nontrivial finite group acting linearly and freely on
$S^n$. Then 

\noindent {\bf 1.} If $n=2m$, $\Gamma$ is isomorphic to $\Z_2$ and the only 
nontrivial even-dimensional
spherical forms are the real projective spaces $\R P^{2m}$.
 Let $M=\R P^m$, so that 
$\pi_1(M)=\Z_2=\{\pm 1\}$. Then the covering space is 
$\widetilde{M}=S^m$ and the deck group action is the usual 
action of $\{\pm 1\}$ on $S^m$ given by multiplication. The nontrivial 
element $-1$ acts as the antipodal map and the trivial element 
$+1$ acts as the identity. The orthogonal representation of 
the fundamental group 
\(
\varrho : \pi_1(\R P^m)=\Z_2 \to O(1)=\Z_2
\label{rrpm}
\)
is given by $\varrho (\pm 1)=\pm 1$.

\vspace{2mm}
\noindent {\bf 2.} If $n=2m+1$, then 
\begin{itemize}

\item if $\Gamma$ is {\it abelian}, then $\Gamma$ is cyclic; these correspond to lens spaces.
 The cyclic group $\Z_n$, the group of $n$th
roots of unity, acts on the unit sphere $S^{2k-1}$ in $\C^k$ by complex 
multiplication. The lens space is the quotient manifold
\(
L^{2k-1}:=S^{2k-1}/\Z_n\;.
\label{eq lens}
\)
The irreducible representations of $\Z_n$ are parametrized by 
$\{\varrho_s\}$, $0 \leq s <n$, where $\varrho_s (\lambda)=\lambda^s$.

\item if $\Gamma$ is {\it nonabelian}, then it has the following equivalent 
properties:
\begin{enumerate}
\item all the abelian subgroups of $\Gamma$ are cyclic;
\item the $p$-Sylow subgroups $\Gamma_p$ of $\Gamma$ are of two types:
either all of them are cyclic, or they are cyclic for $p\neq 2$ and generalized quaternion
groups for $p=2$. The latter corresponds to the quaternionic spherical 
space form $S^{2m+1}/Q_r$. The canonical inclusion $i_p: \Gamma_p \to \Gamma$
induces a ring homomorphism $i_p^*: RO(\Gamma) \to RO(\Gamma_p)$, so that
$\Gamma_p$ acts freely on $S^n$ and the resulting quotient
$M_p^n=S^n/\Gamma_p$ is the  spherical $p$-form associated to $M^n$,
and $\i_p: M_p^n \to M^n$ is a covering fibration.
\end{enumerate}
\end{itemize}
 There is a difference 
between abelian and nonabelian fundamental groups. 

\paragraph{Nonabelian fundamental group.}
A manifold can have a {\it nonabelian} fundamental group yet have only
one Spin structure.
For example \cite{Mc2}: $\Gamma$ is the nonabelian group of order 12 generated by $a$, $b$,
$c$ satisfying 
\(
a^2=b^2=c^3=1, \qquad a b=b a, \qquad c a c^{-1}=b, \qquad c b c^{-1}=a b\;.
\)
The relation ${\hat{c}}~{\hat{a}}~{\hat{c}}^{-1}={\hat{b}}$ shows that ${\hat{a}}$
can be replaced by $-{\hat{a}}$ only if $\hat{b}$ is replaced by 
$-{\hat{b}}$, but this is inconsistent with ${\hat{c}}~{\hat{b}}~{\hat{c}}^{-1}={\hat{a}}~{\hat{b}}$.
Therefore, $M=\widetilde{M}/\Gamma$ has only {\it one} Spin structure
in this case.

\paragraph{Group representation ring of $\Z_n$.} The integers modulo $n$,
$\Z_n :=\{\lambda \in \C ~|~\lambda^n=1\}$, is the group of $n$th roots of unity. 
As above, let $\varrho_s(\lambda):=\lambda^s$. The $\{\varrho_s\}$ for $s\in \Z_n$ parametrize the 
irreducible unitary representations of $\Z_n$, so the group representation ring
$
RU(\Z_n)=\oplus_{s\in\Z_n} \varrho_s \cdot \Z
$
is the free abelian group generated by these elements. The ring structure
is given by 
$
\varrho_s \cdot \varrho_t = \varrho_{s+t}$.
Since $\varrho_1^n=1$, then 
\(
RU(\Z_n)=\Z [\varrho_1]/(\varrho_1^n=1)\;.
\)

\paragraph{Generalized lens spaces.}
The cyclic group $\Z_n$ is a spherical space form group. Let
$\vec{a}=(a_1, \dots, a_k)$ be a collection of integers and let 
$\tau' = \oplus_{\nu} \varrho_{a_\nu}$. Then $\tau'$ is fixed-point-free iff
all the $a_\nu$ are coprime to $n$. 
 The generalized lens spaces 
\(
L(n;\vec{a})=M(\Z_n, \tau')=S^{2k-1}/\varrho (\Z_n)
\)
are the spherical space forms with cyclic fundamental groups. 
 If $k=1$ or $k=3$ then $TM$ is trivial so that $M$ is always Spin.
For $k \geq 5$, \cite{Gi}
 $\varrho$ can be lifted to Spin$(2k)$, so that $M(\Gamma,\varrho)$ is Spin,
 if 
  $|\Gamma|$ is odd, or if $|\Gamma|$ is even and $k$ is even.
Note that $\varrho$ automatically 
lifts to Spin$(2k)$ if $|\Gamma|$ is even and if $k$ is odd.

\vspace{3mm}
Now consider Spin structures on the space bounding a lens space.
Such an example is important for the M-theory partition function.
In general, if $Z^{12}$ is a Riemannian Spin manifold, then 
a Spin structure on $Z^{12}$ induces a Spin structure on 
$Y^{11}=\partial Z^{12}$. 
The frame bundle $P_{SO}Y^{11}$ can be considered as a subbundle 
of the frame bundle $P_{SO}Z^{12}$ restricted to the boundary by adding the 
unit normal vector to the set of orthonormal frames of the former. 
The inverse image of $P_{SO}Y^{11}$ under the covering map 
$P \to P_{SO}Z^{12}$ defines a Spin structure on $Y^{11}$. 
Any Spin structure on $Z^{12}(p, \vec{a})$ is obtained by twisting the 
canonical Spin${}^c$ structure $\sigma_{\rm can}^c$ (see section
\ref{basic spin c})
by the square root 
$K^{1/2}$
of the
canonical line bundle $K$ of $Z^{12}(p, \vec{a})$. 
The canonical line bundle $K$ corresponds to the representation
(cf. \cite{Fu2}) 
\(
a_1 + a_2 + \cdots + a_n \in \Z_p \cong R(\Z_p, U(1))\;.
\)
The square root $K^{1/2}$ corresponds to an element $c \in  \Z_p \cong R(\Z_p, U(1))$
satisfying 
$
2c\equiv a_1 + a_2 + \cdots + a_n$ (mod $p$).
Therefore, 

\textbullet~$Z^{12}(p, \vec{a})$ is Spin iff $p$ is odd, or 
$p$ and the sum $a_1 +\cdots a_n$ are both even. 

\textbullet~If $p$ is odd then the Spin structure $\sigma=c$ is unique.

\textbullet~If $p$ and  $a_1 +\cdots a_n$ are both even then there are
two Spin structures $\sigma_1=c$ and $\sigma_2=c + p/2$. 
\\
Let $\vec{a}\in \Z^n$ be an $n$-tuple of integers coprime to an integer $p$. 
Then the following properties hold  \cite{Fu2}: 

(1) The lens space $L(p;\vec{a})$ admits a Spin structure iff $p$ is odd, or $p$ is even
 and $\sum_{k=1}^n a_k$ is even.

(2) In case the conditions in (1) hold for $(p;\vec{a})\in \Z^{n+1}$ there is a one-to-one
correspondence between the set of all Spin structures on $L(p;\vec{a})$ and the set
of all elements $c \in \Z_p$ satisfying 
\(
2c =a_1 + \cdots + a_n\in \Z_p\;.
\)

\paragraph{Killing spinors in non-simply connected case.} 
Supersymmetry involves (generalized) Killing or parallel spinors.
 Let $(M,g)$ be a complete $n$-dimensional Riemannian Spin 
manifold, and let $SM$ be the Spin bundle of $M$ and $\psi$ 
a smooth section of $SM$. Then $\psi$ is a Killing spinor 
if
\(
\nabla_{\sf X} \psi=\alpha~ {\sf X}\cdot \psi\;, \qquad \forall~ {\sf X} \in \Gamma (TM)\;,
\)
where $\nabla$ is the Levi-Civita connection of $g$ and 
${\sf X}\cdot \psi$ denotes the Clifford multiplication of ${\sf X}$ and 
$\psi$. We say that $\psi$ is {\it imaginary} when 
the Killing constant $\alpha \in {\rm Im}(\C^*)$, $\psi$ is {\it parallel} if 
$\alpha=0$ and 
$\psi$ is real if $\alpha \in {\rm Re}(\C^*)$.
Now let $(M^n,g)$ be a complete 
Riemannian Spin manifold with nontrivial fundamental group,
$\pi_1(M)\neq 0$,
and a nontrivial Killing spinor with $\alpha>0$ or $\alpha<0$. If $n=2m+1$,
$m\geq 2$ even, then there are two possibilities \cite{Wa95}: 

(1) $M$ is a Spin spherical space form.

(2) $M$ is Sasakian-Einstein with 
Hol$(\overline{g})=SU(m+1)$.

\vspace{3mm}
The above spaces are important in classical considerations in M-theory
(and we will consider the partition function later in this paper). 
Superconformal Chern-Simons theories in three dimensions are 
obtained by taking a factor in the eleven-dimensional manifold to be an
eight-dimensional transverse cone with base
the quotient $S^7/\Gamma$, with $\Gamma < SO(8)$
a finite group acting freely on $S^7$ such that the 
quotient is Spin and the space of Killing spinors 
has dimension $N$. 
The classification for $N>3$ is obtained in \cite{DFGM}.
Let $S^{2m-1}/\Gamma$ be a spherical space form with Spin 
structure given by $\epsilon: \Gamma \to {\rm Spin}(2m)$. 
Then the dimension of the space of Killing spinors 
with Killing constant $\alpha=\pm \frac{1}{2}$ is given by 
\cite{Ba1}
$
\frac{1}{|\Gamma|}\sum_{\gamma \in \Gamma}
\chi^\pm (\epsilon (\gamma))
$,
where $\chi^\pm$ are 
the characters of the two irreducible half-spin 
representations of ${\rm Spin}(2m)$.


\paragraph{Homology and higher connectedness.} 
If $\pi_1$ is the first nonvanishing homotopy group of a manifold
$M$, then by the Hurewicz theorem the first homology group is
isomorphic to the abelianization of $\pi_1$. 
Beyond that, 
$n$-connected manifolds, with $n\geq 2$,
 carry unique Spin structures. Examples include:
homotopy spheres (e.g. squashed spheres), Stiefel manifolds, and simply-connected
Lie groups. The case $n=2$ will be considered in section 
\ref{pie2} in connection to Spin${}^c$ structures. The case
$n=3$ is considered in \cite{tcu} in connection to String structures.

\subsubsection{Holonomy and flat vector bundles}
Let $\widetilde{M}$ be the universal cover of $M$. The fundamental group 
$\pi_1(M)$ of $M$ acts on $\widetilde{M}$ by deck transformations 
$
\tilde{x} \to g\cdot \tilde{x}
$,
for $\tilde{x}\in \widetilde{M}$ and $g \in \pi_1(M)$. 
There is a natural identification between representations 
$\varrho$
of the fundamental 
group and flat vector bundles. 
Start with a vector bundle $V$ and a Riemannian connection on $V$ which 
has zero curvature. Let $\mathbb{F}^k_0$ be the fiber of $V$ over the basepoint 
$x_0$, and where
$\mathbb{F}$ is $\R$ or $\C$. Parallel transport around a closed path $\gamma$ 
defines the holonomy representation of $\pi_1(M)$. 
The bundle $V$ is naturally isomorphic to 
the associated flat vector bundle $V_\varrho$, given by
\(
V_\varrho := \left( \widetilde{M} \times_\varrho \mathbb{F}^k\right)/\sim\;,
\label{vrho}
\)
where $(\tilde{x}, v) \sim (g\cdot \tilde{x}, \varrho (g) v)$ for all $g \in \pi_1(M)$.
The associated
connection agrees with the original connection. 
Denote by $\mathcal{O}_{\mathbb{F}}^k$  the trivial real
or complex vector bundle with fiber $\mathbb{F}^k$. Define a flat
connection on this bundle by
$
\widetilde{\nabla} \tilde{f}= d \tilde{f}
$, 
where $\tilde{f}$ is a smooth section of $\mathcal{O}_{\mathbb{F}}^k$. The section 
descends to a section on $V_\varrho$ and also  $\widetilde{\nabla}$
descends to a connection on $V_\varrho$ with zero curvature. 
For example, consider the case around equation \eqref{rrpm}.
The associated vector bundle is given by
$
L:= S^m \times \R/\sim
$,
 where the equivalence is given by $(\xi, x) \sim (-\xi, -x)$. The bundle
 $L$ is a real line bundle which is isomorphic to the classifying line
 bundle over $\R P^m$.
Next, considering the lens space  \eqref{eq lens}, 
let $L_s$ be the complex line bundle 
defined by the representation $\varrho_s$. Then 
$
L_s=S^{2k-1} \times \C/\sim
$,
where the equivalence if given by 
$(\xi, z) \sim (\lambda \xi, \lambda^s z)$ for $\lambda \in \Z_n$, $\xi \in S^{2k-1}$,
and $z \in \C$.

\paragraph{Principal bundles with finite structure group.}
 Let $\Gamma$ be a finite group. A $\Gamma$-structure on a connected manifold $X$ is a principal 
$\Gamma$ bundle $P$ over $X$. The holonomy of the bundle $P$ defines 
a group homomorphism from the fundamental group $\pi_1(X)$ of $X$ to $\Gamma$.
On the other hand, let $\Gamma$ act without fixed points on a compact manifold
$P$. Then $\Gamma \buildrel\pi\over\to P \to P/\Gamma$ is a principal
$\Gamma$ bundle over $X=P/\Gamma$. 
The study of principal $\Gamma$ bundles is equivalent to the study of 
fixed-point-free $\Gamma$ actions (see \cite{GLP}).

\paragraph{Lattices of Lie groups.}
This is yet another rich source of Spin structures. 
Let $\Gamma \subset G$ be a lattice in a Lie group $G$. 
Spin structures on $M=G/\Gamma$ correspond to 
homomorphisms 
$
\epsilon: \Gamma \to \Z_2=\{-1,1\}
$.
The corresponding Spin structure is given by 
\(
P_{{\rm Spin},\epsilon}(G/\Gamma)= G \times_\Gamma {\rm Spin}(n)\;,
\)
where $g_0 \in \Gamma$ acts on $G$ by left multiplication 
and on Spin$(n)$ by multiplication with the central element 
$\epsilon (g_0)$. Spinor fields on $M$ can be identified with
$\epsilon$-equivariant maps to the Spinor module 
$\psi : G \to \Delta_n$, i.e. 
$\psi(g_0 g)= \epsilon(g_0)\epsilon (g)$ for all 
$g\in G$, $g_0\in \Gamma$.

\subsection{Effect of Spin structures on geometric and analytical entities}
\label{ana eff}

\paragraph{The spectrum.} Consider the example of the circle $S^1$
from section \ref{multi spin}.
The Dirac operator is simply $D=i\frac{d}{dt}$, with parameter $t$. 
Spinors on the circle corresponding to the trivial Spin structure 
are complex valued functions. The spectrum corresponding 
to the trivial Spin structure consists of the eigenvalues $\lambda_k=k$ with 
eigenfunctions 
$t \mapsto e^{-ikt}$, $t \in \Z$. On the other hand, spinors corresponding 
to the nontrivial 
Spin structure are $2\pi$-anti-periodic complex-valued functions on 
$\R$, with eigenvalues $\lambda_k=k+\frac{1}{2}$, $k\in \Z$, with 
eigenfunction $t \mapsto e^{-i(k+\frac{1}{2})t}$.
This example illustrates dependence of the spectrum of the Dirac 
operator on the choice of Spin structure. This phenomenon persists
in higher dimensions as well. Many examples, such as spherical space 
forms, can be found in \cite{Ba}.

\paragraph{The noncommutative case.}
In the presence of a Neveu-Schwarz $B$-field, spacetime 
can become noncommutative (see \cite{SW2}).
Spin structures on noncommutative spaces are delicate. They
are closely related to the 
reality structure $J$ appearing in spectral triples, but further require 
invoking spectral properties of the Dirac operator. 
The noncommutative two-sphere, like the classical one, has one admissible 
(real) Spin structure. There are a priori others, but these are ruled out by 
spectral properties of the 
Dirac operator \cite{P}.
The noncommutative two-torus has, similarly to the classical case, 
four inequivalent Spin structures. The spectra of the corresponding 
Dirac operators depend on the Spin structure \cite{PS}.  However, 
interestingly, there are examples of spherical space forms 
where different Spin structures
give rise to 
the {\it same} spectral action \cite{MPT}.

\paragraph{Eta invariants.} As a result of asymmetry in the spectrum, 
the values of the $\eta$-invariants depend on the
choice of Spin structures. For example, for the real projective space 
$\R P^n$, $n \equiv 3$ mod 4, admitting two Spin structures, the $\eta$-invariant for the 
Dirac operator  is given by  $\eta=\pm 2^{-m}$, $n=2m-1$, where the
sign depends on the Spin structure chosen \cite{Ba}. The differences 
are more drastic in the case of Bieberbach manifolds $B^n=T^n/\Gamma$;
for example, for $\Gamma=\Z_4$, $B^n$ has 4 Spin structures and the values 
of the eta invariant are: $\eta=0$ for two of them, $\eta=\frac{3}{2}$
for one, and $\eta=-\frac{1}{2}$ for the fourth Spin structure \cite{Pf}.

\paragraph{Harmonic spinors.}
The space of solutions of a spinor field equation usually depends on the 
Spin structure. For example, the dimension of the kernel of the Dirac operator 
corresponding to the two Spin structures on the circle are: one for the trivial 
Spin structure and zero for the nontrivial Spin structure on $S^1$. 
The case of harmonic spinors -- i.e. the `bare' Killing
spinors-- is explained extensively in \cite{Hit} \cite{BSS}.

\paragraph{Killing and parallel spinors.} 
In the case of multiple Spin structures, extending a local parallel spinor
depends on the choice of Spin structure \cite{Mc2}.
There are compact irreducible Ricci-flat
manifolds with nongeneric holonomy in dimension 4 and 8 
which admit two Spin structures, each of which carries
a parallel spinor \cite{MoS}; an example is a Calabi-Yau 
fourfold with holonomy $SU(4) \rtimes \Z_2$ obtained 
via an involution 
as 
a complete intersection in $\C P^9$. 
Specifying a Spin structure as part of the data defining a supergravity
background was highlighted in \cite{FG}. For example, it was shown that 
the same geometry of four-dimensional anti-de Sitter space times a lens 
space $S^7/\Z_4$ preserves a different 
amount of supersymmetry depending on the choice of Spin structure. 
Such a phenomenon is expected to persist for more general spherical 
space forms.

\paragraph{Spin holonomy and preferred Spin structure(s).}
 Consider an $SO(n)$ bundle $P$ over an oriented Spin 
manifold $(M, g)$ with Levi-Civita connection $\nabla$. 
Choose a Spin structure $(P_{\rm Spin}, \rho)$ on $M$, so that $\nabla$ is
lifted to a Spin connection ${\nabla^s}$ on
the Spin bundle $S= P_{\rm Spin} \times_{{\rm Spin}(n)} \Delta_n$ corresponding 
to $P_{\rm Spin}$. Here $\rho$ is the projection $\rho : {\rm Spin}(n) \to SO(n)$.
The holonomy group Hol$(\nabla^s)$ is a subgroup of Spin$(n)$, and coincides
with Hol$(g)$ under the projection map $\rho$. The projection 
$
\rho : {\rm Hol}(\nabla^s) \to {\rm Hol}(g)
$
is either an isomorphism or a double cover. This depends on the choice
of Spin structure  \cite{Jo}. 
However, if $M$ is simply connected then both Hol$(g)$ and 
Hol$(\nabla^s)$ are connected, so that the latter is the identity component 
of $\rho^{-1}({\rm Hol}(g))$ in Spin$(n)$. Therefore, for simply connected 
Spin manifolds, the Riemannian holonomy groups and the Spin 
holonomy groups coincide. 

\vspace{3mm}
Let $M$ be an $n$-dimensional manifold, $n \geq 3$,
 that admits a $G$-structure $P_G$ with $G$ a connected, simply connected
 subgroup of $SO(n)$. Then $M$ is Spin and has a natural Spin structure
 $P_{\rm Spin}$ induced by $P_G$. 
Since all of the Ricci-flat holonomy groups, $SU(n)$, $Sp(n)$, $G_2$, and 
Spin$(7)$ are connected and simply connected, this gives
the following result \cite{Jo}: Let $(M, g)$ be a Riemannian manifold and suppose
that Hol$(g)$ is one of the Ricci-flat holonomy groups above. Then $M$ is Spin,
with a {\it preferred Spin structure}. With this Spin structure, the spaces of parallel
spinors on $M$ are nonzero. 
Thus, an irreducible metric has one of the Ricci-flat holonomy groups iff
it admits a nonzero constant spinor.

\paragraph{Effect of orientation-reversal on spinors on $Y^{11}$.}
Let $Y^{11}$ be a Riemannian manifold and let $P_{SO(11)}$ be 
the $SO(11)$-principal bundle of oriented orthonormal frames. Let 
$P_{{\rm Spin}(11)} \to Y^{11}$ and 
$\sigma : P_{{\rm Spin}(11)} \times_{{\rm Spin}(11)} SO(11) 
\cong P_{SO(11)}$ represent the Spin structure. Now let 
$P_{SO(11)}^{\rm op}\to Y^{11}$ be the $SO(11)$-principal bundle of 
oriented orthonormal frames of $Y^{11}_{\rm op}$, where the latter 
refers to the manifold $Y^{11}$ with the reversed orientation.  
Define an isomorphism of $SO(11)$-principal bundles 
$\theta : P_{{\rm Spin}(11)} \to P_{{\rm Spin}(11)}^{\rm op}$
which maps the frame $({\sf Y}_1, \cdots, {\sf Y}_{11})$ to 
 $(-{\sf Y}_1, \cdots, -{\sf Y}_{11})$. Then the opposite Spin structure is 
 given by 
$P_{{\rm Spin}(11)} \to Y^{11}$ and 
$P_{{\rm Spin}(11)} \times_{{\rm Spin}(11)} SO(11) 
\buildrel{\sigma}\over{\cong} P_{SO(11)}
\buildrel{\theta}\over{\cong} 
 P_{SO(11)}^{\rm op}$.

\paragraph{Effect of orientation-reversal on spinors on $X^{10}$.}
Let $X^{10}$ be a Riemannian manifold and let $P_{SO(10)}$ be 
the $SO(10)$-principal bundle of oriented orthonormal frames. Let 
$P_{{\rm Spin}(10)} \to X^{10}$ and 
$\sigma : P_{{\rm Spin}(10)} \to X^{10} \times_{{\rm Spin}(10)} SO(10) 
\cong P_{SO(10)}$ represent the Spin structure. Let 
$P_{SO(10)}^{\rm op}\to X^{10}$ be the $SO(10)$-principal bundle of 
oriented orthonormal frames of $X^{10}_{\rm op}$, where the latter 
refers to the manifold $X^{10}$ with the reversed orientation.  
The image $E_1\subset O(10)$ of an element $e_1 \in \R^{10} \subset {\rm Pin}(10) \subset 
C\ell (\R^{10})$ acts as $(x_1, x_2 \cdots, x_{10})
\mapsto  (x_1, -x_2 \cdots, -x_{10})$. Define the map
$\theta : P_{SO(10)} \to P_{SO(10)}^{\rm op}$
by
$({\sf X}_1, {\sf X}_2, \cdots, {\sf X}_{10})
\mapsto  ({\sf X}_1, -{\sf X}_2, \cdots, -{\sf X}_{10})$. 
This becomes an isomorphism of $SO(10)$-principal 
bundles if the action of $SO(10)$ is twisted by $E_1$,
leading to  $\widetilde{P}_{SO(10)}$,  the twisted bundle 
$P_{SO(10)}$. Let 
$P_{{\rm Spin}(10)}^{\rm op}\to Y^{11}$ be the ${\rm Spin}(11)$-principal
bundle given by $P_{\rm Spin(11)}\to Y^{11}$ with the 
${\rm Spin}(11)$-action twisted by $e_1$. As in \cite{Bu},
this gives a 
homomorphism of $SO(11)$-principal bundles 
$P_{{\rm Spin}(11)}^{\rm op} \times_{{\rm Spin}(11)} SO(11)
\buildrel{\tilde{\sigma}}\over{\cong}
\tilde{P}_{SO(11)} \buildrel{\theta}\over{\cong}
P_{\rm SO(11)}^{\rm op}$, where 
$\tilde{\sigma} [s, v]:=\sigma [s E_1, vE_1]$.

\subsection{Circle actions and Spin structures} 
\label{circle actions}

We now consider the situation in the setting of M-theory. 
We take $Y^{11}$ to be a Riemannian eleven-dimensional manifold with metric $g_Y$,
carrying a free isometric and geodesic action of the circle $S^1$.  
Then the 
orbit space $X^{10}=Y^{11} /S^1$ is a manifold and there is a metric $g_X$ 
on $X^{10}$ 
such that the quotient map $\pi : Y^{11} \to X^{10}$ becomes a principal $S^1$ bundle 
and a Riemannian submersion.

\vspace{3mm}
The $S^1$ bundle has a unique connection 1-form $i\omega : TY^{11} \to i\R$
such that ker\hspace{0.5mm}$\omega|_y$ is orthogonal to the fibers with respect to 
$g_Y$
for all $y\in Y^{11}$. The $S^1$-action induces a Killing vector field $v$. 
The size of the fibers is measured by the dilaton field. 
There are two cases:

\noindent {\bf 1.} The dilaton $\phi$ is constant on $X^{10}$, i.e. the length $|v|$ is constant on $Y^{11}$,
then the fibers of $\pi$ are totally geodesic. 

\noindent {\bf 2.} If the dilaton is not constant, then we have a nonconstant length function  
$
e^\phi : X^{10} \to \R^+
$.
Rescaling the metric on $Y^{11}$ is completely characterized by
the connection 1-form $i\omega$, the fiber length $2\pi e^\phi$, and the
metric $g_X$ on $X^{10}$; the Dirac operator can be expressed
in terms of $\omega$, $e^\phi$, and $g_X$. 

\vspace{3mm}
The $S^1$-action on $Y^{11}$ induces an $S^1$-action on $P_{SO}(Y^{11})$. 
A Spin structure $\sigma_Y: P_{\rm Spin}(Y^{11}) \to P_{SO}(Y^{11})$ 
is called (cf. \cite{AB})

1. {\it projectable} or {\it even} if this $S^1$-action on $P_{SO}(Y^{11})$ lifts 
to $P_{\rm Spin}(Y^{11})$.

2. {\it nonprojectabe} or {\it odd} if the $S^1$-action does not lift.

\vspace{3mm}
By a theorem of \cite{Her} all fibers of a Riemannian submersion with totally
geodesic fibers are isometric, so there is a positive real number $r$
such that all the fibers of $\pi$ are isometric to $S^1_{r} \hookrightarrow \C$,
a circle of radius $r$ in $\C \cong \R^2$ with its standard metric. There is the  
disk bundle $\pi_D : Z^{12} = Y^{11} \times_{S^1} \mathbb{D}^2 \to X^{10}$ of 
the associated complex
line bundle $\pi_{\C}: \mathcal{L}=Y^{11} \times_{S^1} \C \to X^{10}$ to $\pi$. The 
disk $\mathbb{D}^2 \subset \C$ 
can be endowed with a metric such 
that $\partial \mathbb{D}^2$ is isometric to $S^1_{r}$.

\vspace{3mm}
Let us consider some details. Denote by 
${\rm pr}: {\rm Fr}(Y^{11}) \to Y^{11}$ 
the bundle of oriented
frames, whose sections are the vierbeins, and by $p: P_{\rm Spin}(Y) \to Y^{11}$ the principal 
${\rm Spin}(11)$-bundle over $Y^{11}$, with covering map $u: P_{\rm Spin}(Y^{11}) \to 
{\rm Fr}(Y^{11})$
corresponding to the double covering $\rho_{\rm Spin}: {\rm Spin}(11) \to SO(11)$.
A given Spin structure is denoted $\sigma_Y=(P_{\rm Spin}(Y^{11}),u)$. 
 Let $\chi : S^1 \times Y^{11} \to Y^{11}$ be a smooth
circle action. Then $\chi$ induces a circle action on ${\rm Fr}(Y^{11})$ that can be viewed as 
a one-parameter family of maps $\chi_t : {\rm Fr}(Y^{11}) \to {\rm Fr}(Y^{11})$, $ t\in \R$. The special values
$\chi_0$ and $\chi_1$ coincide with the identity $id_{\rm Fr}$ on ${\rm Fr}(Y^{11})$. 
Since $P_{\rm Spin}(Y^{11})$ 
is the double cover of ${\rm Fr}(Y^{11})$, $\chi_t$ lifts to 
$\tilde{\chi}_t: P_{\rm Spin}(Y^{11}) \to P_{\rm Spin}(Y^{11})$ with (cf. \cite{AH}):

\begin{enumerate}
\item {\it Even type}: $\tilde{\chi}_0=\tilde{\chi}_1=id_{P_{\rm Spin}(Y^{11})}$, in which case $\tilde{\chi}_t$ induces a
circle action on $P_{\rm Spin}(Y^{11})$ which commutes with the right action of ${\rm Spin}(11)$
on $P_{\rm Spin}(Y^{11})$ and is compatible with $u$. In this case $\chi$ is said to be of {\it even} type.

\item {\it Odd type}: $\tilde{\chi}_0=id_{P_{\rm Spin}(Y^{11})}$ and $\tilde{\chi}_1$ is multiplication by 
$-1 \in \Z_2=\ker \{ {\rm Spin}(11) \to SO(11)\} \subset {\rm Spin}(11)$, in which case
$\tilde{\chi}_t$ induces an action of the connected double covering $\hat{S}^1$ of $S^1$
on $P_{\rm Spin}(Y^{11})$ which, again, commutes with the right action of 
${\rm Spin}(11)$
on $P_{\rm Spin}(Y^{11})$ and is compatible with $u$. In this case $\chi$ is said to be of {\it odd} type.
\end{enumerate}

\noindent Let $\pi_S: P_{\rm Spin}(Y^{11}) \to P_{\rm Spin}(Y^{11})/S^1$ and $\pi_{\rm Fr}: {\rm Fr}(Y^{11}) 
\to {\rm Fr}(Y^{11})/S^1$ be 
the projections of the 
corresponding bundles to the quotient space. Note that there is an isomorphism 
$\pi_{\rm Fr}^*({\rm Fr}(Y^{11})/S^1) \cong {\rm Fr}(Y^{11})$. 

\vspace{3mm}
\noindent The two situations for circle actions above 
are summarized in the following table
\(
\begin{tabular}{|c||c|c|}
\hline
$S^1$-action& $Y^{11}$ & $X^{10}$
\\
\hline
\hline
even & Spin & Spin 
\\
\hline
odd & Spin & ${\rm Spin}^c$
\\
\hline
\end{tabular}
\)
and will be treated separately and in detail in section 
\ref{proj} and section
\ref{nonproj}, respectively. 

 \paragraph{Cobordism of free circle actions.}
Let $F_{11}^{\rm Spin}$ denote the cobordism group of free circle actions on 
closed $11$-dimensional Spin manifolds. This splits, according to whether the
circle action is even or odd, as a direct sum
\(
F_{11}^{\rm Spin}=F_{11}^{\rm Spin, ev} \oplus F_{11}^{\rm Spin, odd}\;,
\)
i.e. into bordism groups of free circle actions of even and odd type,
respectively. The above discussion can be recast into cobordism language as 
follows (cf. \cite{Bo}). For circle actions of even type we have the
isomorphism
$F_{11}^{\rm Spin, ev} \cong  \Omega_{10}^{{\rm Spin}}(K(\Z,2))$,
mapping  
$\left[Y^{11}, \chi\right]$ to  $\left[ Y^{11}/S^1, f \right]$,
where $f: Y^{11}/S^1 \to K(\Z, 2)=\C P^{\infty}$ classifies the complex line bundle
$\mathcal{L}$ associated with $\pi$.
For circle actions of odd type we have the isomorphism 
$F_{11}^{\rm Spin, odd} \cong  \Omega_{10}^{{\rm Spin}^c}$,
mapping $\left[ Y^{11}, \chi \right]$ to $\left[ Y^{11}/S^1\right]$.
See section \ref{cobo} for details on Spin and Spin${}^c$ cobordism.



\subsection{Projectable Spin structures}
\label{proj}

A projectable Spin structure 
$\sigma_Y: P_{\rm Spin}(Y^{11}) \to P_{SO}(Y^{11})$ 
on $Y^{11}$ induces a Spin structure 
on $X^{10}$ as follows (see \cite{AB} \cite{Sz} for the general formalism). Let
the eleventh frame vector be $ve^{-\phi}$. 
The tangent bundle to the quotient 
which factorizes as $T(Y^{11})/S^1=T(Y^{11}/S^1) \oplus \mathcal{O}$  
is given as 
$T(Y^{11}/S^1)=P_{\rm Spin}(Y^{11})/S^1 \times_{{\rm Spin}(11)} \R^{11}$.
The ten-dimensional 
frame bundle $P_{SO}(X^{10})$ can be identified with 
the quotient $P_{SO(10)}(Y^{11})/S^1$ and 
$\sigma_Y^{-1} \left( P_{SO(10)}(Y^{11})/S^1\right)$
is a Spin$(10)$ bundle over $X^{10}$. Thus 
$\sigma_Y$ induces the corresponding Spin 
structure on $X^{10}$.

\vspace{3mm}
The above Spin structure, in fact, induces the Spin structure on $Y^{11}$ via $\pi: Y^{11} \to Y^{11}/S^1$; 
any Spin structure on $X^{10}$ canonically induces a projectable
Spin structure on $Y^{11}$ via pullback. Let
$\sigma_X: P_{\rm Spin}(X^{10}) \to P_{SO}(X^{10})$ be a Spin 
structure on $X^{10}$, and let $\rho_{10}: {\rm Spin}(10)\to SO(10)$ is the 
two-fold covering map. Then 
\(
\pi^* \sigma_X: \pi^* P_{\rm Spin}(X^{10}) \to \pi^* P_{SO}(X^{10})
:= P_{SO(10)}(Y^{11})
\)
is a $\rho$-equivariant map. A Spin structure on $Y^{11}$ is now
obtained by enlarging the structure group from Spin$(10)$ to 
Spin$(11)$
\(
\sigma_Y:=\pi^* \sigma_X \times_{\rho_{10}}\rho_{11}: 
\pi^* P_{\rm Spin}(X^{10}) 
\times_{{\rm Spin}(10)} {\rm Spin}(11)
\longrightarrow P_{SO(10)}(Y^{11})\times_{SO(10)} SO(11)\;.
\)
Thus we conclude that an even type free circle action leads to a Spin structure on 
the quotient ten-dimensional manifold which induces the original one on $Y^{11}$ in eleven dimensions
(cf. \cite{Bo}).

\vspace{3mm}
We can associate to the $S^1$ bundle $Y^{11}\to X^{10}$
the complex line bundle $L:=Y^{11}\times_{S^1}\C$
with the natural connection given by $i\omega$. 
Consider the twisted Spin bundle $SX^{10}\otimes L^{-k}$,
where $L^{-k}$ is the tensor power of $k$ copies of the line bundle
$L^{-1}$.
Note that there is an equality of spinor modules $\Delta_{10}=\Delta_{11}$.

\paragraph{The dilatino.}
The $S^1$-action on $P_{\rm Spin}(Y^{11})$ induces an action of 
$S^1$ on the associated Spin vector bundle 
$SY^{11}=P_{\rm Spin}(Y^{11})\times_{{\rm Spin}(11)} \Delta_{11}$
which we denote $\kappa$. The action $\kappa (e^{it})$ maps a
 spinor with base point $y$ to one with base point $y \cdot e^{it}$. 
The Lie derivative of a smooth spinor $\psi$ in the direction of the 
Killing field $v$ is 
\(
\mathsterling_v(\psi)(y)=\frac{d}{dt}|_{t=0}
\kappa (e^{-it}) (\psi (y\cdot  e^{it}))\;.
\)
Let $e_i$ be an orthonormal basis for the tangent  space. 
Any $r$-form $F_r$ acts on a spinor $\psi$ by
\begin{eqnarray}
\gamma (F_r) \psi&=& F_r(e_{i_1}, \cdots, e_{i_r})
\gamma (e_{i_1}) \cdots \gamma (e_{i_r}) \psi
\nonumber\\
&=& F_r(e_{i_1}, \cdots, e_{i_r}) \gamma (e_{i_1} \wedge 
\cdots \wedge e_{i_r})\psi\;.
\end{eqnarray}
The Christoffel symbols involving one 11-dimensional 
component are proportional to $\frac{1}{4}e^\phi \gamma (d\omega) \Psi$, 
so that the difference between the covariant derivative and the 
Lie derivative $\mathsterling_v \Psi$,
 with 
respect to the Killing vector field $v$ of the
spinor $\Psi$ is
\(
\nabla_v \Psi - \mathsterling_v \Psi= \frac{1}{4}e^{2\phi} \gamma (d\omega)
\Psi\;.
\label{kil psi}
\)

\paragraph{Fourier modes.} With $A$ a spinor corresponding to the principal 
Spin bundle and $\sigma_X$ a Spin structure, let 

\noindent 1. $[A, \sigma_X]$ denote the equivalence class of $(A,\sigma_X)$ in 
$SX^{10}=P_{\rm Spin}(X^{10})\times_{{\rm Spin}(10)} \Delta_{10}$. 

\noindent 2. $[\pi^*A, \sigma_X]$ denote the equivalence class of 
$(\pi^*A, \sigma_X)$ in 
$SY^{11}=\pi^* P_{\rm Spin}(X^{10}) \times_{{\rm Spin}(10)} \Delta_{10}$.

\noindent 3. $[y,1]$ denote the equivalence class of $(y,1)$ in $L=Y^{11}\times_{S^1}\C$.

\noindent Define a vector bundle map 
\begin{eqnarray}
\Pi_k: SY^{11} &\longrightarrow & SX^{10} \otimes L^{-k}
\nonumber\\
\left( y, [\pi^*A, \sigma_X] \right)
&\longmapsto &
\left( \pi( y), [A, \sigma_X] \otimes [y,1]^{-k}\right)\;.
\end{eqnarray}
This is a fiberwise vector space isomorphism preserving Clifford 
multiplication. 
Therefore, for any 
such ten-dimensional spinor coupled to powers of the M-theory 
line bundle 
$\psi: X^{10}\to SX^{10}\otimes L^{-k}$ 
we get an eleven-dimensional spinor $\Psi=Q_k(\psi)$,
with a homomorphism of Hilbert spaces 
$Q_k: L^2(SX^{10} \otimes L^{-k}) \to L^2(SY^{11})$
(injective since $\pi$ is surjective), 
such that we have a commutative diagram
\(
\xymatrix{
Y^{11} 
\ar[rr]^{Q_k (\psi)}
\ar[d]^\pi
&&
SY^{11}
\ar[d]^{\Pi_k}
\\
X^{10}
\ar[rr]^{\hspace{-5mm}\psi}
&&
SX^{10}\otimes L^{-k}\;. 
}
\) 
For any section $\Psi \in \Gamma (SY^{11})$, as in \cite{AB},
\(
\Psi \in {\rm Im}(Q_k) ~~~\Longleftrightarrow~~~
\mathsterling_v \Psi =ik \Psi\;.
\)
Since $\mathsterling_v$ is the differential of a representation 
of the Lie group $S^1=U(1)$ on $L^2(SY^{11})$, we get 
a decomposition 
$
L^2(SY^{11})=\bigoplus_{k\in \Z} V_k
$
into eigenspaces $V_k$ of $\mathsterling_v$ for the eigenvalue 
$ik$, $k\in \Z$. This decomposition is preserved by the Dirac operator
on $Y^{11}$ because the $S^1$-action commutes with that operator. 
So the image of $Q_k$ is precisely $V_k$ and the map $Q_k$ 
is an isometry up to the factor $e^\phi$. In fact, using \cite{AB},
there is a homothety of Hilbert spaces 
$Q_k: L^2(SX^{10}\otimes L^{-k}) \to V_k$ such that the 
horizontal covariant derivative is given by 
\(
\nabla_{\tilde{X}}Q_k(\Psi) = Q_k (\nabla_X\psi) + \frac{1}{4}e^{2\phi}
\gamma(e^{-\phi}v) \gamma(\tilde{V}_X) Q_k(\Psi) \;,
\)
where $\tilde{V}_X$ is the vector field on $X^{10}$ satisfying 
$d\omega (\tilde{X}, \cdot)=\langle \tilde{V}_X, \cdot\rangle$, and 
such that Clifford multiplication is preserved 
\(
Q_k(\gamma(X) \psi) = \gamma (\tilde{X}) Q_k(\Psi)\;.
\)
This can be viewed as dimensional reduction.

\paragraph{Relating spectra.}
Let $D$ be the Dirac operator on $SX^{10}$. Let $E \to X^{10}$ be a Hermitian 
vector bundle with a metric connection $\nabla^E$ and let 
$D^E$ be the twisted Dirac operator on $SX^{10}\otimes E$. 
This splits into three parts (see \cite{AB}). 

\begin{enumerate}
\item The {\it horizontal Dirac operator} is defined as the unique closed linear 
operator $D_h: L^2(SY^{11}) \to L^2(SY^{11})$ on each $V_k$ 
given by 
$
D_h:= Q_k \circ D \circ Q_k^{-1}
$,
where $D$ is the (twisted) Dirac operator on $SX^{10}\otimes L^{-k}$. 

\item The {\it vertical Dirac operator} is defined to be 
$
D_v:=\gamma(e^{-\phi} v)\gamma(d\omega)
$.

\item The {\it zeroth order term} 
$
D_0:= -\frac{1}{4} \gamma (e^{-\phi} v) \gamma (d\omega)
$.
\end{enumerate}
The Dirac operator $\widetilde{D}^\phi$ decomposes as
\(
\widetilde{D}^\phi = \sum_{i=0}^{10} \gamma(e_i) \nabla_{e_i}
= e^{-\phi} D_v + D_h + e^\phi D_0\;.
\label{D split}
\)
Let $\mu_1, \mu_2,\cdots$ be the eigenvalues of 
$D^E$. The eigenvalues of the twisted Dirac operator 
$\widetilde{D}^\phi$ for $\tilde{g}_Y^\phi$ on 
$SY^{11} \otimes \pi^* E \to Y^{11}$ depend
continuously on $e^\phi$ and such that for $e^\phi \to 0$ \cite{AB}:

\noindent 1. For any $j\in \N$ and $k\in \Z$, $e^\phi \cdot \lambda_{j,k}(e^\phi) \to k$.
In particular, $\lambda_{j,k}(e^\phi) \to \pm \infty$ if $k\neq 0$. 

\noindent 2. $\lambda_{j,0}(e^\phi) \to \mu_j$. The convergence of the eigenvalues
$\lambda_{j,0}(e^\phi)$ is uniform in $j$. 

\noindent We interpret this physically as the behavior of the eigenvalues in the 
weak coupling limit, since we view $e^\phi$ as essentially
the string coupling parameter. 

\paragraph{Example: Berger sphere.}
Consider the Hopf fibration $S^{11}\to \CP^5$. The Spin 
structure on $S^{11}$ is projectable since $\CP^5$ is Spin. 
The Dirac operator of $S^{11}$ with Berger metric includes the 
eigenvalues 
$
\lambda (\phi) =e^{-\phi} (m+3) + \frac{5}{2}e^\phi
$
with $m\in \N_0$, with multiplicity 
$\binom{m+5}{m}$. Note that 
$\lim_{e^\phi \to 0} e^\phi \cdot \lambda (\phi)=\pm(m+3)$. 
For the complete set of eigenvalues see \cite{AB}.

\paragraph{The set of Spin structures on $Y^{11}$.}
If $Y^{11}$ is an eleven-dimensional Spin manifold then $w_2(Y^{11})=0$. The 
Gysin sequence of $\pi$ gives that $w_2(X^{10})=0$ or $w_2(X^{10})= w_2(\pi)=c_1(\pi)$
mod 2. Note that, from the point of view of modes on the circle, this corresponds 
to even vs. odd modes, i.e. the Kaluza-Klein level $k$ being even or odd. 

\begin{enumerate}
\item The Spin structure $\sigma_Y$ on $Y^{11}$ is {\it equivariant} if and only if
$X$ is a Spin manifold with Spin structure $\sigma_X$ and $\sigma_Y=\pi^* \sigma_X$. 
Such a Spin structure does not extend to a Spin structure on the 
associated disk bundle $Z^{12}$. The induced ${\rm Spin}^c$-Dirac structure
\footnote{A Spin structure induces a ${\rm Spin}^c$-structure (cf. section \ref{sec spinc}).}
 is {\it strictly 
equivariant.}

\item If $\sigma_Y$ is not equivariant then there is a Spin structure $\sigma_Z$ on 
$Z^{12}$ with $\sigma_Y=\partial \sigma_Z$:
$\sigma_Y$ induces a Spin structure 
on $\pi^* P_{SO} (X^{10})$.
\end{enumerate}
The set Spin$(Y^{11})$ of isomorphism classes of Spin structures on $Y^{11}$, when 
$X$ and $\pi$ are both Spin, is given by: 
Spin$(Y^{11})=\pi^* {\rm Spin}(X^{10}) \bigcup \partial {\rm Spin} (Z^{12})$.

\subsection{Nonprojectable Spin structures}
\label{nonproj}
Now assume that the circle action 
$\chi$ is of odd type. Here Spin${}^c$ structures make an 
appearance (see section \ref{basic spin c} for basic definitions and 
examples).
We will continue to make use of the construction in \cite{AB}.
Let $\sigma_Y: P_{\rm Spin}(Y^{11}) \to P_{SO}(Y^{11})$ be a
nonprojectable Spin structure on $Y^{11}$. In this case, $X^{10}$ 
is not necessarily Spin. We can form a principal Spin$(10)$ bundle
$P:=\sigma_Y^{-1}\left(P_{SO(10)}(Y^{11} )\right)$ by restricting 
the principal $SO(11)$ bundle $P_{SO(11)}(Y^{11})$ to a principal
$SO(10)$ bundle $P_{SO(10)}(Y^{11})$. 
The action of $S^1\cong \R/2\pi \Z$ does not lift to $P$, but the double 
covering of $S^1$, i.e. $S^1\cong \R/4\pi \Z$ acts on $P$.
A free ${\rm Spin}^c(11)= {\rm Spin}(11)\times_{\Z_2} \hat{S}^1$
action on $P_{\rm Spin}(Y^{11})$ is induced from the free action 
$\hat{\chi}: P_{\rm Spin}(Y^{11}) \times ({\rm Spin}(11) \times \hat{S}^1) \to P_{\rm Spin}(Y^{11})$, 
defined by
$\hat{\chi} (\psi, (g, \lambda)) = \lambda^{-1} \psi g$, where 
$\psi \in P_{\rm Spin}(Y^{11})$, $g \in {\rm Spin}(11)$
and $\lambda \in \hat{S}^1$.  A ${\rm Spin}^c(11)$ bundle over $Y^{11}/S^1$ is the 
composition $P_{\rm Spin}(Y^{11}) \to Y^{11} \to Y^{11}/S^1$. 
Now we obtain a ${\rm Spin}^c$ structure on 
${\rm Fr}(Y^{11})/S^1$ and on $T(Y^{11}/S^1)$ as follows. Define the map 
$\tau:  P_{\rm Spin}(Y^{11}) \to {\rm Fr}(Y^{11})/S^1 
\times Y^{11}$ by $\tau (\psi)= ({\pi}_{\rm Fr}(u(\psi)), p(\psi))$, which is equivariant so that 
the map $P_{\rm Spin}(Y^{11}) \to Y^{11}/S^1$ is the composite map $\rho_F \tau$, where 
$\rho_F: {\rm Fr}(Y^{11})/S^1 \times Y^{11} \to Y^{11}/S^1$. Thus, for a circle action of odd type, a Spin structure on 
$Y^{11}$ gives a ${\rm Spin}^c$ structure on the quotient ten-dimensional 
manifold $Y^{11}/S^1$, with
$\pi$ is a principal $U(1)$ bundle, such that the principal ${\rm Spin}^c(11)$-bundle is
given by the composition of the principal ${\rm Spin}(11)$-bundle over $Y^{11}$ and $\pi$.

\vspace{3mm}
Define ${\rm Spin}^c(10)={\rm Spin}(10)\times_{\Z_2} S^1$ where 
$-1 \in \Z_2$ identifies $(-A,c)$ with $(A,-c)$. The Lie group
${\rm Spin}^c(10)$ acts on $\Delta_{10}$. The actions of 
${\rm Spin}(10)$ and $S^1$ on $P$ induce a free action of 
${\rm Spin}^c(10)$ on $P$, so that $P$ can be viewed as 
a principal ${\rm Spin}^c(10)$ bundle over $X^{10}$.
The associated vector bundle is $P\times_{{\rm Spin}^c(10)} \Delta_{10}$,
which is 
$SX^{10}\otimes L^{1/2}$ when $X^{10}$ is Spin.
When $X^{10}$ is not Spin, then neither factor 
exist, but the product $SX^{10}\otimes L^{1/2}$ exists. 
See section \ref{sec using} for more on how the M-theory 
circle is used to construct a Spin${}^c$ structure. 
This is one of the main motivations for 
 considering Spin${}^c$ structures in ten dimensions.

\vspace{3mm}
For the current case of nonprojectable Spin structure on $Y^{11}$, we get 
a splitting
\(
L^2(SY^{11})= \bigoplus_{k \in \Z +\frac{1}{2}} V_k
\) 
into eigenspaces $V_k$ for $\mathsterling_v$ corresponding to the 
eigenvalue $ik$. Therefore, taking the base ten-dimensional manifold
$X^{10}$ to be Spin${}^c$, while $Y^{11}$ is Spin, leads to 
half-integral Fourier modes for the spinors on $Y^{11}$. This seems to 
be a new phenomenon in the Fourier decomposition of spinor
modes in M-theory reduction on the circle and hence
complements the discussion in 
\cite{DMW}.

\paragraph{Relating spectra.}
Let the Spin structure on $Y^{11}$ be nonprojectable. The eigenvalues 
$\left( \lambda_{j,k}(\phi)\right)_{j\in \N, k \in \Z+\frac{1}{2}}$
of the 
twisted Dirac operator $D^\phi_Y$ 
depend continuously on $\phi$ and for $e^\phi \to 0$
\(
e^\phi \cdot \lambda_{j,k}(\phi) \to k \qquad {\rm for~all~} j=1,2, \cdots.
\)
In particular, $\lambda_{j,k} (\phi)\to \pm \infty$.
As in the case of projectable Spin structures in the last section, 
we interpret this physically as the behavior of the eigenvalues in the 
weak coupling limit.

\paragraph{Spin structures on $Y^{11}$ from Spin${}^c$ structures on $X^{10}$.}
Since $U(1)=SO(2)$, there is a natural map 
$SO(10) \times U(1) \to SO(12)$ which extends, via Whitney sum, to 
a map of bundles. Then we can define $P_{{\rm Spin}^c}(X^{10})$
as the pullback by this map of this covering map
\(
\xymatrix{
P_{{\rm Spin}^c}(X^{10}) 
\ar[d]
\ar[rr]
&&
P_{\rm Spin}({L})
\ar[d]
\\
SO(10)\times U(1) 
\ar[rr]
&&
SO(12)
}\;.
\)
Therefore, a Spin${}^c$ structure on $TX^{10}$ consists of a complex line 
bundle $L$ and a Spin structure on $TX^{10}\oplus L$. 
In fact, every Spin${}^c$ structure on $X^{10}$ induces a canonical 
Spin structure on $Y^{11}$. From \cite{Mor2},
by enlargement of the structure groups, the 
two-fold covering $\theta: P_{{\rm Spin}^c(10)}(X^{10}) \to P_{SO(10)}X^{10} \times P_{U(1)}(X^{10})$
gives a two-fold covering $\theta: P_{{\rm Spin}^c(11)}(X^{10}) \to P_{SO(11)}(X^{10})$ which,
by pullback through $\pi$ gives rise to a Spin${}^c$ structure on $Y^{11}$
\(
\xymatrix{
P_{{\rm Spin}^c(11)}(Y^{11})
\ar[rr]^\pi
\ar[d]^{\pi^*\theta}
&&
P_{{\rm Spin}^c(11)}(X^{10})
\ar[d]^\theta
\\
P_{SO(11)}(Y^{11}) \times P_{U(1)}(Y^{11})
\ar[d]
&&
P_{SO(11)}(X^{10}) \times P_{U(1)}(X^{10})
\ar[d]
\\
Y^{11} 
\ar[rr]^\pi 
&&
X^{10}
}\;.
\)
Since the pullback of a principal $G$-bundle with respect to its own projection 
map
is always trivial then this gives a Spin structure on $Y^{11}$.

\subsection{Dependence of the phase on the Spin structure}
\label{dep pha}

Consider the principal $U(1)$ bundle $\pi : Y^{11} \to X^{10}$. Fixing 
a basis vector of the Lie algebra $\frak{u}(1)$ trivializes the 
vertical tangent bundle $T^v \pi$.
Choose a Spin structure on $T^v\pi$ which restricts to the 
trivial (i.e. bounding) Spin structure on each fiber.
Since $T^v\pi$ is trivial, 
The Gysin sequence for the circle bundle gives
\(
\xymatrix{
0 
\ar[r]
&
H^1(X^{10};\Z)
\ar[r]
&
H^1(Y^{11};\Z)
\ar[r]^{\rm \hspace{-2mm} res} 
&
H^0(X^{10};\Z)
\ar[r]^{\hspace{-2mm} d_2}
&
H^2(X^{10};\Z)\;.
}
\label{gys}
\)
The map ${\rm res}: H^1(Y^{11};\Z) \to H^0(X^{10};\Z)
\cong \Z \cong H^1(S^1;\Z)$ is the restriction to the fiber.
Acting on the generator $1\in H^0(X^{10};\Z) \cong \Z$,
$d_2$ gives $d_2(1)=-c_1(Y^{11})$, so that the reduction 
modulo two of (\ref{gys}) gives
\(
\xymatrix{
0 
\ar[r]
&
H^1(X^{10};\Z_2)
\ar[r]
&
H^1(Y^{11};\Z_2)
\ar[rr]^{r_2({\rm res})} 
&&
H^0(X^{10};\Z_2)
\ar[rr]^{\hspace{-2mm}r_2(c_1(Y^{11}))}
&&
H^2(X^{10};\Z_2)\;.
}
\)
A Spin structure of $T^v \pi$ corresponding to 
$x \in H^1(Y^{11};\Z_2)$ restricts to the trivial 
Spin structure on the fibers iff the mod 2 reduction 
$r_2({\rm res})(x) \neq 0$. Since 
$H^0(X^{10};\Z_2) =\Z_2$, the condition is 
$r_2(c_1(Y^{11}))=0$. 
Therefore, the vertical tangent bundle $T^v \pi$ admits  a Spin structure 
which restricts to the trivial Spin structure on the fibers 
iff the reduction modulo 2 of the Euler class $c_1(Y^{11})$  vanishes.

\paragraph{The dependence of the (exponentiated) eta invariant on the Spin structure.}
Let $\sigma_Y$ be a Spin structure on $Y^{11}$. We will study the 
dependence of the phase of the M-theory partition function, via
the eta invariant, on the Spin structure. To indicate which Spin 
structure is being considered, we will use a label for the Spin bundle,
e.g. $S_{\sigma_Y}$ to denote the Spin bundle over $Y^{11}$ corresponding
to $\sigma_Y$. The eta invariant for the twisted 
Dirac operator $D^E$ acting on sections
of $S_{\sigma_Y} \otimes E$ with a given Spin structure $\sigma_Y$ is 
$\eta (\sigma_Y;E)$. As we will see, $E$ for us will be either an
$E_8$ bundle or $TY^{11}-3{\cal O}$. 
Since the set of Spin structures ${\rm Spin} (Y^{11})$  on 
$Y^{11}$ is a torsor over $H^{1}(Y^{11};\Z_2)$, a new Spin structure
$\sigma'_Y$ is obtained from an old one $\sigma_Y$ by acting with 
$H^1(Y^{11};\Z)$ on that set:
\begin{eqnarray}
{\rm Spin} (Y^{11}) \times H^1(Y^{11};\Z_2) &\longrightarrow & {\rm Spin} (Y^{11})
\nonumber\\
(\sigma_Y~,~ \delta)\hspace{1.55cm} &\longmapsto & \sigma_Y + \delta\;.\
\end{eqnarray}
We would like to compare the eta invariant corresponding to $\sigma_Y$, 
$\eta (\sigma_Y;E)$,
to that corresponding to $\sigma'_Y$, $\eta (\sigma_Y + \delta;E)$, i.e.
\(
\Delta \eta (\sigma_Y, \delta;E)= \eta (\sigma_Y + \delta;E) - 
\eta (\sigma_Y;E)\;,
\label{eta difference}
\)
which is a special case of Atiyah-Patodi-Singer (APS)
relative eta invariant \cite{APSII}.

\vspace{3mm}
The effect of $\delta$ will be seen via a line bundle, whose curvature will
essentially  
appear in the APS index formula. To illustrate our point it is enough to consider
the case when the complex line bundle $L_{\delta}$ is topologically trivial,
and hence admits a section. 
The section $\ell$ defines $\beta=\frac{1}{\pi i} \ell^{-1} d\ell$, which is a closed
real-valued one-form on $Y^{11}$. The Dirac operator $D'$ twisted by $\ell$ 
is related to $D$ via 
\(
D'= \ell^{-1} D \ell = D + \ell^{-1} {\rm grad}~ \ell\;.
\)
Both $D$ and $D'$ are twisted by $E$ to give $D^E$ and $D'^E$, respectively.  
It is shown in \cite{Dahl} that $D'$ acting on $\Gamma(S_{\sigma_Y}\otimes E)$ 
has the same spectrum as $D$ acting on $\Gamma(S_{\sigma'_Y} \otimes E)$, so that
the difference of their $\eta$-invariants 
\(
\eta_{D'^E} -\eta_{D^E}= \eta (\sigma_Y + \delta;E) - \eta (\sigma;E)=\Delta \eta 
(\sigma_Y, \delta;E)\;,
\label{d d'}
\)
 the difference between the $\eta$-invariants corresponding to 
 different Spin structures. The $\eta$-invariant part of the APS index
 theorem in this case will involve exactly the above combination, as 
 we see in what follows. 
 
 \vspace{3mm}
Let $Z^{12}=Y^{11} \times [0,1]$. The Spin structure on $Y^{11}$ induces a Spin structure on 
$Z^{12}$ with associated Spin bundle $S^Z:=S_{\sigma_Y} \oplus S_{\sigma_Y}$.
The Spin structure line bundle $L_{\delta}$ gives rise to a 
trivial complex line bundle $L_{tr}$ with connection $\nabla^L$ 
over $Z^{12}$. The main goal will be 
to identify the contribution to the index from this line bundle. 
Near the boundary components of $Z^{12}$, the 
corresponding Dirac operator $\overline{D}^E$ acting on sections of 
$S^Z \otimes E \otimes L_{tr}$ will have the form
\begin{eqnarray}
\overline{D}^E=\frac{\partial}{\partial t} + D^E \qquad {\rm near} \quad Y^{11} \times \{0\}\;,
\nonumber\\
\overline{D}^E=\frac{\partial}{\partial t} + D'^E \qquad {\rm near} \quad Y^{11} \times \{1\}\;.
\end{eqnarray}
The Atiyah-Patodi-Singer index theorem in this case is \cite{Dahl}
\(
{\rm Ind}(\overline{D}^E)=\int_{Z^{12}} \widehat{A}(Y^{11}) \wedge {\rm ch}(\nabla^E)
\wedge {\rm ch} (\nabla^L) -
\eta_{D^E} + \eta_{D'^E}\;.
\label{aps int}
\)
The eta factors, by the fact that $D'^E$ acting on $\Gamma (S_{\sigma_Y}\otimes E)$
is isospectral to $D^E$ acting on $\Gamma (S_{\sigma_Y+ \delta}\otimes E)$, are 
the differences (\ref{eta difference}), so that
\(
\Delta \eta (\sigma_Y, \delta;E)=
-{\rm Ind}(\overline{D}^E) +
\int_{Z^{12}} \widehat{A}(Y^{11}) \wedge {\rm ch}(\nabla^E)
\wedge {\rm ch} (\nabla^L)\;.
\label{delta integ}
\)
Now we relate the integral, involving a polynomial 
$I_{4m+2}$ in characteristic classes, over the manifold with boundary 
$Z^{12}$ to an integral over the manifold with no boundary $Y^{11}\times S^1$
as
\(
\int_{Z^{12}} I_{4m+2} \wedge {\rm ch}(\nabla^L)=-\frac{1}{2}
\int_{Y^{11}} I_{4m+2} \wedge \beta
=
-\frac{1}{2}
\int_{Y^{11} \times S^1} I_{4m+2} \wedge e^{dt \wedge \beta}\;.
\label{relate}
\)
For this we need two smooth scalar functions. The first is $\chi: I \to I$ given by 
$\chi (t)=0$ for $t \leq 1/3$ and $\chi(t)=1$ for $t \geq 2/3$,
with
${\rm ch}_1(\nabla^L)=-\frac{1}{2}d\chi \wedge \beta$. 
The second is $dt$, which measures the size of the circle.
Since $dt \wedge \beta \in H^2(Y^{11} \times S^1;\Z)$,
the exponential in (\ref{relate}) can be viewed as 
the Chern character of a complex line bundle 
$L_S$ on $Y^{11}\times S^1$.  Then the integral in 
(\ref{aps int}) is $-\frac{1}{2} {\rm Ind}(D^{E \otimes L_S})$.
Therefore, from (\ref{delta integ}), we get 
\(
\Delta \eta (\sigma_Y, \delta;E)=
-{\rm Ind}(\overline{D}^E) 
-\frac{1}{2} {\rm Ind}(D^{E \otimes L_S}) \quad \in ~\frac{1}{2} \Z\;.
\label{del eta d}
\)
The above analysis for the 
$E_8$ bundle can be repeated for the 
Rarita-Schwinger bundle  $RS=TY^{11}- 3 \mathcal{O}$,
giving 
\(
\Delta \eta (\sigma_Y, \delta;RS)=
-{\rm Ind}(\overline{D}^{RS}) 
-\frac{1}{2} {\rm Ind}(D^{RS \otimes L_S}) \quad \in ~\frac{1}{2} \Z\;.
\label{var et}
\)
We are now ready to look at the effect on the partition function. 
Using the notation in this section and restoring the Spin structure labels,
we have that the phase 
\eqref{Phi}
of the partition function 
varies with the change in Spin structure as
\begin{eqnarray}
\Phi_{\sigma_Y + \delta} &=& 
\Phi_{\sigma_Y}
\exp \left[ 2\pi i \left( \frac{1}{2} \Delta \overline{\eta}
(\sigma_Y +\delta;E)
+ 
 \frac{1}{4} \Delta \overline{\eta}
 (\sigma_Y +\delta;RS)
\right)\right]
\nonumber\\
&=&
\Phi_{\sigma_Y}
\exp \left[
2\pi i\left( \frac{1}{4}(\Delta h_{E_8}+ \Delta \eta (\sigma_Y, \delta; E))
+\frac{1}{8}(\Delta h_{RS}+ \Delta \eta (\sigma_Y, \delta; RS)\right)\right]\;,
\end{eqnarray}
We have two variations, one from the zero modes $\Delta h$ and one 
from the (unreduced) $\eta$-invariants. We have seen in section 
\ref{ana eff} that the spectrum and the number of zero modes of the 
Dirac operator depend crucially on the Spin structure chosen. 
Hence, we do not expect $\Delta h$ to be zero, nor to be 
a multiple of 4 or 8. On the other hand, we have seen above 
in \eqref{var et} that the variation of the $\eta$-invariant $\Delta \eta$
 is in general
only half-integral and does not necessarily have divisors. 
Hence, in general, the contribution to the phase from the variation 
of the Spin structure is not unity and therefore
we view such possible multi-valuedness as 
an anomaly. We call this anomaly a {\it Spin structure anomaly.}

\paragraph{Special case: flat connections.}
Let us consider the case when the bundle, say $E$, admits a flat connection
$\nabla^E$. In this case, 
\begin{eqnarray}
\Delta \eta (\sigma_Y, \eta;E)
& =& - {\rm Ind}(\overline{D}^E) -
\frac{1}{2} \int_{Y^{11}} \widehat{A} (Y^{11}) \wedge 
 {\rm ch}(\nabla^E) \wedge \beta
\nonumber\\
&=&
- {\rm Ind}(\overline{D}^E) -
\frac{1}{2} \int_{Y^{11}} \widehat{A} (Y^{11}) 
\wedge \beta
\nonumber\\
&=& -{\rm Ind}(\overline{D}^E) \quad \in ~\Z \;,
\end{eqnarray}
where we have used the fact that the integrand in the middle
line is of degree $4m+1$, $m\geq 0$, which cannot match
11, thus giving value zero for the integral.  
The situation in the flat case is a bit better than the general case. 
However, it is not enough to ensure the absence of potential 
anomalies in all such situations. In dimension twelve the
$\widehat{A}$-genus of a Spin manifold is even (see section
\ref{mod 2 ind}) so at least for trivial $E$ the index will be in 
$2\Z$, but that would still lead to a discrepancy in the phase
 of  $1/2$ and $\frac{1}{4}$ for $E$ and $TY^{11}-{\cal O}$, 
 respectively. We will need further divisibility of the 
 index. While this is not true in general, it is certainly 
 not unattainable in special cases.

\paragraph{Example: Flat manifolds with holonomy groups of finite prime order.}
Consider flat manifolds $Y^{11}$ whose holonomy 
groups have prime order. Let $e_1, \cdots, e_{11}$
be a basis of $\R^{11}$. 
Consider the map $J(x)=A(x) + \frac{1}{11}e_{11}$,
with $A: \R^{11} \to \R^{11}$ given by 
\(
A(e_j)=
\left\{
\begin{array}{ll}
e_{j+1} & {\rm for~~~} j\leq 9,
\\
-(e_1 + \cdots + e_{10}) & {\rm for~~~} j=10,
\\
e_{11} & {\rm for~~~} j=11.
\end{array}
\right.
\label{flat et}
\)
Then $Y^{11}=\R^{11}/\Gamma$ with 
$\Gamma=\langle e_1, \cdots, e_{10}, J \rangle$. 
The linear part $A$ of $J$ has 
two lifts $s_\pm \in {\rm Spin}(11)$ such that 
$(s_+)^{11}={\rm id}$ and 
$(s_{-})^{11}=-{\rm id}$. This defines two 
Spin structures $\sigma_{\pm}$ on $Y^{11}$.
For every $\epsilon =(\epsilon_1, \cdots, \epsilon_5) \in \Z_2^5$, consider
$\mu_\epsilon=\sum_{j=1}^5 j\epsilon_j$ and consider elements 
of product 1, i.e. elements such that $\nu=\epsilon_1 \cdots \epsilon_5=1$, and let
$\mathcal{C}_+=\{ \epsilon \in \Z_2^5 ~:~ \nu (\epsilon)=1\}$.
Following \cite{SSz}, the values of the eta invariants corresponding to 
$s_+$ and $s_{-}$, respectively are
\(
\eta_{Y^{11}, \sigma_+} = \sum_{r=1}^{10} A_r^+ \left( 1- \frac{2r}{11}\right)\;,
\qquad \qquad
\eta_{Y^{11}, \sigma_{-}} = \sum_{r=0}^{10} A_r^- \left( 1- \frac{2r+1}{11}\right)\;,
\)
where $A_r^\pm \in 2\Z$ is twice the number of elements 
in $\mathcal{C}_+$ which satisfy 
\(
\frac{\mu_\epsilon +11}{2} \equiv r ~~{\rm mod~} 11\;,
\qquad \qquad \qquad
\frac{\mu_\epsilon +21}{2} \equiv r ~~{\rm mod~} 11\;,
\)
for $\sigma_+$ and $\sigma_{-}$, respectively, for $r\in \{1, 2, \cdots, 10\}$. 
The above eta invariants (\ref{flat et}) are integral 
\cite{SSz}. This can 
be seen as follows. 
From \cite{Sto}, $2^s$ copies of $Y^{11}$, for some $s\in \mathbb{N}$,
 is a boundary of a Spin twelve-manifold $Z^{12}$. From the APS index
 theorem \cite{APSI} $\int_{Z^{12}}\widehat{A}(Z^{12})- \frac{1}{2}2^s \eta_{Y^{11}} \in \Z$.
From \cite{Hir}, the integral can be written as 
$\frac{1}{q_1 \cdots q_r}C_{Z^{12}}$, where $C_{Z^{12}} \in \Z$ and 
$q_1, \cdots q_r \in \{2,3,\cdots, 10\}$ are primes numbers.
Then from (\ref{flat et}), $\eta_{Y^{11}}=\frac{1}{11}C_{Y^{11}}$ for some 
$C_{Y^{11}} \in \Z$. Since $\frac{1}{q_1\cdots q_r}C_{Z^{12}} -2^{s-1}\eta_{Y^{11}}\in \Z$ then 
$\frac{2^{s-1}q_1 \cdots q_r}{11}C_{Y^{11}}\in \Z$, which implies that 
$\eta_{Y^{11}}\in \Z$. 
Now the difference of eta invariants 
$\Delta \eta_{Y^{11}}=\eta_{Y^{11},\sigma^+}- \eta_{Y^{11},\sigma^{-}}
$ takes values in $\frac{1}{2}\Z$, by (\ref{del eta d}).
Now $A_r^\pm \in 2\Z$ so that $\Delta \eta_{Y^{11}}=(\frac{2}{c})(11)$ for 
some $c\in \Z$. Summing up, gives $\frac{11}{2}\Delta \eta_{Y^{11}}\in \Z$, so that
$\Delta \eta_{Y^{11}} \in 2\Z$. 

\vspace{3mm}
The dimension of the space of harmonic spinors can also be obtained from 
the above data \cite{SSz}. For $\dim M=2k+1$, 
\begin{eqnarray}
h(M, \sigma^+)&=& 2 \# \left\{ \epsilon \in \mathcal{C}_+ ~|~ \frac{\mu_\epsilon}{2}
+ c(k) n \equiv 0 {\rm ~mod~} 2k+1\right\}\;,
\nonumber\\
h(M, \sigma^-)&=&0\;. 
\end{eqnarray}
We work out examples in three, seven and eleven dimensions. The first two
(already considered in \cite{SSz}) are relevant for considering them as
fibers of bundles of dimension eleven. We work out the latter case, which is the
most important for our discussion.

\vspace{3mm}
\noindent {\it 1. Three dimensions:} In the case of a Spin three-manifold $M^3$, we have 
$\mu_\epsilon$ and $\nu (\epsilon)$ both equal to 1 for the only element 
$\epsilon$. From $\frac{\mu_\epsilon}{2}+ \frac{1}{2}\dim M^3= 2$ we get that 
the only nonzero component of $A_r^+$ is $A_2^+=2$. 
Then, 
\(
\eta_{M^3, \sigma^+}=\sum_{r=1}^2A_r^+(1-\frac{2r}{3})= -\frac{2}{3}\;.
\)
On the other hand, from $\frac{\mu_\epsilon}{2} + \frac{1}{2}\dim M^3 + 1=3$
we get that the only nonzero component of $A_r^{-}$ is 
$A_0^{-}=2$. Then 
\(
\eta_{M^3, \sigma^-}=\sum_{r=0}^2A_r^-(1-\frac{2r+1}{3})= \frac{4}{3}\;.
\)
Note that here that $\eta_{M^3, \sigma^\pm}$ are not integral. However, 
the difference 
$
\Delta \eta_{M^3}=\eta_{M^3, \sigma^+} - \eta_{M^3, \sigma^-}=-2 ~\in 2\Z$.
Here there are no harmonic spinors, i.e. $h(M^3, \sigma^\pm)=0$.

\vspace{3mm}
\noindent {\it 2. Seven dimensions:}
In order to consider elements in $\mathcal{C}_+$, we need an even number of 
$-1$'s in $\epsilon$. For $\sigma^+$ we have the relation 
$r \equiv \frac{\mu_\epsilon}{2}$ mod 7, while for $\sigma^-$ we 
have $r'\equiv \frac{\mu_\epsilon}{2} +3$ mod 7. 
The values of the parameters are summarized in this 
table
\(
\begin{tabular}{|c||r|c|c|c|}
\hline
$\epsilon$ & $\frac{\mu_\epsilon}{2}$ & $r$ & $\frac{\mu_\epsilon}{2} +3$ & $r'$
\\
\hline
$(1~,1~,1)$& 3 & 3 & 6 & 6 
\\
\hline
$(1,-1,-1)$ & -2 & 5 & 1 & 1
\\
\hline
$(-1,1,-1)$ & -1 & 6& 2 & 2
\\
\hline
$(-1,-1,1)$ & 0 & 0 & 3 & 3
\\
\hline
\end{tabular}
\)
We see that $A_j^+=2$ for $j=0,3,5,6$ and $A_j^+=0$ for all other values of $j$. 
Also, $A_j^{-}=2$ for $j=1,2,3,6$ and $A_j^+=0$ for all other values of $j$. 
From this we get
\(
\eta_{M^7, \sigma^+}= -2\;, \qquad \qquad 
\eta_{M^7, \sigma^{-}}=0\;.
\)
This gives $\Delta \eta_{M^7}= \eta_{M^7, \sigma^+} -
\eta_{M^7, \sigma^{-}}
=-2 \in 2\Z$. The last row in the above table gives that the dimension of 
the space of harmonic spinors corresponding to $\sigma^+$ is 
$h(M^7, \sigma^+)=2\cdot 1=2$. 

\vspace{3mm}
\noindent  {\it 3. Eleven dimensions:}
Again, in order to get elements in $\mathcal{C}_+$, we need 
$\epsilon$'s with an even number (i.e. 0, 2, or 4) of minuses. 
The parameters then are encoded in this table
$$
\begin{tabular}{|c||r|c|c|c|c|}
\hline
$\epsilon$ & $\frac{\mu_\epsilon}{2}$ & $\frac{\mu_\epsilon+11}{2}$ & $\frac{\mu_\epsilon+11}{2}$ mod $11$ 
& $\frac{\mu_\epsilon+ 21}{2}$ & $\frac{\mu_\epsilon +21}{2}$ mod $11$
\\
\hline
$(1,~1,~1,~1,~1)$ & ${\frac{15}{2}}$ & 13 & 2 & 18 & 7
\\
\hline
$(-1, -1, ~1,~1,1)$ & $\frac{9}{2}$ & 10 & 10 & 15& 4
\\
\hline
$(-1, ~1, -1,~1,~1)$ & $\frac{7}{2}$ & 9 & 9 &14 & 3
\\
\hline
$(-1, ~1,~1, -1, ~1)$ & $\frac{5}{2}$ & 8 & 8 & 13 & 2
\\
\hline
$(-1,~1,~1,~1,-1)$ & $\frac{3}{2}$ & 7 & 7 & 12 & 1
\\
\hline
$(1, -1, -1,~ 1,~1)$ & $\frac{5}{2}$ & 8 & 8 & 13 & 2
\\
\hline
$(1,-1,~1,-1,~1)$ & $\frac{3}{2}$ & 7 & 7 & 12 & 1
\\
\hline
$(1,-1,~1,~1,-1)$ & $\frac{1}{2}$ & 6 & 6 & 11 &0
\\
\hline
$(1,~1,-1,-1,~1)$ & $\frac{1}{2}$ & 4 & 4 & 9 & 9
\\
\hline
$(1,~1,-1,~1,-1)$ & $-\frac{1}{2}$ & 5 & 5 & 10 & 10
\\
\hline 
$(1,~1,~1,-1,-1)$ & $-\frac{3}{2}$ & 4 & 4 & 9& 9
\\
\hline
$(-1,-1,-1,-1,1)$ & $-\frac{5}{2}$ &  3 & 3&  8& 8
\\
\hline
$(1,-1,-1,-1,-1)$ & $-\frac{13}{2}$& -1& 10 &4 &4 
\\
\hline 
$(-1,1,-1,-1,-1)$ & $-\frac{11}{2}$ & 0& 0& 5& 5
\\
\hline
$(-1,-1,1,-1,-1)$ & $-\frac{9}{2}$ & 1& 1 & 6 &6
\\
\hline
$(-1,-1,-1,1,-1)$ & $-\frac{7}{2}$ & 2 & 2 & 7 & 7
\\
\hline 
\end{tabular}
$$
This gives the following values for $A_i^+$:
2, 2, 4, 2,  2, 2, 4, 4, 4 2, 4 for $i=0,1, \cdots 10$. It also 
gives 
the following values for $A_i^-$:
4, 4, 4, 2, 4, 2, 2, 4, 2, 2, 2.
Then a straightforward computation shows that
the values of the eta invariants are 
\(
\eta_{Y^{11},\sigma^+}=-2\;, \qquad \qquad 
\eta_{Y^{11},\sigma^-}=4\;.
\)
The difference between the two eta invariants is
$\Delta \eta_{Y^{11}}=
\eta_{Y^{11},\sigma^+}-
\eta_{Y^{11},\sigma^-}=-6$, which is indeed an element 
of $2\Z$. Here the space of harmonic spinors corresponding to 
$\sigma^+$ can be read off from the third row from the bottom of the
table, which gives 
$h(Y^{11}, \sigma^+)=2\cdot 1=2$. 

\vspace{3mm}
More examples can be 
straightforwardly
constructed  
 using the properties of the $\eta$-invariant and 
the $\widehat{A}$-genus under products and bundles (see section \ref{a genu}).

\paragraph{Exponentiated eta invariants.}
When dealing with phases and partition functions, it is 
more appropriate to use exponentiated eta invariants
such as in \cite{DF}. This is applied in 
 \cite{integrand} for the case of the M-theory 
 effective action, where it is called
``setting the quantum integrand". Since we are
dealing with adiabatic limits of the eta invariants, and 
also with eta-forms, we found it more transparent 
to use the `unexponentiated' form. Of course, in the end, 
we have exponentiated the results that
we got in order to study the phase of the partition 
function. Therefore, our results hold at the level of 
exponentiated effective actions.

\section{ M-Theory on Eleven-Dimensional ${\rm Spin}^c$ Manifolds} 
\label{sec spinc}

\subsection{Basic definitions and properties of Spin${}^c$ structures} 
\label{basic spin c}
We recall some facts about ${\rm Spin}^c$ 
structures, mainly following \cite{ABS} \cite{GGK} (Appendix D) and \cite{LM} (Appendix D).
The Spin group ${\rm Spin}(n)$ is the connected double cover of the orthogonal 
group $SO(n)$. Form the product ${\rm Spin}(n) \times U(1)$.
The group ${\rm Spin}^c(n)$ is defined to be the quotient of the above product  
by the group $\Z_2$ composed of the elements $(\epsilon, -1)$, where 
$\epsilon$ is the non-trivial element of the kernel of the double covering map
$\rho_{{\rm spin}}: {\rm Spin}(n) \to SO(n)$. This means that elements of 
${\rm Spin}^c(n)$ are equivalence classes $[s, \gamma]$, where 
$g \in {\rm Spin}(n)$ and $\lambda \in U(1)$, so that 
$[g \epsilon, \lambda]=[g, - \lambda]$. There are two natural homomorphisms
coming from the projection to the two factors, i.e.
$
\rho^c : {\rm Spin}^c(n) \longrightarrow  SO(n)
$
sending  
$\left[g, \lambda \right]$ to  $\rho_{{\rm spin}}(g)$
and
$
{\rm det}: {\rm Spin}^c(n) \longrightarrow  U(1)
$
taking
$\left[g, \lambda \right]$ to $\lambda^2$
which give rise to the short exact sequence
\(
\xymatrix{
1 
\ar[r] 
&
~U(1)~  
\ar[r]
 &
 {\rm Spin}^c(n) 
\ar[r]^{~~~\rho^c} 
&
SO(n)
\ar[r]
&
1
}
\label{U(1)}\;,
\)
\(
\xymatrix{
1 
\ar[r]
&
{\rm Spin}(n)  
\ar[r]
 &
 {\rm Spin}^c(n) 
\ar[r]^{~~{\rm det}}
&
U(1)
\ar[r]
&
1\;.
}
\label{det}
\)
Let $M$ be a Riemannian manifold and $\zeta$ a real oriented vector bundle over $M$ 
of rank $n$ with oriented orthonormal frame bundle $\zeta : P_{SO} (\zeta) \to M$. Let 
$ \rho: {\rm Spin}(n) \to SO(n)$ be the nontrivial double covering, as above,
and define
the induced representation $\rho^c: {\rm Spin}^c (n):= {\rm Spin}(n) \times_{\Z_2} U(1)
\to SO(n)$, 
which is a nontrivial principal $U(1)$ bundle over $SO(n)$. Thus a ${\rm Spin}^c$ structure
on $\zeta$ is a principal $U(1)$ bundle $\sigma^c: P_{{\rm Spin}^c} (\zeta) \to  P_{SO}(\zeta)$, 
whose restriction to any fiber of $P_{SO}(\zeta)$ is the canonical principal $U(1)$
bundle $\rho^c$. This $\sigma^c$ is called a ${\rm Spin}^c$-structure on $X^{10}$. 
The canonical $U(1)$ bundle of $\sigma^c$ is 
$
\xi (\sigma^c) : P_{U(1)} (\sigma^c) : = P_{{\rm Spin}^c} (\zeta) \times_{{\rm Spin}^c(n)} U(1) \to M 
$,
so that the following diagram is commutative
\(
\xymatrix{
P_{SO}(\zeta) 
\ar[d]_{\zeta}
&
P_{SO}(\zeta) \times_M P_{U(1)} (\sigma^c) 
\ar[l]_{\hspace{-1cm} {\tilde{\xi}}}
\ar[d]
&
P_{{\rm Spin}^c}(\zeta)
\ar[l]_{\hspace{1cm} {\sigma^c}'}
\\
M
&
P_{U(1)}(\sigma^c)
\ar[l]_{\xi (\sigma^c)}
&
}\;,
\)
where ${\sigma^c}'$ is a twofold covering, $\sigma^c= \tilde{\xi} \circ {\sigma^c}'$, and the square is 
a pullback diagram. When $\zeta=TM$ then we talk about ${\rm Spin}^c$ structures 
on the manifold $M$ and write $P_{SO}(M)$.
When $M$ has a boundary then on 
$N = \partial M$ the boundary ${\rm Spin}^c$-structure $\partial \sigma^c$ is the 
restriction of $\sigma^c$ to $P_{SO} (N) \hookrightarrow P_{SO} (M)$.  Note 
that the reduction modulo 2 of the first Chern class $c_1 (\sigma^c) :=c_1( \xi (\sigma^c)) \in H^2 (M;\Z)$
is the second Stiefel-Whitney class $w_2( \zeta)$ and ${\rm Spin}^c$ structures on $\zeta$
exist if $w_2(\zeta)$ is the reduction modulo 2 of an integral class $c \in H^2(M; \Z)$.

\vspace{3mm}
Following \cite{BS1} \cite{BS2}, 
a ${\rm Spin}^c$-Dirac structure $(\sigma^c, \omega^{\sigma^c})$ on $M$ is a ${\rm Spin}^c$-structure
$\sigma^c$ together with a connection on $P_{{\rm Spin}^c}(M) \to M$ which is compatible with
the Levi-Civita connection.

\paragraph{The line bundles.}
There are two line bundles corresponding to the homomorphisms (\ref{U(1)}) 
and (\ref{det}). 
The {\it determinant line bundle} associated with the ${\rm Spin}^c$-structure
via the homomorphism (\ref{det}) is 
the complex line bundle
\(
{\sf L}= P \times_{\rm det} \C
\label{sf l}
\)
over $M$. 
As $P$ can be thought of as a circle bundle over $SO(E)$, the associated bundle
$
P \times_{U(1)} \C \to SO(E)
$
is a square root of the pullback to $SO(E)$ of the determinant  bundle
${\sf L}$.

\paragraph{Existence and obstruction.} Existence of Spin${}^c$ structures on a principal bundle $Q$
over a space $M$ can be characterized in several equivalent ways (see \cite{Fried})
\begin{enumerate}
\item $Q$ has a Spin${}^c$ structure;
\item there exists an $S^1$ bundle $P_1$ such that the 
Stiefel-Whitney class of the fiber product $w_2(Q{\times}_MP_1)=0$;
\item there exists an $S^1$ bundle $P_1$ such that $w_2(Q)\equiv c_1(P_1)$ mod 2;
\item there exists a cohomology class $z\in H^2(M;\Z)$ such that 
$w_2(Q)\equiv z$ mod 2. 
\end{enumerate}
Then $Q$ has a Spin${}^c$ structure if and only if the Stiefel-Whitney class 
$w_2(Q)\in H^2(M;\Z_2)$ is the $\Z_2$-reduction of an integral class
$z\in H^2(M;\Z)$. Therefore, if $H^2(M;\Z) \to H^2(M;\Z_2)$ is surjective then
every $SO(n)$-bundle over $M$ admits a Spin${}^c$ structure.

\subsection{Why ${\rm Spin}^c$-structures in eleven dimensions?}
\label{why c}

{\bf The gravitino.}
${\rm Spin}^c$ manifolds are relevant in M-theory and type II string theory. 
The gravitino (in eleven dimensions) a priori requires a Spin structure
since it is a section of a twisted Spin bundle. However, 
due to the coupling to the $C$-field, via its field strength $G_4$, the fermion 
can be charged under a $U(1)$ group, whose field strength is the abelian 
2-form obtained as a result of the dimensional reduction of $G_4$ on the 
torus. Thus, one can define the gravitino on a ${\rm Spin}^c$-manifold $M$
provided that one imposes the following consistency condition,
$w_2(TM)= 2Q c_1(F)$
and, since the second Stiefel-Whitney class $w_2(TM)$ is the mod 2 
reduction of an integral class, $c_1$ (the first Chern class of the corresponding 
circle bundle $F$), 
the fermion charge $Q$ must be half-integral 
\cite{DLP} \cite{BEM}. The main example is type IIA string theory on 
$AdS_5 \times \C P^2 \times S^1$ lifting to M-theory on $AdS_5 \times \C P^2 \times T^2$,
both spaces being ${\rm Spin}^c$ but not Spin. The fact that these spaces
have ${\rm Spin}^c$ structure plays an important role in the discussion of the supersymmetry of the 
Kaluza-Klein massive modes in \cite{LS}.

\paragraph{Parallel and Killing spinors.}  
Manifolds with ${\rm Spin}^c$ structure are also relevant in  supersymmetric 
compactification since they admit parallel and Killing spinors, and hence lead 
to unbroken supersymmetry. For the first, there is the following result \cite{Mor}.
A simply connected ${\rm Spin}^c$ manifold $M$ carries a parallel spinor 
if and only if it 
is isometric to the Riemannian product $M_1 \times M_2$ 
of a simply connected 
K\"ahler manifold $M_1$ and a simply connected Spin manifold 
$M_2$ carrying 
a parallel spinor, and the ${\rm Spin}^c$ structure of $M$ is the 
product of the canonical ${\rm Spin}^c$ structure of 
$M_1$ and the Spin structure of $M_2$. For real Killing spinors
one has \cite{Mor}

\begin{itemize}
\item The cone over ${\rm Spin}^c$ manifolds with real Killing spinors inherits 
 a canonical ${\rm Spin}^c$ structure such that the Killing spinor on the 
 base induces a parallel spinor on the cone. 
 
\item  Every Sasakian manifold $M^{2k+1}$ carries a canonical ${\rm Spin}^c$ 
 structure. If $M$ is Einstein then the auxiliary bundle of the canonical
 ${\rm Spin}^c$ structure is flat, so if in addition $M$ is simply connected
 then it is Spin. 
 
\item  The only simply connected ${\rm Spin}^c$ manifolds admitting 
 real Killing spinors other than the Spin manifolds are the 
 non-Einstein 
 \footnote{Note that a manifold does not necessarily have
 to be Einstein in order for it to satisfy the low-energy equations of string theory.}
 Sasakian manifolds $M^{2k+1}$. 
 \end{itemize}
 
For example, the 3-sphere $S^3$ admits two left-invariant 
non-Einstein Sasakian metrics with scalar curvature $\mathcal{R}=1\pm \sqrt{5}$
 admitting weak Killing spinors \cite{KF}.


\paragraph{Relating Spin${}^c$ on $Z^{12}$ and Spin${}^c$ on $Y^{11}$.}
Spin structures on the bounding manifold $Z^{12}$ have been considered at the end of 
section \ref{proj}. Now
start with $Z^{12}$ being a Spin${}^c$ manifold with boundary $Y^{11}$.
Since $TZ^{12}|_{Y^{11}}=TY^{11}\oplus \mathcal{O}$ and $\mathcal{O}$
obviously has a Spin${}^c$ structure, then we can induce a Spin${}^c$
structure on $Y^{11}$. 
Conversely, given $Y^{11}$ a Spin${}^c$ eleven-dimennsional manifold we can 
take $Z^{12}=[0,\infty)\times Y^{11}$ and induce a natural 
Spin${}^c$ structure on $Z^{12}$.

\subsection{Relevant examples of eleven-dimensional Spin${}^c$ manifolds}
\label{sec rel}


\subsubsection{Contact eleven-dimensional manifolds}
 In eleven (and in fact in all odd) dimensions, contact manifolds have a Spin${}^c$ structure essentially 
for the same reason as for the case of almost complex structures, namely that
the first obstruction is the third integral Stiefel-Whitney class $W_3$. 
Take the $C$-field to be of the form $C_3=\varpi \wedge d\varpi$, where 
$\varpi$ is a non-closed one-form on $Y^{11}$. Then the Chern-Simons 
integrand $C_3 \wedge G_4 \wedge G_4$ in the M-theory action would be 
$\varpi \wedge (d\varpi)^{\wedge 5}$, which is not zero (as it can be taken to be 
the volume form). Such a form 
$\varpi$ is called a contact form and the condition on the above eleven-form 
defines a contact structure on $Y^{11}$, i.e. the hyperplane bundle $H$ of
$TY^{11}$ given by $H=\ker \varpi$. See for instance \cite{Bl} for a more detailed 
description of contact structures. If we take the M-theory circle direction
to be along the first factor $\varpi$ in $C_3$, then the vector field $v$
evaluated on $\varpi$ is a constant, which we normalize to the value 1. 
Such a $v$ for which $\varpi ( v)=1$ is called the Reeb field. 
\footnote{It is obvious but implicit that the resulting $B$-field evaluated on $v$ is zero.}
Now a contact metric structure on $Y^{11}$ produces a reduction of the 
structure group $SO(11)$ of the tangent bundle $TY^{11}$ to the subgroup 
$U(5)$. Homotopy classes of such reductions, for the case of the sphere,
 are classified by (see \cite{Mas})
$
\pi_{11} \left(SO(11)/U(5) \right) \cong \Z
$.

\vspace{3mm}
The pair $(H, d\varpi_H)$ is a symplectic vector bundle, and we have an
almost  complex structure $J$ on $H$ compatible with $d\varpi_H$, so that
we get a Hermitian metric on $H$ given by 
$g_{\varpi, H}(X, Y)=d\varpi (X, JY)$. The trivial extension of the 
complex structure $J$ to $TY^{11}$, via $J v=0$, leads to an extension of the 
metric on $H$ to one on $TY^{11}$ called the {\it Webster metric} by setting 
\(
g_{\varpi}(X, Y)= d \varpi (X, JY) + \varpi (X) \varpi (Y)\;.
\)
Note that $g_{\varpi}(v, X)= \varpi (X)$ and $g_{\varpi}(JX, Y)=d \varpi (X, Y)$, for all
$X, Y \in TY^{11}$.

\vspace{3mm}
Let ${\cal V}_{\C}$ be the subbundle of $T^{\C}Y^{11}$ given by ${\cal V}_\C=\C v$.
Then we have a decomposition 
\(
T^\C Y^{11}= T^{1,0} Y^{11}\oplus T^{0,1} Y^{11} \oplus {\cal V}_\C\;,
\)
where $T^{1,0} Y^{11}$ (respectively $T^{0,1} Y^{11}$) is the subbundle given by the 
eigenspace of the complex extension $J$ on $H^{\C}$ to the eigenvalue $i$ (
respectively -$i$). There is an isomorphism 
\(
\Lambda^{11} T^{\C} Y^{11}= \bigoplus_{p+q+r=11} \Lambda^p T^{1,0} Y^{11}
\otimes \Lambda^q T^{0,1} Y^{11} \otimes \Lambda^r \langle v \rangle \;.
\)
Note that $\Lambda^r \langle v \rangle=0$ for $r >1$. Let us form the following bundles:
$\Lambda_H^{p,q} (Y^{11}):= \Lambda^p T^{1,0}(Y^{11})^* \otimes 
 \Lambda^q T^{0,1}(Y^{11})^*
 $
and then 
\(
 \Lambda^{p,0}(Y^{11}):= \Lambda_H^{p,0} (Y^{11})
 \otimes
 \Lambda_H^{p-1,0} (Y^{11}) \wedge \varpi\;.
\)
The top exterior power bundle
$
K:= \Lambda^{6,0} (Y^{11}) = \Lambda_H^{5,0}(Y^{11})
$
is called the canonical bundle. Its dual $K^{-1}$, called the anticanonical 
bundle, in fact, from \cite{Pe}, 
serves as the determinant line bundle for a Spin${}^c$ structure
on $Y^{11}$.  
This Spin${}^c$ structure is called the canonical Spin${}^c$ structure. Any
other Spin${}^c$ structure with determinant bundle $L$ will have an associated
principal Spin${}^c$ bundle that differs from the canonical principal Spin${}^c$ bundle
by tensoring with some $U(1)$ principal bundle. Now let us take this latter bundle to
be the M-theory circle bundle. The spinor bundle is then 
\(
SY^{11,c}= \Lambda_H^{0, *} (Y^{11}) \otimes {\L}\;, 
\)
where $\L$ is the complex line bundle corresponding to the M-theory circle bundle.
Then we have $\L ^2 = K \otimes L$. 
The Clifford multiplication by $i v$ induces the decomposition
\(
SY^{11,c}= S^+ Y^{11,c} \oplus S^- Y^{11,c}
\)
into eigenspaces $S^{\pm} Y^{11,c}$ of $iv$ with eigenvalue $\pm 1$. 
There is also the Clifford multiplication by $i d \varpi$ which induces the
decomposition
$
SY^{11,c}= \bigoplus_{q=0}^5 S_{5-2q} Y^{11,c}
$, 
where $S_k Y^{11, c}$ is the eigenspace of $i d\varpi$  
with eigenvalue $k$. From \cite{Pe}, we have isomorphisms
\begin{eqnarray}
S^{\pm} Y^{11,c} &\cong & \Lambda_H^{0, {\rm even}/{\rm odd}} (Y^{11}) \otimes \L
\\
S_{5-2q}Y^{11, c} &\cong & \Lambda_H^{0,q} (Y^{11}) \otimes \L\;.
\label{2nd cliff}
\end{eqnarray}

\paragraph{The $B$-field and the K\"ahler form.}
 In the above example, the resulting $B$-field will be $d \varpi$. So the Clifford multiplication 
leading to (\ref{2nd cliff}) is really multiplication by (the lift of ) 
$iB$ from the symplectic manifold whose tangent bundle is $H$. The 
correspondence is provided by the Boothby-Wang theorem (cf. \cite{Bl}):
every compact, regular contact manifold is a principal circle bundle over 
a symplectic manifold whose symplectic form determines an integral cocycle.
The appearance of $i$ multiplying $B$ is compatible with the way it appears
as part of the complexified K\"ahler form $\omega + iB$ in the K\"ahler case.

\paragraph{The Chern-Simons form.}
The condition $\varpi \wedge (d \varpi)^{\wedge 5}\neq 0$, from the beginning 
of this section, is independent of the 
choice of $\varpi$ and is a property of $v=\ker \varpi$. Any other 1-form defining 
the same hyperplane field must be of the form $\lambda \varpi$ for some smooth
function $\lambda : Y^{11} \to \R^{\times}$ so that
\(
(\lambda \varpi) \wedge (d (\lambda \varpi))^5= 
\lambda \varpi \wedge ( \lambda d \varpi + d\lambda \wedge \varpi)^5=
\lambda^{6} \varpi \wedge (d \varpi)^5 \neq 0\;,
\)
as this is proportional to the volume form. In this case we see that the sign of this volume
form depends only on $v$ and not on the choice of $\varpi$, so the contact structure 
$v$ induces a natural orientation of $Y^{11}$. We assume $Y^{11}$ to be oriented 
with a specific orientation so that we can have positive and negative contact structures
depending on the relative sign. With $C_3= \pm \varpi \wedge d \varpi$, and using $dC_3=G_4$,
 we get for the
Chern-Simons term $\frac{1}{6}C_3\wedge G_4 \wedge G_4=\pm \frac{1}{6} 
\varpi \wedge (d\varpi)^{\wedge 5}$, which is proportional to the volume form with 
a possible sign difference. If we take the standard volume form to be 
${\rm vol}_{\rm st}=\frac{1}{5!} \varpi \wedge (d\varpi)^{\wedge 5}$ then the Chern-Simons 
term is $\pm 20 {\rm vol}_{\rm st}$, a cosmological  `constant'.

\subsubsection{Spherical space forms}
\label{ssf}
The representation $\varrho$ from section  
\ref{sph sp}
can always be lifted 
to Spin${}^c(2k)$, so that $M(\Gamma,\varrho)$ is Spin${}^c$. 
Assume that there exists a fixed-point-free representation $\varrho$
 of a finite group $\Gamma$ in the unitary group $U(k)$, for $k\geq 2$. 
 Let $M=M(\varrho):=S^{2k-1}/\varrho(\Gamma)$ be the resulting quotient manifold. 
 The stable tangent bundle $TM \oplus 1$ is naturally isomorphic to 
 the underlying real vector bundle of the complex vector bundle 
 defined by $\varrho$ over $M$ and thus $TM$ admits a stable
 almost complex structure. 
 There exists a unique group homomorphism $f$ from 
 $U(k)$ to Spin${}^c(2k)$, so that the associated determinant 
 line bundle is det$(\varrho)$. This provides $M$ with a natural
 Spin${}^c$ structure. 
 This structure can be reduced to a Spin
 structure iff there exists a square root of the linear representation
 det$(\varrho)$. We will use the following in section 
\ref{fund eleven}.

 \begin{enumerate}
\item  If $|{\Gamma}|$ is odd then the square root of any linear representation exists.
 
\item If $|\Gamma|$ is even then the square root of det$(\varrho)$ exists iff $k$ is even. This
 implies that the dimension of $M$ is congruent to 3 mod 4. In this case, there
 are inequivalent Spin structures on $M$ and the choice of a Spin structure is 
 equivalent to a choice of a square root of det$(\varrho)$. 
 \end{enumerate}

\paragraph{Projective Spaces.}In addition to Spin manifolds being Spin${}^c$,
there are of course Spin${}^c$ manifolds that do not have a Spin structure. 
Let $L$ be the nontrivial flat line bundle over $\RP^m$ given by 
$L:=S^m \times \R/\sim$, with the identification $(\xi, x) \sim (-\xi, -x)$. Let 
$x:=w_1(L) \in H^1(\RP^m;\Z_2)$. The integral second cohomology group is 
$H^2(\RP^m;\Z)=x^2 \cdot \Z_2$. The bundle 
$rL:=L\oplus \cdots \oplus L$, $r$ times, has Stiefel-Whitney 
class
$w(rL)=(1+x)^r$, so that $w_1(rL)=rx$ and $w_2(rL)=\frac{1}{2}r(r-1)x^2$. 
We see that $rL$ admits a Spin${}^c$ structure for $r=4k+2$, since
for these values $w_2=x^2$ lifts to $H^2(\RP^2;\Z)$. Since 
$T(\RP^m)\oplus 1=(m+1)L$, we see that $\RP^m$ admits 
a Spin${}^c$ structure for $m=4k+1$. Therefore, in our range of dimensions
we have:
$\RP^1=S^1$, $\RP^5$ and $\RP^9$. We get eleven-dimensional 
Spin${}^c$ manifolds by taking products (or bundles) with the 
above projective spaces. The Stiefel-Whitney classes satisfy
\(
w_2(X \times Y):=w_2(TX \oplus TY)=
w_2(X) + w_2(Y) + w_1(X) w_1(Y)\;,
\)
so that for oriented manifolds, and with one of the 
factors being Spin and another being Spin${}^c$, the 
product will be Spin${}^c$. 
For the case of bundles, let
$
0 \to V_1 \to V_2 \to V_3 \to 1
$
be a short exact sequence of real vector bundles. Let $\{ i, j, k\}$
be a permutation of $\{ 1,2,3\}$. If $V_i$ and $V_j$ admit Spin${}^c$
structures 
 then $V_k$ admits a natural Spin${}^c$ structure and the 
 canonical line bundles corresponding to the vector bundles 
 satisfy
 $L(V_i)=L(V_j)\otimes L(V_k)$.

\paragraph{Example: Lens spaces (and their bounding spaces).}
Consider the lens spaces from section \ref{sph sp}.
Choose an integer $p$ and an $n$-tuple of integers 
$\vec{a}=(a_1, \cdots, a_n)$ coprime to $p$. 
The quotient of the $(2n-1)$-dimensional sphere $S^{2n-1}\subset \C^{2n}$ 
by the $\Z_p$ action on $\C^{2n}$ is given by 
\(
\zeta \cdot (z_j) = (\zeta^{a_j} z_j) \quad {\rm with~~} \zeta^p=1\;.
\)
The resulting quotient space is the $(2n-1)$-dimensional lens space $L^{2n-1}(p,\vec{a})$.
The lens space $L^{2n-1}(p,\vec{a})$ is diffeomorphic to 
the boundary of the $2n$-manifold
$Z^{2n}(p, \vec{a})$, which is the quotient of the $2n$-dimensional 
ball $\mathbb{B}^{2n}\subset \C^{2n}$ by the $\Z_p$ action. The space
$Z^{2n}(p, \vec{a})$ has a resulting isolated singularity at the origin 
$[(0, \cdots, 0)]$.

\subsection{Multiple Spin${}^c$ structures}
\label{mult c}
\paragraph{The set of Spin${}^c$ structures.}
The group $H^2(M;\Z)= {\rm Vect}_1^{\C}(M)= {\rm Prin}_{U(1)}(M)$ of isomorphism 
classes of principal $U(1)$ bundles over $M$ acts transitively and effectively on the set
${\rm Spin}^c(\zeta)$ of isomorphism classes of ${\rm Spin}^c$ structures on 
the bundle $\zeta$. 
This is analogous to the Spin case, considered in section \ref{multi spin}.

\vspace{3mm}
Given two Spin${}^c$ structures $\sigma^c_1$ and $\sigma^c_2$ we can define their 
difference as the unique line bundle $L$ via $\sigma^c_2=\sigma^c_1 \otimes L$, so that
the collection of Spin${}^c$ structures is (non-canonically) isomorphic to 
$H^2(M;\Z)$. This is an $H^2(M;\Z)$-torsor, i.e. an affine space modeled on
$H^2(M;\Z)$ in the sense that the difference between two Spin${}^c$ structures
is an element in $H^2(M;\Z)$ but there is no distinguished origin of this space.

\paragraph{Example: Lens space and its bounding space.}
\footnote{The following discussion can be given in more general terms; however, 
we are interested in the application to our specific dimensions.} 
Any Spin${}^c$ structure $\sigma^c$ on $Z^{12}(p, \vec{a})$ can be obtained by twisting 
the canoncial Spin${}^c$ structure $\sigma^c_{\rm can}$ by the unique line bundle $L$ on 
$Z^{12}(p, \vec{a})$.
Any topological line bundle $L$  
on $Z^{12}(p, \vec{a})$ can be obtained as the quotient 
of the trivial bundle $\mathbb{B}^{12}\times \C$ by
a $\Z_p$ action given by 
\(
\zeta_p^l\cdot (z, u)=(\zeta_p^l z, \varrho (\zeta_p)^l \cdot u)
\) 
for some $U(1)$ representation $\varrho: \Z_p \to U(1)$ of $\Z_p$. 
Now a Spin${}^c$ structure on a manifold $M$ induces one on 
$M \times \R$ and vice versa. 
Then
there is a one-to-one correspondence between the set of all 
Spin${}^c$ structures ${\rm Spin}^c(Z^{12}(p, \vec{a}))$ on $Z^{12}(p, \vec{a})$
and the set of representations
$R(\Z_p, U(1))\cong \Z_p$ from $\Z_p$ to $U(1)$, which we can see as follows \cite{Fu2}.
 The set ${\rm Spin}^c(L(p;\vec{a}))$ of all Spin${}^c$ structures on $L(p;\vec{a})$
is in one-to-one correspondence with the set Pic${}^t(L(p;\vec{a}))$ of all topological 
line bundles on $L(p;\vec{a})$. Similarly, the set 
 ${\rm Spin}^c(Z^{12}(p;\vec{a}))$ of all Spin${}^c$ structures on $Z^{12}(p;\vec{a})$
is in one-to-one correspondence with the set Pic${}^t(Z^{12}(p;\vec{a}))$ of all topological 
line bundles on $Z^{12}(p;\vec{a})$. 
The idea is to show that the map $i^* : {\rm Pic}^t(Z^{12}(p;\vec{a})) \to {\rm Pic}^t(L(p;\vec{a}))$,
induced from the inclusion $i : L(p;\vec{a}) \hookrightarrow Z^{12}(p;\vec{a})$, is bijective. 
Any line bundle in Pic${}^t(Z^{12}(p;\vec{a}))$ is by definition the quotient of $\mathbb{B}^{12}\times \C$
divided by some $\Z_p$-action with representation 
$\Z_p \to U(1)$. Hence Pic${}^t(Z^{12}(p;\vec{a})) \cong \Z_p$.
On the other hand, Pic${}^t(L(p;\vec{a})) \cong H^2(L(p;\vec{a}))\cong \Z_p$.
Now consider  any $L \in {\rm Pic }^t(L(p;\vec{a}))$ with connection $A$ and curvature 
$F_A$. 
The first Chern class $c_1(L)$ is torsion, i.e.
$c_1(L)\in H^2(L(p;\vec{a}))\cong \Z_p$. By Chern-Weil theory, the de Rham cohomology class
$c_1 (L)=[-F_A]$ vanishes, so that $F_A=da$ for some 1-form $a$ on $L(p;\vec{a})$. This gives
that $A- a$ is a flat connection. The holonomy $\varrho: \Z_p \to U(1)$
of this connection gives an isomorphism of $L$ with the quotient of $S^{12}\times \C$ by 
the $\Z_p$ action given by the holonomy $\varrho$. 
Then the line bundle $(\mathbb{B}^{12} \times \C)/\Z_p$ corresponds to $L$, thus 
establishing surjectivity of $i^*$. 
Since Pic${}^t(L(p;\vec{a}))$ and Pic${}^t(Z^{12}(p;\vec{a}))$ are isomorphic to 
$\Z_p$ and have the same order, the map $i^*$ is bijective. 
 Therefore, the set of all Spin${}^c$ structures on $L(p;\vec{a})$ can be parametrized by
\(
\Z_p\cong R(\Z_p, U(1)) \cong {\rm Spin}^c (Z^{12}(p;\vec{a})) \cong {\rm Spin}^c (L(p;\vec{a}))\;.
\)

\subsection{Spin and Spin${}^c$ cobordism and extension to twelve dimensions}
\label{cobo}

\paragraph{Spin${}^c$ characteristic classes.}
The map $B{\rm Spin}^c \to BSO \times BU(1)$ is an isomorphism on rational 
cohomology. One has Pontrjagin classes $p_i \in H^{4i}(B{\rm Spin}^c;\Q)$ and 
the class $c_1 \in H^2(B{\rm Spin}^c;\Q)$ of the canonical bundles so that 
one has \cite{Sto} the ring $H^*(B{\rm Spin}^c;\Q)=\Q [c_1, p_i, e]$. Every 
${\rm Spin}^c(n)$-bundle over a manifold is a pullback of the universal 
${\rm Spin}^c(n)$-bundle $E{\rm Spin}^c(n) \to B{\rm Spin}^c(n)$ 
through the map to the classifying space. The class $e$ corresponds to the 
Euler class of the bundle $E$, and $c_1$ corresponds to the first Chern 
class of the determinant line bundle ${\sf L}$ in (\ref{sf l}).

\paragraph{Cobordism.} 
Let $F$ denote either  Spin or  ${\rm Spin}^c$.
Consider two $11$-dimensional manifolds $Y^{11}_1$ and $Y^{11}_2$. 
A sum operation $Y^{11}_1+ Y^{11}_2$ is defined by disjoint union
and  $-Y^{11}$ is defined by reversing the orientation of $Y^{11}$ and by taking the appropriate 
$F$ structure.  We say that $Y^{11}_1$ is $F$-bordant to $Y^{11}_2$ if there exists a smooth compact manifold 
$Z^{12}$ so that $\partial Z^{12}=Y^{11}_1 - Y^{11}_2$, and so the $F$-structure on $\partial Z^{12}$ extends 
over $Z^{12}$. 
Bordism defines an equivalence relation. 
\begin{enumerate}
\item Since
$
\partial (Y^{11} \times I)= Y^{11} + (-Y^{11})
$,
$Y^{11}$ is bordant to itself.
\item Reversing orientation of $Z^{12}$ gives that $Y^{11}_1$ is bordant to $Y^{11}_2$, 
implying that $Y^{11}_2$ is bordant to $Y^{11}_1$. 
\item Let $\partial (Z^{12}_1)=Y^{11}_1 - Y^{11}_2$ and $\partial (Z^{12}_2)=Y^{11}_2 - Y^{11}_3$,
so that $Y^{11}_1$ is bordant to $Y^{11}_2$ and $Y^{11}_2$ is bordant to $=Y^{11}_3$. We can find a 
collar neighborhood of $Y^{11}_2$ in $Z^{12}_1$ and form $Y^{11}_2 \times (-\epsilon, 0]$ and a collar 
neighborhood of $Y^{11}_2$ in $Z^{12}_2$ of the form $Y^{11}_2 \times [0, \epsilon)$ for some 
$\epsilon >0$. Gluing $Z_1$ to $Z^{12}_2$ along $Y^{11}_2 \times \{0\}$ creates a smooth 
manifold $Z^{12}_3$ with a natural $F$ structure which provides the desired bordism 
from $Y^{11}_1$ to $Y^{11}_2$. 
\end{enumerate}
Let $\Omega_{11}^F$ be the set of equivalence 
classes. Disjoint union makes
$\Omega_{11}^F$ into an abelian group and 
$
\partial (Y^{11} \times I)=Y^{11} + (-Y^{11})$,
where $-Y^{11}$ is the inverse of $Y^{11}$. 
 Let $\Omega_*^F$ be the associated graded abelian group. This is a
ring under Cartesian product. 
The forgetful functor defines ring morphisms
\(
\Omega_*^{\rm Spin} \longrightarrow  \Omega_*^{{\rm Spin}^c}\;,
\qquad
\Omega_*^{\rm Spin} \longrightarrow  \Omega_*^{{SO}}\;,
\qquad
\Omega_*^{{\rm Spin}^c} \longrightarrow  \Omega_*^{{SO}}\;.
\)
 The characteristic classes define characteristic numbers which are
bordism invariants. 

\begin{enumerate}

\item {\it The $O$ characteristic numbers:} The Stiefel-Whitney classes $w_i \in H^i(-;\Z_2)$.

\item {\it The $SO$ characteristic numbers:} The Pontrjagin classes $p_i\in H^{4i}(-;\Q)$.

\item {\it The Spin characteristic numbers:} The $KO$-characteristic numbers
and the Stiefel-Whitney numbers.

\item {\it The Spin${}^c$ characteristic numbers:}
These are the Chern-Pontrjagin numbers and the Stiefel-Whitney numbers. Define the
Chern-Pontrjagin classes as follows. Let
$
{\rm det} : {\rm Spin}^c \longrightarrow U(1)
$
define a line bundle $L\in {\rm Vect}_{\C}(B{\rm Spin}^c)$. Then
$
H^*(B{\rm Spin}^c;\Z)/{\rm torsion}=\Z [c_1(L), p_i]
$.
There is a natural evaluation (see \cite{Gi})
\(
\mu : \Omega^{{\rm spin}^c} / {\rm torsion} \otimes \Z [c_1(L), p_i] 
\to \Z\;.
\)

\end{enumerate}



\vspace{3mm}
\noindent{\bf Some properties of Spin cobordism.} 

\noindent {\it 1}. Every Spin cobordism class is represented by a connected manifold. 

\noindent {\it 2}.  For $n \geq 3$ every Spin cobordism class is represented by a simply-connected manifold. 

\noindent {\it 3}. For $n \geq 5$ every Spin cobordism class is represented by a 2-connected manifold. 

\noindent {\it 4}. For $n\geq 1$, 
$\Omega_n^{\rm Spin} \cong \Omega_{n+8}^{\rm Spin}$, so there is a mod 8 periodicity 
in the Spin cobordism groups. 

\noindent {\it 5}. $\Omega_n^{\rm Spin}=0$ for $n=3,5,6,7$ mod 8. 

\noindent {\it 6}. For $n=0$: $\Omega_0^{\rm Spin} \cong \Z$. 

\noindent {\it 7}. For $n=1$: A Spin structure is defined to be a 2-fold covering of $P_{SO}(S^1)=S^1$. 
Then
$\Omega_1^{\rm Spin} \cong \Z_2$,
generated by $S^1$ with $2S^1=0$.
See section \ref{multi spin}.

\noindent {\it 8}. For $n=2$: $\Omega_2^{\rm spin}\cong \Z_2$, generated by $S^1 \times S^1$ with
$2(S^1 \times S^1)=0$. 

\noindent {\it 9}. For $n=4$: $\Omega_4^{\rm spin} \cong \Z$, generated by the Kummer surface
$V^2(4)$, which is the variety of degree 4 in $\C P^3$ with vanishing first Chern class.

\noindent {\it 10}. For $n=8$: $\Omega_8^{\rm spin} \cong \Z \oplus \Z$, generated by 
$\mathbb{H}P^2$ and a manifold $L^8$ such that $4L^8$ is Spin cobordant
to $V^2(4) \times V^2(4)$, the product of two Kummer surfaces. 

\noindent {\it 11}. To study Spin cobordism invariant quantities it is enough to check them
against the corresponding  generators, since any other Spin manifold of a given dimension
is `built' out of such generators.

\paragraph{${\rm Spin}^c$-structures on $Y^{11}$.}


Here we apply the construction in \cite{BS1} \cite{BS2} and continue the discussion from 
section \ref{nonproj}. In the non-equivariant case,
$\sigma_Y$ induces a Spin structure 
on $\pi^* P_{SO} (X^{10})$
 which gives an equivariant ${\rm Spin}^c$ structure
$\sigma_Y^{c}$ if we endow $\xi (\sigma_Y^{c}): Y^{11} \times S^1 \to Y^{11}$ with the 
diagonal action of $S^1$ on $Y^{11} \times S^1$. The canonical $U(1)$ bundle
of the quotient ${\rm Spin}^c$ structure $\sigma_X^{c}$ on $X^{10}$ of $\sigma_Y^{c}$
if $\xi (\sigma_X^{c})= \xi (\sigma_Y^{c})/S^1 = -\pi$, i.e. $\pi$ with the $U(1)$-action
reversed. Therefore $\sigma_X^{c}$ induces a ${\rm Spin}^c$ structure $\sigma_Z^{c}$
on $Z^{12}$ with $\xi (\sigma_Z^{c}) = \pi \otimes (-\pi)$ trivial. Given a connection $\omega^{\pi}$ 
on $\pi$ 
we also get an induced ${\rm Spin}^c$-Dirac structure $(\sigma_X^{c}, \omega^X)$  on 
$X^{10}$ with $\xi (\sigma_X^{c})=-\pi$ and connection $\omega_X= - \omega^{\pi}$. 
In this case $Z^{12}$ is a Spin manifold and we must have $w_2(X^{10})= c_1(\pi)$ mod 2. 
Thus the set ${\rm Spin}(Y^{11})$ of isomorphism classes of Spin 
structures on $Y^{11}$ is 
given by
\begin{enumerate}
\item $X^{10}$ is Spin and $\pi$ is not:  ${\rm Spin}({Y^{11}})=\pi^* {\rm Spin} (X^{10})$.
\item Both $X^{10}$ and $\pi$ are not Spin: ${\rm Spin}({Y^{11}})=\partial {\rm Spin} (Z^{12})$.
\end{enumerate}

\vspace{3mm}
We consider two ${\rm Spin}^c$ structures on a fixed equivariant ${\rm Spin}^c$ structure
$\sigma_Y^{c}$ on $Y^{11}$. Such ${\rm Spin}^c$ structures on $Y^{11}$ are obtained from
${\rm Spin}^c$ structures on $X^{10}$ and vice-versa. A ${\rm Spin}^c$ structure 
$\sigma_X^{c} : P_{{\rm Spin}^c} (X^{10}) \to P_{SO}(X^{10})$ on $X^{10}$ induces a 
${\rm Spin}^c$ structure $\sigma_Y^{c} : P_{{\rm Spin}^c} (Y^{11}) \to P_{SO}(Y^{11})$ on $Y^{11}$.
 The equivariant ${\rm Spin}^c$ structure $\sigma_Y^{c}$ extends to a 
${\rm Spin}^c$ structure $\sigma_Z^{c}$ on the disk bundle
$Z^{12}$ which is induced from the ${\rm Spin}^c$ structure
$\sigma_X^{c}$ and the canonical ${\rm Spin}^c$ structure
$\sigma_{\pi}^c : P_{{\rm Spin}^c} (\pi) = Y^{11} \times_{U(1)} {\rm Spin}^c (2) \to Y^{11}=P_{SO}(\pi)$
of the principal $S^1$ bundle $\pi$, 
$
\sigma_Z: 
P_{{\rm Spin}^c} (Z^{12})
\to
P_{SO}(Z^{12})$, where 
\begin{eqnarray}
P_{{\rm Spin}^c} (Z^{12})&=& \pi^* \left( P_{{\rm Spin}^c} (X^{10}) \times_{X^{10}}  P_{{\rm Spin}^c}(\pi)\right)
\times_{{\rm Spin}^c(10) \times {\rm Spin}^c(2)} {\rm Spin}^c(12)
\\ 
P_{SO}(Z^{12})&=&\pi^* \left( 
P_{SO}(X^{10}) \times_{X^{10}} P_{SO}(\pi)
\right)
\times_{SO(10) \times SO(2)} SO(12)\;.
\end{eqnarray}
Its canonical bundle is $\xi (\sigma_Z^{c})=\pi^*_D ( \xi (\sigma_X^{c}) \otimes \pi)$.
Putting $\omega_Z = \pi_D^* (\omega_X^{c} \otimes \omega^{\pi})$ we get an equivariant 
${\rm Spin}^c$-Dirac structure $(\sigma_Z^{c}, \omega_Z)$ on $Z^{12}$. Restriction to $Y^{11}$ 
of such ${\rm Spin}^c$-Dirac structures is {\it boundary $Spin^c$ structures}. They are
equivariant but not strictly equivariant.

\paragraph{Extension to twelve dimensions for ${\rm Spin}^c$}
We need to check that the ${\rm Spin}^c$ eleven-dimensional manifold extends to twelve 
dimensions. We also need to check that the corresponding $E_8$ bundle 
also extends. The former involves checking whether and when the 
cobordism group $\Omega_{11}^{{\rm Spin}^c}$ is zero, and the 
latter involves $\Omega_{11}^{{\rm Spin}^c}(K(\Z, 4))$, since in our 
range of dimensions, $E_8$ has the homotopy type of 
the Eilenberg-MacLane space $K(\Z, 3)$, whose classifying space is 
then of type $K(\Z, 4)$. 



\vspace{3mm}
Rationally \cite{Sto}, 
\(
\Omega_{11}^{{\rm Spin}^c} \otimes \Q \cong \Omega_{11}(K(\Z, 2)) \otimes \Q
\cong   \Omega_{11}^{\rm Spin}(K(\Z, 2)) \otimes \Q
\)

\subsection{The APS index in the Spin${}^c$ case}
\label{sec apsc}
The complex representations of Spin${}^c(n)$ are the
same as those of Spin$(n)$. So, when $n$ is even we
have a $\Z_2$-graded Spin bundle $S=S^+ \oplus S^{-}$. 
The grading is provided by the chirality operator 
$\Upsilon$, which is defined by the Clifford multiplication
$i^{n/2} e_1 \cdots e_n$, where $e_i$ constitute a
local orthonormal frame for $TZ^{12}$. 

\vspace{3mm}
The Spin${}^c$ Dirac  operator is defined in the same way
as the Spin Dirac operator, i.e. by composing the covariant 
derivative with Clifford multiplication 
\(
D^c= \sum_{i=1}^n e_i \cdot \nabla_{e_i}: \Gamma (Z^{12},S) \to 
\Gamma (Z^{12},S)\;,
\)
which decomposes as 
$\ttsmat{0~}{~D^{-}}{D^+}{0}\Big.$
since $D^c$ anticommutes with $\Upsilon$. Here 
$D^\pm$ are the restrictions of $D^c$ to 
$S^\pm$. Unitarity of  $\nabla$ implies that 
$D^\pm$ is the adjoint of $D^\mp$. 
The index of the Spin${}^c$ Dirac operator is 
$
{\rm Ind} (D^c)=\dim \ker D^+ - \dim \ker D^{-}
$. 
Given a Hermitian vector bundle $E$ over $Z^{12}$ 
with unitary connection
$\nabla^E$, the twisted Spin${}^c$ Dirac operator is 
$
D^c_E: \Gamma (Z^{12}, S \otimes E) \to
\Gamma (Z^{12}, S \otimes E)
$.
Assuming $Z^{12}$ has boundary $\partial Z^{12}=Y^{11}$, the restriction 
of the Spin${}^c$ Dirac operator to the boundary is 
\(
D_{Y,E}^c: \Gamma ((S\otimes E)|_{Y^{11}})
\to 
 \Gamma ((S\otimes E)|_{Y^{11}})\;,
\)
which is formally self-adjoint and elliptic. 
Imposing global APS boundary conditions and requiring product type structures
near the boundary, the Atiyah-Patodi-Singer
 index theorem states that (see \cite{LM})
 \(
{\rm Ind}_{\rm APS} (D^c_E) =
 \int_{Z^{12}} \widehat{A}(\nabla^{TZ, LC}) {\rm ch}(\nabla^{L^{1/2}}) {\rm ch}(\nabla^E) 
 - \xi (D_{Y,E}^c)\;,
 \)
where 
\(
\xi (D_{Y,E}^c)=\frac{1}{2}\eta (D_{Y,E}^c) + \frac{1}{2} \dim \ker 
(D_{Y,E}^c)\;. 
\)

\subsection{Effect of the ${\rm Spin}^c$ condition on the $C$-field}
\label{sec eff c}
Here we consider the M2-brane and the M5-brane partition 
functions in the Spin${}^c$ case.
Consider first the Stiefel-Whitney classes in the presence of the 
Spin${}^c$ line bundle with first Chern class $c$. 
\begin{enumerate}
\item $w_2(M)\equiv c$ mod 2, so that the manifold $M$ is Spin if 
$c$ is even. Hence in order for $M$ to be only Spin${}^c$, 
the first Chern class $c$ has to be odd. 

\item  The fourth Stiefel-Whitney class satisfies
\(
w_4(M) \equiv \frac{c^2-p_1(M)}{2} ~~{\rm mod}~ 2.
\label{w4}
\)
We see that if $c$ is even, i.e. when $M$ is Spin, then 
we are back to the condition $w_4(M)\equiv \frac{1}{2}p_1$ mod 2.
However, when $M$ is only Spin${}^c$, so that 
$c$ is odd, $p_1(M)$ can also be odd and still we 
could have $w_4=0$.  
\end{enumerate}


\paragraph{The M5-brane partition function.}
The expression (\ref{w4}) is used in the M5-brane partition function (section 5.2 of \cite{W-duality}). 
Consider the M5-brane on a six-manifold $W^6$. Take a product with a circle and 
extend $W^6 \times S^1$ to a bounding Spin${}^c$ 8-manifold $M^8$ , 
$\partial M^8=W^6\times S^1$.
Witten makes use of the Spin${}^c$ structure on $W^6$ to formulate the partition function. 
The phase is given by the expression
\(
\Omega_{W^6}(x):=\exp \left[ i \pi \int_{M^8} (z^2 + \lambda z)\right]\;,
\)
where $2\lambda =p_1 -c^2$ and the reduction modulo 2 of $c$ is $r_2(c)=w_2(M^8)$.

\paragraph{The M2-brane partition function.}
The Spin${}^c$ index on a manifold $M$ 
coincides with the $\widehat{A}$-genus 
if $c_1(L)$ is torsion and coincides with the index of the 
Dirac operator if the Spin${}^c$ structure is induced from 
a Spin structure on $M$. Thus, in order to go beyond
the standard Spin discussions we have to take 
$c=c_1(L)$ not to be torsion and in particular to 
be odd. 
When $Y^{11}$ is not Spin, $\lambda=\frac{1}{2}p_1$ is not guaranteed to be 
divisible by two, so that the quantization condition $G_4 + \frac{1}{4}p_1(Y^{11})
\in H^4(Y^{11}; \Z)$ of \cite{Flux}
would require careful consideration. In particular, $G_4$ cannot be set to zero.
We now consider the Spin${}^c$ case, i.e. take the membrane worldvolume 
to wrap/be a three-manifold with a Spin${}^c$ structure. Let $D^c$ be the corresponding
Dirac operator and $\sf L$ the corresponding determinant line bundle. 
The effective action involves two factors:
\begin{enumerate}

\item The `topological term' $\exp i \int_{M^3} C_3$, 

\item The fermion term $\exp i \int_{M^3} \overline{\psi} D^c \psi$.

\end{enumerate}
The first factor will simply give $\exp i \int_{X^4} G_4$. 
Now consider the second factor. Corresponding to the map
$\phi: X^4 \to Z^{12}$ we have the index of the Spin ${}^c$ 
Dirac operator for
spinors that are sections of $S^c(X^4) \otimes \phi^* TZ^{12}$ given
via the index theorem by the degree two expression
\begin{eqnarray}
{\rm Ind} D^c &=& \left[ \int_{X^4} \widehat{A}(X^4) \wedge {\rm Ph}(\phi^*TZ^{12})  
e^{\frac{1}{2}c_1 ({\sf L})}
\right]_{(4)}
\nonumber\\
&=& \int_{X^4} \left[ 1 - \frac{1}{24} p_1(TX^4) \right] 
\left[ {\rm rank}(TZ^{12})|_{X^4} + {\rm Ph}_2(\phi^*TZ^{12}) \right]
\left[1 + \frac{1}{8} c_1^2 ({\sf L}) \right]
\nonumber\\
&=& \int_{X^4} \frac{1}{2} p_1(\phi^*TZ^{12}) - \frac{1}{2} p_1(X^4) + 
 \frac{1}{2}
  c_1^2 ({\sf L})\;.
\end{eqnarray} 
Here Ph is the Pontrjagin character, which is the composition of the 
Chern character of a complex bundle with the realification map. 
On the other hand we have a decomposition of the restriction of the
tangent bundle $TZ^{12}$ to $X^4$ as
$
TZ^{12}=TX^4 \oplus NX^4
$,
with $NX^4$ the normal bundle of $X^4$ in $Z^{12}$.
Taking the characteristic class $\lambda$ of both sides of 
the above decomposition we get that the index is equal to $\lambda(NX^4)$.
The effective action involves the square root of the index so that
the contribution from the second factor in the effective action is
\(
\exp 2 \pi i \left[ \frac{1}{2} \lambda(NX^4) + \frac{1}{4} c_1^2 ({\sf L})  \right]\;.
\)
Now taking 
$
N( M^3 \hookrightarrow Y^{11}) \cong N(X^4 \hookrightarrow Z^{12})
$,
we get 
\begin{enumerate}
\item If $M^3$ admits a Spin structure then, assuming that $Y^{11}$ admits a Spin 
structure, then the normal bundle does and so $\lambda$ is divisible by 2. Also,
$c_1 {(\sf L})$ is even in this case, so that $\frac{3}{2} c_1^2({\sf L})$ is integral.
Hence the whole expression is integral and we are essentially back to the 
result of \cite{Flux}.

\item Now if both $Y^{11}$ and $M^3$ are only Spin${}^c$, then we get the 
quantization condition 
\(
G_4 + \frac{1}{2}\lambda + 
\frac{1}{4}
c_1^2({\sf L}) \in H^4(Y^{11};\Z)\;.
\label{spin c quant}
\) 
This is the condition in the Spin${}^c$ case. 
In \cite{DMW}, such a condition is essentially and implicitly 
obtained using 
transformation properties of the $C$-field 
in the presence of the RR field, defined via the 
line bundle with first Chern class $c$.


\end{enumerate}

\subsection{The phase of M-theory in the ${\rm Spin}^c$ case.}
\label{bdry case}
We start by recalling the setting and the results that we need from
\cite{DMW}. 
The partition function of the $C$-field
and the Rarita-Schwinger field is factored into a modulus and a phase.
These can be studied in regards to any of the three dimensions: eleven, 
as well as twelve and ten. 
The two bundles that play an important role are the $E_8$ vector bundle
$E$ (and subbundles thereof) and the Rarita-Schwinger bundle, i.e. the Spin bundle 
$S(Y^{11})$ coupled to the virtual tangent bundle $TY^{11}- 3{\cal O}$. 
 Thus, corresponding to the two bundles there are spinors and hence 
 Dirac operators $D$ acting on them. 
Let $h$ be the number of
zero modes of $D$.

\vspace{3mm}
Now, suppose that $Y^{11}$ is the boundary of a Spin
manifold $Z^{12}$, over which any data, such as an $E_8$ bundle, used in
defining $D$
are extended via cobordism arguments, and let $I(D)$ be the index of the extended operator
$D$ on $Z^{12}$, defined with Atiyah-Patodi-Singer (APS) global boundary 
conditions.  
Let $I_{E_8}$ be the index of the Dirac operator $D_{E}$
on $Z^{12}$, coupled to the
$E_8$ bundle $E$, with APS global boundary conditions.
Let $I_{RS}$ be the index of the Rarita-Schwinger operator 
$D_{RS}$ i.e. of the Spin bundle coupled to the virtual 
tangent bundle $TZ^{12} -4{\cal O}$. Denote the latter by $\Phi (Z^{12})$. Then \cite{DMW}
\begin{enumerate}
\item $\Phi (Z^{12})$ is independent of $Z^{12}$ when $Z^{12}$ has no boundary.  
\item $\Phi (Z^{12})$ is independent of $Z^{12}$ when $Z^{12}$ has a boundary. 
\end{enumerate}
The relation back to eleven dimensions is provided by the APS index theorem. 
This
asserts that $I(D)=\int_Z i_D- \frac{1}{2}{h+\eta}$,
where $i_D$ is the twelve-form whose integral on a closed
twelve-manifold would equal $I(D)$. 
Then the phase of the effective
action on $Y^{11}$ is given by \eqref{Phi}.
Assuming a supersymmetric (nonbounding) Spin structure on $S^1$, the index
of the Dirac operator coupled to $E \otimes {\cal L}^k$ is \cite{DMW}
\(
{\rm Ind} (D_{E \otimes {\cal L}^k}) = \int_{X^{10}}
\widehat{A}(X^{10}) {\rm ch}(E) e^{k c}\;,
\)
where $c$ is the first Chern class $c_1(\pi)$ of the circle bundle $\pi : Y^{11} \to X^{10}$,
and $k$ is the mode number corresponding to powers of the associated line bundle 
${\cal L}^k$ as $c_1({\cal L}^k)= k c_1 ({\cal L})$.  
In what follows we interpret this as corresponding to  ${\rm Spin}^c$ 
structures and study the corresponding expressions for the phase.

\paragraph{The data in the Spin${}^c$ case.} The setting we have is the following:

\noindent 1. $Y^{11}$ is identified with the boundary of the disk bundle $Z^{12}$. 

\noindent 2. $g_Z$ is a metric on $Z^{12}$
such that $Y^{11} \times I_{\epsilon}$ with the product metric $g_Y \oplus dt^2$ 
 isometric to 
a collar neighborhood of $(Z^{12}, g_Z)$ for some $\epsilon >0$. 

\noindent 3. Let $(\sigma^c_Y, \omega_Y)$ be a strictly equivariant or a boundary 
${\rm Spin}^c$ structure on $Y^{11}$ and $\Omega_Y$ the curvature form of $\omega_Y$. 

\noindent 4. $(X^{10}, \sigma^c_X, \omega_X)$ is the induced ${\rm Spin}^c$-Dirac structure
on $X^{10}$. $\Omega$ is the curvature of $\omega_X$ if $\omega_Y$ is strictly equivariant 
and the curvature of $\omega_X \otimes \omega^{\pi}$ otherwise.

\noindent 5. If $(\sigma^c_Y, \omega_Y)$ is an equivariant boundary ${\rm Spin}^c$-Dirac 
structure. We have $\omega_Y= \pi^* \Omega$.

\noindent 6. Now we introduce structure on the $E_8$ vector bundle $E$. This is a 
 vector bundle over $X^{10}$ equipped with a connection
$\nabla^E$ with curvature form $\Omega^E$. 

\noindent 7. $\nabla^Y$ is the induced connection
over $Y^{11}$ and $\nabla^{Z}$ is a connection on the bundle $\pi^* E$ over $Z^{12}$, and
$(\sigma^c_Z, \omega_Z)$ is a ${\rm Spin}^c$-Dirac structure on $Z^{12}$ extending  
$(\sigma^c_Y, \omega_Y)$ to $Z^{12}$. 

\noindent 8. $D_{\pi^* E}$ is the ${\rm Spin}^c$-Dirac operator on $(Z^{12}, \sigma^c_Z, \omega_Z)$
twisted by the coefficient bundle $(\pi^* E, \nabla^Z)$ acting on spinors over $Z^{12}$ 
satisfying the APS boundary conditions. The twisted Dirac operator $D_{\pi^* E}^Y$
is the corresponding operator on $Y^{11}$.

\subsubsection{The adiabatic limit when $(\sigma^c_Y, \omega_Y)$ is a boundary 
${\rm Spin}^c$-Dirac structure}
\label{bdry}
In this case, applying the results of \cite{BS1} \cite{BS2} \cite{Z1} \cite{Z2} gives 
\(
{\lim}_{t \to 0}
 {\overline{\eta}}(D_{\pi^* E^Y, g_Y^t})
= \left\langle 
{\widehat A}(X^{10}) e^{c_1(\sigma^c_X)/2} 
{\rm ch}(E) 
\left( 
\frac{e^{c_1(\pi)/2}}{2{\rm sinh}(c_1(\pi)/2)} - 
\frac{e^{c_1(\pi)/2}}{c_1(\pi)}
\right)
,~
[X^{10}]
\right\rangle\ \mod \Z\;.
\label{eta bound}
\)
Let $a$ be the characteristic class of the $E_8$ bundle $E$. Then the 
Chern character of $E$ is expanded, up to dimension eight, as
\(
{\rm ch}(E)=248 + 60a + 6a^2\;.
\)
What we are really doing is looking at the Chern character of the real bundle, 
i.e. the Pontrjagin character Ph$(E)$. 
The $\widehat{A}$-genus is expanded in degree $4k$ terms as $\widehat{A}=1 + 
{\widehat A}_4 + {\widehat A}_8$, where 
\(
\widehat{A}_4=-\frac{1}{2^3\cdot3}p_1\;, ~~~~~~~~~~~~~~~~~~~~~
\widehat{A}_8=-\frac{7p_1^2- 4p_2}{2^7\cdot 3^2 \cdot 5}\;. 
\)
The expansion of the expression of the adiabatic limit (\ref{eta bound})
gives
\begin{eqnarray}
{\lim}_{t \to 0} {\overline{\eta}}(D_{\pi^* E^Y, g_Y^t})
&=& 
\frac{248}{2^8\cdot 3}
\left[
\frac{11c_1(\pi)^5}{2^2\cdot 3\cdot 5\cdot 7} -
\frac{c_1(\pi)^4 c_1(\sigma^c_X) }{2\cdot 5} 
- \frac{23c_1(\pi)^3 c_1(\sigma^c_X)^2}{2^2\cdot 3\cdot 5}
- \frac{c_1(\pi)^2 c_1(\sigma^c_X)^3}{3} 
-\frac{c_1(\pi) c_1(\sigma^c_X)^4}{2^2\cdot 3} 
\right]
\nonumber\\
&-&
\frac{1}{2^3} \left( \frac{248}{2^2\cdot 3} \widehat{A}_4(X^{10}) + 5a\right)
\left[ 
\frac{ c_1(\pi) c_1(\sigma^c_X)^2}{2} + c_1(\pi)^2 c_1(\sigma^c_X)
+ \frac{23c_1(\pi)^3}{2^2\cdot 3\cdot 5} 
\right]
\nonumber\\
&-&
\frac{1}{2^2} \left[ \left(
a^2 + \frac{248}{2\cdot 3} \widehat{A}_8(X^{10})
\right) c_1(\pi) 
\right]
\end{eqnarray}

\paragraph{Special cases. }
\noindent {\bf 1.} When the circle bundle $\pi$ is trivial, $c_1(\pi)=0$, then the 
adiabatic limit is zero.

\noindent {\bf 2.} When the ${\rm Spin}^c$ structure is trivial, i.e. 
$c_1(\sigma^c_X)=0$, the right-hand side of the expression simplifies to 
\( 
\frac{31\cdot 11}{2^7\cdot 3^2\cdot 5\cdot 7}c_1(\pi)^5
-
\frac{23}{2^5\cdot 3\cdot 5} \left( \frac{248}{2^2\cdot 3} \widehat{A}_4(X^{10}) + 5a\right)
c_1(\pi)^3
-
\frac{1}{2^2} \left[ \left(
a^2 + \frac{248}{2\cdot 3} \widehat{A}_8(X^{10})
\right) c_1(\pi) 
\right]
\label{simp1}
\)
{\it Example. }
 Consider the ten-dimensional manifold $S^{10}/\Gamma$ or 
$S^8/\Gamma \times T^2$, where $\Gamma$ is chosen so that
the resulting manifold is Spin${}^c$; for instance, take $|\Gamma|$ to be odd. 
Then $\widehat{A}(S^8/\Gamma) \widehat{A}(T^2)$ can be arranged to be
1 with an appropriate choice of Spin structure (see section 
\ref{mod 2 ind} for more details on the $\widehat{A}$-genus 
in these dimensions). 
Consider the trivial $E_8$ vector bundle 
$S^{8}/\Gamma \times T^2 \times \R^{248}$; then, using $\widehat{A}(X^{10})=1$
and $a(E)=0$ in this case, the expression (\ref{simp1}) 
simplifies considerably to
\(
\frac{31\cdot 11}{2^7\cdot 3^2\cdot 5\cdot 7}c_1(\pi)^5\;.
\)
If the Rarita-Schwinger index does not contribute then this expression 
is nontrivial when exponentiated. Requiring the phase to be integral 
\(
\exp 2\pi i \left( \frac{1}{4}\cdot \frac{11\cdot 31}{2^7\cdot 3^2\cdot 5\cdot 7} c_1(\pi)^5 \right) 
\)
then imposes the integrality condition
\(
c_1(\pi)^5 \equiv 0~\mod~ 2^7\cdot 3^2\cdot 5\cdot 7
\label{mod 1}
\) 
on the Euler class of the M-theory circle.



\subsubsection{The adiabatic limit when $(\sigma^c_Y, \omega_Y)$ is a strictly equivariant 
${\rm Spin}^c$-Dirac structure}
\label{str eq}
In this case, applying the results of \cite{BS1} \cite{BS2} \cite{Z1} \cite{Z2} gives 
\(
{\lim}_{t \to 0} {\overline{\eta}}(D_{\pi^* E^Y, g_Y^t})
= \left\langle 
{\widehat A}(X) e^{c_1(\sigma^c_X)/2} 
{\rm ch}(E) 
\left( 
\frac{e^{c_1(\pi)/2}}{2{\rm sinh}(c_1(\pi)/2)} - 
\frac{1}{c_1(\pi)}
\right)
,~
[X^{10}]
\right\rangle\ \mod \Z;.
\label{eta equiv}
\)
The expansion of the expression of the adiabatic limit (\ref{eta equiv})
gives
\begin{eqnarray}
{\lim}_{{t \to 0}} {\overline{\eta}}(D_{\pi^* E^Y, g_Y^t})
&=& 
\frac{248}{2^5\cdot 3}\left[ \frac{c_1(\sigma^c_X)^5}{2^4\cdot 5} 
+ \frac{c_1(\pi) c_1(\sigma^c_X)^4}{2^4\cdot 3}
-  \frac{c_1(\pi)^3 c_1(\sigma^c_X)^2}{2^2\cdot 3\cdot 5}
+  \frac{c_1(\pi)^5}{3\cdot 5\cdot 7} 
\right]
\nonumber\\
&+&
\left( \frac{60 a}{2^4\cdot 3} + \frac{31}{2\cdot 3} \widehat{A}_4(X^{10}) \right)
\left[ 
 \frac{c_1(\sigma^c_X)^3}{2} +  \frac{c_1(\pi) c_1(\sigma^c_X)^2}{2^2} 
 -  \frac{c_1(\pi)^3}{3\cdot 5}  
\right]
\nonumber\\
&+&
\left( \frac{6a^2}{2^2} + 15a \widehat{A}_4(X^{10}) + 62 \widehat{A}_8(X^{10})\right)
\left[ c_1(\sigma^c_X)+ \frac{c_1(\pi)}{2\cdot 3}  \right]
\end{eqnarray}

\paragraph{Special cases.}
\noindent {\bf 1.}
When the circle bundle $\pi$ is trivial, $c_1(\pi)=0$, then the expression simplifies to
$$
{\lim}_{t \to 0} {\overline{\eta}}(D_{\pi^* E^Y, g_Y^t})
=
\left[
\frac{31}{2^6\cdot 3\cdot 5} c_1(\sigma^c_X)^4 
+ \frac{1}{2^2} \left( \frac{5a}{2} + \frac{31}{3} \widehat{A}_4\right) c_1(\sigma^c_X)^2
+ \left( \frac{3a^2}{2} + 15a\widehat{A}_4+ 62 \widehat{A}_8\right) 
\right]
c_1(\sigma^c_X)\;,
$$
where $\widehat{A}_i$ in the equation mean $\widehat{A}_i(X^{10})$.

\noindent {\bf 2.} When the ${\rm Spin}^c$ structure is trivial, i.e. 
$c_1(\sigma^c_X)=0$, the right-hand side of the expression simplifies to 
\(
\frac{31}{2^4\cdot 3^2\cdot 5\cdot 7}c_1(\pi)^5 -
\frac{1}{2\cdot 3\cdot 5} \left(\frac{5}{2}a +\frac{31}{3}  \widehat{A}_4(X^{10}) \right)
c_1(\pi)^3
+
\frac{1}{2.3} \left( \frac{3a^2}{2} + 15a \widehat{A}_4(X^{10}) + 62 \widehat{A}_8(X^{10})\right)
c_1(\pi)
\label{simp2}
\)
{\it Example 1.}
Consider the same example as in the previous section, i.e. 
the product $E_8$ bundle 
$S^{8}/\Gamma \times T^2\times \R^{248}$ or 
$S^{10}/\Gamma \times \R^{248}$, then, using $\widehat{A}(X^{10})=1$
and $a(E)=0$ in this case, the expression (\ref{simp2}) reduces to
\(
\frac{31}{2^4\cdot 3^2\cdot 5\cdot 7}c_1(\pi)^5\;.
\)
If the Rarita-Schwinger index does not contribute then this expression 
is nontrivial when exponentiated. Requiring the phase to be integral 
\(
\exp 2\pi i \left( \frac{1}{4}\cdot \frac{31}{2^4\cdot 3^2\cdot 5\cdot 7} c_1(\pi)^5 \right) 
\)
then imposes the integrality condition
\(
c_1(\pi)^5 \equiv 0~\mod~ 2^2\cdot 3^2\cdot 5\cdot 7
\label{mod 2}
\) 
on the Euler class of the M-theory circle.


\vspace{3mm}
\noindent {\it Example 2.}
Note that in both examples we could also use $\R P^5 \times \R P^5$ or  $\R P^9 \times \R P^1$.
This would require $G_4$ to be trivial, but that is consistent with requiring the 
$E_8$ bundle to be trivial, as above.

\paragraph{The adiabatic limit for the Rarita-Schwinger operator:}
The Rarita-Schwinger operator can also be written down using the results 
in \cite{Z1} \cite{Z2} and a similar formula holds. The same is true for the 
boundary case of section \ref{bdry case}.

\section{ Ten-Dimensional Spin and Spin${}^c$ Manifolds}

\subsection{Even dimensions: Almost complex and K\"ahler manifolds.}
\label{sec eve}

 In ten 
(and any even real) dimensions,  
almost complex manifolds are Spin${}^c$. The first 
obstruction to having an almost complex structure is the third integral Stiefel-Whitney
class $W_3$, which is exactly the only obstruction to having a Spin${}^c$ 
structure.  

\paragraph{Why Spin${}^c$ structures in ten dimensions?}
The main point of including even-dimensional ${\rm Spin}^c$ manifolds that are not necessarily 
Spin is to allow for K\"ahler manifolds such as complex projective spaces and
Fano varieties. Such compactifications appear in eleven-dimensional 
supergravity \cite{PV} and in M-theory \cite{BeBe}. In the Lorentzian case, 
${\rm Spin}^c$ manifolds also allow for a K\"ahler factor in the decomposition 
of the eleven- or ten-dimensional manifold and also allows for a factor having special 
holonomy (see \cite{Ik}). K\"ahler manifolds are especially interesting if we
want the M-theory manifold $Y^{11}$ to be Spin but the type IIA manifold
$X^{10}$ to be ${\rm Spin}^c$ as then we would be in the situation 
of nonprojectable Spin structures discussed in section
\ref{nonproj}. We can have a factor such as Witten spaces $M^{p,q,r}$ 
(see \cite{PP})  in $Y^{11}$ 
with the circle bundle identified with the M-theory circle so that 
the base in $X^{10}$ is a product of complex projective spaces.

\paragraph{ Spin${}^c$ structures on complex vector bundles over $X^{10}$.}
A complex vector bundle $E_\C$ and a complex line bundle $L$ over $X^{10}$
determine a Spin${}^c$ structure on $E_\C$ which is unique up to homotopy. 
The determinant line bundle of the Spin${}^c$ structure is isomorphic to 
$K^* \otimes L^2$, where $K^*$ is the anticanonical bundle 
of $E_\C$. Homotopic (respectively, bundle equivalent) complex structures on 
$E_\C$ give rise to homotopic (respectively, bundle equivalent) 
Spin${}^c$ structures (see e.g. Appendix D in \cite{GGK}).
We thus can use a complex $SU(5)$ bundle on $X^{10}$ 
to define a Spin${}^c$ structure on that bundle. In fact, we 
can consider the decomposition $E_8 \supset \left(SU(5) \times SU(5)\right)/\Z_5$
to define Spin${}^c$ structures on the corresponding
product bundles. These encode the Ramond-Ramond fields in 
the reduction of M-theory to type IIA string theory \cite{DMW}. 
Note that $SU(5)$ structures in relation to the geometry of Killing spinors in 
eleven dimensions are discussed in \cite{GP}.

\subsubsection{K\"ahler manifolds}
 K\"ahler manifolds
are Spin${}^c$. There is no homomorphism $U(5) \subset SO(10) \to {\rm Spin}(10)$
but there is a homomorphism $U(5) \to {\rm Spin}^c(10)$ that covers the 
inclusion $U(5) \to SO(10) \times U(1)$ which maps $A$ to $(A, \det A)$. 

\vspace{3mm}
\noindent {\it Complex projective space.} The complex projective space 
$\C P^m$ has a canonical Spin${}^c$ structure for all $m$. The twisting line 
bundle in this case is the canonical line bundle $H$, since
\(
H \otimes \left( T\C P^m \oplus 1\right) \cong (m+1) (H^* \otimes H) \cong m+1
\)
is trivial. 
 Recall also the following situation in the real case. Let $L$ be the nontrivial flat line bundle over $\R P^n$. Let 
$x=w_1 (L) \in H^1(\R P^m;\Z_2)$. Then $(4k+2)L$ admits a Spin${}^c$ structure since
for this line bundle $w_1=0$ and $w_2=x^2$ lifts to 
$H^2(\R P^2;\Z)$. From $T (\R P^m) \oplus 1 = (m+1)L$ we see that 
$\R P^{4k+1}$ admits a Spin${}^c$ structure.

\vspace{3mm}
\noindent {\it Weighted complex projective space.}
The weighted complex projective space of dimension $n$ is the quotient by the $\C^*$-action 
on the punctured complex space of dimension $n+1$,
\(
\C P (a_0, \cdots, a_n)=\frac{\C^{n+1}\backslash \{(0,\cdots, 0)\} }{\C^*}
\)
where the action of $\C^*$ on $\C^{n+1}\backslash \{(0,\cdots, 0)\}$
is given by
$
t\cdot (z_j)=(t^{a_j} z_j)$, 
with $(z_j)\in \C^{n+1}$.
The set of all complex line bundles over $\C P (a_0, \cdots, a_n)$ is the topological 
Picard group Pic${}^t(\C P (a_0, \cdots, a_n))$. 
This is isomorphic to $\Z$,
\begin{eqnarray}
{\rm Pic }^t(\C P (a_0, \cdots, a_n)) &\buildrel{\cong}\over{\longrightarrow}& \Z
\nonumber\\
\left[L^m\right] & \longmapsto & m\;,
\end{eqnarray}
where $L$ is the hyperplane line bundle.
Therefore, there is a $\Z$ worth of Spin${}^c$ structures 
on a weighted complex projective space (cf. \cite{Fu1}).

\subsubsection{Almost complex structures}

An even-dimensional manifold $M^{2n}$ is said to have an almost complex structure (acs)
if there exists a complex $n$-plane bundle $E_{acs}$ 
on $M^{2n}$ whose underlying real $2n$-plane bundle
is isomorphic to $TM^{2n}$. The manifold
$M^{2n}$ is said to have a {\it stable acs} if $TM^{2n}$ 
is stably isomorphic to the underlying real bundle of some
complex vector bundle over $M^{2n}$. 
 In \cite{He} necessary and sufficient conditions for the 
existence of acs in terms of the cohomology ring and 
characteristic classes of $M^{2n}$ are determined. 
A result of E. Thomas \cite{Th2} shows that a stable almost complex structure 
$E_{acs}$ on $M$ induces an almost complex structure iff $c_n(E_{acs})$
is equal to the Euler class of $M$:
$
c_n (E_{acs})= \chi (M^{2n})
$.

\paragraph{The index theorem.}
The index theorem 
takes the 
following form. If $M^{2n}$ is a manifold, 
$x \in H^2(M^{2n};\Z)$ a class such that 
the mod 2 reduction of $x$ is $r_2(x)=w_2(M^{2n})$ and $\xi$ a complex vector bundle 
over $M^{2n}$ then 
\(
\langle 
{\rm ch}(\xi) \cdot e^{x/2} \cdot \widehat{A} (M^{2n}),
[M^{2n}]
\rangle 
\)
is an integer. This is a special instance of the Spin${}^c$
index theorem for a manifold without boundary. 

\paragraph{ How many acs?} 
Let $M$ be a $2n$-dimensional manifold with almost complex structure $\tau$.
The stable almost complex structures of $M$ are given by the coset
$\tau + \ker (r)$, where $r : K(M) \to KO(M)$ denotes the realification map. 
From \cite{Th2} the almost complex structures on $M$ are given by
$\tau + F$, where $F \in \ker (r)$ satisfies 
$c_n(\tau + F)=c_n(\tau)$, i.e. satisfies
\(
c_n(F) + c_{n-1}(F) \cdot c_1(\tau) + \cdots +
c_1(F) \cdot c_{n-1}(\tau)=0\;.
\)
This shows that the set of almost complex structures only depends on 
$c(\tau)$ and the homotopy type of $M$ \cite{De98}.

\begin{enumerate}
\item Let $M$ be a $2n$-dimensional almost complex manifold. 
Assume there exists $x\in H^2(M;\Q)$ such that $x^n\neq 0$ (e.g. 
$M$ is a symplectic manifold). Then $M$ admits
infinitely many almost complex structures if 
$n \neq 1$, 2, 4 \cite{De98}.

\item Let $G$ be a compact connected simple 
Lie group and $U$ a maximal closed connected 
subgroup of maximal rank. Among the homogeneous 
manifolds $M=G/U$ 
there are many that are almost complex.

\item Let $M=G/U$ as above. Then $M$ admits 
only finitely many almost complex structures iff $M$ is isomorphic to 
\cite{De98}
\(
S^2, \quad S^6, \quad \C P^2, \quad \C P^4, \quad \frac{SO(6)}{SO(4)\times SO(2)}, {\rm ~or~~}
\frac{U(4)}{U(2)\times U(2)}
\)


\end{enumerate}

\paragraph{Spherical space forms.} Consider a spherical space form $M(\Gamma,\varrho)$.
 Let $m=2k-1$ be odd and let $\varrho : \Gamma \to O(m)$ be fixed-point-free. Recall
 that $\varrho$ is conjugate to a unitary fixed-point-free representation \cite{Wo}
\(
\varrho : \Gamma \to U(k) \subseteq SO(2k) \subset O(2k)\;.
\)
This shows that the stable tangent space
$
T(M(\Gamma, \varrho))\oplus 1
$
is the underlying real bundle of the complex vector bundle $\pi (\varrho)$ corresponding to 
the representation $\varrho$ so $M(\Gamma, \varrho)$ admits a {\it stable complex structure} \cite{Gi}.

\paragraph{Dimension eight.} 
A manifold $M^8$ has an acs iff \cite{He}
\begin{enumerate}
\item $w_8(M^8)\in Sq^2 H^6(M^8;\Z)$\newline
and there exist cohomology classes $u \in H^2(M^8;\Z)$ and 
$v\in H^6(M^8;\Z)$ such that 
\item $r_2(u)=w_2(M^8)$, $r_2(v)=w_6(M^8)$
\item $\chi (M^8) + u \cdot v \equiv 0$ mod 4
\item $8 \chi (M^8) = 4p_2(M^8) + 8u \cdot v - u^4 
+ 2u^2 \cdot p_1(M^8) -p_1(M^8)^2$.

(In case $Sq^2H^6(M^8;\Z)=0$ then the third is implied
by the first and second).
\end{enumerate}
Let $M^8$ be a manifold with a stable acs. Let $(u,v)$ be 
a pair of classes in $H^2(M^8;\Z)\times H^6(M^8;\Z)$ such that
$r_2(u)=w_2(M^8)$ and $r_2(v)=w_6(M^8)$. 
Then $M^8$ has a stable acs $E_{acs}$ such that 
$c_1(E_{acs})=u$ and 
$c_3(E_{acs})=v$ iff
\(
2\chi (M^8) + u \cdot v \equiv 0 ~~{\rm mod~}4.
\)

\paragraph{Dimension ten and orientation in generalized cohomology.} 
The existence of an almost complex structure in 
dimension ten is close to both the existence of Spin${}^c$ structure and 
the existence of an $EO(2)$-orientation of \cite{KS1}. 
If $X^{10}$ is a manifold such 
that $H_1(X^{10};\Z_2)=0$ and $w_4(X^{10})=0$
then $X^{10}$ has a stable acs
iff $W_3(X^{10})=0$ \cite{Th2}.
Furthermore, if $x\in H^4(X^{10};\Z)$
then
\(
Sq^4 x= Sq^2 (x \cup w_2 (X^{10}))\;.
\label{4.9}
\)
 Let us consider some consequences of this for the partition 
function. Applying the Bockstein $\beta =Sq^1$ and 
using the Cartan formula in (\ref{4.9}), we get
(abbreviating $w_2(X^{10})$ by $w_2$ in the third and fourth lines)
\begin{eqnarray}
\beta Sq^4 x &=&
Sq^3(x \cup w_2(X^{10}))
\nonumber\\
&=&
\sum_{i+j=3} Sq^i(x) \cup Sq^j (w_2(X^{10}))
\nonumber\\
&=&
Sq^0(x) \cup Sq^3(w_2)
+
Sq^1(x) \cup Sq^2(w_2)
+ 
Sq^2(x) \cup Sq^1(w_2)
+ 
Sq^3(x) \cup Sq^0(w_2)
\nonumber\\
&=&
x \cup 0 
+
Sq^1 (x) \cup w_2^2
+
Sq^2 (x) \cup W_3
+
Sq^3(x) \cup w_2\;.
\end{eqnarray}
Now if $X^{10}$ is Spin, i.e. if $w_2(X^{10})=0$, then 
the right-hand side is not zero in general. 
However, if $w_2(X^{10})\neq 0$  but
$X^{10}$ is only Spin${}^c$, i.e. $W_3(X^{10})=0$, the right hand side
will be equal to $Sq^3(x) \cup w_2(X^{10})$. Take 
$x=r_2(G_4)$. Then if $W_7(X^{10})=0$ 
(take $Sq^3(x)=W_7$), then the right hand side
is zero. Having $W_7(X^{10})=0$ while $w_2(X^{10}) \neq 0$ 
is equivalent to orientation with respect to Morava E-theory
\cite{KS1}. 

\vspace{3mm}
Suppose $X^{10}$ is a manifold such that
$H_1(X^{10};\Z_2)=0$ and $w_4(X^{10})=0$. 
Then if $u \in H^2(X^{10};\Z)$ is a class such that 
$r_2(u)=w_2(X^{10})$, there exists 
a stable acs $E_{acs}$ on $X^{10}$ such that
$
c_1(E_{acs})=u$ and $c_3(E_{acs})=2u^3$.
Let $(u, v)$ be a pair of classes in 
$H^2(X^{10};\Z)\times H^6(X^{10};\Z)$. Then there 
exists a stable acs $E_{acs}$ on $X^{10}$ such that
$c_1(E_{acs})=u$ and $c_3(E_{acs})=v$ iff
$
r_2(u)=w_2(X^{10})$ and $v=2u^3 + 2x$,
where $Sq^2 x=0$. 

\vspace{3mm}
Let $X^{10}$ be a 10-manifold with $H_1(X^{10};\Z)=0$, 
$H_i(X^{10};\Z)$ contains no torsion for $i=2$, 3, and
$H^2(X^{10};\Z)$  is generated by $h$ and $h^2 \equiv 0$ mod 2. 
Then $X^{10}$ admits a stable acs iff \cite{De98}
$
w_2(X^{10}) \cdot w_4(X^{10})^2=0
$.
Thus the existence of an acs only depends on the 
homotopy type of $X^{10}$. 
We see from the above discussion that there are plenty of examples.

\paragraph{Dependence on almost complex structures and a potential anomaly.}
In the presence of 
the RR field $F_2$, we have a line bundle with first Chern class $e$. 
The phase of the M-theory partition function in this case is given by 
\cite{DMW}
\(
\Omega_M(e,a)=(-1)^{f(a)} \exp 
\left[ 
2\pi i 
\left\langle 
\frac{e^5}{60} + \frac{e^3 a}{6} - \frac{11e^3 \lambda}{144}
-\frac{e\lambda a}{24}
+\frac{e\lambda^2}{48}
-\frac{e\widehat{A}_8}{2}
, [X^{10}]
\right\rangle 
\right]\;.
\) 
When the characteristic classes $a$, $\lambda$ and $p_2$ vanish, the 
expression for the phase reduces to 
\(
\Omega_M(e, 0)=
\exp \left[
2\pi i 
\left\langle 
\frac{e^5}{60}
,
[X^{10}]
\right\rangle
\right]
\label{e5 action}
\)
Take $e=c_1(X^{10})$. 
There are examples of two 5-dimensional algebraic varieties which 
are $C^\infty$-diffeomorphic, but have different 
Chern numbers \cite{BH}, 
namely the ten-dimensional flag manifold 
\(
X^{10}=U(4)/U(2) \times U(1) \times U(1)\;,
\)
whose points are the ordered triples of one 
2-dimensional and two 1-dimensional linear subspaces of the 
standard hermitian spaces $\C^4$ which are pairwise Hermitian 
orthogonal. 
The manifold $X^{10}$ carries two homogeneous
complex structures:
\begin{enumerate}
\item the 5-dimensional complex manifold 
$X^{10}_1$ consisting of the flags in $\C^4$ of type
$(0) \subset (1) \subset (3) \subset (4)$, i.e. the 
origin is contained in 
a one-dimensional linear subspace, contained in a three-dimensional
linear subspace, contained in $\C^4$.
\item the 5-dimensional complex manifold $X^{10}_2$ 
consisting of flags of type 
$(0) \subset (1) \subset (2) \subset (4)$.
\end{enumerate}
In \cite{BH} it was shown that that the first Chern classes and 
the Chern numbers of $X^{10}_1$ and $X^{10}_2$ are different:
$c_1(X^{10}_1)  {\rm~is ~divisible~by~}3~{\rm while~}
c_1(X^{10}_2)  {\rm ~is~not~divisible~by~}3$, and the particular 
Chern numbers are
\begin{eqnarray}
c_1^5[X^{10}_1]&=&2^2\cdot 3^5\cdot 5=4860\;, 
\\
c_1^5[X^{10}_2]&=&2^2\cdot 3^3\cdot 5^3=4500\;. 
\end{eqnarray}
Metrics $g_1$ and $g_2$ on $X_1^{10}$ and $X_2^{10}$, respectively, 
are equivalent when the 
the two manifolds are 
considered as differentiable manifolds but {\it not} when the manifolds 
are considered 
as complex analytic manifolds. 
This implies that
if we consider a family of metrics $g_t=tg_1 + (1-t)g_0$ through which 
a coordinate transformation takes $g_0$ to $g_1$ does not necessarily 
leave invariant the phase of the partition function (\ref{e5 action}); instead the 
change is given by $\exp 2\pi i \Delta S$, where $\Delta S= 360$.  However,
in the particular example, the effective action (or phase) 
is invariant. Of course, one might
be able to come up with other examples where the action does not change 
by a multiple of 60 and then the phase would 
pick up a potential anomaly. This provides a concrete physical realization
of speculations given in \cite{Nash} on the hope for a
 role of such different values of Chern 
numbers in anomalies.

\subsection{Using the M-theory circle to define a Spin${}^c$ square root} 
\label{sec using}

Let $X^{10}$ be an almost complex manifold and let $K_X=\Lambda^{5,0}(X^{10})$
be the canonical bundle of $X^{10}$, where $\Lambda^{r,0}$ denotes the bundle of 
$r$-forms of holomorphic  type. The canonical line bundle does not admit a square root
as the space $X^{10}$ is not necessarily Spin. However, this bundle admits 
a square root when tensored with an appropriate power of the complex line bundle
$\L$. We will take this line bundle to be associated to the M-theory circle bundle. Hence 
we will use the M-theory
circle to define spinors appropriately. Note that line bundles related to other entities 
can be used to define Spin${}^c$ structures, e.g. for
the $B$-field $B_2$ introduced in the context of D-branes \cite{FW}. 
Note also that in $A$- and $B$-twisted topological field theories \cite{W91}
the nature of spinors is changed by tensoring the spin bundle (which is the square root
of the canonical class in this case) with a line bundle.

\vspace{3mm}
In what follows
we make use of some constructions from \cite{HMU}.
Isomorphism classes of complex line bundles or $S^1$-principal bundles on $X^{10}$ 
are parametrized by the isomorphism of cohomology groups
\begin{eqnarray}
c_1 : H^1(X^{10}, \underline{U(1)}) &\cong & H^2(X^{10};\Z)
\nonumber\\
L &\longmapsto & c_1(L)\;.
\end{eqnarray}
where $\underline{U(1)}$ is the sheaf of germs of circle-valued functions. 
Suppose that $c_1(K_X)$ is a nonzero cohomology class. Let $p \in \mathbb{N}$
be the greatest number such that $c=\frac{1}{p} c_1(K_X) \in H^2(X^{10};\Z)$.
When $X^{10}$ is K\"ahler, $p$ is called the {\it index} of the manifold, and it is called
the {\it minimal Chern number} of $X^{10}$ when the latter is symplectic.   
If we take the M-theory circle bundle $\mathcal{L}$ such that 
$c_1(\L)= c = \frac{1}{p} c_1(K_X)$ then $K_X = \L^p$, i.e. $\L$ is the 
$p$-th root of the canonical bundle $K_X$, and so $\L$ is the $p$-fold covering
space of $K_X$. 

\paragraph{Example:} The index of $\C P^5$ with the usual complex structure is 6. 
The only line bundle $\L$ with first Chern class $\frac{1}{6}c_1(K_{\C P^5})$ is the universal bundle over
$\C P^5$ \cite{GH} and the corresponding principal $S^1$-bundle $\L$ is the Hopf bundle
$S^{11} \to \C P^5$. Then $K_{\C P^5}=\L^6 = S^{11}/\Z_6$ so that 
$(K_{\C P^5})^2= \L^{12}= S^{11}/\Z_{12}$. We could also deduce that $c_1(K_{\C P^5})$ 
is divisible by 3 from the fact that $\C P^5$ is a complex contact manifold, i.e. it carries
a codimension one holomorphic subbundle of $T^{1,0}\C P^5$ which is maximally 
non-integrable. 

\vspace{3mm}
\noindent {\bf Spin and Spin${}^c$ structures on $Y^{11}$ from Spin${}^c$ structures
on $X^{10}$.}
Next we consider the situation in eleven dimensions.
In going from ten to eleven dimensions, we could consider pullback of structures. 
Given a ${\rm Spin}^c$ structure on $X^{10}$ with determinant line bundle ${\sf L} \in H^2(X^{10};\Z)$
the oriented Riemannian submersion $\pi : \L \to X^{10}$, associated to 
$\pi: Y^{11}\to X^{10}$, induces a ${\rm Spin}^c$ structure 
on $\L$ whose determinant line bundle is $\pi^*({\sf L}) \in H^2(\L; \Z)$.
Now, from the Gysin sequence for the circle bundle 
\(
\xymatrix{
0 
\ar[r]
&
H^0(X^{10};\Z)
\cong \Z
\ar[rr]^{~~\cup c_1(\L)=\alpha}
&&
H^2(X^{10};\Z)
\ar[r]^{~\pi^*}
&
H^2(\L;\Z)
\ar[r]
&
0
}
\)
we have that $\ker \pi^* = \Z (\alpha)$. This implies the following.
\begin{enumerate}
\item ${\rm Spin}^c$ structures on 
$X^{10}$ whose determinant line bundles are tensor powers $\L^q$, for $q \in \Z$,
are exactly those inducing on $\L$, through the projection $\pi$, the unique Spin
structure.  
\item The ones that induce a ${\rm Spin}^c$ structure, however, are the ones
for which 
$
c_1(\L^q) \equiv c_1(X^{10})$ mod 2.
\end{enumerate}

\paragraph{ Examples.}
Now we specialize to K\"ahler manifolds. 
Using $c_1(\L)=\alpha$ and $c_1(X^{10})=-c_1(K_X)=-p\alpha$, with $p$ the index
of $X^{10}$,  we get 
\(
( q +p) c \equiv 0 ~~{\rm mod}~2 \in H^2(X^{10};\Z)\;,
\)
which in turn means that 
$
p + q \in 2 \Z
$
since $c$ is an indivisible class in $H^2(X^{10};\Z)$. 
For example if $X^{10}=\C P^5$ then $p=6$ is already even and so 
in order for $\L^q \otimes K_{\C P^5}$ to have a square root, $q$ has to 
be even. That is, we are considering only even Kaluza-Klein modes in the 
Fourier mode spectrum. 
The argument for the other (perhaps more phenomenologically realistic) 
Fano varieties is similar.

\paragraph{Dimensional reduction of $\L^q$.}
Now let us consider the odd case, i.e. $Y^{11}$ is Spin and $X^{10}$ is ${\rm Spin}^c$. 
Let us also specialize to the class of examples where $X^{10}$ is K\"ahler. Then we have
a K\"ahler form $\Omega^X$ on $X^{10}$. We would like to consider how spinors and 
spinor bundles on $X^{10}$ and $\L^q$ are related. 

\vspace{3mm}
{\it The Spin bundles:} Denote by $S^q X^{10}$ the spinor bundle of the corresponding 
Spin${}^c$ structure with determinant line bundle $\L^q$ with $q \in -p + 2\Z$.  
All these bundles pull back to the spinor bundle $S\L$, 
\(
\pi^* (S^q X^{10})=S\L~~~ \forall q\in \Z\;.
\)

\vspace{3mm}
{\it The Levi-Civita connections:} The Levi-Civita connection $\nabla^X$ on $X^{10}$
gives a corresponding $S^1$-connection on the principal $S^1$ bundle 
$\L \to X^{10}$ given by a 1-form $i \theta \in \Gamma ( \Lambda^1 (\L) \otimes i\R)$.
The metrics are related, via $\pi$ and $\theta$, as
$
g_{{}_{\L}} = \pi^* (g_X) + \frac{p^2}{36} \theta \otimes \theta 
$
such that the vector field $v$ defined by the free circle action on $\L$ 
has constant length $p/6$, i.e. the fibers are totally geodesic circles of length
$p\pi/3$. Note that the above quantities have normalized values for 
$\C P^5$, which has index $p=6$. 
For each integer $ q \in \Z$, there is a Levi-Civita connection on the line bundle 
$\L^q$ given by $i\theta^q \in \Gamma (\Lambda^1 (\L^q)\otimes i\R)$, which is mapped 
via the pullbacks of the $q$-fold covering $pr_{1,q}: \L \to \L^q$ as
$pr_{1,q}^* \theta^q= q \theta$. Hence the metric on $\L^q$ is
\(
g_{{}_{\L^q}} = \pi^* (g_X) + \frac{q^2p^2}{36} \theta \otimes \theta 
\)
Obviously, the larger $q$ is the larger is the part of the metric coming from the
circle part. This limit is in a sense the inverse of that of the adiabatic limit, 
for which the volume of the base is taken to be very large compared to that of
the fiber (cf. section \ref{equiv e}).

\paragraph{Correspondence with Spin structures.}
A Spin${}^c$ structure with trivial canonical line bundle is canonically
identified with a Spin structure. Since the corresponding $U(1)$ bundle 
is trivial then it admits a section $s$. Then the inverse image 
$P_{{\rm Spin}(11)}(Y^{11})$ of $P_{SO(11)}Y^{11} \times s$ defines 
the desired Spin structure on $Y^{11}$. Furthermore, if the connection 
$\nabla^L$ of $L$ is flat, then the Spin${}^c$ connection corresponds to 
the connection on the Spinor bundle. Thus, in using the M-theory circle to 
define the canonical line bundle, if that circle bundle admits a flat connection
 then we have a Spin structure.

\subsection{Spin${}^c$ ten-dimensional manifolds with $G$-actions}
Let $X^{10}$ be a Spin${}^c$-manifold on which a compact 
(not necessarily connected) Lie group $G$ acts, and let $S^1$ denote
a fixed subgroup of $G$. 
Let $V$ (respectively $W$) be a complex (respectively Spin) vector bundle over $X^{10}$.
The question is whether the $S^1$-action lifts to the Spin${}^c$ structure and the 
vector bundles $V$ and $W$.
Let $X_G:=EG\times_G X$ denote the Borel construction, where $EG$ is 
the classifying space for $G$.
 Let 
$L$ be a complex line bundle over $X$. Then the 
$G$-action lifts to $L$ iff $c_1(L)$ is in the image of the forgetful
homomorphism $H^2(X_G;\Z) \to H^2(X;\Z)$
\cite{HY}.

\paragraph{Circle action.}
Let $Q_c$ denote the Spin${}^c$ structure on $X^{10}$. This 
induces two complex line bundles:
\begin{enumerate}
\item $L_c$, a complex line bundle over $X^{10}$ defined by the $U(1)$ 
principal bundle 
$
Q_c/{\rm Spin}(10) \to Q_c/{\rm Spin}^c(10) \cong X^{10}
$
using the standard embedding of Spin$(10)$ in Spin${}^c(10)$.
The class $c_1(L_c)$, also denoted $c_1(Q_c)$ or $c_1(X^{10})$,
 will be called the first Chern class of $X^{10}$.

\item The group $U(1)$ acts on $Q_c$ via the embedding $U(1) \hookrightarrow 
{\rm Spin}^c(10)$. The quotient $Q_c/U(1)$ is the $SO(10)$ principal bundle 
$P$ of orthonormal frames, for the metric induced by $Q_c$. 
The projection $\xi : Q_c \to P$ is a $U(1)$ principal bundle and defines 
a second complex line bundle $\xi$. 

\end{enumerate}
\noindent The two line bundles are related: The pullback of $L_c$ to $P$ is isomorphic to $\xi^2$. 

\vspace{3mm}
We now consider the possible lift of the circle action.
The $S^1$ action on $X^{10}$ lifts to $P$ via differentials. 
If the $S^1$-action lifts to the principal $U(1)$ bundle 
$\xi: Q_c \to P$ then for a modified lift the Spin${}^c$ structure 
$Q_c \to X$ is $S^1$-equivariant  \cite{P72}.
Furthermore, If the first Betti number $b_1(X)$ vanishes or 
$c_1(X)$ is a torsion element then the $S^1$-action lifts to the
Spin${}^c$ structure $Q_c$. This can be shown as follows
 \cite{De99}. The $S^1$-action on $P$ lifts to the principal 
$U(1)$ bundle $\xi: Q_c \to P$ iff $c_1(\xi)$ is in the image of
$H^2(P_{S^1};\Z) \to H^2(P;\Z)$. 
Consider the Leray-Serre spectral sequence $\{E_r^{p,q}\}$ for 
$P_{S^1} \to BS^1$ in integral cohomology. Since 
$H^*(BS^1;\Z)$ is a polynomial ring in one generator of
degree 2, the only possibly nontrivial differential in the spectral 
sequence restricted to the subgroup of bidegree $(0,2)$ is
\(
d_2 : E_2^{0,2} \to E_2^{2,1} \cong H^2(BS^1; H^1(P;\Z))\;.
\)
The Leray-Serre spectral sequence for $P \to X$ shows that 
$b_1(P)$ vanishes if $b_1(X)$ does. In this case, 
$H^2(BS^1; H^1(P;\Z))=0$ and 
$d_2$ is the zero map. 

\vspace{3mm}
If $c_1(X)$ is a torsion class then $c_1(\xi)$ is also torsion since 
the pullback of $L_c$ to $P$ is isomorphic to $\xi^2$. Since 
$E_2^{2,1} \cong H^2(BS^1; H^1(P;\Z))$ is always torsion-free,
the image of $c_1(\xi)$ under $d_2$ is zero. 
Therefore, the class $c_1(\xi)$ survives and all differentials 
vanish on it, implying that it lies in the image of 
$H^2(P_{S^1};\Z) \to H^2(P;\Z)$. Thus, the $S^1$-action on 
$P$ admits a lift to $Q_c$ for which the Spin${}^c$ structure 
$Q_c\to X$ is equivariant.

\subsection{Category of Spin representations and the Atiyah-Bott-Shapiro 
construction}
\label{sec abs}

\vspace{3mm}
The Atiyah-Bott-Shapirro (ABS) construction \cite{ABS} relates the 
Grothendieck groups of real Clifford 
algebras to the KO-theory of spheres. 
Consider the following algebraic objects over a manifold $X$:

\noindent $\circ$ $V_\R(X)$, the set of isomorphism classes of real vector 
bundles over $X$. It is an abelian semigroup with the addition operation
being the Whitney (direct) sum; 

\noindent $\circ$ $F_\R(X)$, the free abelian group generated by elements of 
$V_\R(X)$;

\noindent $\circ$ $E_\R(X)$, the subgroup of $F_\R(X)$ generated by 
elements of the 
form 
$
[V]+ [W]-([V]\oplus [W])$,
where $+$ denotes addition in $F_\R(X)$ and 
$\oplus$ denotes addition in $V_\R(X)$.
Then the real K-theory of $X$ is defined to be the abelian group
$KO(X)=F_\R(X)/E_\R(X)$ whose elements are virtual bundles.

\vspace{3mm}
The tensor product of two Clifford algebras
$C\ell_n \otimes C\ell_m$ is in general not a Clifford algebra. Thus, to find
a multiplicative structure in the representations of Clifford algebras it is natural
to consider the category of $\Z_2$-graded modules \cite{LM}. 
A $\Z_2$-graded module for $C\ell_{11}$ is a module $W$ with a decomposition
$W=W^0 \oplus W^1$ such that 
\begin{eqnarray}
C\ell_{11}^0\cdot W^0&\subseteq &W^0\;, \qquad \qquad
C\ell_{11}^0\cdot W^1 \subseteq W^1\;,
\nonumber\\
C\ell_{11}^1\cdot W^0&\subseteq &W^1\;, \qquad \qquad
C\ell_{11}^1\cdot W^1 \subseteq W^0\;.
\label{prop 20}
\end{eqnarray}
The category of $\Z_2$-graded modules over $C\ell_{11}$ is equivalent to the 
category of ungraded modules over $C\ell_{10}\cong C\ell_{11}^0$ by passing from the
graded module
\footnote{e.g. $S^+ \oplus S^{-}$ to $S^{+}$.}
 $W^0 \oplus W^1$ to the module $W^0$. 
Define $\mathcal{M}_n$ to be the Grothendieck group of 
equivalence classes of irreducible representations of $C\ell_n$
and let
$\widehat{\mathcal{M}}_n$ be the Grothendieck 
group of real $\Z_2$-graded modules over $C\ell_n$.  
Thus, there is an isomorphism  $\widehat{\mathcal{M}}_{11}=\mathcal{M}_{10}$.
(See the end of the third paragraph in section \ref{proj} for an 
instance of this).

\vspace{3mm}
A natural $\Z_2$-graded tensor product of $\Z_2$-graded modules 
$W=W^0 \oplus W^1$ and $V=V^0\oplus V^1$ over 
$C\ell_n$ and $C\ell_m$ respectively,
is defined as 
follows. Set 
\begin{eqnarray}
(W \hat{\otimes}V)^0&=&W^0 \otimes V^0 + W^1 \otimes V^1
\nonumber\\
(W \hat{\otimes}V)^1&=&W^0 \otimes V^1 + W^1 \otimes V^0\;.
\end{eqnarray}
The action of $C\ell_n \hat{\otimes}C\ell_m$ on $W\hat{\otimes}V$ is given by 
$
(\varphi \otimes \psi)\cdot (w \otimes v)= (-1)^{pq}(\varphi w) \otimes (\psi v)
$,
where ${\rm deg}(\psi)=p$ and ${\rm deg}(w)=q$. 
The $\Z_2$-graded tensor product induces a natural associative pairing
$
\widehat{\mathcal{M}}_n \otimes_\Z \widehat{\mathcal{M}}_m \longrightarrow 
\widehat{\mathcal{M}}_{n+m}
$
which gives $\widehat{\mathcal{M}}_*=\bigoplus_{n\geq1}\widehat{\mathcal{M}}_n$
the structure of a graded ring. 
This multiplication gives
\(
\left(\widehat{\mathcal{M}}_*/i^*\widehat{\mathcal{M}}_{*+1} \right)
\equiv \bigoplus_{n\geq 0}
\left(\widehat{\mathcal{M}}_n/i^*\widehat{\mathcal{M}}_{n+1} \right)\;.
\)

Let $W=W^0\oplus W^1$ be a $\Z_2$-graded module over the 
Clifford algebra $C\ell_{10}\equiv C\ell(\R^{10})$. Let $\mathbb{D}^{10}= \{
x \in \R^{10} ~:~||x||\leq 1\}$ be the unit disk and 
set $S^{9}=\partial \mathbb{D}^{10}$. 
Let $E_0=\mathbb{D}^{10}\times W^0$ and $E_1=\mathbb{D}^{10}\times W^1$
be the trivial product bundles and let $\mu: E_0 
\buildrel{\simeq}\over{\to} E_1$ be the isomorphism over $S^{9}$ 
given by Clifford multiplication 
$
\mu(x,w)\equiv (x, x\cdot w)
$. 
Associate to the graded module $W$ the element
$
\varphi (W)=\left[ E_0, E_1;\mu\right] \in K(\mathbb{D}^{10},S^{9})$,
which depends only on the isomorphism class of the graded module
$W$, with the map $W \mapsto \varphi (W)$ being an additive 
homomorphism. Thus, this gives a homomorphism 
\(
\varphi : \widehat{\mathcal{M}}_{11} \to KO(\mathbb{D}^{10}, S^{9})=\widetilde{KO}(S^{10})=
KO^{-10}({\rm pt})\;.
\)  
Then there is a graded ABS isomorphism 
\(
\varphi_* : \left(\widehat{\mathcal{M}}_*/i^*\widehat{\mathcal{M}}_{*+1} \right)
\to KO^{-*}({\rm pt})\;,
\) 
giving a Clifford algebra definition of the $\alpha$-invariant. Further definitions 
and discussions are given in section \ref{mod 2 ind}.

\vspace{3mm}
The ABS isomorphism provides explicit generators for $KO^{-*}({\rm pt})$ 
defined via representations of
Clifford algebras. Let $S=S^+\oplus S^{-}$ be the fundamental 
graded module for $C\ell_{4n}$ where $S^{\pm}=(1\pm \omega)S$, where
$\omega$ is the volume element. Then 
$
\sigma_{4n}\equiv [S^+, S^{-};\mu]
$
is a generator of the group $KO^{-4n}({\rm pt})\cong KO_{\rm cpct}(\R^{4n})\cong \Z$, 
where $\mu: S^+\to S^{-}$ denotes Clifford multiplication by $x\in \R^{4n}$.
Clifford algebra relations give
the relation 
$
4 \sigma_8=(\sigma_4)^2$.

\paragraph{The Clifford linear Dirac operator.}
We utilize a unified approach which involves general Clifford 
modules, i.e. takes into account all possible vector bundles
coming from a given principal Spin bundle. 
The starting point is a principal Spin bundle 
$P_{\rm Spin}(X)$. Then in order to do analysis we need an 
associated vector bundle. This can be associated to one of 
the spinor representations 
$\Delta$, $\Delta^+$, 
$\Delta^{-}$, etc., depending on the dimension. However,
there is a formulation via Clifford algebras 
which automatically takes care of all 
cases without having to worry about making such 
(dimension-dependent) choices. Furthermore, it allows for
generalization to coupling to other vector bundles. This is the 
concept of a Clifford-linear Dirac operator (see \cite{LM})
 associated to the bundle
\(
\mathcal{S}= P_{\rm Spin}(X) \times_{{\rm Spin}(10)} C\ell_{10}\;.
\)
In constructing the bundle we use the left action of the 
Clifford algebra and we can make use of the right action 
to define the Dirac operator. This makes $\ker D$ not just
a vector space but a $\Z_2$-graded Clifford module.

\vspace{3mm}
Over a compact 10-manifold $X$ any $C\ell_{10}$-linear Dirac operator $D$ 
has an analytic index ${\rm Ind}_{10}(D)\in KO^{-10}({\rm pt})$ defined 
by applying the Atiyah-Bott-Shapiro isomorphism to the 
residue class of the Clifford module $[{\rm ker} D]$ as follows (cf. \cite{LM}). 
Consider a Clifford bundle $C\ell(X)$, which is the bundle of Clifford 
algebras over $X$ whose fiber at $x \in X$ is the Clifford algebra
$C\ell(T^*_xX)$ of the Euclidean spaces $T_x^*X$. 
Define a canonical bundle map 
$
\lambda_\omega : C\ell(X) \to C\ell(X)
$
by setting $\lambda_\omega (\varphi)=\omega \cdot \varphi$. In ten 
dimensions  this
satisfies $\lambda_\omega^2=-1$, so that if we complexify 
$S$ and consider the operator $i\lambda_\omega$ we get a splitting
$
S\otimes \C=(S\otimes \C)^+ \oplus (S \otimes \C)^{-}
$,
i.e. $\lambda_\omega$ defines a complex structure on $S$
with the above decomposition corresponding to the 
$(1,0)$, $(0,1)$ decomposition for the complex structure. 
Let $\mathcal{S}=\mathcal{S}^0 \oplus \mathcal{S}^1$ be a 
$\Z_2$-graded $C\ell_{10}$-Dirac bundle over a compact 10-manifold
$X$. Then the analytic index of the Dirac operator 
$
\mathcal{D}
$
of $\mathcal{S}$ 
 is the residue
class
\(
[\ker \mathcal{D}] \in 
\widehat{\mathcal{M}}_{10}/i^* \widehat{\mathcal{M}}_{11}\cong KO^{-10}({\rm pt})=\Z_2\;,
\)
or, equivalently, via $\mathcal{D}^0: \Gamma (\mathcal{S}^0) \to 
\Gamma (\mathcal{S}^1)$,
\(
[\ker \mathcal{D}^0] \in 
{\mathcal{M}}_{0}/i^* {\mathcal{M}}_{10}\cong KO^{-10}({\rm pt})=\Z_2\;.
\)

\paragraph{Examples.}
Let $S=S^+ \oplus S^{-}$ be the ordinary complex
$\Z_2$-graded Dirac bundle over a compact ten-dimensional manifold 
$X$. 
\begin{itemize}
\item First, take $V$ to to be the complexified tangent bundle $T_\C X$,
which splits into the $\Z_2$-graded decomposition
$T_\C X=T \oplus \overline{T}$, where $T$ is the holomorphic 
and $\overline{T}$ is the antiholomorphic tangent bundle. 
Then a $\Z_2$-graded $C\ell_{10}$-bundle $\mathcal{S}$ is
\begin{eqnarray}
\mathcal{S}=S \otimes T_\C X&=&
(S^+ \oplus S^{-}) \otimes (T \oplus \overline{T})
\nonumber\\
&=&
(S^+ \otimes T \oplus  S^- \otimes \overline{T}) 
\oplus
(S^+ \otimes \overline{T} \oplus  S^- \otimes {T}) 
\nonumber\\
&:=&
\mathcal{S}^+ \oplus \mathcal{S}^{-}\;.
\end{eqnarray}
\item Next, take  $V$ to be an $E_8$ bundle. The  
inclusion $\left(SU(5)\times SU(5)\right)/\Z_5 \subset E_8$ allows us
to break $V$ into a pair of $SU(5)$ bundles as in
\cite{DMW}, $V \supset E \oplus \overline{E}$. Then, as in 
the first case, we get the $\Z_2$-graded $C\ell_{10}$-module
\begin{eqnarray}
\mathcal{S}=S \otimes V&=&
(S^+ \oplus S^{-}) \otimes (E \oplus \overline{E})
\nonumber\\
&=&
(S^+ \otimes E \oplus  S^- \otimes \overline{E}) 
\oplus
(S^+ \otimes \overline{E} \oplus  S^- \otimes {E}) 
\nonumber\\
&:=&
\mathcal{S}^+ \oplus \mathcal{S}^{-}\;.
\end{eqnarray}
\end{itemize}

\paragraph{Topological definition of the mod 2 index.}
We follow \cite{LM}.
 Consider the graded ring structure of $KO^*({\rm pt})$, generated over 
$\Z$ by elements $\eta$, $\omega$, $\mu$ of degrees 1, 4, and 8, 
respectively, with relations
\(
2\eta=\eta \omega = \eta^3 =0\;, \qquad \omega^2=4\mu\;.
\)
That is, 
\begin{eqnarray}
KO^{-1}({\rm pt})&=&\Z_2 \eta\;, \quad \hspace{5mm} KO^{-4}({\rm pt}) = \Z \omega\;,
\nonumber\\
KO^{-2}({\rm pt})&=&\Z_2 \eta^2\;, \qquad KO^{-8}({\rm pt}) = \Z \mu\;,
\end{eqnarray}
with the periodicity $KO^{*\pm 8}({\rm pt})\cong KO^*({\rm pt})$.
  Consider the embedding 
$i: X^{10} \to S^{10 +8k}$, where $k$ is large, 
and let $NX^{10}$ be the normal bundle of $X^{10}$ inside 
$S^{10+8k}$. Let $U(NX^{10})$ be the 
$KO$-Thom class of $NX^{10}$, and let $j$ be the natural 
isomorphism $\mathbb{D}N/SN\simeq S^{10+8k}$, with $\mathbb{D}N$ and 
$SN$ the disk and the sphere bundle of the normal bundle,
respectively.
Then, for any $E \in KO(X^{10})$, 
define the Gysin morphism 
\(
f_! : KO(X^{10}) \to KO^{-10}({\rm pt})
\)
defined as 
\(
f_!(E)=j^*(E\cup U(NX^{10})) \in \widetilde{KO}(S^{10+8k}) \cong KO^{-10}({\rm pt})\;,
\)
which is independent of the embedding $i$. 
The mod 2 index of the Dirac operator is then defined as \cite{AS5}
\(
{\rm Ind}_2(D\otimes E) \equiv \pi (f_! E) ~~{\rm mod}~2\;,
\)
where $\pi : KO^{-1}({\rm pt}) \buildrel{\simeq}\over{\to}\ \Z_2$, $KO^{-2}({\rm pt})\buildrel{\simeq}\over{\to}\Z_2$
are the isomorphisms sending $\eta$ and $\eta^2$ to 
$1 \in \Z_2$.

\vspace{3mm}
Consider the classical genera
\(
\widehat{\mathcal{U}}=
\left\{
\begin{array}{ll}
{\rm Ind}_2D,& X {\rm ~has ~dimension~} 1,2,9, {\rm or~} 10\\
{\rm Ind} D,& X {\rm ~has~ dimension~} 4 {\rm ~or~} 8.
\end{array}
\right.
\)
The mod 2 index on $X^{10}$ can be related to the index 
on the twelve-manifold $\mathcal{T}(X^{10})$,
the mapping torus of $X^{10}$,
 via the theorem 
proved in \cite{Liu}:
\(
{\rm Ind}_2 D\equiv \frac{1}{2}\int_{\mathcal{T}(X)}\hat{\mathcal{U}}(\mathcal{T}(X))~~~
{\rm mod} ~2\;.
\)

\paragraph{When is the mod 2 index nonzero.}
The mod 2 invariant given by the index map
\(
{\rm Ind}_{2} : \Omega_{10}^{\rm Spin} 
\to KO^{-10}({\rm pt})
\)
is a ring homomorphism. 
The nonzero element 
$\eta \in KO^{-1}({\rm pt})$ has the property that 
$\eta x$ and $\eta^2 x$ are nonzero whenever 
$x$ is an odd multiple of the generator in degree 
8 \cite{LM}.

\paragraph{ Example 1:  the circle $S^1$.} Consider the nontrivial Spin structure on
$S^1$. The $\Z_2$-grading on $C\ell_1\cong \C=\R 
\oplus i \R$, $C\ell_1^0\cong \R$, 
 gives the corresponding grading on the 
Clifford module $\mathcal{S}=\mathcal{S}^0 \oplus 
\mathcal{S}^1=(S^1 \times \R) \oplus (S^1 \times i\R)$.
Sections of $\mathcal{S}$ are complex-valued functions 
on the circle. The Dirac operator 
$\mathcal{D}=i\frac{d}{dt}$ splits into $\mathcal{D}^0$ and $\mathcal{D}^1$,
and the kernel of 
$\mathcal{D}: \Gamma (\mathcal{S}^0) \to \Gamma (\mathcal{S}^1)$ is the 
set of real-valued constant functions on $S^1$, a space whose real dimension
is 1. Thus, ${\rm ind}_2(S^1)\neq 0$ is the generator of $KO^{-1}({\rm pt})\cong \Z_2$. 
On the other hand, Ind${}_2=0$ for the trivial Spin structure on $S^1$. 

\paragraph{ Example 2 : the torus $T^2=S^1 \times S^1$.}
Choose the Spin structure on $T^2$ which is given by squaring 
the nontrivial Spin structure on $S^1$. This is the covering 
\(
\xymatrix{
P_{\rm Spin}(T^1)=T^2 \times S^1 
\ar[rr]^{{\rm Id} \times z^2} 
&&
T^2 \times S^1=P_{SO}(T^2)
}\;.
\)
The Clifford algebra $C\ell_2 \cong \H$ and $C\ell_2^0 \cong \C$ give that 
$\mathcal{S}(T^2)=T^2 \times \H$ and 
$\mathcal{S}^0(T^2)=T^2 \times \C$. The kernel of $\mathcal{D}^0$ is 
the complex-valued constant functions. Then 
\(
{\rm Ind}_2(T^2)\neq 0 \quad {\rm for~nontrivial~Spin~structure~ on~}S^1\;.
\) 

\paragraph{Multiplicative properties of the index.}
Let $X_1$ and $X_2$ be two Riemannian Spin manifolds of dimensions
$n_1$ and $n_2$, with $n_1 + n_2=10$. Let 
 $\mathcal{D}_k:
\Gamma (\mathcal{S}_k) \to \Gamma (\mathcal{S}_k)$ be the 
Atiyah-Singer operator for the 
$C\ell_{n_k}$-Dirac bundle $\mathcal{S}_k=\mathcal{S}(X_k)$, for $k=1,2$. 
Let $\mathcal{D}: \Gamma (\mathcal{S}) \to \Gamma (\mathcal{S})$ 
be the operator for $X^{10}=X_1 \times X_2$ with the product Riemannian
and Spin structure. For this structure,
$
P_{\rm Spin}(X_1 \times X_2) \supset P_{\rm Spin}(X_1) \times_{\Z_2}
P_{\rm Spin}(X_2)
$,
where $\Z_2$ acts on the two factors by $(-1,1)$. Then 
\(
\mathcal{S}=\mathcal{S}_1 \hat{\otimes}~ \mathcal{S}_2\;,
\label{z2 g}
\)
the exterior $\Z_2$-graded tensor product, which is a 
$C\ell_{n_1 + n_2} = (C\ell_{n_1} \hat{\otimes} C\ell_{n_2})$-Dirac
bundle. This gives \cite{LM}
$
\ker (\mathcal{D})=\ker (\mathcal{D}_1) \hat{\otimes} 
\ker (\mathcal{D}_2)$,
with a $\Z_2$-graded tensor product inherited from the 
$\Z_2$-grading of the Clifford modules (\ref{z2 g}) and is 
the multiplication in 
$
KO^{-*}=\widehat{\mathcal{M}}_*/i^* \widehat{\mathcal{M}}_{*+1}
$.
Then ${\rm Ind}_2^*$ is a ring homomorphism
\(
{\rm Ind}^{n_1 + n_2}_2 (\mathcal{D})= 
{\rm Ind}^{n_1}_2 (\mathcal{D}_1)~
{\rm Ind}^{n_2}_2 (\mathcal{D}_2)\;.
\label{n1 n2}
\)

\paragraph{Extension to the case of coupling to a vector bundle.}
Let $E$ be a real vector bundle over $X$ with an orthogonal 
connection. The bundle 
$\mathcal{S}(X) \otimes E$ is naturally a
$\Z_2$-graded $C\ell_{10}$-Dirac bundle with a Dirac operator 
$\mathcal{D}_E$. Take the Spin bundle $SX$ and tensor it with
$E$ to get $SX\otimes E$ with Dirac operator $D_E$. 
Let $h_E(X)\equiv \ker (D_E)$ denote the space of real 
harmonic $E$-valued spinors. Then the above result on 
multiplicativity \eqref{n1 n2} extends to this case of 
twisted Dirac operators \cite{LM}. 


\vspace{3mm}
Note that there is also a Clifford-module approach to twisted (complex) 
K-theory, which is explained for instance in \cite{Ka}.

\section{Anomalies in the Partition Function and Geometry}
\label{sec geo}

\subsection{Effect of scalar curvature}
\label{sec sca}

\paragraph{Scalar curvature for products and bundles.}
Let $M$ and $N$ be closed Riemannian 
 manifolds with scalar curvatures $\mathcal{R}_M$ and $\mathcal{R}_N$, 
 respectively.    
 Let $E:=M \times N$ and let $\pi(x,y)=y$ define a trivial Riemannian submersion
from $E$ to $N$. Consider the canonical variation given by
$
g_t := tds_M^2 + ds_N^2
$
on $E$ so that the scalar curvature corresponding to $g_t$
is
$\cR_t=t^{-1} \cR_M + \cR_N + O(t)$. Assuming $\cR_M$ is 
positive, $\cR_t$ tends to $\infty$
as $t$ goes to zero. 
Thus, if $M$ admits a metric of positive scalar curvature
then so does $M\times N$ (see e.g. \cite{GLP}).

\vspace{3mm}
\noindent Next let $\pi : Y \to X$ be a Riemannian submersion with totally 
geodesic fibers. For any $t$, $\pi : (Y, g_Y) \to (X, g_X)$ is a Riemannian submersion with totally geodesic 
fibers.  Let $\cR_t$ be the scalar curvature of the metric $g_t$, let 
$\cR_F$ be the scalar curvature of the original metric on the 
fiber $F$, and let $\cR_X$ be the scalar curvature of the metric on $X$. 
Then 
$
\cR_t = t^{-1} \cR_F + \cR_X + O(t)$.
This is proved using O'Neill's formula \cite{Be}. Then a similar conclusion as for the 
case of products follows.

\paragraph{Scalar curvature vs. Spin.} Scalar curvature is closely related 
to (non)existence of Spin structure as well as to Killing spinors. 
Let $M$ be a closed simply connected manifold of dimension at least
5. If $M$ does not admit a Spin structure, then $M$ carries a metric of positive scalar curvature
\cite{GL2}.
Let $(M,g)$ be an $n$-dimensional connected complete Riemannian Spin 
manifold with a nontrivial Killing spinor with $\alpha\neq 0$. 
Then (see \cite{Fried} \cite{BFGK})

\noindent {\bf 1.} $(M,g)$ 
is locally irreducible. 

\noindent {\bf 2.} $M$ is Einstein with 
Einstein constant $\lambda=4(n-1)\beta^2$, i.e. $M$
is a space
of constant sectional curvature equal to $4\beta^2$. 
In particular, when $\beta$ is a nonzero real number, $M$ is compact
of positive scalar curvature.

\paragraph{$G$-action and scalar curvature.}
Let $Y^{11}$ be a closed manifold equipped with an $S^1$ action. Then, from \cite{BB},
 the following 
are equivalent:
\begin{enumerate}
\item $Y^{11}$ admits an $S^1$-invariant metric of positive scalar curvature.
\item $Y^{11}/S^1$ admits a metric of positive scalar curvature.
 \end{enumerate}
The above equivalence is not true for $S^1$ replaced by a connected nonabelian 
group $G$. For example, consider $Y^{11}=G \times T^n$, with $G=SU(2)=Sp(1)$, 
$SO(3)$, 
$SO(4)$, $SU(3$), $SO(5)=Sp(2)$, and $n$ is respectively, 8, 8, 5, 3, and 1. 
The bi-invariant Riemannian metric on $G$ has positive scalar metric, and $G$ 
acts freely on the first factor in $Y^{11}$. The quotient $T^n$ obviously does not 
admit a metric of positive scalar curvature.    

\vspace{3mm}
Let $Y^{11}$ be a closed oriented free $S^1$-manifold which is simply connected and does not 
admit a Spin structure. Then $Y^{11}$ carries an $S^1$-invariant metric of positive scalar 
curvature. This can be proved as follows \cite{Ha}. The long exact sequence on homotopy of the
$S^1$-fibration $S^1 \hookrightarrow Y^{11} \buildrel{\pi}\over{\longrightarrow} 
X^{10}=Y^{11}/S^1$ shows that $X^{10}$ 
is also simply connected. Furthermore, 
$
TY^{11} \cong \pi^* \left( 
TX^{10} \oplus \mathcal{O}_{\R}
\right)$,
where $\mathcal{O}_\R$ is a trivial real line bundle. Calculating $w_2$, and using the 
fact that $w_2(\mathcal{O}_\R)=0$ gives that $X^{10}$ does not admit a Spin structure. 
By the Gromov-Lawson theorem \cite{GL1} \cite{GL2}, 
$X^{10}$ admits a metric of positive scalar curvature.
By \cite{BB}, the manifold $Y^{11}$ admits an $S^1$-metric of positive scalar curvature. 

\paragraph{Positive curvature and the kernel of the Dirac operator.}
If $D$ is the Dirac operator on a Spin manifold $M$, $\nabla$ is the 
covariant derivative on the Spin bundle $SM$, and $\nabla^*$ is the adjoint
of $\nabla$,
then 
\(
D^2=\nabla^* \nabla + \frac{1}{4}\mathcal{R}\;,
\label{lich wen}
\)
where $\mathcal{R}$ is the scalar curvature of $M$. Since the first term on the
right hand side
 is
non-negative then the Dirac operator cannot have any kernel when 
$\mathcal{R}$ is positive \cite{Li}. This classical result has a refinement to $KO$-theory 
via the $\alpha$-invariant. See section \ref{mod 2 ind}.

\paragraph{Motivation for positive scalar curvature for $Y^{11}$.}
Let $D^+(Z^{12}, g_Z)$ be the (chiral) Dirac operator 
with respect to a Riemannian metric $g_Z$ that coincides
with the product metric on $Y^{11}\times I$ in a collar neighborhood 
of the boundary $\partial Z^{12}=Y^{11}$. This is a Fredholm 
operator when taking APS boundary conditions. 
Now consider a continuous family of metrics $g_Z(t)$ on 
$Z^{12}$. Then in this case the corresponding family of projections
$P(t)$ is {\it not} continuous for those values of the parameter $t$
 where an eigenvalue of $D(Y^{11}, g_{Y})$ crosses the
 origin. 
 If $g_Y$ has positive scalar curvature then from the Lichnerowicz
 argument \cite{Li} using the Weitzenb\"ock formula (\ref{lich wen}) shows that
 the kernel of $D(Y^{11}, g_{Y}(t))$ is trivial. 
Hence $D^+(Z^{12}, g_Z(t))$ is a continuous family of Fredholm 
operators and therefore ${\rm Ind}\hspace{0.5mm}D^+(Z^{12}, g_Z(t))$ is independent 
of $t$. 

\vspace{3mm}
One motivation for positive scalar curvature for $Z^{12}$ is the following.
If $g_Z$ has positive scalar curvature then 
 ${\rm Ind} D^+(Z^{12}, g_Z(t))$ vanishes \cite{APSII}.

\paragraph{Effect of the scalar curvature in the Spin${}^c$ case.}
Let $(M, g, \sigma^c, \omega^{c})$ be a ${\rm Spin}^c$-Dirac manifold with
metric $g$ of scalar curvature $\cal R$, ${\rm Spin}^c$ structure $\sigma^c$ 
and a connection
$\omega^{c}$ on the canonical $U(1)$ bundle $\xi (\sigma^c)$. 
Let $D_{\zeta}$ be 
the Dirac operator on $(M, g, \sigma^c, \omega^{c})$ twisted with a bundle
$(\zeta, \nabla^{\zeta})$. Let $\Omega^{\zeta}$ and $\Omega^{\alpha}$ be the principal 
curvature forms 
of $\nabla^{\zeta}$ and $\omega^c$, respectively. 
If we write the Ricci curvature as $\Omega^{c}= \sum_{i \leq (n-1)/2} \lambda_i e_i \wedge
e_{[i + (n-1)/2]}$, using the basis $\{ e_1, \cdots, e_n \}$, then 
a `norm' is defined as \cite{Hit}
$||\Omega^{c} \otimes 1||= \sum_{i \leq (n-1)/2} |\lambda_i|$. Hence this 
is, in a sense, a measure of the scalar functions in front of the Ricci curvature written 
in a Cartan basis.  

\vspace{3mm}
If 
$\frac{1}{4} {\cal R} - || \Omega^{c} \otimes 1 + 1 \otimes \Omega^{\zeta} ||$
is positive somewhere and nonnegative everywhere on $M$ then
\cite{Hit} \cite{Li}
${\rm Ind}\hspace{0.5mm}D_{\zeta}^+ =0$; on closed manifolds, 
${\rm ker}\hspace{0.5mm}D_{\zeta}=0$.
Let ${\cal R}_Y$ and ${\cal R}_X$ be the scalar curvatures of $g_Y$ and $g_X$
on $Y^{11}$ and $X^{10}$, respectively. 

\vspace{3mm}
\noindent {\it 1. Effect of scalar curvature of $Y^{11}$.}
If ${\cal R}_Y (y) > 4 || \Omega_Y \otimes 1 + 1 \otimes \pi^* \Omega^E || (y)$
for all $y \in Y^{11}$ then the index of the Dirac operator $D^+_{\pi^*E}$ vanishes. 

\vspace{3mm}
\noindent{\it 2. Effect of scalar curvature of $X^{10}$.}
Since $\lim_{t \to 0} {\cal R}_{(Y, g_Y^t)}={\cal R}_X$ and 
${\cal R}_X \geq {\cal R}_{(Y, g_Y^t)}$ then 
if ${\cal R}_X (x) > 4 || \Omega_Y \otimes 1 + 1 \otimes  \Omega^E || (x)$ for all $x \in X^{10}$
then $\lim_{t \to 0} {\rm Ind}\hspace{0.5mm}D_{\pi^* E}^+(g_Z^t)=0$.
In the case when the  ${\rm Spin}^c$-Dirac structure on $Y^{11}$ is 
bounding then, from \cite{BS1} \cite{BS2}, (\ref{eta bound}) holds without the 
mod $\Z$ requirement. 
Similarly, for (\ref{eta equiv}) when the  ${\rm Spin}^c$-Dirac structure on $Y^{11}$ is 
strictly equivariant. This is viewed as the classical analog (i.e. at the level of
Lagrangians and actions) of the 
semi-classical statements in section \ref{bdry} and section \ref{str eq}
 (which are given there, at least implicitly, at the level of partition functions).
This is in harmony with the fact that large scalar curvature takes us further 
and further into the classical regime. 

\paragraph{Ricci curvature and genera.}
Conditions on Ricci curvature are generically stronger
than conditions on the scalar curvature. One example is 
the Stolz-H\"ohn Conjecture \cite{St96}, which is 
 the following statement.
Let $M$ be a String manifold, i.e. a manifold admitting a lift of the 
Spin bundle to a String bundle. If $M$ admits a metric of positive
Ricci curvature then the Witten genus $\varphi_W(M)$ 
vanishes. 
Now any Spin manifold of positive Ricci curvature has positive scalar 
curvature and, hence, vanishing $\widehat{A}$-genus. This implies the conjecture 
in dimension $4k< 24$ or dimension 28. 
As seen above, the 
main sources  of positive Ricci curvature include K\"ahler geometry 
(any K\"ahler manifold with positive first Chern class carries a metric of 
positive Ricci curvature), Lie groups, and  
homogeneous spaces (a compact homogeneous space admits
an invariant metric of positive Ricci curvature if its fundamental 
group $\pi_1$ is finite \cite{Be95}).

\paragraph{Scalar curvature and the ten-dimensional partition function.}
The partition function in ten dimensions is well-defined 
when $\langle  x\otimes \overline{x}, [X^{10}] \rangle
=0 \in KO_{10}({\rm pt})\cong \Z_2$ (see section \ref{mod 2 ind}).
In the case when $X^{10}$ is a connected simply connected Spin manifold, 
we use \cite{Ros3} to deduce that the 
KO fundamental class $[X^{10}]$ is zero in $KO_{10}({\rm pt})\cong\Z_2$ if and
only if $X^{10}$ admits a metric of positive scalar curvature. 
Therefore, the partition function
is well-defined if and only if $X^{10}$ admits a metric 
of positive scalar curvature. Hence, 
in the simply connected case,
anomalies arise only for manifolds
not admitting metrics of positive scalar curvature. 
However, a refinement is needed when the fundamental group is nontrivial
(see section \ref{nonsim}).

\subsection{The mod 2 index, the $\widehat{A}$-genus, and the $\alpha$-invariant}
\label{mod 2 ind}

The K-theoretic description of the partition function of the 
Ramond-Ramond fields in type IIA string theory, reduced 
from eleven dimensions on a circle, leads to an anomaly
given by the mod 2 index of the Dirac operator \cite{DMW}. 
We provide a characterization of this within our context.

\vspace{3mm}
We first recall some setting from \cite{DMW}.
 Consider the product case, $Y^{11}=X^{10} \times S^1$ with the 
	             C-field $C_3$ with field strength $G_4$, both pullbacks from $X^{10}$. In this case, there is an 
	             orientation-reversing symmetry under which 
	             the term $I_{CS}=\int_{Y^{11}} G_4 \wedge G_4 \wedge C_3$ reverses sign so that
	             the phase, which contains the factor $\exp(iI_{CS})$,  is complex-conjugated.
	             This implies that the phase is $\Z_2$-valued. 
	             In terms of the index theorem, the Dirac operator changes sign under the reflection of 
	             one coordinate so that the nonzero eigenvalues appear in pairs $(\lambda, -\lambda)$, implying that 
	             the eta invariants are zero. The corresponding analysis for the zero modes
	             shows that the $E_8$ part of the phase 	             
	             \eqref{Phi}
	             reduces to $\Phi_a=(-1)^{f(a)}$, where $f(a)$ is the mod 2 index of the Dirac operator.
	             Here $a$ is the class of the $E_8$ bundle.  
The breaking of this bundle on $Y^{11}$ via $E_8 \supset (SU(5) \times SU(5))/\Z_5$ allows for relating 
to complex bundles on $X^{10}$, and this is extended to K-theory.

\paragraph{The mod 2 index from twelve dimensions}
The K-theoretic partition function in \cite{DMW} depended on 
a mod 2 index with values in ${KO}(X^{10})$. The ten-dimensional
description of the KO-theoretic class is as the tensor product 
$V \otimes \overline{V}$, where $V$ is a complex vector bundle 
and $\overline{V}$ its conjugate, corresponding in the homological 
setting, respectively, to D-branes and anti-D-branes. The mod 2 index
is just ${\rm Ind}(D_{V \otimes \overline{V}})$ mod 2, which is 
a topological invariant in ten dimensions \cite{AS5}. 

\vspace{3mm}
We provide a description of the mod 2 index entering the partition function 
using twelve-manifolds which are closed, i.e. without the use of the lift to M-theory.
Let $M^{12}$ be a {\it closed} Spin${}^c$ twelve-manifold. Our $X^{10}$ will
be considered as a submanifold of $M^{12}$ as follows. Consider the projection
${\rm Spin}^c(12) \to SO(2)$ given by 
the determinant (\ref{det}). At the level of bundles we get the determinant bundle of 
$M^{12}$. The ten-dimensional manifold 
$X^{10}$ is taken to be a (codimension two) 
submanifold dual to the determinant bundle with the inclusion being 
$i : X^{10} \to M^{12}$. Since the dimension of $X^{10}$ is 
divisible by 2 mod 8, then we can apply a result of \cite{FO}. For every 
$\psi \in KO(M^{12})$, the KO-characteristic number 
\(
\langle i^* \psi , [X^{10}] \rangle _{KO} \in \Z_2 
\) 
is the reduction mod 2 of the index in twelve dimensions
\(
\left\langle 
{\rm Ph}(\psi) \exp (e/2) \widehat{A}(TM^{12}) , [M^{12}]  
\right\rangle \in \Z\;,
\)
where Ph is the Pontrjagin character (composition of realification with the
Chern character) and $e \in H^2(M^{12};\Z)$ is the Euler class of the determinant 
bundle. We can then take $E \otimes \overline{E} = i^* \psi$.



\subsubsection{The $\widehat{A}$-genus}
\label{a genu}
The $\widehat{A}$-genus is usually defined for manifolds of dimension
$4k$. However, there is a lift to KO-theory, which allows an extension to 
dimensions $8k+1$ and $8k+2$. Thus, for us, this genus will be important 
in dimensions 1, 2, 9, and 10.  Furthermore, the value of the 
$\widehat{A}$-genus in these dimensions will in general depend on the 
Spin structure chosen, whenever the manifold admits more than one
Spin structure.

\vspace{3mm}
Let $D=D(M, \sigma)$ be the Dirac operator defined by a Spin structure $\sigma$ on a closed
manifold $M$ of dimension $m$. Then, following \cite{GLP}, we consider
the following cases.
\begin{enumerate}
\item Suppose that $m\equiv 0$ mod 4. Decompose $D=D^+ + D^{-}$ into the chiral Dirac 
operators and define $\widehat{A}(M, \sigma)={\rm ind} (D^+)$. Inequivalent Spin structures are
twisted by flat real line bundles (as in section \ref{spinpie}). 
This does not affect the index density and hence
$
 \widehat{A}(M, \sigma)= \widehat{A}(M)
$ 
is independent of the Spin structure. 
Furthermore, ${\rm dim~ker} (D)$ is even in this case.

\item Suppose that $m \equiv 1$ mod 8. Let  $\widehat{A}(M, \sigma) \in \Z_2$ be the mod
2 reduction of dim ker$(\widehat{A}(M, \sigma))$.

\item Suppose  $m\equiv 2$ mod 8. The index of the Spin complex is zero. Therefore,
dim ker$(\widehat{A}(M, \sigma))=2 {\rm dim}(D^+(M,s))$ is even. Let
 $\widehat{A}(M, \sigma)\in \Z_2$ be the mod 2 reduction of 
$\frac{1}{2} {\rm dim~ ker}(\widehat{A}(M, \sigma))$.

\item Set $\widehat{A}(M, \sigma)=0$ otherwise.
\end{enumerate}
\noindent The first case is the `usual' $\widehat{A}$-genus, while the second and third
are $\alpha$-invariants (which will be discussed in more detail in section \ref{alp in}).
Hence, the above can also be called {\it generalized} $\widehat{A}$-genera.

\vspace{3mm}
We start with examples of the first case and then recall the examples from section \ref{multi spin}.

\paragraph{Example 1: Hypersurfaces in projective space.}
Let $V^{n}(d)$ denote the nonsingular complex hypersurface of degree $d$ in 
$\C P^n$. In homogeneous coordinates $[Z_0, \cdots, Z_{n+1}]$, $V^{n}(d)$ is given as
the zeros of a homogeneous polynomial $P(Z_0, \cdots, Z_{n+1})$ of degree $d$. 
The diffeomorphism class of $V^{n}(d)$ is uniquely determined by the integers $n$ and $d$.
The first Chern class of $V^{n}(d)$ for $n>1$ is $c_1= (n+2-d)x$, 
where $x$ is the canonical generator of $H^2(V^{n}(d);\Z)$, i.e. the K\"ahler form induced 
from $\C P^{n+1}$. The Spin condition means that $w_2$ is the mod 2 reduction 
of the integral class $c_1$, so that in order for $V^{n}(d)$ to be Spin, $c_1$ should be even.
This is equivalent to saying that $n+d$ should be even.
When $n=2m$ is even then \cite{LM}
\(
\widehat{A} (V^{2m}(d))=\frac{2^{-2m}d}{(2m+1)!}\prod_{k=1}^m 
\left( d^2 - (2k)^2\right)\;.
\)
Thus, each of the Spin manifolds $V^{2m}(2d)$, for $d>n$, 
has nonzero $\widehat{A}$. 
\begin{itemize}
\item For $m=1$, i.e. for $\C P^3$:
$
\widehat{A} (V^2(2d))=\frac{d}{3}(d^2-1)
$.

\item For $m=2$, i.e. for $\C P^5$:
$
\widehat{A} (V^4(2d))=\frac{d}{60}(d^2-1)(d^2-4)
$.
\item In general
$
\widehat{A} (V^{2k}(2k+2))=2
$.
\end{itemize}

\paragraph{Example 2: Bott manifold.}
Let $b$ be a generator of $KO_8({\rm pt})=KO_8 (\R)\cong \Z$. One can find 
a simply-connected manifold $B^8$ of dimension 8
 which represents Bott periodicity in $KO_*$,
 with $\alpha (B^8)=b$ (see Section \ref{alp in}). 
There are many possible choices for $B^8$. One such is the 
Bott manifold, which is a simply-connected Spin manifold with $\widehat{A}(B^8)=1$ \cite{RS}.
An example of this is a Joyce manifold with ${\rm Spin}(7)$ holonomy. 
Such examples are important in compactifications of  
 M-theory (see e.g. \cite{AG} for a survey).

\vspace{3mm}
We now consider examples of the first and second cases. Here, 
dependence on the Spin structure emerges. This discussion will be naturally
continued in section \ref{alp in}.

\paragraph{ Example 3: The circle $M=S^1$.} Let $\theta$ be the usual periodic parameter 
on the circle, with $\partial_\theta$ providing a 
global trivialization of the tangent bundle $TS^1$ and defining a Spin structure
$\sigma_1$. 
The associated Spin bundle is trivial and the Dirac operator is 
$D=-i \partial_\theta$. Thus dim ker$(D)=1$ and 
$
\widehat{A}(S^1, s\sigma_1)=1$.
There is another Spin structure $\sigma_2$ which is defined by twisting $\sigma_1$
with the M\"obius bundle. There are no harmonic spinors for $\sigma_2$ and 
$
\widehat{A}(S^1, \sigma_2)=0$.

\vspace{3mm}
Note the following {\it terminology}: The $\widehat{A}$-genus is a bordism invariant (as will
be indicated in the properties below). 
The first Spin structure $\sigma_1$ does not bound.  
The second Spin structure $\sigma_2$ is induced by regarding the circle
as the boundary of the disk. This second Spin structure, corresponding to 
the nontrivial Spin bundle, is called the ``trivial" structure since it bounds.

\paragraph{Example 4: The torus $M=T^2$.}  The two-dimensional 
torus $T^2=S^1 \times S^1$ admits 4 Spin structures
$\sigma_i$, $i=1, \cdots, 4$. The product Spin structure is $\sigma_0$ gives the Spin
bundle $T^2 \times \C^2$. This has ${\rm dim~ker}(D)=2$, so that
$
\widehat{A}(T^2, \sigma_0)=1
$.
For the other three Spin structures on $T^2$ we have
$
\widehat{A}(T^2, \sigma_j)=0$, for $j=1,2,3$.

\paragraph{Some properties of the $\widehat{A}$-genus.}
The $\widehat{A}$-genus satisfies the following properties. 
\begin{enumerate}
\item {\it Multiplicative behavior of the $\widehat{A}$-genus.} 
Let $N$ be a closed Spin manifold of dimension $n=4k$ and let $M$
be a closed Spin manifold of dimension $m$. Then
\(
\widehat{A} (N \times M)= \widehat{A} (N) \cdot \widehat{A} (M)\;.
\)
\item{\it Relation to the eta-invariant.}
If $\varrho \in RU(\pi_1(M))$ and if $M$ is odd-dimensional 
then, from Atiyah-Patodi-Singer \cite{APSI},
we get
\(
\eta (N \times M)(\varrho)=\widehat{A}(N)\cdot \eta(M) (\varrho)\;.
\)
For example, for the dimensions of interest to us we have: 
\begin{eqnarray}
\widehat{A}(X^{10})&=& \widehat{A}(M^{8}) \cdot \widehat{A}(N^{2}) 
\nonumber\\
\widehat{A}(X^{10})&=& \widehat{A}(M^{4}) \cdot \widehat{A}(Y^{4}) \cdot \widehat{A}(N^{2})
\nonumber\\
\eta (N^4 \times M^7)  (\varrho)&=& \widehat{A}(N^8) \cdot \eta (M^3)(\varrho)
\nonumber\\
\eta (N^4 \times M^7)  (\varrho)&=& \widehat{A}(N^4) \cdot \eta (M^7)(\varrho)\;.
\end{eqnarray}
This allows us to extend the evaluation of the eta invariant 
in eleven dimensions to many cases in which the eleven-dimensional manifold is
reducible. 
\item {\it $\widehat{A}$ is a bordism invariant.}
If $M$ is the boundary of a compact Spin manifold $N$
then $\widehat{A}(M, \sigma)=0$. Furthermore, the $\widehat{A}$ genus is independent of 
the Riemannian metric.
\item {\it Behavior in fiber bundles.}
The $\widehat{A}$-genus vanishes on any smooth fiber bundle of closed
oriented manifolds provided that the fiber is a Spin manifold and the 
structure group is a compact connected Lie group which acts smoothly
and non-trivially on the fiber \cite{De08}.
\end{enumerate}

\subsubsection{The $\alpha$-invariant}
\label{alp in}

We saw in the previous section that the definition of the $\widehat{A}$-genus 
can be extended beyond manifolds of dimension $4k$. We now consider 
in more detail how that is done.
The $\widehat{A}$-genus, being the index of an elliptic operator, 
admits a lift to some $K$-group. Indeed, Milnor found a surjective 
homomorphism $\alpha$ from the Spin cobordism ring 
$\Omega_*^{\rm Spin}$ onto $KO^{-*}({\rm pt})$ such that 
$\alpha (M)=\widehat{A}(M)$ if $n=8m$, and which 
captures additional $\Z_2$-information in dimensions
$n \equiv 1,2$ mod 8.
The $\alpha$-invariant associates to any $n$-dimensional Spin 
manifold $(M, \sigma)$ an element $\alpha(M, \sigma)\in KO^{-n}({\rm pt})$,
where 
\(
KO^{-n}({\rm pt}) \cong 
\left\{
\begin{tabular}{ll}
$\Z$ & if $n$ is divisible by 4
\\
$\Z_2$ & if $n \equiv 1,2$ mod 8\\
0& otherwise.
\end{tabular}
\right.
\) 
The map $\alpha$ defines a surjective ring homomorphism from the Spin 
cobordism ring $\Omega^{\rm Spin}_*$ to $\bigoplus_{k\in \N}KO^{-n}({\rm pt})$.

\paragraph{The index of the Dirac operator.}
The Dirac operator on $X^m$ has an index in $KO^{-m}({\rm pt})$, given by 
$f_!(1)$, where $f: X\to {\rm pt}$ is the collapsing map. Depending on 
$m$, one has \cite{AS69}
\(
\begin{tabular}{lll}
$KO^{-8m}({\rm pt})\cong \Z$ && $f_!(1)=\widehat{A}(X)$
\\
$KO^{-(8m+4)}({\rm pt})\cong \Z$ && $f_!(1)=\frac{1}{2}\widehat{A}(X)$ 
\\
$KO^{-(8m+1)}({\rm pt})\cong \Z_2$ && $f_!(1)=h_D$ mod 2
\\
$KO^{-(8m+2)}({\rm pt})\cong \Z_2$ && $f_!(1)=h_{D^+}$ mod 2
\end{tabular}
\)
Define $\alpha(X)=f_!(1) \in KO^{-n}(X)$ for a Spin manifold 
$X$ of dimension $n$. Then 
\(
\alpha(X \times Y)=\alpha(X) \cdot \alpha(Y)\;.
\)
Thus, if $\alpha(M,\sigma)\neq0$ then the Dirac operator has 
a nontrivial kernel.

\paragraph{Some properties of $\alpha$.}
\begin{enumerate}


\item The Spin cobordism class of a manifold is completely determined by its 
Stiefel-Whitney and $KO$-characteristic numbers \cite{ABP}. A fundamental such 
$KO$-invariant is the $\alpha$-invariant. 

\item The homomorphism $\alpha$ is an isomorphism for $n \leq 7$. 
Thus it is not an isomorphism for $n=8, 9, 10$. 

\item Let the fundamental group of a connected compact Spin manifold 
$M$ of dimension
$m \geq 5$ be a spherical space form group. Then $M$ 
admits a metric of positive scalar curvature iff $\alpha (M)=0$ \cite{BGS}.

\end{enumerate}

\paragraph{Example: $\alpha$-invariant of a Riemann surface.}
In the case of two dimensions, the $\alpha$-invariant is 
\(
\alpha(\Sigma_g) \in KO^{-2}({\rm pt})=\Z_2=\{0,1\}\;.
\)
Associate to any Spin structure $\sigma$ on a Riemann surface $\Sigma_g$
a quadratic
function $q_\sigma: H_1(\Sigma_g; \Z_2)\to \Z_2$. Let $V$ be a finite-dimensional 
$\Z_2$-vector space. 
For any quadratic map $q_\sigma: V \to \Z_2$ associated to a nondegenerate symmetric 
bilinear form one defines the Arf invariant 
\(
{\rm Arf}(q):=\frac{1}{\sqrt{\# V}} \sum_{\alpha \in V} (-1)^{q_{\sigma}(\alpha)}\;,
\)
which has values $\pm$, by virtue of quadraticity of $q$. 
For a compact surface of genus $g$, there are
\begin{itemize}
\item $2^{2g-1}+2^{g-1}$ Spin structures for which the Arf invariant of $q_\sigma$ is $+1$.
\item $2^{2g-1}-2^{g-1}$ Spin structures for which the Arf invariant of $q_\sigma$ is $-1$.
\end{itemize}
An alternative interpretation of the Arf invariant is 
$
{\rm Arf}(q_\sigma)= (-1)^{\dim \mathcal{H}_s}
$,
where $\mathcal{H}_s$ is the space of holomorphic sections of the Spin bundle $S(\Sigma_g)$
\cite{ACGH}.
The $\alpha$-invariant  $\alpha(\Sigma_g ,\sigma)$
can be defined via ${\rm Arf}(q_{\sigma})=(-1)^{\alpha(\Sigma_g,\sigma)}$
\cite{AHu}.
The index theorem gives that for any compact Riemann surface $\Sigma_g$
with Riemannian 
metric $g$ and Spin structure $\sigma$ 
\(
\alpha(\Sigma_g, \sigma) =\frac{1}{2}\dim_\C \ker D \quad {\rm mod}~2.
\)

{\bf  Case genus $g=0$: The two-sphere $S^2$.} $\Sigma_0=S^2$ with Spin structure $\sigma_0$. Then 
$q_{\sigma_0}:\{0 \} \to \Z_2$, $q_{\sigma_0}(0)=0$, and the 
$\alpha$-invariant vanishes: $\alpha (S^2,\sigma_0)=0$.

\vspace{3mm}
{\bf Case genus $g=1$: The torus.} $\Sigma_1=T^2=\R^2/\Gamma$ with a Euclidean metric for a lattice 
$\Gamma \subset \R^2$, $\Gamma = \pi_1(\Sigma_1)=H_1(\Sigma_1;\Z)$. 
Any group homomorphism $\gamma: \Gamma \to \Z_2\subset \ker ({\rm Spin}(2)
\to SO(2)) \subset {\rm Spin}(2)=S^1$ defines an orthogonal action of $\Gamma$
on $\R^2 \times {\rm Spin}(2)$. A ${\rm Spin}(2)$ principal bundle is defined by 
factoring the $\Gamma$-action 
\(
P_{\rm Spin}(\Sigma_1, g)_\Gamma:=\R^2 \times_\gamma {\rm Spin}(2)\;.
\label{spin sig 1}
\)
A Spin structure $\sigma_\gamma$ on $\Sigma_1$ is defined via
the principal bundle (\ref{spin sig 1}) and the natural map
\begin{eqnarray}
\sigma_\gamma: P_{\rm Spin}(\Sigma_1, g)=\R^2 \times_\gamma {\rm Spin}(2) &\longrightarrow &
P_{SO}(\Sigma_1, g)=(\R^2/\Gamma) \times SO(2)
\nonumber\\
\left[ (x,z)\right]_\gamma &\longmapsto & 
(x + \Gamma, z^2)\;.
\end{eqnarray}
Two Spin structures $\sigma_{\gamma_1}$ and $\sigma_{\gamma_2}$ are equivalent 
if and only if $\gamma_1=\gamma_2$. 
Let $\overline{v}$ denote the image of 
$v\in \Gamma$ in $H_1(\Sigma_1;\Z_2)=\Gamma\otimes_\Z \Z_2$. The quadratic form 
$q$ of the Spin structure associated to $\gamma$ satisfies
\(
q(\overline{v})\cong
\left\{
\begin{tabular}{ll}
$\gamma(v) +1$ & for $\overline{v}\neq 0$
\\
$\gamma(v)=0$ & for $\overline{v}=0$,
\end{tabular}
\right.
\)
so that 
\(
\alpha(\Sigma_1, \sigma)=
\left\{
\begin{tabular}{ll}
1 & if $\sigma$ is the Spin structure associated to the trivial map $\gamma$,
\\
0 & otherwise. 
\end{tabular}
\right.
\)
One can be more explicit in this, using elliptic curves \cite{KuS}.
There are 4 Spin structures on the torus. Let $\C / \{ 2\omega_1, 2\omega_3\}={\rm Jac} (T^2)$
be the Jacobian for $T^2$, and let $e_i=\wp (\omega_i)$, $(i=1,2,3)$, where $\omega_2=\omega_1 + \omega_3$.
Then $h(u)=(\wp (u), \wp' (u))$ is a conformal diffeomorphism from the Jacobian to the Riemann surface
$\Sigma_1$ defined by 
\(
\omega^2=4(z-e_1)(z-e_2)(z-e_3)\;.
\)
Then the four distinct Spin structures are defined by the four differentials
\begin{eqnarray}
du &=& dz/\omega 
\nonumber\\
(\wp (u) - e_i)du &=& (z-e_i) dz/\omega\;.
\label{4diffs} 
\end{eqnarray}
With $a_i$ the generator of $H_1(T^2;\Z_2)$ defined by 
$a_i: [0,1]\to {\rm Jac}(T^2)$, $a_i(t)=2t\omega_i$, 
then the values of the Arf invariant for the last three of the
four Spin structures in (\ref{4diffs}) are 1, whereas, for 
$du$:
\(
q_\sigma (0)=0, \qquad q_\sigma (a_i)=+1 \quad {\rm for~} i=1,2,3\;,
\)
so that Arf$(q_\sigma)=-1$. This gives an $\alpha$-invariant which 
is odd in this case, $\alpha (T^2, \sigma_4)=1$.

\subsubsection{Ten-dimensional manifolds: products and bundles}
\label{pr bu}

\paragraph{Ten-dimensional manifolds which are products.}
Manifolds with free $S^1$-action and nonvanishing $\alpha$-invariant 
can be obtained by taking the product of 
$S^1$, having the nontrivial Spin structure,
with any 8-dimensional 
Spin manifold with an odd $\widehat{A}$-genus \cite{Mil2}.
(Simply connected examples are given in \cite{LY}).
Thus we have the following examples of (reducible) ten-dimensional manifolds for 
which the $\alpha$-invariant is possibly nonzero:
\begin{eqnarray}
\alpha(X^{10}) &=& h_{D^+} ~ {\rm mod}~2:=r_2(h_{D^+})
\nonumber\\
\alpha(X^9 \times S^1)&=& \alpha(X^9) \alpha(S^1)= (h_D(X^9) ~{\rm mod~} 2) (h_D(S^1)~{\rm mod~}2)
\nonumber\\
&=&r_2(h_D(X^9)) r_2(h_D(S^1))
\nonumber\\
\alpha(X^8 \times \Sigma_{g }) &=& \widehat{A}(X^8) (h_{D^+}(\Sigma)~{\rm mod~} 2)
=\widehat{A}(X^8) r_2(h_{D^+}(\Sigma))
\nonumber\\
\alpha(X^4 \times Y^4 \times \Sigma_{g})
&=&
\frac{1}{4} \widehat{A}(X^4) \widehat{A}(Y^4) r_2(h_{D^+}(\Sigma))
\nonumber\\
\alpha(X^4 \times \Sigma_{g_1} \times \Sigma_{g_2} \times \Sigma_{g_3})
&=&
\frac{1}{2} \widehat{A}(X^4) 
r_2(h_{D^+}(\Sigma_{g_1}))
r_2(h_{D^+}(\Sigma_{g_2}))
r_2(h_{D^+}(\Sigma_{g_3}))
\nonumber\\
\alpha( \Sigma_{g_1} \times \Sigma_{g_2} \times \Sigma_{g_3}
 \times \Sigma_{g_4} \times \Sigma_{g_5}
)
&=&
r_2(h_{D^+}(\Sigma_{g_1}))
r_2(h_{D^+}(\Sigma_{g_2}))
r_2(h_{D^+}(\Sigma_{g_3}))
r_2(h_{D^+}(\Sigma_{g_4}))
r_2(h_{D^+}(\Sigma_{g_5}))\;.
\nonumber
\end{eqnarray}
In addition, we can have factors with products of circles 
\(
\alpha \left( M^{4n} \times \prod_{i=1}^{10-4n} (S^1)^i \right)
= \epsilon' \widehat{A}(M^{4n}) \prod_{i=1}^{10-4n}
(h_D(S_i^1)~{\rm mod~} 2)^i\;,
\)
where $\epsilon'$ equals $\frac{1}{2}$ for $n=4$ and equals 
1 for $n=8$, and where $S_i^1$ denotes the $i$th circle. 

\vspace{3mm}
For example, for $X^{10}=M^8 \times (S^1)^2$, where 
$S^1$ is the circle with the nonbounding Spin structure, 
$\alpha (X^{10})=0$ is equivalent to the condition 
$\widehat{A}(M^8)\equiv$ 0 mod 2. Examples of $M^8$ are
the quaternionic projective plane $\H P^2$, for which 
$\widehat{A}(\H P^2)_8=0$. Thus $\alpha (\H P^2 \times S^1 \times S^1) = 0$.
Similarly for the Milnor manifold $M_0^8$ \cite{Mil2}, which  is a Spin 
manifold with $\widehat{A}(M_0^8)=1$.


\paragraph{Ten-dimensional manifolds which are bundles.}
 Let
$
0 \to V_1 \to V_2 \to V_3 \to 0
$
be a short exact sequence of real vector bundles. let $\{ i, j, k\}$
be a permutation of $\{ 1,2,3\}$. If $V_i$ and $V_j$ are Spin
 then $V_k$ has a natural Spin structure. Thus consider the 
 following situation. 
Let $F^n \to X^{10} \buildrel{p}\over{\to}M^{10-n}$ 
be a differentiable fiber 
bundle with Spin base and total space. The
Spin structures on the base and the total space
induce a Spin structure on the tangent bundle along the fibers
$TF^n$. For $f: M^{10-n} \to {\rm pt}$, the $\alpha$-invariant of the
ten-dimensional manifold is defined as
\(
\alpha (X^{10})= (f_p)_! (1)=f_! (p_!(1)) \in KO^{-10}({\rm pt})\cong \Z_2\;,
\) 
where the direct image homomorphisms are taken relative to the 
Spin structures on the base and the fiber. This means that if $p_!(1)=0$, then
$\alpha(X^{10})=0$. Thus, we have examples with nonzero $\alpha$ 
(cf. \cite{Hit})
\begin{eqnarray}
F^9 &\to & X^{10} \to S^1
\nonumber\\
F^8 &\to & X^{10} \to \Sigma_g\;.
\end{eqnarray}
The fiber in the each of the above bundles admits harmonic spinors with respect to some 
metric.

\paragraph{Cobordism generators for ten-dimensional manifolds which are bundles with $\alpha=0$.}
Recall from section \ref{cobo} that one of the two generators of 
$\Omega_8^{\rm spin}$ is the quaternionic projective plane
$\H P^2$.
The space ker~$\alpha$ can be represented by total spaces of bundles with 
fibers given as $\H P^2$ \cite{St92}. Note that
\begin{eqnarray}
\ker \alpha &=& 0 \qquad {\rm in~dim~} \leq 7
\nonumber\\
\ker \alpha &=& \Z \qquad {\rm in~dim~} 8\;,
\end{eqnarray}
where the generator is represented by $\H P^2$. 
Any closed
simply connected spin manifold $M$ with $\alpha(M) = 0$ is 
Spin bordant to the total
space of a bundle with $\H P^2$ as fiber and 
structural group Isom$(\H P^2)=P Sp(3)=Sp(3)/\Z_2$, the group of isometries of 
$\H P^2$ \cite{St92}. Given such a fiber bundle,
one can use the O'Neill formulas to produce a metric of positive scalar curvature on
the total space from the standard positive scalar curvature metric on $\H P^2$ and an
arbitrary Riemannian metric on the base. Note that the base of such a fiber bundle
is compact, so one can squeeze the fibers so that at each point the scalar curvature
contribution coming from the fiber dominates the contribution coming from the
base. Thus any $\H P^2$-bundle can be equipped with a metric of
positive scalar curvature on its total space.
Let $G=PSp(3)$ and $H=P(Sp(2)\times Sp(1))$. 
Given a manifold $X$ and a map $f: X \to BG$, let $\widehat{X} \to X$
be the pullback of the fiber bundle 
\(
\H P^2 
\to
 G/H 
 \to
  BH 
  \buildrel{\pi}\over{\longrightarrow}  BG
\)
via $f$. 
Consider the homomorphism $\Psi: \Omega_2^{\rm Spin}\to \Omega_{10}^{\rm Spin}$
by mapping the bordism class of $f$ to the bordism class of $\widehat{X}$. 
Then \cite{St92}
\(
\ker \alpha = {\rm Im} \Psi\;,
\)
that is, if $X$ is a closed Spin manifold with $\alpha(X)=0$ then $X$ is 
Spin cobordant to the total space of a fiber bundle with fiber $\H P^2$ and 
structure group the projective symplectic group ${\rm PSp}(3)$. Thus, the kernel
of $\alpha$ forms a subgroup $T_{10}$ consisting of bordism classes represented 
by total spaces of $\H P^2$ bundles.

\vspace{3mm}
Any Spin manifold of dimension $\geq 8$ with $\widehat{A}(M)=0$
is rationally cobordant to an $\H P^2$-bundle with 
nontrivial $S^3$-action along the fibers. 
Kreck-Stolz show the much deeper result that any Spin
manifold $M$ of dimension $\geq 8$ with $\alpha (M)=0$ is {\it integrally}
cobordant to the total space of an 
$\H P^2$-bundle with structure group $PSp(3)$ \cite{KS93}.
There is a canonical isomorphism 
\(
KO_n(X)\cong 
\left( 
\bigoplus_k \Omega_{n+8k}^{\rm spin} (X) 
\right)/ \sim\;,
\)
where the equivalence is generated by 
\begin{enumerate}
\item $[E, f \circ p] \sim 0$ if $p: E \to M$ is an $\mathbb{H}P^2$ bundle over a Spin
manifold $M$ and $f$ is a map from $M$ to $X$. 
\item $[M, f]\sim [M \times B, f \circ {\rm pr}_1]$, where $B$ is the Bott manifold 
and ${\rm pr}_1$ denotes the projection to the first factor.
\end{enumerate}

\vspace{3mm} Given a space $M$, let $T_*(M)$ be the subgroup of
$\Omega_*^{\rm Spin}(M)$ represented by pairs $(X^{10}, f\circ p)$
where $X^{10} \to \Sigma_g$ is an $\H P^2$ bundle and $f: \Sigma_g \to M$ is 
a map. Let 
$b\in \Omega_8^{\rm Spin}({\rm pt})/T_8({\rm pt})\cong {KO}_8({\rm pt})\cong \Z$
be the generator, i.e., the class represented by the Bott manifold. Then the 
homomorphism $\rho \circ D_*: \Omega_*^{\rm Spin}(M) \to {KO}_*(M)$ 
induces an isomorphism 
\(
\Omega_*^{\rm Spin}(M)/T_*(M)[b^{-1}] \cong {KO}_*(M)\;.
\)
%
%
This makes a connection between 2 and 10 dimensions.
In particular, we can take a type IIA string wrapping a Riemann surface,
which we take as the base of the above fibration. 

\paragraph{Eleven-dimensional manifolds which are total spaces of $\H P^2$ bundles.}
Consider a Spin eleven-dimensional manifold $Y^{11}$ which is an $\H P^2$ bundle 
$\pi$ over a Spin three-manifold $M^3$. Let there be a map from $Y^{11}$ 
to some other space (e.g. classifying space) $\mathcal{M}$ which 
factors through $M^3$, so that we have the following diagram
\(
\xymatrix{
\H P^2 
\ar[r]
&
Y^{11} 
\ar[d]_\pi
\ar[drr]
&&
\\
&
M^3 
\ar[rr]^{\hspace{-3mm} f}
&&
\mathcal{M}
}
\)
Let $T_{11} (\mathcal{M})$ be the subgroup of $\Omega_{11}^{\rm, spin}(\mathcal{M})$ 
consisting of bordism classes $[Y^{11}, f \circ \pi]$. 
Let $\widetilde{T}_{11}(\mathcal{M})$ be 
the subgroup with the additional assumption that
$[M^3, f]=0 \in \Omega_{3}^{\rm Spin} (\mathcal{M})$.

\begin{itemize}

\item {\bf Case $\mathcal{M}={\rm pt}$:} Here we have $\Omega_{11}^{\rm Spin}({\rm pt})$,
which is zero. So there are no obstructions in
this case.

\item {\bf Case $\mathcal{M}=BE_8$:} We have an $E_8$ bundle on $Y^{11}$ 
which we are asking to extend to twelve dimensions. However, since 
by a result of Stong, $\Omega_{11}^{\rm Spin}(K(\Z, 4))=0$, the above subgroups 
will be trivial and so there is not much to do in this case. 

\item {\bf Case $\mathcal{M}=B\pi_1(Y^{11})$:}
Since the fiber $\H P^2$ is simply connected, then $\pi_1(Y^{11})$ and 
$\pi_1(M^3)$ are isomorphic. This implies that there is indeed a map 
$f: M^3 \to B\pi_1(Y^{11})=B\pi_1(M^3)$, which is the classifying map
for the fundamental group. 
\end{itemize}
As in the ten-dimensional case, this provides a link between dimension
eleven (M-theory) and three (which we can take to be where the membrane 
wraps or lives).

\paragraph{Ten-dimensional manifolds which are total spaces of $\H P^2$ bundles.}
Consider a Spin ten-dimensional manifold $M^{10}$ which is an $\H P^2$ bundle 
$\pi$ over an oriented Riemann surface  $\Sigma_g$. 
We take a map from $M^{10}$ to $\mathcal{M}$ which 
factors through $\Sigma_g$, so that we have the following diagram
\(
\xymatrix{
\H P^2 
\ar[r]
&
M^{10} 
\ar[d]_\pi
\ar[drr]
&&
\\
&
\Sigma_g 
\ar[rr]^{\hspace{-3mm} f}
&&
\mathcal{M}
}
\)
Let $T_{10} (\mathcal{M})$ and $\widetilde{T}_{10}(\mathcal{M})$
be the subgroups as in the eleven-dimensional case.  
We consider ten-dimensional Spin cobordism 
in the setting where M-theory is taken on an eleven-dimensional 
Spin manifold $Y^{11}$ with boundary $M^{10}=\partial Y^{11}$
as in \cite{DFM}. 

\begin{itemize}

\item {\bf Case $\mathcal{M}={\rm pt}$:} Here we have $\Omega_{10}^{\rm Spin}({\rm pt})=\Z_2 
\oplus \Z_2 \oplus \Z_2$. The $\alpha$-invariant has this as a domain. 
Here $T_{10}({\rm pt})=\Z_2 \oplus \Z_2$.

\item {\bf Case $\mathcal{M}=BE_8$:} We have an $E_8$ bundle on $X^{10}$ 
which we wish to extend to eleven dimensions. Unlike the case $n=11$, here
 $\Omega_{10}^{\rm Spin}(K(\Z, 4))=\Z_2 \oplus \Z_2$.
 
\item {\bf Case $\mathcal{M}=B\pi_1(M^{10})$:}
Since the fiber $\H P^2$ is simply connected, then $\pi_1(M^{10})$ and 
$\pi_1(\Sigma_g)$ are isomorphic. This implies that there is indeed a map 
$f: \Sigma_g \to B\pi_1(X^{10})=B\pi_1(\Sigma_g)$, which is the classifying map
for the fundamental group. 
\end{itemize}

\paragraph{Homotopy spheres.} Let $\Sigma^n$ be an
 $n$-dimensional homotopy sphere, i.e.
a compact differentiable manifold which is homotopy equivalent to the $n$-sphere $S^n$. 
Then $\Sigma^n$ is cobordant to zero but not necessarily Spin cobordant to zero. 
In fact, the homotopy $n$-spheres form an abelian group $\Theta_n$ under the operation 
of connected sum, and 
\(
\alpha : \Theta_n \to KO^{-n}({\rm pt})
\)
is a homomorphism. For $n\equiv 1$ or 2 (mod 8) and $n>8$, the homomorphism
$
\alpha : \Theta_n \to \Z_2
$
is surjective (see \cite{LM}).
This means that we can find exotic spheres in dimensions nine and ten for which the 
alpha invariant is $1$, the nontrivial element in $\Z_2$. 
This can serve as a source of an anomaly for the partition function.

\paragraph{Exotic spheres.}
Among the manifolds for which $\alpha$ is nonzero is half the exotic spheres 
in dimensions 9 and 10. 
In these dimensions every compact Spin manifold is
homeomorphic to a manifold which does not carry a positive scalar curvature. This can 
be seen as follows (see \cite{LM}). Let $X$ be a Spin $k$-manifold with $k=9$ or 10, and let
$\Sigma$ be an exotic $k$-sphere with $\alpha (\Sigma)\neq 0$. Since 
$\alpha (X \# \Sigma)= \alpha (X) + \alpha (\Sigma)$, then
$X$ and $X \# \Sigma$ 
are compact Spin manifolds with positive scalar curvature, so that their 
$\alpha$-invariant vanishes.
Thus {\it every} Spin manifold of dimension 9 or 10 is homeomorphic to one 
with nontrivial $\alpha$-invariant. 
This means that there is an abundance of manifolds which could lead to an 
anomaly. 
 
 \vspace{3mm}
\noindent {\it  Which exotic spheres have a non-trivial $\alpha$-invariant?}
The $\alpha$-invariant constitutes a surjective group homomorphism from
the group $\Theta_n$ of homotopy $n$-spheres (with the addition 
induced by the connected sum operation) onto the group $\Z_2$, provided 
$n \equiv 1,2$ mod 8 and $n \geq 9$. Roughly
speaking, Adams has shown \cite{Ad} that a non-trivial $\alpha$-invariant in
dimensions $n\equiv 1, 2$ mod 8 can always be realized through a stably framed closed
manifold, while Milnor has shown \cite{Mil} that one can alter
an accordingly framed manifold to a homotopy sphere by a sequence of surgeries
without changing its $\alpha$-invariant, provided $n \geq 9$ (and $n \equiv 1,2$ mod 8).

\paragraph{Interpretation of $\alpha$ in terms of connective real K-theory.}
The $\alpha$-invariant $\alpha (X)$ can be interpreted as the image 
of the class $[X]$ under a natural transfomation of generalized cohomology theories 
\cite{RS2}. Let $KO_*(X)$ and $ko_*(X)$ denote the periodic and 
connective real K-theory of a space $X$, respectively. Then there are 
natural transformations
\begin{eqnarray}
\Omega_*^{\rm Spin} 
&\buildrel{D}\over{\longrightarrow}& 
~{ko}_*(X)~~
\buildrel{\rm per}\over{\longrightarrow}~
{KO}_*(X)
\nonumber\\
\left[ X,f \right] &\longmapsto & f_*([X]_{ko}) 
\longmapsto 
{\rm per}\circ D(X)=\alpha (X)\;,
\label{ko}
\end{eqnarray}
where $[X]_{ko} \in {ko}_*(X)$ denotes the ko$-$fundamental class of $X$ determined
by the Spin structure. The natural transformation per induces an isomorphism 
${ko}_*({\rm pt})\cong {KO}_n({\rm pt})$ for 
$n\geq 0$.

\subsection{Types of metrics and boundary conditions}
\label{sec typ}
 We have considered M-theory on an eleven-dimensional manifold $Y^{11}$.
This is a total space of a circle bundle over $X^{10}$, in relating 
to type IIA string theory in ten dimensions. On the other hand, this 
is the boundary of a twelve-manifold $Z^{12}$, in studying 
the topological aspects of the partition function. Of course, we 
could use both at the same time and consider $Z^{12}$ as the total space
of a two-disk bundle over $X^{10}$. The task then is to 
study effects of the geometry, and in particular, the effects of choice of 
metrics on each one of the above spaces. Furthermore, one could ask
whether there are preferred metrics which are favored by both the 
fibration structure on $Y^{11}$ as well as by (a variation on)
the Atiyah-Patodi-Singer index theorem. We will make use of 
the results in \cite{HHM} on metrics and those in \cite{LMP}
on the corresponding index theorems.

\vspace{3mm}
Let $Z^{12}$ be a smooth twelve-dimensional Spin manifold 
with boundary $\partial Z^{12}=Y^{11}$. Let $z$ be a boundary 
function such that on the boundary we have $z=0$ and $dz\neq 0$.
Let $g_Y$ be a smooth metric on $Y^{11}$. Assume that both $Z^{12}$ and
$X^{10}$ are Spin, and fix a Spin structure on each. There is an 
induced structure on the fibers $\pi^*(X^{10}):=S^1_x \subset Y^{11}$. The
Spin bundles on $Z^{12}$, $Y^{11}$, and $S^1_x$ are denoted
$SZ=S^+(Z) \oplus S^{-}(Z)$, $SY$ and $S{S^1_x}$,
respectively. Let $E \to Z^{12}$ be a Hermitian 
complex vector bundle corresponding to an $E_8$ principal 
bundle endowed with a unitary connection. 
Fixing a metric $g_Z$ on $Z^{12}$, we have a twisted  Dirac operator 
\(
D_{g_Z}^+: C^\infty (Z^{12}, E\otimes S^+(Z)) \to
C^\infty (Z^{12}, E\otimes S^-(Z))\;.
\label{dir z 4}
\)
The boundary operator $D^Y$ induces a family of operators 
$\{D_x^Y\}_{x\in X}$, where each $D_x^Y$ acts on the space of sections
$C^\infty (\pi^{-1}(x), E \otimes S_x^Y)$. 
See section \ref{fam} for more on families.

\paragraph{Case $Z^{12}$ arbitrary.} Here we have the following two types of metrics:

\begin{enumerate}
\item {\it Exact $b$-metric} $g^b_Z$ on the interior of $Z^{12}$: This takes the
form
$g^b_Z=\frac{1}{z^2}dz^2 + g_Y$\; 
 in the neighborhood of the boundary. Setting $z=e^{-t}$ gives 
 the metric $dt^2 + g_Y$ on $\R^+\times \partial Z^{12}$, so that it 
 has cylindrical ends. The index theorem corresponding to the 
 Dirac operator $D_{g^b_Z}$, defined using the metric $g_Z^b$,
  is the Atiyah-Patodi-Singer
 index theorem. 
 
  \item {\it Exact cusp $c$-metric} or {\it scattering metric} $g^c_Z$ on $Z^{12}$:
$g^c_Z=\frac{1}{z^4}dz^2 + \frac{1}{z^2}g_Y$. Taking 
$z=\frac{1}{r}$ gives $dr^2 + r^2g_Y$ with $r\to \infty$, which is the standard 
metric of the infinite end of a cone and corresponds to the asymptotically 
locally Euclidean (ALE) class of gravitational 
instantons, such as the Eguchi-Hanson metric. This is thus appropriate for 
the Kaluza-Klein monopole in M-theory. The corresponding index theorem 
in this case, the index theorem corresponding 
to the Dirac operator $D_Z^c$, defined using the $c$-metric,
reduces to the Atiyah-Patodi-Singer index theorem in a simple way. 

\end{enumerate}

\paragraph{Case $\partial Z^{12}=Y^{11}$ is a fibration.}
Consider the case when the eleven-dimensional boundary 
is the total space of a fibration 
$\pi : Y^{11} \to B^n$, with $(11-n)$-dimensional fiber $F$.
In this case we have the index theorem relating 
$Z^{12}$ to $X^{10}$, i.e. respecting the 
fibration structure of $Y^{11}$.
The tangent bundle decomposes as
\begin{eqnarray}
T(\partial Z^{12})&=&T_V(\partial Z^{12}) \oplus 
T_H(\partial Z^{12})
\nonumber\\
&=& 
T(\partial Z^{12}/X^{10})\oplus \pi^*(TX^{10})\;.
\end{eqnarray}
Let $k_Y$ be a symmetric two-tensor on $\partial Z^{12}$ which 
restricts to a metric on each fiber $F$.  In this case, there are two natural types of metrics:

\begin{enumerate}

\item {\it Fibered boundary metric} $g^\Phi_Z$ on $Z^{12}$:
$g^\Phi_Z=\frac{1}{z^4}dz^2 + \frac{1}{z^2}\pi^*g_X+k_Y$.
Letting $z=1/r$ gives the metric $dr^2 + r^2 \pi^*g_X + k_Y$,
which is the ALF (asymptotically locally flat) and 
the ALG classes 
\footnote{The terminology ALG in the literature is meant to 
`mimic' ALE, that is alphabetically G is the letter right after F.}
of gravitational 
instantons such as Taub-NUT space. the index of the Dirac operator 
using this metric, from \cite{LMP}, is
\(
{\rm Ind}(D_\Phi^+)=\int_{Z^{12}} \widehat{A}(Z^{12}, g_Z^\Phi)
\wedge {\rm ch}E -\frac{1}{2}
\int_{X^{10}} \widehat{A}(X^{10}, g_X)\wedge \widehat{\eta}\;,
\)
 where $\widehat{\eta}\in \Omega^*(X^{10})$ is the Bismut-Cheeger
 eta form for the boundary family $(D_x^\partial)_{x\in X^{10}}$. 
Eta forms will be discussed in detail in section \ref{fam}.

 \item {\it Cusp fibered $d$-metric} $g^d_Z$ on $Z^{12}$:
$g^d_Z=\frac{1}{z^2}{dz^2} + \pi^*g_X+ z^2k_Y$.
Letting $z=e^{-t}$ gives the metric $dt^2+ \pi^*g_X
+ e^{-2t} k_Y$, which is the standard form for 
$\partial Z^{12}$ a torus bundle over a torus. 
The $d$-metric is related to the $\Phi$-metric by 
conformal transformations $g_Z^d=z^2g_Z^\Phi$, so that
$\widehat{A}(Z^{12}, g_Z^\Phi)=\widehat{A}(Z^{12}, g_Z^d)$ pointwise. 
Then, the results of \cite{Mor} \cite{Va} \cite{LMP} imply 
that the index for the $d$-metric coincides with that of the 
$\Phi$-metric
\(
{\rm Ind}(D_d^+)=\int_{Z^{12}} \widehat{A}(Z^{12}, g_Z^d)
\wedge {\rm ch}E -\frac{1}{2}
\int_{X^{10}} \widehat{A}(X^{10}, g_X)\wedge \widehat{\eta}\;.
\)

\end{enumerate}

\noindent The above results can be generalized to 
the case when $Y^{11}$ is a general fiber bundle 
$F^{11-n} \to Y^{11} \to B^{n}$.

\paragraph{Dependence of space of harmonic spinors on the metric in 10 and 11 dimensions.}
From \cite{Hit}, one can get examples of Spin $10$-manifolds for which the space of harmonic
spinors depends on the metric:
$S^7 \times S^3$. Of course we can also take quotients by finite groups
$S^7/\Gamma_1 \times S^3/ \Gamma_2$ and/or consider nontrivial 
bundles.  
In eleven dimensions one has $S^8 \times S^3$ and similarly their 
quotients and nontrivial bundles.

\section{Non-simply Connected Manifolds}
\label{nonsim}

\subsection{Non-simply connected ten-dimensional manifolds $\pi_1(X^{10})\neq 1$}
\label{pi 1 10}
If $X^{10}$ is a Spin manifold with nontrivial fundamental group then
we can consider a refinement to a cobordism class which takes 
$\Gamma=\pi_1(X^{10})$ into account. Consider maps 
$X^{10} \to B\Gamma \times B{\rm Spin}$.

\paragraph{$C^*$-algebras.}
 Let $\Omega_n^{\rm Spin}(B\Gamma)$ be the bordism group of closed Spin manifolds
$M$ of dimension $n$ with a reference map to $B\Gamma$. 
Let $C_r^*(\Gamma;\R)$ be the real reduced group $C^*$-algebra of $\Gamma$,
and let $KO_n(C_r^*(\Gamma;\R))$ be its topological K-theory.
 Given an element $[u: M \to B\Gamma]\in \Omega_n^{\rm Spin}(B\Gamma)$,
we can take the $C_r^*(\Gamma;\R)$-valued index of the equivariant Dirac
operator associated to the $\Gamma$-covering
$\widetilde{M} \to M$. 
Recall that a Bott manifold is any simply connected closed Spin manifold $B^8$ of
dimension 8 with $\widehat{A}$-genus $\widehat{A}(B^8)=1$. This manifold 
is not unique. $B^8$ geometrically represents Bott periodicity in 
KO-theory \cite{RS2}. 
 A generator is
$
{\rm Ind}_{C_r^* (\{1\};\R)} (B^8) \in KO_8(\R)\cong \Z
$
and the product with this element induces the Bott periodicity
isomorphism 
\(
KO_n(C_r^*(\Gamma;\R)) \buildrel{\cong}\over{\longrightarrow} KO_{n+8} (C_r^*(\Gamma;\R))\;.
\label{per cstar}
\)
 In particular
\(
{\rm Ind}_{C_r^* (\Gamma;\R)} (M)=
 {\rm Ind}_{C_r^* (\Gamma;\R)} (M\times B^8)
\)
if we use the identification (\ref{per cstar}) via Bott periodicity
(see e.g. \cite{KL}).
 Relevant examples for us are, as in section \ref{pr bu}, 
when $M$ is a Riemann surface $\Sigma_g$ or a Spin 
3-manifold $M^3$ for type IIA and M-theory, respectively. 
This again is a way to relate phenomena in 2 or 3 dimensions 
to ones in 10 or 11 dimensions, via a sort of dimensional 
reduction/lifting governed by Bott periodicity when the internal 
space is, for instance, an eight-manifold of Spin$(7)$ holonomy.

\paragraph{Flat vector bundles.} Consider the Mishchenko-Fomenko bundle \cite{MF}
\(
\mathcal{V}:= \widetilde{X} \times_\Gamma C^*\Gamma\;,
\)
which is a bundle of right $C^*\Gamma$-modules. 
This is a flat bundle, so that it has no effect on characteristic 
classes. 
Consider the Spin bundle $SX$ coupled to $\mathcal{V}$ via
$SX \otimes \mathcal{V}$, and form the 
space of sections $\Gamma (X;SX\otimes \mathcal{V})$, on 
which the Dirac operator acts on the left and 
$C^*\Gamma$ acts on the right, so that the two actions commute. 
This gives $\ker D$ as a $C^*\Gamma$-module, and 
$[\ker D]$ as a finitely generated projective module, and hence 
represents a class $[\ker D]\in KO(C^*\Gamma)$.

\vspace{3mm}
In general, the mod 2 index of the $C\ell$-linear Dirac operator 
twisted by the Mishchenko-Fomenko bundle $\mathcal{V}$ is an 
element of $KO^{-n}(C^*\Gamma)$. This is the desired mod 2 index when
the fundamental group of $X$ is nontrivial. 
 The above reduces to the known formula when the fundamental 
group is trivial. This is because of the equalities in that case
\(
KO(C^*\Gamma)=KO_{\rm cpct}(\R)=KO({\rm pt})\;.
\)
Here $\alpha$ is the composition 
\(
\xymatrix{
\Omega_n^{\rm Spin} (B\Gamma) 
\ar[r]^D
&
{ko}_n (B\Gamma)
\ar[r]^p
&
{KO}_n(B\Gamma)
\ar[r]^{\hspace{-2mm}A}
&
{KO}_n(C_r^*\Gamma)\;,
}
\)
where $A$ is the {\it assembly map}, whose target is the 
KO-theory of the (reduced) real group $X^*$-algebra 
$C_r^*\Gamma$, which is the norm completion of the real
group ring $R\Gamma$ (equal to it if $\Gamma$ is finite).

\paragraph{The Rosenberg-Stolz construction.}
Let $\xi$ be an auxiliary real vector bundle over the classifying space $B\Gamma$. 
 If $\Gamma=\Z_n$, any complex line bundle over $B \Z_n$ is given by a representation 
of $\Z_n$. Also, any element of $H^2(B\Z_n;\Z_2)$ lifts to $H^2(B\Z_n;\Z)$.  
To define equivariant Spin bordism we take $\xi$ to be the trivial line bundle.
Taking nontrivial $\xi$ allows one to define twisted Spin bordism groups. 
Consider triples $(M, f, \sigma)$, where $f$ is a $\Gamma$ structure on $M$ and 
$\sigma$ is a Spin structure on $TM \oplus f^* \xi$. 
Introduce the {\it bordism relation}: $[(M, f, \sigma)]=0$ if there exists a compact manifold $N$ with boundary 
$M$ so that the structures $f$ and $\sigma$ extend over $N$. This induces a suitable 
equivalence class and the {\it twisted bordism group} 
$\Omega_m^{\rm Spin}(B\Gamma, \xi)$ consists of bordism classes of these triples. 
Disjoint union defines the group structure. Cartesian product makes 
$\Omega_m^{\rm Spin}(B\Gamma, \xi)$ into an $\Omega_*^{\rm Spin}$-module. 

\vspace{3mm}
 A Spin structure $\sigma$ on $TM \oplus f^* \xi$ and a Spin${}^c$ structure 
$\sigma^c_\xi$ on $\xi$ define a natural Spin${}^c$ structure $\sigma^c_M$ on $M$.
If $\sigma^c_\xi$ is a Spin structure $\sigma_\xi$ then we get a Spin structure on $M$. 
So there is a homomorphism \cite{RS}
\(
\Omega_m^{\rm Spin} (B\Gamma, \xi) \to \Omega_m^{{\rm Spin}^c}(B\Gamma) \qquad {\rm if} ~
\sigma^c_\xi ~{\rm is~ a~}{\rm Spin}^c ~{\rm structure.}
\)
If $\xi$ is trivial , then $\Omega_m^{\rm Spin}(B\Gamma, \xi)$ can be identified with the ordinary Spin
bordism group groups $\Omega_m^{\rm Spin}(B\Gamma)$. 
The map that sends $(M, f, \sigma)$ to $(M, \sigma)$
which forgets the $\Gamma$ structure induces a natural forgetful map 
\(
\Omega_m^{\rm Spin}(B\Gamma) \to \Omega_m^{\rm Spin}\;.
\label{forg fun}
\)
Conversely, to any Spin manifold $M$ can be associated a trivial $\Gamma$ structure. Therefore, there
is a direct sum decomposition
$
\Omega _m^{\rm Spin}(B\Gamma)=\Omega_m^{\rm Spin}\oplus \widetilde{\Omega}_m^{\rm Spin}(B\Gamma)
$,
where the {\it reduced bordism groups} $\widetilde{\Omega}_m^{\rm Spin}(B\Gamma)$ carry the essential 
new information.
Note that the even-dimensional reduced bordism groups for $\Gamma=\Z_n$ vanish \cite{GLP}
$\widetilde{\Omega}_{2k}^{\rm Spin}(B\Z_n)=0$.

\vspace{3mm}
 Let $\ell=2^q \geq 4$, let $f$ be a $\Z_{\ell}$ structure on $X^{10}$, and let 
$\sigma$ be a Spin structure on $TX^{10} \oplus  f^* \xi$. 
Let $\sigma_X$ be a Spin structure on $X^{10}$, and let $\hat{\sigma}$ be the
 Spin structure $\sigma_X$ twisted by the representation $\varrho_{\ell/2}$. Since 
 $10\equiv 2$ mod 8, then \cite{GLP}
 \(
 \alpha (M, f, \sigma) := \widehat{A}(M, \sigma) \oplus  \widehat{A}(M, \hat{\sigma})
 \subset \Z_2 \oplus \Z_2\;.
 \)
\noindent {\it Example: the circle and the two-torus. }
Consider the simpler case of dimension two.
Let $S^1$ be the circle with the trivial $\Z_\ell$ structure and nonbounding Spin structure. 
Let $\overline{S}^1$ be the circle with nontrivial $\Z_\ell$ structure and nonbounding Spin structure. 
The $\alpha$-invariants are
\(
\alpha (S^1)= 1 \oplus 1\;,
\qquad
\alpha (\overline{S}^1)=1 \oplus 0\;.
\)
Next consider the torus. Let $T^2=S^1 \times S^1$ and 
$\overline{T}^2=\overline{S}^1 \times \overline{S}^1$. The $\alpha$-invariants
are
\(
\alpha (T^2) = 1 \oplus 1\;,
\qquad
\alpha (\overline{T}^2)=1 \oplus 0\;.
\)
Further examples leading to ten-dimensional manifolds can be constructed using the above spaces
as parts (i.e. fibers or base spaces) of ten-dimensional bundles. 
 The $\alpha$-invariant on the bordism groups is given by \cite{GLP}
\begin{eqnarray}
\alpha \left( \Omega_{8k+1}^{\rm Spin} (B\Z_\ell)\right)&=&\Z_2 \oplus \Z_2\;,
\qquad
\alpha \left( \Omega_{8k+2}^{\rm Spin} (B\Z_\ell)\right)=\Z_2 \oplus \Z_2\;,
\nonumber\\
\alpha \left( \widetilde{\Omega}_{8k+1}^{\rm Spin} (B\Z_\ell)\right)&=&\Z_2\;,
\qquad \qquad~
\alpha \left( \widetilde{\Omega}_{8k+2}^{\rm Spin} (B\Z_\ell)\right)=\Z_2 \;.
\end{eqnarray}

\paragraph{Equivariant Spin and Spin${}^c$ cobordism.}
 Let $X^{10}_1$ and $X^{10}_2$ be smooth compact $10$-dimensional 
manifolds without boundary and with $({\rm Spin}, \Gamma)$ structure, and 
similarly for $({\rm Spin}^c, \Gamma)$ structure. 
$X^{10}_1$ is $({\rm Spin}, \Gamma)$ bordant to $X^{10}_2$ if there exists a smooth
compact eleven-dimensional manifold $N^{11}$ so that $\partial N^{11}=X^{10}_1 - X^{10}_2$ such that the 
$({\rm Spin}, \Gamma)$ structure extends over $N^{11}$. 
 Let $\Omega_{10}^{\rm Spin} (B\Gamma)$ be the set of $({\rm Spin}, \Gamma)$
bordism equivalence classes. 

\textbullet~Disjoint union gives $\Omega_{10}^{\rm Spin} (B\Gamma)$ the structure of an abelian group.

\textbullet~$\Omega_*^{\rm Spin} (B\Gamma)$ is a right $\Omega_*^{\rm Spin}$-module. 
 
 \textbullet~If $M \in \Omega_*^{\rm Spin} (B\Gamma)$ and $N \in \Omega_*^{\rm Spin}$ then 
$M \times N$ has a natural Spin structure. 

\vspace{3mm}
\noindent We have seen above, in (\ref{forg fun}), that there is a functor that forgets
the $\Gamma$-structure.
By forgetting the map to the classifying space of the 
fundamental group, 
one can define the forgetful functor
$
\Omega_*^{\rm Spin} (B\Gamma) \to \Omega_*^{\rm Spin}$.
This forgetful functor can be split by associating to any element of 
$\Omega_*^{\rm Spin}$ the trivial Principal $\Gamma$ bundle 
$M \times \Gamma$. Let the reduced equivariant bordism groups
$\widetilde{\Omega}_*^{\rm Spin}(B\Gamma)$ be the kernel of 
the forgetful functor. This leads to the decomposition 
\(
\Omega_*^{\rm Spin}(B\Gamma)=\Omega_*^{\rm Spin} \oplus 
\widetilde{\Omega}_*^{\rm Spin}(B\Gamma)\;.
\)
The above can all be extended to Spin${}^c$ in a parallel way. 
One can also construct many classes of examples (see \cite{Gi}). 
For instance, the ten-dimensional Spin${}^c$ cobordism group
has rank ${\rm Rank}_\Z (\Omega_{10}^{{\rm Spin}^c})=4$.


\paragraph{Cobordism and positive scalar curvature.}
 For any space $M$,  $\Omega_n^{\rm Spin, +}(M)$
is the subgroup of $\Omega_n^{\rm Spin}(M)$ defined by
$$
\Omega_n^{\rm Spin, +}(M)=\left\{
[N,f]  : N~{\rm is~an~}n{\rm- dimensional ~Spin~manifold~
 with~ psc~ metric,~} 
f: N \to M
\right\}\;.
$$
For a finitely presented group $\Gamma$, \cite{DSS}
\begin{eqnarray}
{\rm Gromov}-{\rm Lawson~conjecture} &\Longleftrightarrow & \Omega_n^{\rm Spin,+}(B\Gamma)=\ker ({\rm per} \circ D)
\\
{\rm Gromov-Lawson-Rosenberg~conjecture} &\Longleftrightarrow & \Omega_n^{\rm Spin,+}(B\Gamma)=\ker \alpha\;.
\end{eqnarray}
Consider the example of cyclic fundamental group.
Let $X^{10}$ be a connected compact manifold 
 with cyclic fundamental group. Assume that the universal
cover $\widetilde{X}^{10}$ of $X^{10}$ is Spin.
\begin{enumerate}
\item If $X^{10}$ is Spin, then $X^{10}$ admits a metric of positive scalar curvature if and 
only if $\widehat{A}(X^{10},\sigma)$ for all Spin structures on $X^{10}$ 
\cite{BGS} \cite{KwS1}. 

\item  If $X^{10}$ is orientable but not Spin, then $X^{10}$ 
admits a metric of positive scalar curvature if and only if 
$\widehat{A}(\widetilde{X}^{10})=0$, where $\widetilde{X}^{10}$ is the 
universal cover of $X^{10}$
\cite{BG} \cite{KwS2}. 
\end{enumerate}

\paragraph{Consequences for the partition function.}
Let $X^{10}$ be a Spin manifold of dimension 
$n \geq 5$ 
with fundamental group $\Gamma$, and let $u: X^{10} \to B\Gamma$ be the classifying map 
of the universal covering $\widetilde{X}^{10} \to X^{10}$. Then, by the
Gromov-Lawson-Rosenberg (GLR) conjecture, $X^{10}$ has 
a positive scalar curvature metric 
iff the element $\alpha (X^{10}, u)$ in $KO(C_r^* (\Gamma))$ vanishes. 
Since the GLR conjecture is a theorem in our range of dimensions, then 
the spaces for which $\alpha=0$, and hence for which there is no mod 2 anomaly 
in the type IIA partition function, are 
characterized; their $\alpha$-invariant
 has as a source the subgroup $\Omega_n^{\rm Spin,+}(B\Gamma)$ of manifolds of positive scalar curvature admitting a map to the classifying spaces 
of their fundamental group.


\subsection{Non-simply connected eleven-dimensional manifolds $\pi_1(Y^{11})\neq 1$}
\label{fund eleven}

\paragraph{Equivariant Spin and Spin${}^c$ cobordism.}
The discussion in eleven dimensions parallels that in ten 
dimensions from section \ref{pi 1 10}.
Let $\Gamma$ be a spherical space form group and let 
$\varrho: \Gamma \to U(k)$ be fixed-point-free. Since we have a Spin condition:
If $|\Gamma|$ 
is even then we suppose $k$ to be even. $M=M(\Gamma, \varrho)$
 has a Spin structure so that
$
M \in \widetilde{\Omega}_*^{\rm Spin}(B\Gamma)
$.
 The above can all be extended to Spin${}^c$ in a parallel way. 
\begin{enumerate}
\item If $M=S^m/\varrho (\Gamma)$ is Spin then 
$
M \in \widetilde{\Omega}_*^{\rm Spin}(B\Gamma)
$.
\item If $M=S^m/\varrho (\Gamma)$ is Spin${}^c$ then 
$
M \in \widetilde{\Omega}_*^{{\rm Spin}^c}(B\Gamma)
$.
\end{enumerate}
There are many examples of such equivariant cobordism 
groups which are very far from being trivial. For example
(see \cite{Gi} for an extensive list), for cyclic groups
\(
\Omega_{11}^{{\rm Spin}^c}(B\Z_2)= \Z_{64} \oplus \Z_{16} \oplus
2\cdot \Z_4 \oplus \Z_2\;,
\qquad
\Omega_{11}^{{\rm Spin}^c}(B\Z_3)= 2\cdot \Z_{27} \oplus 2\cdot \Z_{9} \oplus
4\cdot \Z_3\;,
 \)
and, for quaternionic groups,
\begin{eqnarray}
\Omega_{11}^{{\rm Spin}^c}(BQ_2)&=& \Z_{128} \oplus \Z_{32} \oplus
4\cdot \Z_8 \oplus 3\cdot \Z_4 \oplus 7\cdot \Z_2\;,
\nonumber\\
\Omega_{11}^{{\rm Spin}^c}(BQ_3)&=& \Z_{256} \oplus \Z_{64} \oplus
2\cdot \Z_{16} \oplus 3\cdot \Z_8 \oplus 3\cdot \Z_4 \oplus 7\cdot \Z_2\;.
\end{eqnarray}
Thus, in extending $Y^{11}$ to $Z^{12}$ in the case of Spin${}^c$ 
with nontrivial fundamental group one encounters various potential 
obstructions. This implies that the study of the partition function 
in terms of index theory in twelve dimensions requires careful 
analysis in this case. We will not pursue that aspect further in this paper and 
leave it for future discussions.

\paragraph{Invariants from K-theory.}
 For any finite group $\Gamma$, There is the standard isomorphism 
$
R(\Gamma) \to K^0(B\Gamma)
$,
which assigns to every representation $\varrho$ of $\Gamma$ an associated 
bundle $V_{\varrho}$ over $B\Gamma$ determined by the representation 
module associated to $\varrho$ for every point of $B\Gamma$. 
By \cite{APSII}, there is a correspondence between the representation ring of 
the finite fundamental group $\Gamma=\pi_1(Y^{11})$ of an eleven-dimensional 
manifold 
 $Y^{11}$ 
bounding a twelve-manifold $Z^{12}$ and the 
equivalence classes of vector bundles on $Y^{11}$, taken modulo the integers
\(
R(\pi_1(Y^{11})) \to K^{-1} (Y^{11};\Q/\Z)\;.
\)
This map sends the class of a representation $\varrho$ to the class 
$[\varrho]$ of the representation vector bundle $V_{\rho}$ associated to 
$\varrho$ over $Y$, and then using the pushforward in K-theory to obtain an 
element in the coefficient group of K-theory modulo the integers. 
This is done by means of the completion homomorphism, which relates
$K_\Gamma^*$ and $K^*(B\Gamma)$, and the pullback of the 
corresponding classifying map $f: Y^{11} \to B\Gamma$.
 The inclusion of the trivial group in $\Gamma$ induces a representation 
between representation rings. The
kernel of this representation is the augmentation ideal $I_\Gamma$. The quotient 
$
\widetilde{R}(\Gamma):=R(\Gamma)/I_\Gamma
$ 
is the reduced representation ring of $\Gamma$.
The completion with respect to $I_\Gamma$ is done by means of 
$c_{I_\Gamma}: I_\Gamma \to \widetilde{K}^0(B\Gamma)
$. 
The Bockstein homomorphism in K-theory
associated to the short exact sequence of coefficients 
$0 \to \Z \to \Q \to \Q/\Z \to 0$ is
$
\delta^{-1} :  \widetilde{K}^0(B\Gamma) \to K^{-1}(B\Gamma;\Q/\Z)
$.
The composition 
\(
\gamma :=\delta^{-1} \circ c_{I_\Gamma} 
: R(\Gamma)/I_\Gamma \to  K^{-1}(B\Gamma;\Q/\Z)
\)
sends the representation 
$\varrho: \Gamma \to U(k)$ of rank $k$ to the class
$\gamma (\varrho -k)$. 
Consider the pullback via $f$ 
with respect to the $\pi_1(Y^{11})$-action on $Y^{11}$;
we get a map
$
f^* : K^{-1}(B\Gamma;\Z) \to K^{-1}(Y^{11};\Q/\Z)
$. 
The composition 
$
f^* \circ \gamma : \widetilde{R}(\Gamma) \to K^{-1}(Y^{11};\Q/\Z)
$
sends the reduced representation $\varrho-k$ 
to a class $[\varrho]$ in $K^{-1}(Y^{11};\Q/\Z)$.

\paragraph{The invariant in the Spin${}^c$ case.}
When 
$Y^{11}$ is Spin${}^c$, the direct image map 
$
K^{-1}(Y^{11};\Q/\Z) \to \Q/\Z
$
sends the class of $[\varrho]$ in $K^{-1}(Y^{11};\Q/\Z)$ 
 to a number in $\Q/\Z$. This number is the eta invariant 
 of a corresponding Dirac operator twisted by the flat 
 line bundle $V_\varrho$ defined by the representation \cite{APSII}. 
 Below, we will take this flat line bundle to be associated to the fundamental 
 group.
Note that associated to every rank $k$ is a 
unitary representation of $\Gamma$ in some $U(k)$.

\paragraph{The eta-invariant associated to the fundamental group.}
Take 
 $\varrho$ to be a unitary representation of $\pi_1 (Y)$ and let $V_\varrho$ be the associated
vector bundle
$
V_\varrho :=(\widetilde{Y}^{11} \times_\varrho \R^k)/\sim
$,
with $(\tilde{y}, v)\sim (g\cdot \tilde{y}, \varrho (g) v)$ for all $g \in \pi_1(Y^{11})$
(cf. (\ref{vrho})). 
Let $D$ be a Dirac operator on a bundle $E$. The transition functions of $V_\varrho$
are locally constant and the operator $D$ extends naturally to an operator
$D_\varrho$ on $C^{\infty} (E \otimes V_\varrho)$. 
Define $\eta (D)(\varrho):=\eta (D_\varrho)$. The map $\varrho \to \eta (D)(\varrho)$ is
additive and extends to a map from $RU(\pi_1(Y^{11}))$ to $\R$. 
Let $\tilde{\varrho}:=\varrho (\lambda^{-1})^t$ be the dual virtual representation.
The above works in any odd dimension.

 \paragraph{Example 1: The circle $S^1$.} Let $n \geq 2$. Give $S^1$ the canonical $\Z_\ell$ structure
and either of the two inequivalent Spin structures. Then (see \cite{GLP})
\(
\eta (S^1) (\varrho_0 - \varrho_s)=-\frac{s}{\ell} \in \R/\Z\;.
\)
Interchanging the Spin structures replaces $\varrho_0 - \varrho_s$ by 
$\varrho_{\frac{\ell}{2}} - \varrho_{\frac{s + \ell}{2}}$ and leaves the
eta invariant unchanged.

\paragraph{Example 2: General odd-dimensional spaces.}
Let $D$ be the Dirac operator on a compact
Spin manifold $Y$ of odd dimension $m$ which admits a metric of positive 
scalar curvature. Let $\varrho$ be a unitary representation of $\pi_1(Y)$. Then
${\rm ker}(D_\varrho)=\{0\}$ and \cite{DLP}
\begin{eqnarray}
\eta (D_\varrho)&=& \eta (D_{\tilde{\varrho}}) \qquad ~~ {\rm if~~} m\equiv 3 ~({\rm mod}~ 4)
\nonumber\\
\eta (D_\varrho)&=& -\eta (D_{\tilde{\varrho}}) \qquad {\rm if~~} m\equiv 1 ~({\rm mod} ~4)\;.
\end{eqnarray}

\paragraph{Scalar curvature and geometric bordism groups.}
In order to include geometric data such as the metric, 
the fundamental group, and the Spin structure, one introduces 
geometric bordism groups. 
The geometric bordism groups $G^+\Omega_{11}^{\rm Spin}(B\Gamma, \xi)$ are defined as follows.
Consider quadruples $(Y^{11}, f, \sigma, g)$, where $f$ is a $\Gamma$ structure 
on a closed manifold $Y^{11}$, where $\sigma$ is a Spin structure on $TY^{11}\oplus f^* \xi$, and 
where $g$ is a metric of positive scalar curvature on $Y^{11}$. 
Introduce the equivalence relation $[(Y^{11}, f, \sigma, g)]=0$ if there exists a compact manifold $Z^{12}$ with
boundary $Y^{11}$ so that the structures $f$ and $\sigma$ extend over $Z^{12}$ and so that the metric $g$ extends
over $Z^{12}$ as a metric of positive scalar curvature which is a product near the boundary. 
Assume that there exists a manifold 
$Y^{11}_1$ which admits a metric $g_1$ of positive scalar curvature and which 
admits structures $(f_1, \sigma_1)$ so that
$[(Y^{11}, f, \sigma)]=[(Y^{11}_1, f_1, \sigma_1)]$ in $\Omega_{11}^{\rm Spin}(B\Gamma, \xi)$. 
Then, by the results of  
\cite{Ga} \cite{Miy} \cite{Ros1} \cite{Ros2}
\cite{Ros3},
$Y^{11}$ admits
a metric of positive scalar curvature $g$ so that 
$[(Y^{11}, f, \sigma, g)]=[(Y^{11}_1,f_1, \sigma_1, g_1)]$ in 
$G^+\Omega_{11}^{\rm Spin}(B\Gamma, \xi)$.

\vspace{3mm}
Let $[(Y^{11}, f, \sigma, g)] \in G^+\Omega^{\rm spin}_{11}(B\Gamma;\xi)$ and 
let $\varrho \in RU_0(\Gamma)$.
Assume that $\xi$ is orientable so that $\sigma_\xi$ is a Spin${}^c$ structure. 
Suppose that $Y^{11}$ is the boundary of $Z^{12}$ and that the structures extend from 
$Y^{11}$ to $Z^{12}$. Then the APS index theorem \cite{APSIII}
gives
\(
{\rm Ind}(D_\varrho)=\int_{Z^{12}} {\rm dim} \varrho \cdot \widehat{A}(Z^{12}) 
+ \eta(Y^{11},\varrho)\;.
\)
Since $Z^{12}$ has a metric of positive scalar curvature, then 
the kernel of the Dirac operator on $Z^{12}$ with spectral boundary conditions
is trivial, so that the index is zero. 
Therefore, the 
map which sends $(Y^{11},f,\sigma)$ to $\eta(Y^{11},\varrho)$ extends to a homomorphism 
from $G^+\Omega_{11}^{\rm Spin}(B\Gamma, \xi)$ to $\R$ (see \cite{GLP}).

\paragraph {Example: Spherical space forms.}
We will use the discussion on Spin${}^c$ structures and representations from the 
end of section \ref{ssf}.
 Let $\delta$ be a linear representation so that $(\delta {\rm det}(\varrho))^{1/2}$ is 
 a well-defined representation of $\Gamma$. Such a representation always exists
 (e.g. $\delta={\rm det}(\varrho)$). Then $M$ admits a Spin${}^c$ structure whose
 associated determinant line bundle is given by 
 the representation $\delta$.  
 Give $M$ a Spin${}^c$ structure with associated 
 determinant line bundle given by the linear representation $\delta$.
 Let $\varrho \in  RU(\Gamma)$. If $2k-1\equiv 3$ mod 4, assume $\varrho \in RU_0(\Gamma)$. Then
 \cite{GLP}
 \(
\eta(M)(\varrho)=|\Gamma|^{-1} \sum_{g\in \Gamma, g\neq 1} 
{\rm Tr}(\varrho (g)) \left\{
\delta (g) {\rm det}(\varrho (g))
\right\}^{1/2}
{\rm det} (\varrho (g) - I)^{-1}\;.
\label{dettau}
 \)
 
\paragraph{A Spin${}^c$ example: The eta invariant of lens space bundles.}
 \cite{BGS} \cite{GLP}
Let $n \geq 2$ and let $\vec{a}:=(a_1, \cdots, a_5)$ be a collection of odd
integers which are coprime to $n$. Let $L_1, \cdots, L_5$ be complex line 
bundles over the Riemann sphere $S^2=\C P^1$. 
Let $S(L_1 \oplus \cdots \oplus L_5)$
be the associated sphere bundle of dimension eleven. 
This manifold is Spin iff the sum of the first Chern classes 
$c_1(L_1) + \cdots + c_1 (L_5)$ is even in $H^2 (S^2;\Z)=\Z$.
Assume this is the case. 
Let 
$
\varrho_{\vec{A}} (\lambda) (\xi_1, \cdots, \xi_5):=
(\lambda^{a_1} \xi_1, \cdots, \lambda^{a_5} \xi_5)$, for $\lambda\in \Z_n$, 
define a fiberwise action of $\Z_n$ on 
$L_1 \oplus \cdots \oplus L_5$. The restriction 
of this action defines a fixed point free action of $\Z_n$ 
on $S(L_1 \oplus \cdots \oplus L_5)$. Let the resulting 
quotient manifold be
\(
Y^{11} (n; \vec{a}; L_1, \cdots, L_5):=
S(L_1 \oplus \cdots \oplus L_5)/\varrho_{\vec{a}} (\Z_n)\;.
\)
$Y^{11}$ admits a metric of positive scalar curvature.
Furthermore, since $11=2k+1$ with $k$ odd, $Y^{11}$ 
admits a Spin${}^c$ structure, with 
associated determinant line bundle defined by the representation 
$\rho_1$. 
Define $
\delta_{\vec{a}}$ to be 
$\frac{1}{2}(1+ a_1 + \cdots + a_5)$.
Let $\varrho \in RU_0(\Z_n)$.
Then, using \cite{GLP},
\(
\eta (Y^{11})(\varrho)=-\frac{1}{2n}
\sum_{\lambda\neq 1} {\rm Tr} (\varrho(\lambda)) \lambda^{\delta_{\vec{a}}}
\left\{ 
\prod_i \frac{1}{\lambda^{a_i}-1} 
\right\}
\left\{ 
\sum_i \int_{\C P^1} c_1 (L_j) \frac{\lambda^{a_j+1}}{\lambda^{a_j}-1}
\right\}\;.
\label{eta fl}
\)

\paragraph{A Spin example: The eta invariant of lens space bundles.}
Consider the nine-dimensional manifold
 \(
X^{9} (n; \vec{a}; L_1, \cdots, L_4):=
S(L_1 \oplus \cdots \oplus L_4)/\varrho_{\vec{a}} (\Z_n)\;.
\)
similarly to $Y^{11}$ above. 
This $X^9$ admits a Spin structure. The formula for the eta invariant 
is similar to (\ref{eta fl}) with $\varrho \in RU(\Z_n)$
and
$\delta_{\vec{a}}=\frac{1}{2}(a_1 + \cdots + a_4)$. Note that in 
dimension seven, this is an $M^{p,q,r}$ manifold, i.e. 
a circle bundle over $\C P^ 1\times \C P^1 \times \C P^1$, which
is important in flux compactification to gauged supergravity
in four dimensions.


\paragraph{Trivial Spin structures: Framed eleven-dimensional manifolds and the Adams $e$-invariant.}
The Pontrjagin-Thom construction gives an identification of stable 
homotopy groups of spheres $\pi^S$ with the cobordism groups of 
framed manifolds. A framing is a trivialization of the stable normal bundle.
Up to homotopy, this is equivalent to a trivialization of the stable tangent 
bundle (because we are embedding in flat $\R^m$ for very large $m$).
A particular form of framing is a parallelism, which is a trivialization 
of the tangent bundle. So, a parallelism $\mathcal{P}$
on $Y^{11}$ induces a framing on $Y^{11}$ and hence 
defines an element 
$
[Y^{11}, \mathcal{P}] \in \pi_{11}^S
$.
Since $\Omega_{11}^{\rm spin}=0$ then 
$Y^{11}=\partial Z^{12}$ for some Spin twelve-manifold 
$Z^{12}$. The Spin structure induced on $Y^{11}$ is the 
trivial Spin structure defined by $\mathcal{P}$. 
The Adams $e$-invariant for $Y^{11}$ is a homomorphism 
$e: \pi_{11}^S \to \Q/\Z$ defined via the relative 
$\widehat{A}$-invariant as (cf. \cite{Sto})
\(
e[Y^{11}, \mathcal{P}]=\frac{1}{2}\widehat{A}(Z^{12}, Y^{11})\;,
\)
in $\Q/\Z$ and is independent of the choice of 
$Z^{12}$.

\vspace{3mm}
Let $g_{\mathcal{P}}$ be the Riemannian metric defined by 
$\mathcal{P}$, 
and consider $\overline{\eta} (g_{\mathcal{P}})=\frac{1}{2}\left(
h(g_{\mathcal{P}}) + \eta(g_{\mathcal{P}})
\right)$. Then the analytic formula for the Adams $e$-invariant
is obtained as follows \cite{APSII}.
Let $\omega^Z$ be a connection on $Z^{12}$ extending 
the product connection defined by the connection 
$\omega^Y$ near $Y^{11}$. Let $p(\omega^Z)$ be 
the total Pontrjagin class on $TZ^{12}$ defined by $\omega^Z$.
A closed one-form 
$\mathcal{W}$ on the space of connections on $Y^{11}$
is $\mathcal{W}=d\mathcal{F}$, where 
$\mathcal{F}(\omega^Z)=\int_{Z^{12}} \widehat{A}(p(\omega^Z))$
mod $\Z$. Then 
\begin{eqnarray}
e[Y^{11},\mathcal{P}]&=& \frac{1}{2}\mathcal{F}(\mathcal{P})
=
\frac{1}{2}\left[ \mathcal{F}(g_{\mathcal{P}}) 
+ \left( \mathcal{F}(\mathcal{P})- \mathcal{F}(g_{\mathcal{P}})
\right)
\right]
\nonumber\\
&=&
\frac{1}{2}\left( 
\overline{\eta} (g_{\mathcal{P}}) -
\int_{\mathcal{P}}^{g_{\mathcal{P}}}
\mathcal{W}
\right) ~~{\rm mod~} \Z\;.
\end{eqnarray}
In the case when 
$\int_{\mathcal{P}}^{g_{\mathcal{P}}} \mathcal{W} \in 2\Z$,
the phase of the partition function (with no nontrivial bundles) 
will be given by the Adams $e$-invariant 
$
e^{\pi i \overline{\eta}}=e^{2\pi i e[Y^{11}, \mathcal{P}]}
$.

\vspace{3mm}
Let $D_\varrho$ be the Dirac operator on $Y^{11}$ twisted by a unitary 
representation $\varrho$ of $\pi_1(Y^{11})$. 
Let $\overline{\eta}_\varrho=\overline{\eta}_{D_\varrho}$ and let 
$\underline{n}=\dim \varrho$ be the trivial $n$-dimensional 
representation for which $\overline{\eta}_{\underline{n}}=n \overline{\eta}_D$. Then
$
\widetilde{\overline{\eta}}_\varrho= \overline{\eta}_\varrho - 
\overline{\eta}_{\underline{n}}$ (mod $\Z$)
is independent of the metric on $Y^{11}$ and is a cobordism invariant
of $(Y^{11}, \varrho)$;
if  $Y^{11}=\partial Z^{12}$ (as Spin${}^c$ manifolds) with 
$\varrho$ extending a unitary representation of 
$\pi_1(Z^{12})$, then $\widetilde{\overline{\eta}}_\varrho=0$.
Thus, the phase of the partition function in this case (with 
non-nontrivial bundles) is one. 

\vspace{3mm}
Now consider a finite covering $\widetilde{Y}^{11} \to Y^{11}$ of degree $n$
with representation $\varrho : \pi_1(Y^{11}) \to O(n)$.
Then 
\begin{eqnarray}
e[\widetilde{Y}^{11}, \tilde{\mathcal{P}}]- n e[Y^{11}, \mathcal{P}]&=&
\frac{1}{2}\left( \overline{\eta}(\tilde{g}_{\tilde{\mathcal{P}}}) - n 
\overline{\eta}( g_{\mathcal{P}})\right)
\quad {\rm mod}~\Z
\nonumber\\
&=&
\frac{1}{2}
\widetilde{\overline{\eta}}_\varrho \quad {\rm mod}~\Z\;.
\end{eqnarray}
In this case, the phase will involve the terms on the left hand side, 
the second of which can be arranged to give a unit phase by 
appropriately choosing the degree of the covering when possible.


\subsection{The second homotopy group and Spin${}^c$ structures}
\label{pie2}
We have seen that the fundamental group provides a vast number of 
examples of Spin manifolds with possibly many Spin structures
(see section \ref{spinpie}).  
The main source of manifolds with Spin${}^c$ structures are manifolds with
nontrivial second homotopy group. This is due to the fact that 
Spin${}^c$ structures are classified by the second integral cohomology
group of the manifold (see section \ref{basic spin c}). If the space is 2-connected 
then, by the Hurewicz theorem, the first nonzero homotopy is the same as the homology in the 
same degree, and Spin${}^c$ structures exist on $M$ if
 $H^2(M;\Z) \to H^2(M;\Z_2)$ is surjective.  

\vspace{3mm}
If the second homotopy group $\pi_2(M)$ is finite, then assuming further 
that the fundamental group $\pi_1(M)$ is trivial, Spin and Spin${}^c$ structures
are equivalent.  This is because of the diagram
\(
\xymatrix{
& \vdots 
\ar[d]
& &  \\
& \pi_2(M) 
\ar[d]^\partial
& &  \\
& \Z_2 \oplus \Z
\ar[d]
& &  \\
\Z\cong \pi_1(S^1)
\ar[r]
\ar[ur]^{\alpha \mapsto (0,\alpha)}
&
\pi_1(Q \buildrel{\sim}\over{\times} P_1)
\ar[r]
\ar[d]
&
\pi_1(Q)
\ar[r]
&
1
\\
& \pi_1(M)=1 &&
}
\)
in which rows and arrows are exact \cite{Fried}. Since 
$\pi_2(M)$ is finite, the image of $\partial$ is contained in the 
subgroup $\Z_2$, so that either $\partial\equiv 0$ or 
im$(\partial)=\Z_2$. The first gives $\pi_1(Q)=\Z_2$ and 
$Q$ admits a Spin${}^c$ structure. The second 
does not give a Spin${}^c$ structure on $Q$.
For example, the five-manifold $M^5=SU(3)/SO(3)$ 
has trivial $\pi_1$ and $\pi_2(M^5)=\Z_2$, and 
the homotopy exact sequence argument shows that 
$\pi_1(Q)=1$, i.e. $Q$ admits no Spin${}^c$ structure. 



\paragraph{Sources for the fields from non-vanishing of $\pi_1$ and $\pi_2$.}
Physical fields are generally taken to be cohomology classes. Hence such fields in a given 
degree can be supported on a manifold if the cohomology of the manifold 
is nonzero in that degree. For a connected manifold $M$, $\pi_1(M)\neq 0$ implies that
$H_1(M) =\pi_1(M)$ and, by Poincar\'e duality, a field in $H^1(M)$ can be supported. 
Similarly for $\pi_2(M)\neq 0$, as we saw at the beginning of this section.
\begin{itemize}
\item Now taking $M=Y^{11}$, then the only fields in cohomological degree 1 or 2
should be related to $E_8$ gauge theory, since the only supergravity 
form field available is $C_3$ (with its curvature $G_4$) and are not of
 that degree.
\item For $M=X^{10}$, there is no field strength of degree 1 but we can 
take flat Ramond-Ramond fields. For $H^2(X^{10})\neq 0$, we have either 
a flat $B$-field or an RR field strength, essentially the curvature of the M-theory 
circle bundle. This latter will be considered in more detail in section \ref{fam}.
\end{itemize}
We summarize the situation for the relation between low degree homotopy 
groups and the fields in ten and eleven dimensions in the following table.
$$
\begin{tabular}{|l||l|l|}
\hline
Dimension& $\pi_1\neq 0$ & $\pi_2\neq 0$\\
\hline
\hline
Ten & Flat RR 1-form connection $C_1$ &
Flat B-field $B_2$ and/or RR field $F_2$\\
\hline
Eleven & Flat connection for $E_8$ bundle &
$E_8$ gauge field $F$\\
\hline
\end{tabular}
$$
The case of $\pi_3\neq 0$ is considered in \cite{tcu}.

\paragraph{Note on the fields in type IIB.}
We briefly consider the case of type IIB string theory.
Here there is a 1-form, the RR field $F_1$,
which is supported by the first nontrivial homotopy or 
homology.   
Let $X^{10}$ be a compact manifold and 
consider its associated Jacobian torus 
$J_X=H^1(X^{10};\R)/H^1(X^{10};\Z)$. 
Let $\widetilde{X}$ denote the universal
covering of $X^{10}$. Define a family of flat complex line
bundles $L$ over $X^{10}$ parametrized by $J_X$ 
by taking the
quotient, as in \cite{LM},  
\(
L:= \left(\widetilde{X} \times H^1(X^{10};\R) \times \C \right)
/ \pi_1(X^{10}) \times H^1(X^{10};\Z)\;,
\)
where the action of  $\pi_1(X^{10}) \times H^1(X^{10};\Z)$is given by 
$\phi_{(g,h)} (\widetilde{x}, v, z)= (\widetilde{g}x, v+ h, e^{2\pi i v(g)}z)$. 
We can in this case also consider $L$ as a twist for spinor 
modules $S \otimes L=(S^+ \otimes L) \oplus (S^- \otimes L)$,
as a pair of vector bundles over $X^{10}$ parametrized by 
$J_X$. We can also extend the flat connection on $L$ to 
define a Dirac operator 
$D_v^+: \Gamma (S^+ \otimes L_v) \to \Gamma (S^- \otimes L_v)$
for $v \in J_X$. Of course we could also do the same for type
IIA string theory, using the fields in the above table.

\section{Families, Eta Forms and Harmonic Forms}
\label{fam}

\subsection{The $B$-field and harmonic representatives for the Spin${}^c$ structure}
\label{B har}
 The $C$-field in eleven dimensions as a harmonic representative of the String structure is 
described in \cite{tcu}.
Here we consider the case of the $B$-field in ten-dimensional type IIA string theory. 

\paragraph{The Euler class and the harmonic representative of the Ramond-Ramond 2-form.}
 Let $\mathcal{L}$ be a complex line bundle over $X^{10}$ equipped with 
a smooth fiber metric and a Riemannian connection $\nabla$. Let
$S(\mathcal{L}):=\{ \xi \in \mathcal{L}~ :~ |\xi|=1\}$ be the associated
circle bundle. $S^1$ acts transitively on the fibers of $S(\mathcal{L})$ by 
complex multiplication, so $S(\mathcal{L})$ is a principal circle bundle. 
Conversely, every principal bundle arises this way.
The Chern form of a principal circle bundle over $X^{10}$ is an invariantly defined 
real closed two-form on $X^{10}$ given by 
$
c_1 (\nabla) := \mathcal{F} (\nabla)$.
The de Rham cohomology class is independent of the connection chosen 
and represents the complexification of $c_1$. This is the description of 
the Ramond-Ramond (RR) 2-form $F_2$, in the absence of a cosmological
constant (i.e. the RR 0-form $F_0$). 

\vspace{3mm}
We could ask for a harmonic representative for $F_2$. 
Indeed, there exists
a unitary connection $\nabla$ on $\mathcal{L}$ so that the curvature $\mathcal{F}$ is harmonic. This can be shown as follows (see e.g. \cite{GLP}).
Let $\nabla'$ be any unitary connection on $\mathcal{L}$ with curvature two-form $\mathcal{F}'$.
Since this is a closed two-form, the Hodge-de Rham theorem ensures the existence of 
a harmonic two-form $\mathcal{F}_{\rm harm}$ with
$[\mathcal{F}']=[\mathcal{F}_{\rm harm}]$
in cohomology, so that $\mathcal{F}_{\rm harm}=\mathcal{F}' + d\upsilon$, where $\upsilon$ is a 
smooth 1-form on $X^{10}$. A unitary connection $\nabla$ can be built out of 
$\nabla'$ and $\upsilon$ as
$
\nabla := \nabla' + i \upsilon$,
so that
\(
\mathcal{F}=dA' + d\upsilon = \mathcal{F}' + d\upsilon= \mathcal{F}_{\rm harm}\;.
\)  

\vspace{3mm}
Next we ask: When is this harmonic representative nontrivial?
Since the first Chern class identifies ${\rm Vect}_{\C}^1 (X^{10})$ with 
$H^2(X^{10};\Z)$ and since 
$0 \neq H^2(X^{10};\R)=H^2(X^{10};\Z)\otimes_{\Z} \R$ we may choose
$\mathcal{L} \in {\rm Vect}_{\C}^1 (X^{10})$ so that 
$0 \neq c_1(\mathcal{L})$ in $H^2(X^{10};\R)$. 
Therefore, 
if $H^2(X^{10};\R)\neq 0$, then there exists a complex line  
bundle $\mathcal{L}$ over $X^{10}$ and a unitary connection on $\mathcal{L}$
so that the curvature $\mathcal{F}$ is harmonic and nontrivial.

\paragraph{The $B$-field as a harmonic representative of the 
Spin${}^c$ structure.}
Next we consider the $B$-field. We can use the line bundle corresponding 
to the M-theory circle to be the line bundle which enters in defining $B$ as
a connection on a gerbe. This brings out the relation
between $B$ and the RR field $F_2$. This is most manifest in the 
presence of the RR 0-form (the cosmological constant), where
$dF_2=F_0 H_3$, so that essentially $F_2=F_0 B_2$.  
Let $P=P_{SO}(X^{10})$ be a principal $SO(10)$ bundle with connection $\omega_X$
over a $10$-dimensional Riemannian manifold $(X^{10},g_X)$. 
Consider a natural one-parameter family of metrics $g_t$ 
on $P$,
$
g_t = g_X + t g_{SO}
$.
The adiabatic limit is given by taking $t \to 0$ (see more in section \ref{equiv e}). 
We use the following cohomological definition of Spin${}^c$ structure.
A Spin${}^c$ structure on an $SO(n)$ bundle $P \to M$ is
a cohomology class $\sigma^c_X \in H^2(P;\Z)$ such that $i^* \sigma^c_X \neq 0\in H^2(SO(n);\Z)\cong \Z_2$,
where $i: SO(n) \hookrightarrow P$ is the fiberwise inclusion. 
 A Spin${}^c$ structure, topologically, is a choice of a particular complex line bundle
$L \to P$ characterized by its first Chern class $\sigma^c_X \in H^2(P;\Z)$. 
To any Spin${}^c$ structure there is an associated line bundle $\cL$ on $X^{10}$. 
This is determined by requiring that $\pi^*c_1 (\cL)=2\sigma^c_X \in H^2(P;\Z)$. 
However, upon introducing a harmonic metric $g_X$ on $X^{10}$ and a connection
on $P$ (which is usually determined by $g$), then taking the harmonic representative
of $\sigma^c_X$ in the adiabatic limit produces a canonical 2-form $B_{g,\sigma^c_X} \in \Omega^2(X^{10})$,
such that
\(
\pi^* c_1(\cL)=2 [B] \in H^2(P;\R)\;.
\)
This gives that $2B_{g,\sigma^c_X}\in \Omega^2(X^{10})$ is the curvature of some connection on 
$\cL$. 
Furthermore, $B_{g,c}$ is harmonic so it picks out a class of connections on $\cL$
with energy-minimizing curvature. 
 The
 harmonic representative in the adiabatic limit of (the real cohomology 
 class given by) $\sigma^c_X$  is the form equal to the pullback of 
 a 2-form from $X^{10}$, using \cite{Red},
 \(
 \lim_{t \to 0} [\sigma^c_X]_{g_t}=\pi^* B_{g, \omega_X, \sigma^c_X} \in \Omega^2(P)\;,
 \)
where the 2-form $B_{g, \omega_X, \sigma^c_X} \in \Omega^2(X^{10})$.
%
Furthermore, if the Spin${}^c$ structure is changed by $\xi \in H^2(X;\Z)$ then
\(
[\sigma^c_X+ \pi^*\xi]_0=[\sigma^c_X]_0 + \pi^*[\xi]_g\;.
\)
 For $\xi \in H^2(X^{10};\Z)$ the adiabatic harmonic representative of the
class pulled back to $P$ is just the pullback of the harmonic representative on
$X^{10}$:
\(
[\pi^* \xi]_0=\pi^* [\xi]_g \in \Omega^2(P)\;.
\)

\subsection{Families of Dirac operators and superconnections over $X^{10}$}
\label{sec fam 10}
\subsubsection{Families of Dirac operators on the M-theory disk}
Consider the two-dimensional disk $\mathbb{D}^2$ (which is compact but not closed), 
with a collection of Dirac operators 
$\{D_x\}$, parametrized by points of the ten-dimensional 
space $X^{10}$. 
As $x$ varies continuously in $X^{10}$, the index 
${\rm Ind}(D_x)$ remains constant. However, 
we have a family of Dirac operators which can be 
nontrivial globally, in an analogous way that a vector 
bundle, which is locally a product of a vector space with
the base space, can be globally nontrivial. 
Let $\pi_D : Z^{12} \to X^{10}$ be a family of disks over $X^{10}$.
A family of Dirac operators on this family of disks 
consists of:

\noindent (i) A $\Z_2$-graded Hermitian vector bundle $E$ over $Z^{12}$.

\noindent (ii) For each $x \in X^{10}$, a Dirac operator $D_x$ on the manifold 
$Z_x=\pi_D^{-1}(x)$, acting on section of the restriction of the bundle $E$ 
to $Z_x$, such that:

\noindent (iii) The operators $D_x$ vary smoothly with $x$, i.e. they are 
restrictions to $\mathbb{D}^2_x$ of a single Dirac operator on $Z^{12}$.

\noindent A family of Dirac operators over $X^{10}$ can also be defined as
the leafwise Dirac operator on the following groupoid (see 
\cite{HR}). A family $G_{\pi_D}$ of pair groupoids parametrized by $X^{10}$
 is the smooth groupoid
given by:

$\circ$ The object space is $Z^{12}$;

$\circ$ The morphism space is 
$\left\{ (z_1, z_2) \in Z^{12} \times Z^{12} ~:~\pi_D(z_1)=\pi_D(z_2) \in X^{10}  \right\}$;

$\circ$ The source and the range maps are 
$s(z_1, z_2)= z_1$, $r(z_1, z_2)=z_2$;

$\circ$ The composition law is $(z_1, z_2) \cdot (z_2, z_3)=(z_1, z_3)$;

$\circ$ The inverse is $(z_1, z_2)^{-1}= (z_2, z_1)$;

$\circ$ The inclusion of identities is $z \longmapsto (z, z)$.

\noindent This is equivalent to a family of Dirac operators in the sense that 
the $C^*$-algebra of this groupoid $G_{\pi_D}$ is Morita equivalent to 
$C(X^{10})$, the algebra of continuous functions on 
$X^{10}$. 
The family index gives rise to an element of the K-theory of 
$X^{10}$, via the isomorphism $K^0(X^{10})=K(C_r^* (G_{\pi_D}))$
with the K-theory of the groupoid $C^*$-algebra (cf. section \ref{pi 1 10}). 
The vertical tangent bundle to the fibers is 
\(
T_{\pi_D} Z^{12}= \ker ({\pi_D}_*: TZ^{12} \to TX^{10})\;.
\)
Equipping this family with a Spin${}^c$ structure gives a family
of Dirac operators on $Z^{12} \to X^{10}$. Then there is 
a map ${\pi_D}_!: K^0(Z^{12}) \to K^0(X^{10})$ associated to the 
$K$-oriented map $\pi_D$.

\vspace{3mm}
There are two forms of the index theorem: K-theoretic and 
cohomological.  The first is more powerful, since the Chern 
character, mapping K-theory to cohomology, loses torsion 
information.  
The cohomological formula is 
\(
{\rm ch}({\rm Ind} D)=\int_{\mathbb{D}^2} {\rm ch}(\sigma_D) 
{\rm Todd}(T_{\pi_D} Z^{12} \otimes \C) \in H^*(X^{10})\;,
\)
where the integral is integration over the fiber, going from 
$H^*(Z^{12})$ to $H^*(X^{10})$. We will consider disk bundles
 further in section \ref{form del}.


\paragraph{Ramond-Ramond (RR) fields via Dirac families.}
Consider Dirac operators with coefficients in a bundle 
$E \to \mathbb{D}^2$ for a family $\mathbb{D}^2 \to X^{10}$.  
A geometric form of the index theorem gives the index of 
the family of Dirac operators is the image of 
$E$ under pushforward in a form of K-theory.
When $E$ is complex then complex K-theory
pushforward gives, via 
$
K(\mathbb{D}^2) \to K(X^{10})$,
an element of the K-theory of the base.
The Dirac operator $D_x$ is taken to be a field 
parametrized by points
$x \in X^{10}$. Take a complex vector bundle $E$ to 
be $\ker D_x$ and its conjugate 
$\overline{E}$ to be ${\rm coker} D_x$. Then
\(
E - \overline{E}=\ker D_x - {\rm coker}\hspace{0.5mm}D_x\;,
\)
which is the index of $D_x$, represents an 
element of K-theory $K(X^{10})$. 
The RR fields satisfy the quantization 
condition \cite{MW} \cite{FH}
\(
F(E)=\sqrt{\widehat{A}(X^{10})} {\rm ch}(E)
\)
corresponding to a complex vector bundle $E$
(or the class of such a bundle). Thus, for the 
tensor product $E\otimes E'$, with another vector
bundle $E'$, we get
\begin{eqnarray}
F(E) \wedge F(E')&=&
\widehat{A}(X^{10}) {\rm ch}(E)~{\rm ch}(E')
\nonumber\\
&=&\widehat{A}(X^{10}) {\rm ch}(E\otimes E')
\nonumber\\
&=& {\rm Ind}_{E\otimes E'}(X^{10})\;.
\end{eqnarray}

\paragraph{The mod 2 index.}
If we take $E'=\overline{E}$ to be the complex 
conjugate of $E$, then we get the tensor bundle 
$E \otimes \overline{E}$, which is a real bundle. 
Thus, in this case, the complex K-theory 
description refines to that of $KO$-theory 
$KO(X^{10})$.

\subsubsection{Families of Dirac operators on the M-theory circle}
\label{sub fam}
In this section we consider the pushforward of bundles in eleven dimensions
to bundles in ten (and lower) dimensions at the level of K-theory. 
Several new features arise, including the appearance of
infinite-dimensional bundles.
 Let $\pi : Y^{11} \to X^{10}$ 
be a smooth fibration of manifolds with a Spin structure
along the fibers, so that $X^{10}$ is Spin with Spin bundle $SX \to X^{10}$.
Form 
a metric along the fibers, i.e., a 
metric on $T(Y^{11}/X^{10})$, and a smoothly varying family of
``horizontal subspaces" transverse to ker $\pi_*$.

\begin{itemize}

\item A family of circles over a ten-dimensional manifold $(Y^{11}_x~|~x \in X^{10})$ 
is a family of fibers $Y^{11}_x=\pi^{-1}(x)$ of a smooth
circle bundle $\pi : Y^{11}\to X^{10}$. 
Let $T(Y^{11}/X^{10})=TS^1 \subset TY^{11}$ be the bundle of 
vertical tangent vectors with dual $T^*(Y^{11}/X^{10}) \cong 
T^*Y^{11}/\pi^*T^*X^{10}$.

\item A family of vector bundles $(E_x~|~x\in X^{10})$ is a smooth 
vector bundle $E \to Y^{11}$, so that $E_x$ is the restriction 
of the bundle $E$ to $Y^{11}_x$. 
To the family $E\to Y^{11}$ we associated the infinite-dimensional 
bundle
$\pi_*E$ over $X^{10}$, whose fiber
$(\pi_* E)_x$
 at
$x \in X^{10}$ is the Fr\'echet  space 
$\Gamma (Y^{11}_x, E_x)$. 
A smooth section of $\pi_*E$ over $X^{10}$ is defined to be 
a smooth section of $E$ over $Y^{11}$:
\(
\Gamma (X^{10}, \pi_* E)=
\Gamma (Y^{11}, E)\;.
\)
\end{itemize}
Then the geometric family of Dirac operators is defined
as a differential geometric version of the 
analytical index
\(
{\rm Ind}\hspace{0.5mm}D=\ker D_x - {\rm coker}\hspace{0.5mm}D_x\;,
\label{ind def}
\)
which is the formal difference of parametrized families of vector 
spaces. If dim(ker\hspace{0.5mm}$D_x$) is constant in $x$, then each term
determines a vector bundle and so ${\rm Ind}\hspace{0.5mm}D$ makes sense
as an element of $K(X^{10})$. Similarly, the discussion can be 
extended to Dirac operators $D_E$ twisted by the bundle $E$.

\paragraph{The tangent bundles and Killing vector.}
Let $X^{10}$ be a ten-dimensional compact connected Spin 
manifold with a fixed Spin structure. Let $g_X$ be a metric on $TX^{10}$,
let $\nabla^X$ be the associated Levi-Civita connection and ${R}_X$ the 
curvature of $\nabla^X$. 
 Let $Z^{12}\to X^{10}$ be a 2-dimensional oriented vector 
bundle over $X^{10}$. Let $g_Z$ be a metric on $Z^{12}$ and let
$\nabla^Z$ be a connection on $Z^{12}$ preserving $g_Z$ and 
with curvature ${R}_Z$. 
Let $Y^{11}$ be the boundary of $Z^{12}$ with metric $g_Y$, so that
$
TY^{11}=TS^1 \oplus \pi^* TX^{10}$,
$
g_Y=g_Z|_{Y^{11}}=g_{S^1} \oplus \pi^* g_X$.
Let $\nabla^{LC}$ denote the Levi-Civita connection of $g_Y$. 
Then $Y^{11}$ is a circle bundle over $X^{10}$ with structure group
$SO(2)$ acting by isometries on the fibers. Furthermore, it carries a canonical 
induced Spin structure induced from the Spin structure on $X^{10}$. 
A Spin structure is also induced on the fiber (see section \ref{proj}).

\vspace{3mm}
Consider the free circle action on $Y^{11}$ and let $\xi$ 
denote  the Killing vector generating this action. 
Locally, $\xi=d\theta$, where $\theta$ is the local coordinate on
$S^1$. 
For every $y\in Y^{11}$, the tangent space to $Y^{11}$ at
$y$ splits into vertical and tangent horizontal spaces 
$
T_yY^{11}=TS^1_y \oplus T^HY^{11}_y
$,
where $TS^1_y$ is the one-dimensional subspace 
spanned by $\xi (y)$ and $T^HY^{11}_y=(TS^1_y)^\perp$
is the orthogonal complement. The projection 
$\pi_*$ defines an isomorphism $T^HY^{11}_y\cong T_{\pi y}X^{10}$
and there is a unique metric $h_X$ on $X^{10}$
for which this is also an isometry.
The horizontal distribution $T^HY^{11}$ defines a one-form 
$\xi^*$ such that
$\ker (\xi^*)=T^HY^{11}$ and is normalized so that 
$\xi^* (\xi)=1$. Locally, $\xi^*=d\theta +A$, where
$A$ is a horizontal one-form, the RR one-form potential. 
Its field strength $F_2$ pulls back to the curvature 
$d\xi^*$ of the principal connection. $F_2$ is 
both horizontal and invariant, hence basic
(see \cite{FS} \cite{MS}).


\paragraph{Geometric representative of the Euler class of the circle bundle.}
 Let $\pi^{S^1}$ be the orthogonal projection on $TS^1$ with respect to 
$g_Y$. Let $\nabla$ be a connection on $TY^{11}$ defined as
(cf. \cite{Bi})
\begin{eqnarray}
\nabla_z v_z&=&\pi^{S^1} (\nabla_z^{LC} v_z)\;,
\qquad \qquad \qquad
\nabla_\mu v_z=\pi^{S^1} (\nabla_\mu^{LC} v_z)\;,
\nonumber\\
\nabla_z v_\mu&=&0\;,
\hspace{3.8cm}
\nabla_\mu v_\nu=\nabla_\mu^X v_\nu\;,
\end{eqnarray}
for $v$ a vector with component $z$ along the eleventh direction or 
$\mu$ along $X^{10}$. 
As indicated in the introduction, this is more general than the usual 
Kaluza-Klein and Scherk-Schwarz settings for dimensional reduction 
in the sense of allowing $\nabla_\mu v_z$ to be nonzero in general.
We define the following:
\begin{enumerate}
\item $S$, a tensor defined by
$
S=\nabla^{LC} -\nabla
$.
Then \cite{Z2}
$
S(\xi)(\xi)=0
$
for $\xi \in TS^1$ the unit vector field determined by $g_Y$ and the Spin structure
on $TS^1$. 

\item $T$, the torsion of $\nabla$ defined by 
$
T_{\mu\nu}=-S_{\mu\nu} + S_{\nu \mu}
$.
\end{enumerate}
As above, let $\xi^* \in T^*S^1$ be the dual of $\xi$. Then, since
$\nabla^{LC}$ is torsion-free, we have the pairing 
\(
\langle T(U,V), \xi \rangle = d\xi^* (U,V)\;, \qquad U, V \in TX^{10}\;.
\label{Tuv}
\)
Therefore, $T$ determines a 2-form in $\Lambda^2(T^*X^{10})$ 
such that $\frac{1}{2\pi}T$ represents the Euler class $e(Z^{12})$
of $Z^{12}$. Thus the above torsion can be viewed as the RR 2-form
$F_2$.
 Let $\{e_1, \cdots e_{10}\}$ be an orthonormal basis of $TX^{10}$
and $\{ e^*_1, \cdots, e_{10}^*\}$ be the dual basis, i.e. a basis of 
$T^*X^{10}$. Set
\(
c(T)=\frac{1}{2}\sum_{a,b} e^*_a e^*_b c(T(e_a, e_b))\;.
\)
We view this as the $\gamma^{\mu \nu}F_{\mu \nu}$ (in the physicist's 
component notation) term 
in the dilatino supersymmetry transformation. We will make more connection 
below.

\paragraph{The Killing spinors.} 
Consider the case when the circle action in M-theory 
lifts to an action on the Spin bundle. This is 
infinitesimally generated by the spinorial Lie derivative 
\(
\mathsterling_\xi \varepsilon =\nabla_\xi \varepsilon +
\frac{1}{4}c(d \xi^*) \varepsilon\;,
\label{spinor lie}
\)
for any spinor $\varepsilon$ (compare to expression (\ref{kil psi})).
An invariant Killing spinor, specified by $D_\xi \varepsilon =0$ and
$\mathsterling_\xi \varepsilon=0$,
gives rise to a IIA Killing spinor \cite{J-max}.
In the presence of the dilaton $\phi$, i.e. for non-constant length vector,
we have that expression (\ref{spinor lie}) is changed to 
\(
(\mathsterling_\xi -\nabla_\xi)\varepsilon= \frac{1}{4}e^{-\phi/2}c(d\xi^*)\varepsilon\;,
\label{tino}
\)
which represents the supersymmetry transformation of the dilatino
in type IIA superstring theory.

\paragraph{Quillen superconnection and differential geometric representatives.}
 Quillen's superconnection \cite{Qu} is the differential geometric 
refinement of an element in K-theory in the same way 
that a connection on a smooth vector bundle is the 
differential geometric refinement of an equivalence of 
topological vector bundles. 
Let $V=V^+\oplus V^{-}$ be a smooth $\Z_2$-graded 
vector bundle over a smooth ten-dimensional manifold $X^{10}$.
A superconnection on $V$ consists of an ordinary connection 
$\nabla=\nabla^+ \oplus \nabla^{-}$ and a linear 
endomorphism 
$
L : V \to V$,
which anticommutes with the grading, i.e. $L(V^{\pm})\subseteq V^{\mp}$.
The superconnection is the operator $\nabla + L$ acting on the space
$\Omega^*(X^{10},V)$ of differential forms on $X^{10}$ with 
values in the bundle $V$. 
The K-theory element corresponding to 
$\nabla +L$  is 
$V^+ \buildrel{L}\over{\longrightarrow} V^{-}$.
Define the supertrace as the
map ${\rm Tr}_s: \Omega^*(X^{10},{\rm End}V) \longrightarrow 
\Omega^*(X^{10})$
taking
$\ttsmat{\alpha}{\beta}{\gamma}{\delta}\Big.$
to
$
{\rm tr}~\alpha - {\rm tr}~\delta
$.
Then the curvature is $(\nabla +L)^2$ and the Chern
character is ${\rm Tr}_s \exp (\nabla+L)^2$, and the latter 
differential form represents the topological Chern character
of the associated K-theory element \cite{Qu}.


\paragraph{Bismut superconnection.} The Bismut superconnection is obtained 
by applying Quillen's construction to a geometric family of 
Dirac operators. In our setting we have the following family
$
\pi : Y^{11} \to X^{10}
$.
The corresponding construction gives an {\it infinite-dimensional} superconnection 
over $X^{10}$. The fiber of the vector bundle at $x\in X^{10}$ is the space of 
spinors on $Y^{11}_x$, the fiber of 
$\pi$ at $x$. The metric on $Y^{11}_x$ gives this space 
an $L^2$ inner product, and the horizontal subspaces 
give it a unitary connection. 
The operator $L$ is the 
  Dirac operator $D$ modified by a degree 2 term which is 
 essentially the curvature $T$ of the field of horizontal planes
 (\ref{Tuv}).
The superconnection represents the associated 
differential geometric 
shriek map $\pi_!$, the pushforward in K-theory. 
 Because $D$ is Fredholm there is an element of 
$K(X^{10})$ associated to the superconnection
which is precisely Ind$D$ in (\ref{ind def}). 
Let $\widetilde{\nabla}$ be the natural lifting of 
$\nabla$ to the infinite-dimensional vector bundle
$
\Gamma \left( S(S^1) \otimes {\sf R}(TY^{11}|_{S^1})\right)
$
over $X^{10}$, where ${\sf R}$ denotes symmetric or antisymmetric 
powers. The connection $\widetilde{\nabla}$ is unitary. 
 The Bismut superconnection for the family $\{ D_{S^1,{\sf R}}\}$
is \cite{Bi}
\(
\mathbb{A}_u =\widetilde{\nabla} + {u} D_{S^1, {\sf R}}
- \frac{c(T)}{4 u}\;.
\label{bis}
\)
The last term, which is the difference of the Bismut superconnection and 
the Quillen superconnection, is the Clifford multiplication with the 
horizontal distribution, which can be considered as a two-form 
on $X^{10}$ with values in the vertical vector fields. 

\vspace{3mm}
{\it The interpretation:} Expression (\ref{bis}), which takes the important effect of the 
family on the M-theory circle into account,  is a families version of the 
dilatino supersymmetry transformation 
(\ref{tino}), where the Bismut 
superconnection replaces the spinorial Lie derivative 
$\mathsterling_\xi$
and the Quillen superconnection replaces the covariant derivative 
on spinors
 $\nabla_\xi$.

\vspace{3mm}
Now consider the Bismut superconnection 
(\ref{bis})
in the limit $u \to 0$. 
In this case,
\begin{itemize}
\item The last term is dominant so that the effect of the 
torsion field takes over. 
\item The adiabatic limit is essentially taking $u \to 0$ in the 
Quillen superconnection, i.e. in the first two terms in (\ref{bis}).
\item Comparing with the expression of the Dirac operator 
(\ref{D split})
we see that $u$ is essentially $e^{-\phi}$, with $\phi$ the dilaton.
In dynamical terms, $e^{\phi}$ represents the string coupling constant, so that
$u$ is the inverse of the string coupling constant  $e^{-\phi}$
(up to possible normalization 
factors). Therefore, the $u \to 0$ limit is the weak string coupling limit.
Viewed from a different angle, this corresponds to $\phi \to \infty$,
which geometrically means that the volume of the fiber is getting 
large. 
\item The behavior as $u \to 0$
leads to a differential geometric Riemann-Roch
formula.
Indeed, the Chern character of the Bismut superconnection approaches
\cite{Bi}
\(
\int_{Y/X} \widehat{A}(\Omega^{(Y/X)})
\label{fam fr}
\)
as $u \to 0$. Here $\Omega^{(Y/X)}$ is the curvature of the connection
on $T(Y/X)$ determined by the geometric data. 
The differential form in (\ref{fam fr}) represents in de Rham cohomology 
the cohomology class in 
$
{\rm ch}({\rm Ind~}D)=\pi_* (\widehat{A}(Y/X))$.

\end{itemize}

%
%

\subsection{Equivariant eta invariants, eta forms, and adiabatic limits}
\label{equiv e}


%

\paragraph{The Spin bundle.} 
Now that we described the integration over the fiber, we can continue
the discussion from section \ref{sub fam} and generalize
to Clifford modules.
Let $\mathcal{E}$ be a Clifford module along the 
fibers of $\pi$ (e.g. an $E_8$ bundle). This means that we 
have a Hermitian 
vector bundle over $Y^{11}$ with a skew-adjoint 
action $c: C(T^*(S^1)) \to {\rm End}(\mathcal{E})$ of the 
vertical Clifford bundle of $\pi$, and a Hermitian connection
$\nabla^{\mathcal{E}}$ compatible with this action,
$
\left[\nabla_{\sf Y}, c(\xi_z) \right]=
c(\nabla_{\sf Y}^{S^1} \xi_z)
$,
for ${\sf Y}\in \Gamma(Y^{11}, TY^{11})$ and 
$\xi_z \in \Gamma (Y^{11}, T^*S^1)$. 
There is a one-to-one correspondence (see e.g. \cite{BGV})
\(
\{{\rm Superconnections ~on ~the ~bundle~} \pi_* \mathcal{E}\}
=\{ {\rm Dirac~ operators~ on~ the~ Clifford~ module~}
SX^{10}\otimes \pi_* \mathcal{E}\}\;,
\)
 that is, an isomorphism
\(
\Gamma (X^{10}, SX^{10} \otimes \pi_*\mathcal{E})
\cong
\Gamma (Y^{11}, \pi^*SX^{10} \otimes \mathcal{E})\;.
\)

\paragraph{ The adiabatic limit for circle bundles.}
 Consider $S^1$-equivariant Dirac operator $D$ on $\cE \to S^1$.
Let $\overline{D}$ on $\overline{\cE}$ be a family of Dirac operators on $X^{10}$ induced by $D$.
Then  $\cE_Y := \overline{\cE} \otimes \pi^* SX^{10}$ becomes a Clifford module over 
$Y^{11}$ and 
$\nabla^{\overline{\cE}}$ induces a Clifford connection on $\cE_Y$. 
Rescale the metric on the fibers by $t >0$. Then the Dirac operator 
$D_{Y,t}$ may be viewed as the Dirac operator on 
$\pi_* {\cE}_Y=\pi_* \overline{\cE} \otimes SX^{10}$
associated to the Bismut Levi-Civita connection $\mathbb{A}_t$ on $\pi_*\overline{\cE}$.
 The results of Bismut-Cheeger \cite{BC} and Dai \cite{Dai} give
\(
\lim_{t \to 0} \eta (D_{Y, t}) \equiv 
2 (2\pi i)^{-\frac{\dim X}{2}} \int_X \widehat{A}(X) \wedge \widehat{\eta} (\overline{D}) \qquad \mod \Z\;,
\label{BCD}
\)
where $\widehat{\eta}$ is an even form on
$X^{10}$ defined by Bismut-Cheeger \cite{BC}
\(
\widehat{\eta}=\frac{1}{\sqrt{\pi}}
\int_0^\infty
{\rm Tr}^{\rm even} \left[ \left( D_{S^1,{\sf R}}+\frac{c(T)}{4u}\right)
\exp (-\mathbb{A}_u^2)
\right]
\frac{du}{2\sqrt{u}}\;.
\)
\begin{enumerate}
\item Since $u\sim e^{-\phi}$ then 
$u=0$ corresponds to $\phi=+\infty$ and 
$u=\infty$ to $\phi=-\infty$.
\item The integral is then over all $\phi$ and resembles
a soliton configuration. Thus we are summing over 
all dilaton configurations, that is, in a sense, integrating over
string coupling parameters. 
\end{enumerate}

\paragraph{Eta forms on circle bundles giving an $H$-field on the base.}
Consider the $S^1$ bundle $\pi : Y^{11} \to X^{10}$, with the $E_8$ vector
bundle $V$ over $Y^{11}$. Then, considering the degree 2 eta-form,
\(
d \widehat{\eta}_2=\int_{S^1} a \in \Omega^3(X^{10})\;.
\)
As considered in \cite{Sgerbe}, this gives the $B$-field
and the $H$-field in type IIA string theory as the differential forms 
$\widehat{\eta}_2$ and $d \widehat{\eta}_2$, respectively. 
Refinement to Deligne cohomology will be considered in 
section \ref{form del}.

\paragraph{Adiabatic limit for bundles with higher-dimensional fibers.}
Consider a fiber 
bundle $M^d \to Y^{11} \buildrel{\pi_M}\over{\longrightarrow} X^{11-d}$ 
 with a
Spin structure on the vertical tangent bundle $TM^d=\ker (d \pi_M)$,
with curvature $R_{TM}$. 
Let $V$ be a vector bundle on $Y^{11}$ (which will be either an 
$E_8$ bundle or the Rarita-Schwinger bundle, or some formal combination)
with a connection and curvature $F^V$. 
The differential of the eta-form will define 
a differential form over the base $X^{11-d}$ obtained by integrating 
an index formula over the fiber $M^d$,
\(
d\widehat{\eta}=\int_{M^d} \widehat{A}(R_{TM}) \wedge {\rm ch}(F^V).
\) 
{\it Example: $S^3$-bundles.}
Consider the following example. Let $M^d=S^3$.
Taking $V$ to be the $E_8$ bundle with characteristic class $a$ of degree 4, then,
with $\widehat{A}(R_{TM})=1$, we have that we get a degree zero eta-form on 
the base $X^8$ obtained by integrating $a$ over $S^3$ (up to a sign)
\(
d \widehat{\eta}_{0}=\int_{S^3} a\;.
\)
Thus, we get an eta-function $\eta=\widehat{\eta}_0$ 
over $M^7$, the  
boundary of the space $X^8$,
\begin{eqnarray}
S^3 \longrightarrow Y^{11} &\longrightarrow & 
X^8 \hookleftarrow M^7
\nonumber\\
a &\longmapsto & d \eta \longleftarrow \eta\;.
\end{eqnarray}
 Such a space is guaranteed to exist because of the 
vanishing of the Spin cobordism group in dimension seven, $\Omega_7^{\rm Spin}=0$.
Also, if we extend an $E_8$ bundle then we would need 
 $\Omega_7^{\rm Spin}(K(\Z, 4))$, which indeed vanishes (see e.g. \cite{HoS}). 
The case of the Rarita-Schwinger bundle is similar.

\vspace{3mm}
Now let $\pi : Y^{11} \to X$ be a totally geodesic fiber bundle with odd-dimensional 
fiber $M$ and structure group $G$. Similar arguments give (\ref{BCD}).
If the dimension of the fiber is even then the infinite-dimensional bundle 
$\pi_*E$ is graded, i.e. it splits as 
$\pi_*E^+ \oplus \pi_*E^{-}$. This case will be considered in section
\ref{form del}.

\paragraph{Equivariant eta invariants.}
Let $G$ act isometrically on $Y^{11}$, a compact oriented Riemannian 11-manifold.
Let $D$ be a $G$-equivariant Dirac operator on a $G$-equivariant Dirac bundle 
$\cE$ over $Y^{11}$. 
One can form equivariant eta invariants with respect to either 
the group or its Lie algebra. The latter is the $\gg$-equivariant 
eta invariant $\eta_{\sf X} (D)$ with respect to an element ${\sf X}\in \gg$
of the Lie algebra $\gg$ \cite{Go1}.   
The former is the $G$-equivariant eta invariant $\eta_{e^{-{\sf X}}}(D)$ 
with respect to an element $e^{-{\sf X}}$ of the Lie group $G$ \cite{Do}.
The two are directly related; in fact, the difference $\eta_{\sf X} (D)
- \eta_{e^{-{\sf X}}}(D)$ is locally computable when $Y^{11}$ is a boundary,
a fact that allows for relating $G$-equivariant $\eta$-invariants and $\eta$-forms
of totally geodesic fiber bundles with structure group $G$ 
\cite{Go1}.

\vspace{3mm}
Let $\C [\gg^*]$ denote the space of formal power series in ${\sf X} \in \gg$. 
Goette \cite{Go1} defines equivariant eta invariants $\eta_{\sf X} (D) \in \C [\gg^*]$
such that when ${\sf X}=\Omega$ this gives the eta-form of any family of 
Dirac operators associated to a $G$-principal bundle over $X$ with 
curvature
 $\Omega \in \cA^2(X;\gg):=\Gamma(\Lambda^2 X) \otimes \gg$, 
 a $\gg$-valued 2-form on  $X$.
For ${\sf X}\in \gg$, let ${\sf X}_Y(y)=\frac{d}{dt}|_{t=0} e^{-t{\sf X}} y$ be the Killing 
field induced by ${\sf X}$, let $c_{\sf X}$ denote Clifford multiplication with 
${\sf X}_Y$, and let $\mathsterling_\gg$ denote the infinitesimal action of 
$\gg$ on $\cE$. Define the $\gg$-equivariant Dirac operator 
$
D_{\sf X}:= D -\frac{1}{2}c_{\sf X}
$
and the equivariant Bismut Laplacian (cf. \cite{BGV})
\(
\mathcal{H}_{\sf X}:= D_{-{\sf X}}^2 + \mathsterling_\gg =
(D +\frac{1}{2}c_{\sf X})^2 + \mathsterling_\gg\;.
\)
This is a generalized Laplacian which depends polynomially on 
${\sf X}\in \gg$ and can be thought of as a quantum 
analog of the equivariant Riemann curvature 
$R_\gg({\sf X})=(\nabla - \iota ({\sf X}))^2 +\mathsterling_\gg$.
The $\gg$-equivariant eta invariant is now defined as 
the convergent power series
\(
\eta_{\sf X} (D):= \int_0^\infty \frac{1}{\sqrt{\pi t}}
{\rm tr} \left( 
D_{\frac{\sf X}{t}} e^{-t\mathcal{H}_{\frac{\sf X}{t}}}
\right)
dt\;.
\)
Note that this encodes the Clifford multiplication in the 
fiber dimensions.

\paragraph{\bf Relating eta form to eta invariant.}
Let $Y^{11} \to X^{11-d}$ be a totally geodesic fiber bundle with 
compact, oriented, odd-dimensional fiber $M^d$ and compact holonomy group 
$G$. Let $\overline{D}_M$ be a family of Dirac operators induced by a $G$-equivariant 
Dirac operator $D_M$ on a $G$-equivariant Clifford module $E_M$. 
Let $V$ be an element of $\gg$, the Lie algebra of the isometry 
group of the fiber. Then the 
$\eta$-form $\widehat{\eta}(\overline{D}_M)$ of the family $\overline{D}_M$ is related to
the $g$-equivariant $\eta$-invariant $\eta_V(D_M)$ of $D_M$ by (\cite{Go1})
\(
\widehat{\eta} (\overline{D}_M) =\frac{1}{2} \eta_{V} (D_M)\;.
\) 
Note that by definition, $\eta_V (D_M)\in \C [\gg^*]$ is
${\rm Ad}_G$-invariant. As the image of $\eta_V (D_M)$
under the Chern-Weil homomorphism, 
$\widehat{\eta} ({\overline D}_M)$ is thus a characteristic form
of the bundle $P \to X^{11-d}$. In particular, it is closed and 
$[\widehat{\eta} ({\overline D}_M)]\in H^*(X^{11-d};\R)$ is independent of the 
connection on $P$. However, a change of the metric 
on $M^d$ will in general affect $[\widehat{\eta} ({\overline D}_M)]$.

\paragraph{Eta form on the circle.}
This is considered in \cite{Go1}.
Let $D=-i\left(\frac{\partial}{\partial \theta}  + c\right)$
be the Dirac operator on the trivial bundle $\cE:=S^1 \times \C$ over
$S^1=\R/2\pi \Z$ with respect to the connection 
$\nabla_{\frac{\partial}{\partial \theta} }=\frac{\partial}{\partial \theta} +ic$. 
Define an $\R$-operation on $\cE$ by
\(
g(\theta, z):=(\theta + g, e^{i\lambda g}z), \qquad {\rm for}~g\in\R\;. 
\)
The Dirac operator with respect to the trivial Spin structure corresponds to 
$c=\lambda=0$ and is $\R/2\pi \Z$-equivariant. 
The Dirac operator with respect to the non-trivial Spin structure corresponds to 
$c=-\lambda=\frac{1}{2}$ and is $\R/4\pi \Z$-equivariant. 
Then $D(e^{i \nu \theta})=(\nu +c)e^{i\nu \theta}$ and $g^*(e^{i\nu \theta})=
e^{i\nu (\theta-g) +i \lambda g}$.
For the untwisted Dirac operators $D_1$ and $D_2$ with
respect to the trivial and nontrivial Spin structures on $S^1$, respectively,
we have
\(
\eta_V(D_1)=i\frac{\frac{x}{2}-{\rm tan}\frac{x}{2}}
{\frac{x}{2}{\rm tan}\frac{x}{2}}
=2\sum_{k=1}^\infty \zeta (-k) \frac{(-ix)^k}{k!}\;,
\label{d1}
\)
and
\(
\eta_V(D_2)=i\frac{\frac{x}{2}-{\rm sin}\frac{x}{2}}
{\frac{x}{2}{\rm sin}\frac{x}{2}}
=2\sum_{k=1}^\infty \zeta (-k, \frac{1}{2}) \frac{(-ix)^k}{k!}\;.
\)


\paragraph{Eta forms over principal circle bundles.}
Let $Y^{11} \to X^{10}$ be an $S^1$-principal bundle. If the associated 
Hermitian line bundle $\cL$ has curvature $F^{\cL} \in i \cA^2(X^{10})$, its
first Chern form is $c_1(\cL)=-\frac{1}{2\pi i} [F^{\cL}]$
so that 
${\rm ch}(\cL)=e^{2\pi i c_1(\cL)}$. Then the curvature 
of $Y^{11}$ itself equals $\Omega=iF\frac{\partial}{\partial \theta}$.
Let $D_{S^1}$ be the equivariant Dirac operator on $S^1$. 
Then $D_{S^1}$ induces a family $\overline{D}_{S^1}$ of fiber-wise 
Dirac operators on $Y^{11}$. Then the $\eta$-form of this family
represents the characteristic class \cite{Go1}
\begin{eqnarray}
[\widehat{\eta}(\overline{D}_{S^1})] &=& \frac{1}{2} [\eta (D_{S^1})]
\nonumber\\
&=& 
\sum_{k=0}^{\infty} 
\frac
{B_{k+1}(\lambda-[-c]) + 
B_{k+1}(1+\lambda+[c]) 
-2 (c+\lambda)^{k+1}
}
{(k+1)!} (2\pi i c_1(\cL))^k\;.
\label{Go s1}
\end{eqnarray}
Here $B_k(r)$ are Bernoulli polynomials, defined by
$
\frac{te^{rt}}{e^t-1}=\sum_{k=0}^\infty B_k(r) \frac{t^k}{k!}
$, 
which coincide with Bernoulli 
numbers when $r=0$, $B_k=B_k(0)$, i.e. 
$B_0(r)=1$,
$B_1(r)=r -\frac{1}{2}$,
$B_2(r)=r^2 -r +\frac{1}{6}$.
For example, for the operator induced by $D_1$ in (\ref{d1}),
\(
[\widehat{\eta}(\overline{D}_1)]=\frac{1}{2}
\left[ 
\frac{\frac{F^{\cL}}{2}-{\rm tanh}\frac{F^{\cL}}{2}}
{\frac{F^{\cL}}{2}{\rm tanh}\frac{F^{\cL}}{2}} \right]
=-\sum_{k=0}^\infty \zeta (-k) \frac{(2\pi i c_1(\cL))^k}{k!}\;.
\)
This was also derived in \cite{Z1} \cite{Z2}.


\paragraph{Example 1: Heisenberg manifolds.}
The $11$-dimensional Heisenberg manifold 
can be described as an $S^1$-bundle $Y^{11}$ over the ten-torus $T^{10}$,
with Chern class
$
c_1(Y^{11})=-(d\theta_1 \wedge d\theta_2 + \cdots + d\theta_{9} \wedge d\theta_{10})
$,
where $\theta\in \R/\Z$ parametrizes the $i$th factor of $T^{10}=\underbrace{S^1 \times \cdots \times S^1}_{10}$.
Since the torus is flat, it has $\widehat{A}(T^{10})=1$, so that the only term 
contributing to the adiabatic limit is
$
c_1(Y)^5=(-1)^5 5! d\theta_1 \wedge \cdots \wedge d\theta_{10}
$.
For the untwisted Dirac operator $D_{Y, t}$ determined
by $P_{\rm Spin}Y^{11}$ on the bundle $Y^{11}$ with fiber of length $t$, 
the 
adiabatic limit is
\(
\lim_{t \to 0} \eta (D_{Y, t})
=
\left\{
\begin{tabular}{ll}
$2\zeta (-5)$  & if the Spin structure restricted to the fiber is trivial, and\\
$2\zeta (-5, \frac{1}{2} )$ & otherwise.
\end{tabular}
\right.
\)
Now consider the seven-dimensional case, i.e. $Y^{7}$ is a circle bundle over the 
six-torus $T^6$. In this case, the adiabatic limit vanishes.

\paragraph{Example 2: $S^3$-bundles.}
Let $D$ be the untwisted Dirac operator on $S^3$. The equivariant 
eta invariant corresponding to 
\(
V=
\left(
\begin{tabular}{cccc}
0&$x$&& \\
$-x$ & 0 &&\\
&&0& $y$\\
&& $-y$ & 0
\end{tabular}
\right)
\in \frak{so}(4) \cong \frak{spin}(4)
\label{mat xy}
\)
is \cite{Go1}
\(
\eta_V=2 \sum_{k,l=1}^\infty (-1)^{k+l}
\left(
\frac{B_{2k+2l}(1/2)}{(2k+2l)!}
-
\frac{B_{2k}(1/2) B_{2l}(1/2)}{(2k)!(2l)!}
\right)
x^{2k-1}y^{2l-1}\;.
\)
Now we would like to consider $S^3$ bundles $Y^{11} \to X^8$. 
The first Pontrjagin class $p_1(E_4)$
and the Euler class $\chi (E_4)$ of the Spin$(4)$ bundle correspond to the polynomials
\(
V \mapsto (-2\pi i)^{-2} (x^2 + y^2)\;,
\qquad \qquad
V \mapsto  (-2\pi i)^2 xy\;.
\)
The $\eta$-form for the induced Dirac operator $\overline{D}$ on $S^3$-bundle is 
(without the factors of $2\pi i$)
\begin{eqnarray}
\widehat{\eta} (\overline{D})&=& 
\widehat{\eta}_4 + \widehat{\eta}_8 
+ \cdots
\nonumber\\
&=&
\frac{\chi(E_4)}{2^7\cdot 3 \cdot 5} -
\frac{\chi(E_4) p_1(E_4)}{2^9\cdot 3 \cdot 5 \cdot 7}\;.
\end{eqnarray}
Now $X^8$ is eight-dimensional so that the possible components of the eta-forms appearing
in (\ref{BCD}) are 0, 4, and 8,
\begin{eqnarray}
\lim_{t \to 0} \eta (D_{Y, t}) &=&
\int_{X^8} \widehat{A}_8(X^8) + \widehat{A}_4 (X^8) \wedge \widehat{\eta}_4 +
\widehat{\eta}_8
\nonumber\\
&=&
\int_{X^8} 
\frac{\left(7p_1(X^8)^2-4p_2(X^8)\right)}{2^7\cdot 3^2 \cdot 5}
-
 \frac{\chi(E_4)p_1(X^8)}{2^{14}\cdot 3^3\cdot 5}
-  \frac{\chi(E_4) p_1(E_4)}{2^9\cdot 3 \cdot 5 \cdot 7}\;.
\end{eqnarray}

\paragraph{Spherical space forms.}
Consider the three-dimensional Spin spherical space form 
$M^3=S^3/\Gamma$.
The value of the equivariant eta invariant depends on the Spin structure.
For example, consider $\R P^3=S^3/\Gamma$, where $\Gamma=\{{\rm id}, -{\rm id} \}
\subset SO(4)$. The finite group
$\Gamma$ acts freely on $S^3$ preserving the orientation. There are
two lifts of $\Gamma$ to Spin$(4)$
\(
\tilde{\Gamma}_1:= \{1, e_1e_2 e_3 e_4\}\;,
\qquad \qquad
\tilde{\Gamma}_2:= \{1,- e_1e_2 e_3 e_4\}\;,
\)
both in Spin$(4)$ $\subset C\ell (\R^4)$. 
When lifted 
to the Spin bundle, these give two different Spin structures 
on the quotient (see section \ref{sph sp}).
Let $D$ be the Dirac operator on $S^3$ and 
let $D_1$ and $D_2$ be the untwisted 
Dirac operators on $\R P^3$. The equivariant eta invariants are
given by (see \cite{Go1})
\begin{eqnarray}
\eta_V(D_1) &=& \frac{1}{2} \eta_V(D) -\frac{1}{4 \cos \frac{x}{2} \cos \frac{y}{2}}\;,
\\
\eta_V(D_2) &=& \frac{1}{2} \eta_V(D) +\frac{1}{4 \cos \frac{x}{2} \cos \frac{y}{2}}\;,
\end{eqnarray}
where $x$ and $y$ are the entries in the matrix (\ref{mat xy}).



\paragraph{Effect of parity on Dirac structures.}
In M-theory, the action and equations of motion admit a parity symmetry
given by an odd number of space or time reflections together with 
$C_3 \mapsto - C_3$ \cite{DNP}. Thus, in the Riemannian case, this is 
orientation reversal. 
In the presence of the one-loop 
term and of $E_8$ bundles, 
this parity symmetry takes the form \cite{Flux} \cite{DMW}
 $G_4 \mapsto -G_4$ and 
$a \mapsto \lambda -a$, where $a$ is the characteristic class of 
the $E_8$ bundle and $\lambda$ is the first Spin characteristic 
class. This parity is also a reflection on the $E_8$ class $a$ if 
M-theory is taken on String eleven-dimensional manifolds such as in \cite{SE8}.

\vspace{3mm}
We saw at the end of section
\ref{ana eff} how the Spin structures in ten and eleven 
dimensions get modified when the orientation of the 
manifold is reversed. 
 Let $-Y^{11}$ denote the eleven-dimensional manifold 
with the opposite orientation and Spin structure. 
Then the Dirac structures on $Y^{11}$ and $-Y^{11}$ can be 
related as \cite{Bu}
\(
\mathcal{S}(-Y^{11}) \cong \mathcal{S}(Y^{11})^{\rm op}\;,
\)
where $\mathcal{S}(-Y^{11})$ is the Clifford module corresponding 
to the orientation-reversed manifold $-Y^{11}$ and 
$\mathcal{S}(Y^{11})^{\rm op}$ is obtained from 
$\mathcal{S}(Y^{11})$ by reversing  the homomorphism 
$c : C\ell (TY^{11}) \to {\rm End}(V)$, with $V$ the vector bundle 
used to give the module structure. For us, this is the 
$E_8$ bundle, or the tangent bundle $TY^{11}$ or a flat 
bundle, capturing the fundamental group. The eta-forms 
satisfy $\widehat{\eta}(\mathcal{E}^{\rm op})=-\widehat{\eta}(\mathcal{E})$,
where $\mathcal{E}$ collectively corresponds to the geometric family 
as in \cite{Bu}.
When the eleven-dimensional manifold $Y^{11}$ is a product or a bundle, we 
can apply the construction to the fiber (which could be a spherical space 
form) and the base. 
An analogous situation in type IIA string theory for the $B$-field 
is considered in \cite{Sgerbe}.


\subsection{The fields via eta forms, gerbes, and Deligne cohomology}
\label{form del}

\paragraph{The disk bundle and the circle bundle.}
Let $\mathbb{D}^2 \to Z^{12} \buildrel{\pi_D}\over{\longrightarrow} X^{10}$ be the disk bundle corresponding to
the circle bundle $S^1=\partial \mathbb{D}^2 \to Y^{11}=\partial Z^{12} \buildrel{\pi_S}\over{\longrightarrow}
 X^{10}$.
Assume that $X^{10}$ is compact and Spin and 
fix a Spin structure on $T\mathbb{D}^2$. 
Let $g_{\mathbb{D}^2}$ be a metric on $T\mathbb{D}^2$ and let $g_{S^1}$ be the
restriction to the circle. 
Assume that there is a neighborhood $[-1,0]\times  Y^{11}$ of 
$Y^{11}$ in $Z^{12}$ such that for any $x\in X^{10}$, on 
$\pi_D^{-1}(x) \cap ([0,1] \times Y^{11})=[-1,0]\times S_x^1$,
$g_{\mathbb{D}^2}$ takes the form 
$
g_{\mathbb{D}^2}|_{[-1,0]\times S_x^1}=dt^2 \oplus g_{S^1_x}$.
\begin{itemize}
\item Let $S\mathbb{D}^2$ be the spinor bundle of $(T\mathbb{D}^2, g_{\mathbb{D}^2})$. Since 
the disk is even-dimensional, the Spin bundle splits canonically 
into positive and negative Spin bundles
$
S\mathbb{D}^2=S^+ \mathbb{D}^2 \oplus S^{-}\mathbb{D}^2
$.
\item Let $E$ be a complex vector bundle over $Z^{12}$. Let $g_E$ be a metric on $E$
such that
$
g_E|_{[-1,0]\times S^1}=\pi^* g_E|_{S^1}
$.
\item Let $\nabla^E$ be a Hermitian connection on $E$ such that
$
\nabla^E|_{[-1,0]\times S^1}=\pi^* \nabla^E|_{S^1}
$.
\item Let $SS^1$ be the Spin bundle of $(TS^1, g_{S^1})$. 
Associated to $g_{S^1_y}$, $g_{E|_{S^1_x}}$, and $\nabla^{E|_{S^1_x}}$ are
canonically defined (twisted) Dirac operators $D_x^{S^1, E}$.
\end{itemize}
For any $x \in X^{10}$, associated to $g_{\mathbb{D}^2_x}$, $g_{E|_{\mathbb{D}^2_x}}$ and $\nabla^{E|_{\mathbb{D}^2_x}}$,
 there is also a twisted Dirac operator 
\(
D_x^{\mathbb{D}^2, E}: \Gamma(S^+ \mathbb{D}^2_x \otimes E|_{\mathbb{D}^2_x}) \to
\Gamma(S^{-} \mathbb{D}^2_x \otimes E|_{\mathbb{D}^2_x})\;.
\)
On $U_x=[-1,0]\times S^1_x$, this takes the form
$
D_x^{\mathbb{D}^2, E}=c (\partial_t ) 
\left(\partial_t + D_x^{S^1, E}
\right)
$.

\paragraph{Fields in Deligne cohomology from eta forms.}
Eta-forms behave nicely under inclusion of open subsets in a space. If 
$\mathcal{E}|_U$ is the restriction of the geometric family 
$\mathcal{E}$ of Dirac operators 
to the subset $U \subseteq X$, then (see \cite{Bu})
\(
\widehat{\eta} (\mathcal{E}|_U)= \widehat{\eta}(\mathcal{E})|_U\;.
\)
Then there is a corresponding Dirac operator $D_\alpha$ 
on each open subset
$U_\alpha$.
This allows for a description in terms of Deligne cohomology 
\cite{Lo} \cite{Bu}. Thus, we will refine the discussion of
section \ref{equiv e}.

\vspace{3mm}
The degree zero component $\widehat{\eta}_0$ is half of the Atiyah-Patodi-Singer
eta invariant of $D_\alpha$. Consider an $S^3$ bundle $Y^{11} \to X^8$.
If $U_\alpha \cap U_\beta \neq \emptyset$ then 
$\widehat{\eta}^0_\beta|_{U_\alpha \cap U_\beta} 
-\widehat{\eta}^0_\alpha|_{U_\alpha \cap U_\beta}$ is an integer-valued
function on $U_\alpha \cap U_\beta$. Define $f_\alpha : U_\alpha \to S^1$
by
$f_\alpha=e^{2\pi i \widehat{\eta}_\alpha^0}$. Then if 
$U_\alpha \cap U_\beta\neq \emptyset$ then 
$f_\alpha|_{U_\alpha \cap U_\beta}
=
f_\beta|_{U_\alpha \cap U_\beta}
$, 
so that the functions $\{f_\alpha\}_{\alpha \in I}$ glue together to form 
a function 
$f: X^{8} \to S^1$ such that $f|_{U_\alpha}=f_\alpha$. 
Let $[S^1]\in H^1(S^1;\Z)$ be the fundamental class of $S^1$. Then, applying
\cite{Lo}, 
$f^*[S^1]\in H^1(X^{8};\Z)$ is represented in real cohomology
by the closed form 
\(
\frac{1}{2\pi i} {\rm ln} f= 
\left( \int_{S^3} \widehat{A}(R_{S^3}) \wedge {\rm ch} (F^V) \right)_{[1]}
\in \Omega^1(X^8)\;.
\)
Next we consider the gerbe on $X^{10}$ via the circle bundle
$\pi : Y^{11} \to X^{10}$. The `phase' 
$\frac{D}{|D|}$ 
of the Dirac operator $D$ has eigenvalues $\pm 1$. Thus on $U_\alpha \cap U_\beta$ 
the operator $\frac{D_\beta}{|D_\beta|}-\frac{D_\alpha}{|D_\alpha|}$
has eigenvalues $0,2$ and $-2$. 
Let $P_0$, $P_2$ and $P_{-2}$, respectively,
 be the images of the projections of the  
corresponding eigenspaces. These are finite-dimensional 
vector bundles on $U_\alpha \cap U_\beta$. 
From \cite{Lo}, this gives a gerbe with connection on $X^{10}$ 
\(
d \widehat{\eta}^2=\left( 
\int_{S^1} \widehat{A}(R_{S^1}) \wedge {\rm ch}(F^V)
\right)_{[3]}
\in \Omega^3(X^{10})
\label{h from eta}
\)
with data:
\begin{enumerate}
\item Line bundle $L_{\alpha \beta}=\Lambda^{\rm max} (P_2) \otimes (\Lambda^{\rm max}(P_{-2}))^{-1}$
with a unitary connection $\nabla_{\alpha \beta}$, inherited
from the projected connections on $P_2$ and $P_{-2}$. 
\item The curvature of $\nabla_{\alpha \beta}$ is 
$F_{\alpha \beta}=\widehat{\eta}^2_\alpha - \widehat{\eta}^2_\beta$
\item a nowhere zero section $\theta_{\alpha \beta \gamma}$ on 
$L_{\alpha \beta} \otimes L_{\beta \gamma} \otimes L_{\gamma \alpha}$ 
if $U_\alpha \cap U_\beta \cap U_\gamma\neq \emptyset$.
\item $F_\alpha$, the 2-form component of the eta-form.
\end{enumerate} 
As a rational cohomology class, (\ref{h from eta}) lies in the image of 
$H^3(X^{10};\Z)\to H^3(X^{10};\Q)$. This is the $H$-field on $X^{10}$. 

\vspace{3mm}
Next we consider examples where the dimension of the fiber is even. 
Here we can start in twelve or in eleven dimensions. In the first case
we can a flat $B$-field in type IIA string theory. Let 
$\pi_D : Z^{12}\to X^{10}$ be a disk bundle with fiber the 
disk $\mathbb{D}^2$. Let $D^\mathbb{D}$ denote the family 
$D^\mathbb{D}=\{D_x\}_{x \in X^{10}}$ of Dirac operators, with 
$D_x$ acting on $C^\infty (\mathbb{D}_x^2;E|_{\mathbb{D}^2_x})$. Then 
\(
d \widehat{\eta}_1=
\left( 
\int_{\mathbb{D}^2} \widehat{A}(R_{\mathbb{D}^2}) \wedge {\rm ch}(F^V)
\right)_{[2]}
\in \Omega^2 (X^{10})\;. 
\)
Next we consider M-theory on the two-torus, i.e. take $Y^{11}$
to be the total space of a torus bundle $\pi': Y^{11}\to X^9$, with 
fiber $T^2$. Let $D'$ be the family of Dirac operators 
$D'=\{D'_x\}_{x \in X^9}$ of Dirac operators, with
$D'_x$ acting on 
$C^\infty (T^2_x; E|_{T^2_x})$. Then we similarly get a flat $B$-field,
but now on $X^9$.

\vspace{3mm}
New contributions occur from the kernel of the Dirac operator.
Assume that that ${\ker}(D^D)$ has a constant rank, i.e. is a 
$\Z_2$-graded vector bundle on $X^{10}$. The connection 
$\nabla^{\pi_*E}$ projects into a connection 
$\nabla^{{\rm ker}(D^D)}$ on ${\rm Ker}(D^D)$
with curvature $F^{{\rm ker}(D^D)}$. Then 
\(
d \widehat{\eta}=\int_{\mathbb{D}^2} \widehat{A}(R_{\mathbb{D}^2}) \wedge {\rm ch}(F^V)
-{\rm ch}(F^{{\rm ker}(D^D)})\;.
\)
Assume further that for each $x \in X^{10}$, the index of $D_x^D$ vanishes in $\Z$, so that 
the bundles ${\rm ker}(D^D)_+$ and ${\rm ker}(D^D)_{-}$ have the same rank. 
The corresponding connections $\nabla^{{\rm ker}(D^D)_{\pm}}$
have curvatures $F^\pm$.  We choose an open covering of $X^{10}$ by open subsets such that
there is an isometric isomorphism $\kappa_\alpha : \ker (D^D)_+|_{U_\alpha}
\to \ker (D^D)_{-}|_{U_\alpha}$.
There is a 1-form $CS_1^\alpha \in \Omega^1(U_\alpha)$ 
such that 
\(
dCS_1^\alpha=
{\rm ch}(F^+)-
{\rm ch}(\kappa_\alpha^{-1} F^{-} \kappa_\alpha)
=
{\rm ch}(F^{{\rm ker}(D^D)})\;.
\)
Then this `Chern-Simons form' combines with the eta-form to give
\(
d(\widehat{\eta}_1 + CS_1^\alpha)=
\left( 
\int_{\mathbb{D}^2} \widehat{A}(R_{\mathbb{D}^2}) \wedge {\rm ch}(F^V)
\right)
\in \Omega^2(X^{10})\;.
\)
This gives an extra contribution to the flat $B$-field, given by the above 
Chern-Simons form coming from the kernel of the Dirac operator. 
The `descent relations' can be found in \cite{Lo}. 

\vspace{3mm}
Similarly we can consider M-theory on a two-torus. In this case, we have
\(
d(\widehat{\eta}_1 + CS_1^\alpha)=
\left( 
\int_{T^2} \widehat{A}(R_{T^2}) \wedge {\rm ch}(F^V)
\right)
\in \Omega^2(X^{9})\;.
\)

\paragraph{Filtration in K-theory.}
The above gerbes have obstructions which can be seen using 
the filtration in K-theory that leads to the Atiyah-Hirzebruch Spectral 
Sequence \cite{Lo} \cite{Bu}.
For a positive integer $p$ and an element in K-theory
$\psi \in K^*(X^{10})$, we say $\psi \in K_p^*(X^{10})$ if
$f^* \psi=0$ for all branes $M$ of dimension $<p$ and 
continuous maps $f: M \to X^{10}$. Then there is 
a natural decreasing filtration 
\(
\cdots \subseteq K_p^*(X^{10}) 
 \subseteq K_{p-1}^*(X^{10})
  \subseteq 
  \cdots 
   \subseteq K_0^*(X^{10})
  =  K^*(X^{10})
\)
which preserves periodicity and the ring structure. 
Consider a class $x\in K_p^r(X)$. 
A class in $E_2^{p,r-p}$ is
$z\in H^p(X;\Z)$. Its image 
$z_\Q \in H^p(X;\Q)$ of $z$ in rational cohomology
satisfies 
$
{\rm ch}_p (x)=z_\Q
$.

\vspace{3mm}
\noindent{\it Example: circle bundle.}
Using the calculation of the eta-form for circle bundles
 \eqref{Go s1},
the Deligne cohomology class corresponding to $z_{2m}$ is 
\cite{Bu}
\(
[\widehat{\eta} (z_{2m})]= a\left[ \frac{B_{m+1}}{(m+1)!} c_1(\omega)^m\right]
\in H^{2m+1}_{Del}(X^{10})\;.
\)
In particular, the curvature of the Deligne class is zero. 
Consider the case $m=1$. The Deligne cohomology class of $z_2$ 
corresponds to a gerbe of the geometric family. 
The index gerbe is then \cite{Bu} 
\(
[\widehat{\eta}(z_2)]=a\left[ \frac{1}{12}c_1(\omega)\right]\;.
\)
For example, take $X=\CP^1$ and let $Y \to X$ be the square of the 
Hopf bundle. Then $c_1(Y)=2$ and 
$[\widehat{\eta}(z_2)]\cong [1/6]_{\R/\Z}$ under the 
isomorphism 
$H^3_{Del}(\CP^1)\cong \R/\Z$.

\paragraph{Kaluza-Klein modes and Adams operations.} 
In \cite{DMW}, the massive modes of the spin $1/2$ fermions were used in the expressions for the 
index of the Dirac operator. Their idea could be summarized as follows. The M-theory circle 
bundle can be viewed as unit vectors in a complex line bundle $\mathcal{L}$. So functions on 
$Y^{11}$ that transform as 
$e^{-ik\theta}$ under rotations of the circle can be viewed as sections of ${\mathcal{L}}^k$ 
over $X$. The space of functions on $Y^{11}$ can be written in terms of the space of sections of 
${\mathcal{L}}^k$ on $X^{10}$ as
${\rm Func}(Y^{11})=\bigoplus_{k\in \ZZ} \Gamma(X^{10}, {\mathcal{L}}^k)$.
Here ${\rm Fun}(Y^{11})$ is the space of functions on $Y^{11}$, and 
$\Gamma(X^{10},{\cal L}^k)$ the space of sections of ${\cal L}^k$.

\vspace{3mm}
One can think of this line bundle as generating the higher modes. One can get from level $n=1$
all the other levels. Mathematically, one could think of using an Adams operation on the line bundle,
\(
\psi^k(\mathcal{L})={\mathcal{L}}^k
\)
so the bundles on $X^{10}$ are tensored with ${\mathcal{L}}^m$. 
This is a cohomology operation on $K(X^{10})$, i.e. 
$\psi^k: K^0(X^{10}) \to K^0(X^{10})$, $k=0, 1, 2, \cdots$. 
The Adams operations satisfy 
$\psi^k \circ \psi^l =\psi^{kl}
=\psi^l \circ \psi^k$. This property is satisfied in our
context of the M-theory line bundle and corresponds to 
additivity and commutativity of the Fourier modes. 
The coupling of this line bundle to
the vector bundles already in $X^{10}$ might give an idea about the 
extension beyond K-theory. We will consider this elsewhere. 


\vspace{7mm}
\noindent {\bf \large Acknowledgements}

\vspace{2mm}
\noindent 
The author thanks Jonathan Rosenberg and Stephan Stolz for 
very useful discussions and remarks. He also thanks 
S. Stolz his kind hospitality at the University of 
Notre Dame. The useful editorial suggestions from  
Arthur Greenspoon on the first draft
 are very much appreciated. The careful reading of the manuscript by the 
 referee and their useful suggestions are also gratefully acknowledged. 
This research is supported in part 
by NSF Grant PHY-1102218.



\begin{thebibliography}{99}

\bibitem{AG}
B. S. Acharya and S. Gukov,
{\it M theory and singularities of exceptional holonomy manifolds},
Phys. Rept. {\bf 392} (2004) 121--189,
[{\tt arXiv:hep-th/0409191}].

\bibitem{Ad}
J.F. Adams, {\it On the groups J(X), IV}, Topology {\bf 5} (1966), 21--71.

\bibitem{AB}
B. Ammann and C. B\"ar,
{\it The Dirac operator on nilmanifolds and collapsing circle bundles},
 Ann. Global Anal. Geom. {\bf 16} (1998), no. 3, 221--253.

\bibitem{AHu}
B. Ammann and E. Humbert, {\it The spinorial $\tau$-invariant and $0$-dimensional
surgery}, J. Reine Angew. Math. {\bf 624} (2008), 27--50.

\bibitem{ABP}
D. W. Anderson, E. H. Brown, and F. P. Peterson, 
{\it SU-cobordism, KO-characteristic numbers, and the Kervaire invariant},
Ann. Math. (2) {\bf 83} (1966), 54-67.



\bibitem{ALT}
L. Andrianopoli, M.A. Lled\'o, and M. Trigiante,
{\it The Scherk-Schwarz mechanism as a flux compactification 
with internal torsion},
J. High Energy Phys. {\bf 0505} (2005), 051,
[{\tt arXiv:hep-th/0502083}].

\bibitem{ACGH}
E. Arbarello, M. Cornalba, P. A. Griffiths, and J. Harris,
{\it Geometry of Algebraic Curves}, Springer-Verlag, New York, 1985. 


\bibitem{At}
M. F. Atiyah,
{\it Riemann surfaces and spin structures},
Ann. Sci. \'Ec. Norm. Sup\'er. (4) {\bf 4} (1971), 47--62.

\bibitem{ABS}
M. Atiyah, R. Bott and A. Shapiro,
{\it Clifford Modules}, 
Topology {\bf 3} (1964) 3--38.

\bibitem{AH}
M. F. Atiyah and F. Hirzebruch, 
{\it Spin-manifolds and group actions}, Essays on
Topology and Related Topics, 18--28, Springer, New York 1970.

\bibitem{APSI}
M. F. Atiyah, V. K. Patodi, and I. M. Singer, 
{\it Spectral asymmetry and Riemannian geometry} I,
Math. Proc. Camb. Philos. Soc. {\bf 77} (1975) 43--69.

\bibitem{APSII}
M. F. Atiyah, V. K. Patodi, and I. M. Singer, 
{\it Spectral asymmetry and Riemannian geometry} II,
Math. Proc. Camb. Philos. Soc. {\bf 78} (1975) 405--432.

\bibitem{APSIII}
M. F. Atiyah, V. K. Patodi, and I. M. Singer, 
{\it Spectral asymmetry and Riemannian geometry} III,
Math. Proc. Camb. Philos. Soc. {\bf 79} (1976) 71--99.


\bibitem{AS69}
M. F. Atiyah and I. Singer, {\it Index theory for skew-adjoint Fredholm 
operators},
Inst. Hautes \'Etudes Sci. Publ. Math. {\bf 37} (1969) 5--26. 

\bibitem{AS5}
M. F. Atiyah and I. M.  Singer, {\it The index of elliptic operators V},
 Ann. Math. (2) {\bf 93} (1971) 139--149. 

\bibitem{ALST}
R. Aurich, S. Lustig, F. Steiner, and H. Then,
{\it Hyperbolic universes with a horned 
topology and the CMB anisotropy},
Class. Quant. Grav. {\bf 21} (2004) 4901--4926,
[{\tt 	arXiv:astro-ph/0403597}].

\bibitem{AuS}
L. Auslander and R. H. Szczarba, 
{\it Characteristic classes of compact solvmanifolds},
Ann. Math. (2) {\bf 76} (1962), 1--8.



\bibitem{Ba1}
C. B\"ar, 
{\it The Dirac operator on space forms of positive curvature},
J. Math. Soc. Japan {\bf 48} (1996), no.1, 69--83.

\bibitem{Ba}
C. B\"ar,
{\it Dependence of the Dirac spectrum on the spin structure},
Global analysis and harmonic analysis (Marseille-Luminy, 1999), 17--33, 
S\'emin. Congr., 4, Soc. Math. France, Paris, 2000.





\bibitem{BSS}
C. B\"ar and P. Schmutz,
{\it  Harmonic spinors on Riemann surfaces},
 Ann. Global Anal. Geom. {\bf 10} (1992), no. 3, 263--273.


\bibitem{BFGK}
H. Baum, T. Friedrich, R. Grunewald, and I. Kath, 
{\it Twistors and Killing Spinors on Riemannian
Manifolds}, Teubner-Texte f\"ur Mathematik, vol. 124, Teubner, Stuttgart, Leipzig,
1991.



\bibitem{BS1}
S. Bechtluft-Sachs,
{\it On the $\eta$-invariants of Dirac operators on manifolds with free 
circle action},
PhD dissertation, U. of Mainz, 1993.

\bibitem{BS2}
S. Bechtluft-Sachs,
{\it The computation of $\eta$-invariants on manifolds with free circle
action},
J. Func. Anal. {\bf 174} (2000) 251--263.

\bibitem{BeBe}
K. Becker and M. Becker,
{\it M-theory on eight-manifolds},
Nucl. Phys. {\bf B477} (1996) 155--167,
[{\tt arXiv:hep-th/9605053}].


\bibitem{BB}
L. B\'erard Bergery, 
{\it Scalar curvature and isometry group}, 
in Spectra of Riemannian Manifolds,
Tokyo, 1983, Kagai Publications, 9Ð28.

\bibitem{Be95}
V. N. Berestovskij, {\it Homogeneous Riemannian manifolds of 
positive Ricci curvature}, Math.
Notes {\bf 58} (1995), no. 3, 905--909.

\bibitem{BGV}
N. Berline, E. Getzler, M. Vergne,
{\it Heat kernels and Dirac operators},
Springer, Berlin, 2004.

\bibitem{Be}
A. Besse, {\it Einstein manifolds},
Springer-Verlag, Berlin, 1987.

\bibitem{Bi}
J.-M. Bismut,
{\it The index theorem for families of Dirac operators: two heat 
equation proofs},
Invent. Math. {\bf 83} (1986), 91--151. 

\bibitem{BC}
J.-M. Bismut and J. Cheeger, 
{\it Remarks on the index theorem for families of Dirac operators on manifolds with boundary},
Differential geometry, Pitman Monogr. Surv. Pure Appl. Math. {\bf 52}, 59--83 (1991).


\bibitem{Bl}
D. E. Blair, 
{\it Riemannian geometry of contact and symplectic manifolds},
Birkh\"auser Boston, Inc., Boston, MA, 2002.

\bibitem{BH}
A. Borel and F. Hirzebruch, 
{\it Characteristic classes and homogeneous spaces III}, 
Amer. J. Math. {\bf 82} (1960), 491--504.


\bibitem{Bo}
L. D. Borsari,
{\it Bordism of semifree circle actions on spin manifolds},
Trans. Amer. Math. Soc. 
{\bf 301} (1987), no. 2, 479--487.

\bibitem{BG}
B. Botvinnik and P. Gilkey, 
{\it The Gromov-Lawson-Rosenberg conjecture: the twisted case},
Houston J. Math. {\bf 23} (1997), 143--160.

\bibitem{BGS}
B. Botvinnik, P. Gilkey, and S. Stolz,
{\it The Gromov-Lawson-Rosenberg conjecture for groups with
periodic cohomology},
J. Diff. Geom. {\bf 46} (1997) 374--405.


\bibitem{BEM}
P. Bouwknegt, J. Evslin, and V. Mathai,
{\it T-duality: Topology change from $H$-flux},
Commun. Math. Phys. {\bf 249} (2004), 383--415,
[{\tt 	arXiv:hep-th/0306062}].

\bibitem{BOPR}
 V. Braun, B. A. Ovrut, T. Pantev, and R. Reinbacher,
 {\it Elliptic Calabi-Yau threefolds with $\Z_3 \times \Z_3$ Wilson lines},
J. High Energy Phys. {\bf 0412} (2004) 062,
[{\tt arXiv:hep-th/0410055}].

\bibitem{BOS1}
 J. D. Breit, B. A. Ovrut, and G. C. Segre, {\it $E_6$ symmetry breaking in the
superstring theory}, Phys. Lett. {\bf B158} (1985), 33--39.
 


\bibitem{Bu}
U. Bunke, 
{\it Index theory, eta forms, and Deligne cohomology},
Mem. Am. Math. Soc. {\bf 928}, 120 p., 2009.


\bibitem{LWRLU}
S. Caillerie, M. Lachi\`eze-Rey, J.-P. Luminet, R. Lehoucq, A. Riazuelo, 
and J. Weeks, 
{\it A new analysis of Poincar\'e dodecahedral space model},
Astron. \& Astrophys. {\bf 476} (2007), 691--696,
	[{\tt arXiv:0705.0217}]  [{\tt astro-ph}].




\bibitem{CG}
J. Cheeger and D. Gromoll, 
{\it On the structure of complete manifolds of nonnegative curvature}, 
Ann. Math. {\bf 96} (1972), 413--443.

\bibitem{CJS}
E. Cremmer, B. Julia and J. Scherk, {\it Supergravity in theory in
11 dimensions}, Phys. Lett. {\bf B76} (1978), 409--412.

\bibitem{Dai}
X. Dai, 
{\it APS boundary conditions, eta invariants and adiabatic limits},
Trans. Am. Math. Soc. {\bf 354} (2002), no.1, 107--122.

\bibitem{DF}
X. Dai and D. S. Freed,
{\it Eta-invariants and determinant lines},
J. Math. Phys. {\bf 35} (1994), 5155--5194; Erratum-ibid. 
{\bf 42} (2001), 2343--2344,
[{\tt 	arXiv:hep-th/9405012}].


\bibitem{DT}
L. Dabrowski and A. Trautman,
{\it Spinor structures on spheres and projective spaces}
J. Math. Phys. {\bf 27} (1986), 2022--2028.

\bibitem{Dahl}
M. Dahl,
{\it Dependence on the spin structure of the eta and Rokhlin invariants},
Top. Appl. {\bf 118} (2002), 345--355.


\bibitem{DM}
J. M. Davis and R. J. Milgram,
{\it A survey of the spherical space form problem},
Math. Rep. Ser. 2, no. 2, 1986.

\bibitem{DSSz}
K. Dekimpe, M. Sadowski, and A. Szczepa\'nski, 
{\it Spin structures on flat manifolds},
Monatsh. Math. {\bf 148} (2006), no. 4, 283--296.

\bibitem{DFGM}
P. de Medeiros, J. Figueroa-O'Farrill, S. Gadhia, 
and E. M\'endez-Escobar,
{\it Half-BPS quotients in M-theory: ADE with a twist},
J. High Energy Phys. {\bf 0910} (2009), 038,
[{\tt arXiv:0909.0163}] [{\tt hep-th}].

 
\bibitem{De98}
A. Dessai,
{\it Some remarks on almost and stable almost complex manifolds}, 
Math. Nachr. {\bf 192} (1998), 159--172.

\bibitem{De99}
A. Dessai,
{\it Spin$^c$-manifolds with Pin$(2)$-action}, 
Math. Ann. {\bf 315} (1999), 511--528.

\bibitem{De08}
A. Dessai,
{\it Some geometric properties of the Witten genus},
 Contemporary Mathematics {\bf 504} (2009), 99--115
 American Mathematical Society, Providence, RI.

\bibitem{DFM}
E. Diaconescu, D. S. Freed, and G. Moore, {\it The M-theory 3-form
and $E_8$ gauge theory}, 
in Elliptic cohomology, 44--88, London Math. Soc. Lecture Note Ser., 
342, Cambridge Univ. Press, Cambridge, 2007,
[{\tt arXiv:hep-th/0312069}].


\bibitem{DMW}
D. Diaconescu, G. Moore, and E. Witten, {\it $E_8$ gauge theory, and
a derivation of K-theory from M-theory}, Adv. Theor. Math. Phys.
{\bf 6} (2003), 1031--1134, [{\tt arXiv:hep-th/0005090}].


\bibitem{DOPR}
R. Donagi, B. A. Ovrut, T. Pantev, and R. Reinbacher, 
{\it SU(4) instantons on
Calabi-Yau threefolds with $\Z_2 \times
\Z_2$ fundamental group}, J. High Energy Phys.  {\bf 01} 
(2004) 022, [{\tt arXiv:hep-th/0307273}].

\bibitem{DOTW1}
 R. Donagi, B. A. Ovrut, T. Pantev, and D. Waldram,
 {\it  Standard-model bundles
on non-simply connected Calabi-Yau threefolds}, J. High Energy Phys.  
 {\bf 08} (2001) 053,
[{\tt arXiv:hep-th/0008008}].
 
 \bibitem{DOTW2} 
 R. Donagi, B. A. Ovrut, T. Pantev, and D. Waldram, 
 {\it Standard-model bundles},
Adv. Theor. Math. Phys. {\bf 5} (2002), 563--615, [{\tt arXiv:math.ag/0008010}].

\bibitem{DOTW3}
 R. Donagi, B. A. Ovrut, T. Pantev, and D. Waldram, 
 {\it Spectral involutions on
rational elliptic surfaces}, Adv. Theor. Math. Phys. {\bf 5} (2002), 499--561,
[{\tt arXiv:math.ag/0008011}].
 

\bibitem{Do}
H. Donnelly,
{\it Eta invariants for $G$-spaces}, 
Indiana Univ. Math. J. {\bf 27} (1978), 889--918.


\bibitem{DNP}
M. J. Duff, B. E. W. Nilsson,
and C. N. Pope, 
{\it Kaluza-Klein supergravity},
Phys. Rep. {\bf 130}, no. 1 \& 2 (1986)  1--142.


\bibitem{DLM}
M. J. Duff, J. T. Liu, and R. Minasian, {\it Eleven dimensional
origin of string/string duality: A one loop test}, Nucl. Phys. {\bf
B452} (1995) 261, [{\tt hep-th/9506126}].

\bibitem{DLP}
M. J. Duff, H. Lu and C. N. Pope, 
{\it ${\rm AdS}^5 \times S^5$ untwisted}, 
Nucl. Phys. {\bf B532} (1998), 181--209, 
[{\tt arXiv:hep-th/9803061}].


\bibitem{DSS}
W. Dwyer, T. Schick, and S. Stolz,
{\it Remarks on a conjecture of Gromov and Lawson},
in Farrell, F. T. (ed.) et al., High-dimensional manifold topology,
World Scientific,  River Edge, NJ, 159--176 (2003).


\bibitem{EO}  
 M. Evans and B. A. Ovrut, {\it Splitting the superstring vacuum degeneracy}, 
 Phys. Lett. {\bf B174} (1986), 63--68.


\bibitem{FO}
J. L. Fast and S. Ochanine, 
{\it On the $K{\rm O}$ characteristic cycle of a $Spin\sp c$ manifold},
Manuscripta Math. {\bf 115} (2004), no. 1, 73--83. 


\bibitem{J-max}
J. Figueroa-O'Farrill,
{\it Maximal supersymmetry in ten and eleven dimensions},
in Proceedings of the workshop on special geometric structures in string theory, Bonn, 
Germany, September 8--11, 2001, 
[{\tt arXiv:math/0109162}] [{\tt math.DG}].



\bibitem{FG}
J. Figueroa-O'Farrill and S. Gadhia,
{\it Supersymmetry and spin structures},
Class. Quantum Grav. {\bf 22} (2005), 
L121--L126.

\bibitem{FS}
J. Figueroa-O'Farrill, J. Sim\'on,
{\it Supersymmetric Kaluza-Klein reductions of M2 and M5-branes},
 Adv. Theor. Math. Phys. {\bf 6} (2003), 703--793,
[{\tt arXiv:hep-th/0208107}].


\bibitem{FH}
D. S. Freed and M. Hopkins,
{\it On Ramond-Ramond fields and $K$-theory},
J. High Energy Phys. {\bf 4} (2000) 044,
[{\tt arXiv:hep-th/0002027}].

\bibitem{FW}
D.~S.~Freed and E.~Witten, {\it Anomalies in string theory with D-Branes},
Asian J. Math. {\bf3} (1999) 819--851, [{\tt arXiv:hep-th/9907189}].

\bibitem{integrand}
D. S. Freed and G. W. Moore,
{\it Setting the quantum integrand of M-theory},
Commun. Math. Phys. {\bf 263} (2006), 89-132,
[{\tt arXiv:hep-th/0409135}].


\bibitem{Fr1}
T. Friedrich, 
{\it Zur Abh\"angigkeit des Dirac-Operators von der Spin-Struktur},
Colloq. Math. {\bf 48} (1984), 57--62.

\bibitem{Fried}
T. Friedrich,
{\it Dirac operators in Riemannian geometry},
American Mathematical Society, Providence, Providence, RI, 2000.

\bibitem{Fu1}
Y. Fukumoto,
{\it The index of the ${\rm Spin}^c$ Dirac operator on the weighted projective space and 
the reciprocity law of the Fourier-Dedekind sum}, 
J. Math. Anal. Appl. {\bf 309} (2005), no. 2, 674--685. 

\bibitem{Fu2}
Y. Fukumoto,
{\it A cancellation formula of Alvarez-Gaum\'e and Witten and a signature 
defect-type invariant of 11-dimensional lens spaces with spin structures},
Pacific J. Math. {\bf 235} (2008), no. 2, 213--234.

\bibitem{Ga}
P. Gajer,
{\it Riemannian metrics of positive scalar curvature on compact
manifolds with boundary},
Ann. Global Anal. Geom. {\bf 5} (1997),  179--191.

\bibitem{GP}
J. P. Gauntlett and S. Pakis,
{\it The geometry of $D=11$ Killing spinors},
J. High Energy Phys. {\bf 0304} (2003) 039,
[{\tt arXiv:hep-th/0212008}].


\bibitem{Gi}
P. Gilkey,
{\it The Geometry of Spherical Space Form Groups},
World Scientific, Singapore, 1989.


\bibitem{GLP}
P. Gilkey, J. Leahy, and J. Park, 
{\it Spectral Geometry, Riemannian Submersions and the Gromov-Lawson 
Conjecture},
 Chapman \& Hall/CRC, Boca Raton, FL, 1999.

\bibitem{GGK}
V. L. Ginzburg, V. Guillemin, and Y. Karshon,
{\it  Cobordisms and Hamiltonian group actions}, American 
Mathematical Society, Providence, RI, 2002.


\bibitem{Go1}
S. Goette, 
{\it Equivariant $\eta$-invariants and  $\eta$-forms},
J. Reine Angew. Math. {\bf 526} (2000), 181--236.


\bibitem{GH}
P. Griffiths and J. Harris, {\it Principles of Algebraic Geometry}, 
John Wiley, New York, 1994.

\bibitem{GL1}
M. Gromov and B. Lawson Jr., {\it Spin and scalar curvature in the presence of the fundamental group}, 
Ann. Math. (2) {\bf 111} (1980), 209--230.

\bibitem{GL2}
M. Gromov and H. B. Lawson, Jr., 
{\it The classification of simply connected
manifolds of positive scalar curvature}, 
Ann. Math. (2) {\bf 111} (1980), 423--434.



%

\bibitem{Ha}
B. Hanke, 
{\it Positive scalar curvature with symmetry},
J. Reine Angew. Math. {\bf 614} (2008), 73--115.

\bibitem{HY}
A. Hattori and T. Yoshida,
{\it  Lifting compact group actions in fiber bundles}, 
Japan. J. Math. {\bf 2} (1976), 13--25.

\bibitem{HHM}
T. Hausel, E. Hunsicker, and R. Mazzeo,
{\it Hodge cohomology of gravitational instantons},
Duke Math. J. {\bf 122} (2004), no. 3, 485--548.

\bibitem{He}
T. Heaps,
{\it Almost complex structures on eight and ten dimensional manifolds},
Topology {\bf 9} (1970), 111--119.


\bibitem{Her}
R. Hermann,
{\it A sufficient condition that a mapping of Riemannian manifolds be a fibre bundle},
Proc. Amer. Math. Soc. {\bf 11} (1960), 236--242.

\bibitem{HMU}
O. Hijazi, O. Montiel, and F. Urbano,
{\it Spin${}^c$ geometry of K\"ahler manifolds and the Hodge Laplacian 
on minimal Lagrangian submanifolds},
Math Z. {\bf 253} (2006), 821--853.

\bibitem{HR}
N. Higson and J. Roe,
{\it Operator K-Theory and the Atiyah-Singer Index Theorem}
to appear through Princeton University Press.


\bibitem{Hir}
F. Hirzebruch, 
{\it Topological Methods in Algebraic Geometry},
Springer-Verlag, 
Berlin-Heidelberg-New York, 1978. 

\bibitem{HiS}
G. Hiss and A. Szczepa\'nski,
{\it Spin structures on flat manifolds with cyclic holonomy},
Commun. Algebra {\bf 36} (2008), no. 1, 11--22.

\bibitem{Hit}
N. Hitchin, {\it Harmonic spinors},
 Adv. Math. {\bf 14} (1974), 1--55.

\bibitem{HoS}
M. J. Hopkins and I. M. Singer,
{\it Quadratic functions in geometry, topology, and M-theory},
J. Diff. Geom. {\bf 70} (2005), 329--452.


\bibitem{HW}
P. Horava and E. Witten, {\it Heterotic and type I string dynamics
from eleven dimensions}, Nucl. Phys. {\bf B460} (1996), 506--524, [{\tt
arXiv:hep-th/9510209}].


\bibitem{Ik}
A. Ikemakhen,
{\it Parallel spinors on Lorentzian Spin${}\sp c$ manifolds},
 Differential Geom. Appl.  {\bf 25}  (2007),  no. 3, 299--308. 


\bibitem{Jo}
D. Joyce,
{\it Compact Manifolds With Special Holonomy},
Oxford University Press, New York, 2000. 

\bibitem{Ka}
M. Karoubi, {\it Clifford modules and twisted K-theory},
Adv. Appl. Clifford Algebr. {\bf 18} (2008), no. 3-4, 765--769.



\bibitem{KF}
E. Kim and T. Friedrich,
{\it The Einstein-Dirac equation on Riemannian Spin manifolds},
J. Geom. Phys. {\bf 33} (2000) 128--172,
[{\tt arXiv:math/9905095}] [{\tt math.DG}].


\bibitem{KL}
M. Kreck and W. L\"uck, 
{\it The Novikov conjecture, Geometry and algebra,}  
Oberwolfach Seminars 33, Birkh\"auser, Basel, 2005. 


\bibitem{KS93}
M. Kreck and S. Stolz, 
{\it $\mathbb{H}{\rm P}^2$-bundles and elliptic homology},
 Acta Math. {\bf 171} (1993), no. 2, 231--261.



\bibitem{KS1}
I. Kriz and H. Sati,
{\it M theory, type IIA superstrings, and elliptic cohomology},
Adv. Theor. Math. Phys. {\bf 8} (2004), 345--395, 
[{\tt arXiv:hep-th/0404013}].


%
%

\bibitem{KuS}
R. Kusner and N. Schmitt,
{\it The spinor representation of surfaces in space},
[{\tt arXiv:dg-ga/9610005}].

\bibitem{KwS1}
S. Kwasik and R. Schultz,
{\it Positive scalar curvature and periodic fundamental groups},
Comment. Math. Helv. {\bf 65} (1990), no. 2, 271--286. 

\bibitem{KwS2}
S. Kwasik and R. Schultz,
{\it Fake spherical space forms of constant positive scalar curvature},
 Comment. Math. Helv. {\bf 71} (1996), no.1, 1--40.


\bibitem{LM}
H.B. Lawson and M.-L. Michelson,
{\it Spin Geometry}, Princeton University Press, Princeton, NJ, 1989.


\bibitem{LY}
H. B. Lawson and S. T. Yau,
{\it Scalar curvature, nonabelian group action and the degree
of symmetry of exotic spheres},
 Comm. Math. Helv. {\bf 49} (1974), 232--244.


\bibitem{LMP}
E. Leichtnam,
R. Mazzeo,
and P. Piazza,
{\it The index of Dirac operators on manifolds with
fibered boundaries},
Bull. Belg. Math. Soc. - Simon Stevin {\bf 13} (2006), no. 5, 845--855. 

\bibitem{Li}
A. Lichnerowicz,
{\it Spineurs harmoniques},
C. R. Acad. Sci. Paris, S\'er. A-B {\bf 257} (1963), 7--9. 



\bibitem{LS}
J. T. Liu and H. Sati,
{\it Breathing mode compactifications and supersymmetry of the brane-world},
Nucl. Phys. {\bf B605} (2001), 116--140,
[{\tt arXiv:hep-th/0009184}].


 \bibitem{Liu}
 K. Liu, {\it On mod 2  and higher elliptic genera},
 Commun. Math. Phys. {\bf 149} (1992), no. 1, 71--95.

\bibitem{Lo}
J. Lott,
{\it Higher degree analogs of the determinant line bundle},
Commun. Math. Phys. {\bf 230} (2002), 41--69. 

\bibitem{MPT}
M. Marcolli, E. Pierpaoli, and K. Teh,
{\it The spectral action and cosmic topology},
Commun. Math. Phys. {\bf 304} (2011), 125--174,
	[{\tt arXiv:1005.2256}] [{\tt hep-th}].


\bibitem{Mas}
W. S. Massey,
{\it Obstructions to the existence of almost complex structures},
Bull. Amer. Math. Soc. {\bf 67} (1961), 559--564. 

\bibitem{MS}
V. Mathai and H. Sati,
{\it Some Relations between twisted K-theory and $E_8$ gauge theory},
J. High Energy Phys. {\bf 0403} (2004) 016,
[{\tt arXiv:hep-th/0312033}].


\bibitem{Mc1}
B. McInnes,
{\it Multiple spin structures in higher dimensional physics},
Class. Quant. Grav. {\bf 13} (1986), 3175--3182.

\bibitem{Mc2}
B. McInnes,
{\it Existence of parallel spinors on nonsimply connected Riemannian manifolds},
J. Math. Phys. {\bf 39} (1998), 2362--2366. 


\bibitem{MP}
R. J. Miatello and R. A. Podest\'a, 
{\it Spin structures and spectra of $\Z_2^k$-manifolds},
Math. Z. {\bf 247} (2004), no. 2, 319--335. 


\bibitem{Mil2}
J. W. Milnor, 
{\it Spin structures on manifolds}, 
Enseignement Math. (2) {\bf 9} (1963), 198--203.


\bibitem{Mil}
J. Milnor, {\it Remarks concerning spin manifolds}, 
Differential and Combinatorial Topology,
55Ð62, Princeton Univ. Press, Princeton, NJ, 1965.




\bibitem{MM}
R.~Minasian and G.~Moore,
{\it K-theory and Ramond-Ramond charge},
J. High Energy Phys. {\bf 11} (1997) 002,
[{\tt arXiv:hep-th/9710230}].

\bibitem{MF}
A. S. Mishchenko and A. T.  Fomenko, 
{\it The Index of elliptic operators over $C^*$-algebras},
 Math. USSR-Izv. {\bf 15} (1980), 87--112.


\bibitem{Miy}
T. Miyazaki,
{\it On the existence of positive scalar curvature metrics on non simply 
connected manifolds},
J. Fac. Sci., Univ. Tokyo, Sect. I A {\bf 30} (1984), 549--561.

\bibitem{MW}
G. Moore and E. Witten,
{\it Selfduality, Ramond-Ramond fields and $K$-theory}
J. High Energy Phys. {\bf 4} (2000), 032,
[{\tt arXiv:hep-th/9912279}].


\bibitem{Mor}
A. Moroianu,
{\it Parallel and Killing spinors on Spin${}^c$ manifolds},
Commun. Math. Phys. {\bf 187} (1997), 417--427.

\bibitem{Mor2}
A. Moroianu,
{\it $\text{Spin}^c$ manifolds and complex contact structures},
Commun. Math. Phys. {\bf 193} (1998), no.3, 661--674.


%

\bibitem{MoS}
A. Moroianu and U. Semmelmann,
{\it Parallel spinors and holonomy groups},
J. Math. Phys. {\bf 41} (2000), no.4, 2395--2402.

\bibitem{Nash}
C. Nash,
{\it A complex anomaly},
Phys. Lett. {\bf B 184} (1987), 239--241.


\bibitem{OPR1}
 B. A. Ovrut, T. Pantev, and R. Reinbacher, {\it Torus-fibered 
 Calabi-Yau threefolds with non-trivial fundamental group}, 
 J. High Energy Phys.  {\bf 05} (2003) 040, [{\tt arXiv:hep-th/0212221}].

\bibitem{OPR2}
B. A. Ovrut, T. Pantev, and R. Reinbacher, {\it 
Invariant homology on standard
model manifolds}, J. High Energy Phys.  {\bf 01} (2004) 059, [{\tt arXiv:hep-th/0303020}].


\bibitem{PP}
D. N. Page and C. N. Pope,
{\it Stability analysis of compactifications of $D = 11$ 
supergravity with $SU(3) \times SU(2) \times U(1)$ symmetry},
Phys. Lett. {\bf B145} (1984), 337--341. 

 

\bibitem{P}
M. Paschke, {\it Von nichtkommutativen Geometrien, ihren 
Symmetrien, und etwas Hochenergiephysik}, 
Ph.D. thesis, Mainz 2001.


\bibitem{PS}
M. Paschke and A. Sitarz,
{\it On Spin structures and Dirac 
operators on the noncommutative torus},
Lett. Math. Phys. {\bf 77} (2006), no. 3, 317--327,
[{\tt	arXiv:math/0605191}]  [{\tt math.QA}].
	
	
\bibitem{Pe}
R. Petit, 
{\it Spin${}^c$-structures and Dirac operators on contact manifolds},
Differential Geom. Appl. {\bf 22} (2005), no. 2, 229--252. 


	
\bibitem{P72}	
T. Petrie,
{\it Smooth $S^1$-actions on homotopy complex projective spaces and related topics},
Bull. Math. Soc. {\bf 78} (1972), 105--153.

\bibitem{Pf}
F. Pfaffle,
{\it The Dirac spectrum of Bieberbach manifolds},
J. Geom. Phys. {\bf 35} (2000), no. 4, 367--385.



\bibitem{Pol}
J. Polchinski,
{\it Dirichlet branes and Ramond-Ramond charges},
Phys. Rev. Lett. {\bf 75} (1995), 4724--4727,
[{\tt arXiv:hep-th/9510017}]. 

\bibitem{PV}
C.N. Pope and P. van Nieuwenhuizen,
{\it Compactifications of $d=11$ supergravity on K\"ahler manifolds}, 
 Commun. Math. Phys. {\bf 122} (1989), 281--292.

\bibitem{Qu}
D. Quillen, 
{\it Superconnections and the Chern character}, 
Topology {\bf 24} (1985), no. 1, 89--95. 

\bibitem{Red}
C. Redden, 
{\it Canonical Metric Connections Associated to String Structures},
 Ph.D. thesis, University of Notre Dame, 2006.
		
\bibitem{Ros1}
J. Rosenberg, 
{\it $C^*$-algebras, positive scalar curvature, and the Novikov conjecture},
Publ. Math., Inst. Hautes \'Etud. Sci. {\bf 58} (1983), 409--424.

\bibitem{Ros2}
J. Rosenberg, 
{\it $C\sp*$-algebras, positive scalar curvature and the Novikov conjecture} II,
Geometric methods in operator algebras, 341--374, 
Pitman Res. Notes Math. Ser. 123, 1986. 

\bibitem{Ros3}
J. Rosenberg,
{\it $C\sp *$-algebras, positive scalar curvature, and the Novikov conjecture} III,
Topology {\bf 25} (1986), 319--336.
		
\bibitem{RS}
J. Rosenberg and S. Stolz, 
{\it A ``stable" version of the Gromov-Lawson conjecture},
Contemp. Math. {\bf 181}, 405--418,
American Mathematical Society,
 Providence, RI, 1995.

\bibitem{RS2}
J. Rosenberg and S. Stolz,
{\it Metrics of positive scalar curvature and connections
with surgery}, in Surveys in surgery theory,
Ann. Math. Stud. 149, 353--386,
 Princeton University Press, Princeton, NJ, 2001.

\bibitem{SSz}
M. Sadowski and A. Szczepa\'nski,
{\it Flat manifolds, harmonic spinors, and eta invariants},
Adv. Geom. {\bf 6} (2006), no. 2, 287--300.

\bibitem{Sgerbe}
H. Sati,
{\it $E_8$ gauge theory and gerbes in string theory},
Adv. Theor. Math. Phys. {\bf 14} (2010), 1--39, 
[{\tt arXiv:hep-th/0608190}].

\bibitem{SE8}
H. Sati,
{\it Anomalies of $E_8$ gauge theory on String manifolds},
 Int. J. Mod. Phys. {\bf A26} (2011), 2177--2197, 
[{\tt arXiv:0807.4940}] [{\tt hep-th}].

\bibitem{tcu}
H. Sati,
{\it Geometric and topological structures related to M-branes},
 Proc. Symp. Pure Math. {\bf 81} (2010), 181--236, 
[{\tt arXiv:1001.5020}] [{\tt math.DG}].

\bibitem{SW}
N. Seiberg and E. Witten,
{\it Spin structures in string theory},
 Nucl. Phys. {\bf B276}  (1986), 272--290.
 
 \bibitem{SW2}
 N. Seiberg and E. Witten,
 {\it String theory and noncommutative geometry}, 
 J. High Energy Phys. 
 {\bf 9909} (1999) 032,
 [{\tt arXiv:hep-th/9908142}].

 \bibitem{Se1}
  A. Sen, {\it The heterotic string in arbitrary background field}, 
  Phys. Rev. {\bf D32}
(1985), 2102--2112.


\bibitem{St92}
S. Stolz, 
{\it Simply connected manifolds with positive scalar curvature}, 
Ann. Math. (2) {\bf 136} (1992), 511--540.

 \bibitem{St96}
 S. Stolz, {\it A conjecture concerning positive Ricci curvature
and the Witten genus}, Math. Ann. {\bf 304} (1996),
785-800.



\bibitem{Sto}
W. Stong,
{\it Notes on cobordism theory},
Princeton University Press, Princeton, NJ, 1968.

\bibitem{Sz}
R. H. Szczarba,
{\it The tangent bundle of fiber spaces and quotient spaces},
Amer. J. Math. {\bf 86} (1964),
685--697.


\bibitem{Th2}
E. Thomas, 
{\it Complex structures on real vector bundles},
Amer. J. Math. {\bf 89} (1967), 887--908.

\bibitem{VW}
C. Vafa and E. Witten, {\it A One-loop test of string duality},
Nucl. Phys. {\bf B447} (1995), 261--270, 
[{\tt arXiv:hep-th/9505053}].

\bibitem{Va}
B. Vaillant, {\it Index and spectral theory for manifolds with generalized fibred
cusps}, Dissertation, Rheinische Friedrich-Wilhelms-Universit\"at Bonn,
Bonn, 2001.

\bibitem{Vas}
A. T. Vasquez, 
{\it Flat Riemannian manifolds},
J. Diff. Geom. {\bf 4} (1970), 367--382.




\bibitem{Wa95}
M. K. Wang, 
{\it On non-simply connected manifolds with non-trivial parallel spinors},
Ann. Global Anal. Geom. {\bf 13} (1995), 31--42.


\bibitem{W0}
 E. Witten, {\it Symmetry breaking patterns in superstring models}, 
 Nucl. Phys. {\bf B258} (1985), 75--100.
 


\bibitem{W91}
E. Witten, 
{\it Mirror manifolds and topological field theory},
 Essays on mirror manifolds, International Press, Cambridge, MA, 120--159 (1992),
[{\tt arXiv:hep-th/9112056}].



\bibitem{Flux}
E. Witten,
{\it On Flux quantization in M-theory and the effective action},
J. Geom. Phys. {\bf 22} (1997), 1--13,
[{\tt arXiv:hep-th/9609122}].


\bibitem{W-duality}
E. Witten, 
{\it Duality relations among topological effects in string theory},
J. High Energy Phys. {\bf 0005} (2000) 031,
[{\tt arXiv:hep-th/9912086}].

\bibitem{Wo}
J. Wolf, 
{\it Spaces of Constant Curvature},
Publish or Perish, Inc., Boston, MA,
1977.




\bibitem{Z1}
W. Zhang,
{\it Eta invariants and Rokhlin congruences},
  C. R. Acad. Sci. Paris, Serie I, {\bf 315} (1992), 305--308.

\bibitem{Z2}
W. Zhang,
{\it Circle bundles, adiabatic limits of eta invariants and Rokhlin congruences},
 Ann. Inst. Fourier {\bf 44} (1994), 249--270.   






\end{thebibliography}
\end{document}